%% file: thesis.tex
\documentclass[12pt,a4paper,twoside]{scrreprt}

\pdfoutput=1

\usepackage{a4wide}
\usepackage[Glenn]{fncychap}
\usepackage{fancyheadings}
\usepackage{graphicx}
\usepackage{amsmath}
\usepackage{bbm}
\usepackage{bm}
\usepackage{epsfig}
\usepackage{epsf}
\usepackage{dsfont}
\usepackage[nottoc]{tocbibind}

\newcommand{\del}{\partial}
\newcommand{\be}{\begin{equation}\begin{aligned}}
\newcommand{\ee}{\end{aligned}\end{equation}}
\newcommand{\D}{\displaystyle}

\newcommand{\arctanh}{\operatorname{arctanh}}
\newcommand{\bea}{\begin{eqnarray}}
\newcommand{\eea}{\end{eqnarray}}
\newcommand{\bdm}{\begin{displaymath}}
\newcommand{\edm}{\end{displaymath}}
\newcommand{\beas}{\begin{eqnarray*}}
\newcommand{\eeas}{\end{eqnarray*}}
\newcommand{\av}[1]{\left< #1\right>}
\newcommand{\intdom}[1]{\frac{1}{V}\int_{\mathcal{D}} #1\sqrt{h}d^3\mathbf{x}}
\newcommand{\dom}[1]{#1_{\mathcal{D}}}
\newcommand{\Qd}{\mathcal{Q}_{\mathcal{D}}}
\newcommand{\Rd}{\mathcal{R}_{\mathcal{D}}}
\newcommand{\Pd}{\mathcal{P}_{\mathcal{D}}}
\newcommand{\Td}{\mathcal{T}_{\mathcal{D}}}
\newcommand{\bkr}{\overline{\rho}}

\makeatletter
\def\ps@headings{\let\@mkboth\markboth
\def\@oddfoot{} \def\@evenfoot{}
\def\@evenhead{\protect\underline{
\hbox to\hsize{ \thepage \it \hfil \leftmark}}}
\def\@oddhead{\protect\underline{
\hbox to\hsize{{\it \rightmark\hfil}\thepage}}}
\def\chaptermark##1{\markboth {\ifnum \c@secnumdepth>\m@ne
\thechapter. \ \fi ##1}{}}
\def\sectionmark##1{\markright{\ifnum \c@secnumdepth >\z@
\thesection \ \fi ##1}}}
\makeatother

\begin{document}

\pagestyle{empty}
\pagenumbering{roman}


\begin{center}   
\begin{sc}
{\Huge Dissertation}\\[0.6cm]
\end{sc}
\begin{Large}
submitted to the\\
Combined Faculties for the Natural Sciences and for Mathematics\\
of the Ruperto-Carola University of Heidelberg, Germany\\
for the degree of\\[0.2cm]
Doctor of Natural Sciences
\end{Large}
\vspace{13.5cm}

\begin{large}
presented by\\[0.1cm]
{\bf Diplom-Physikerin Juliane Behrend}\\[0.1cm]
born in L\"ubeck, Germany\\
\vspace{0.5cm}
Oral examination: April 23, 2008
\end{large}

\end{center}

\newpage

\mbox{}

\newpage
\thispagestyle{empty}
\begin{center}
\begin{large}
\mbox{}\\[2.0cm]

{\sc\Huge Metric Renormalization}\\[0.6cm]
{\sc\Huge in General Relativity}

\vspace{16cm}

{\bf Referees:$~~$ Prof. Dr. Otto Nachtmann}\\[.2cm]
\hspace{1.8cm}{\bf Prof. Dr. Iring Bender}\\

\end{large}
\end{center}

\cleardoublepage
\input{abstract.tex}

\cleardoublepage
\pagestyle{empty}
\vspace*{4.5cm}
\begin{center}
{\large\it to my father}\\[0.5cm]
{\Large\sc Volkmar Behrend}\\[0.2cm]
{\sc Jan 20, 1930 - Aug 28, 2004}
\end{center}

\cleardoublepage
\pagestyle{empty}
\tableofcontents

{\newpage{\pagestyle{empty}\cleardoublepage}}
\pagestyle{headings}
\pagenumbering{arabic}

\input{intro.tex}

{\newpage{\pagestyle{empty}\cleardoublepage}}
\input{problem.tex}

{\newpage{\pagestyle{empty}\cleardoublepage}}
\input{paper.tex}

{\newpage{\pagestyle{empty}\cleardoublepage}}
\input{process.tex}

{\newpage{\pagestyle{empty}\cleardoublepage}}
\input{twodimensions.tex}

{\newpage{\pagestyle{empty}\cleardoublepage}}
\input{twosphere.tex}

{\newpage{\pagestyle{empty}\cleardoublepage}}
\input{threesphere.tex}

{\newpage{\pagestyle{empty}\cleardoublepage}}
\input{gauss.tex}
{\newpage{\pagestyle{empty}\cleardoublepage}}
\input{julelagrangian.tex}
{\newpage{\pagestyle{empty}\cleardoublepage}}

\input{concl.tex}

{\newpage{\pagestyle{empty}\cleardoublepage}}

\input{append.tex}
{\newpage{\pagestyle{empty}\cleardoublepage}}
\bibliographystyle{amsplain}
\bibliography{thesis}

{\newpage{\pagestyle{empty}\cleardoublepage}}
\input{acknowledgement.tex}
\end{document}

%% file: abstract.tex
\begin{center}
{\sc\Large Renormierung von Metriken}\\[0.25cm]
{\sc\Large in der Allgemeinen Relativit\"atstheorie}\\[0.7cm]
{\bf\large  Zusammenfassung}\\[0.7cm]
\end{center}

Das Mittelungsproblem in der Allgemeinen Relativit\"atstheorie besteht in der Definition eines wohldefinierten Mittels \"uber Tensorgr\"o\ss en und wir beleuchten dieses Problem von verschiedenen Seiten. Zun\"achst gehen wir auf die kosmologische R\"uckreaktion ein, die dadurch verursacht wird, dass der gemittelte Einstein-Tensor nicht identisch zu dem Einstein-Tensor der gemittelten Metrik ist. Es gibt Vermutungen, nach denen diese R\"uckreaktion die  Erkl\"arung f\"ur Dunkle Energie sein soll. Wir zeigen numerisch, dass im Buchert Formalis- mus die Korrekturen von (quasi)linearen  St\"orungen nur von der Gr\"o\ss enordnung $10^{-5}$ sind und die Eigenschaften von Dunkler Materie aufweisen. Anschlie\ss end besch\"aftigen wir uns mit der Formulierung eines allgemein kovarianten Mittelungsprozesses, der die Metrik in Vielbeine zerlegt und diese mit Hilfe eines relativistischen Wegner-Wilson Operators an einen gemeinsamen Punkt parallelverschiebt, wo sie anschlie\ss end gemittelt werden. F\"ur die Festlegung des entsprechenden Vielbeinfeldes wird der Lagrange-Formalismus verwendet. Die Funktionsweise des Mittelungsprozesses wird an speziellen Beispielen in zwei und drei Raumdimensionen verdeutlicht. Dazu werden partielle Differentialgleichungen numerisch mit dem Simulationspaket Gascoigne gel\"ost.

\vspace{1.5cm}
\begin{center} 
{\sc\Large Metric Renormalization in General Relativity}\\[0.7cm]
{\bf\large Abstract}\\[0.7cm]
\end{center}

The averaging problem in general relativity concerns the difficulty of defining meaningful averages of tensor quantities and we consider various aspects of the problem. We first address cosmological backreaction which arises because the averaged Einstein tensor is not the same as the Einstein tensor of the averaged metric. It has been suggested that backreaction might account for the dark energy. We show numerically in the Buchert formalism that the corrections from (quasi)linear perturbations are only of the order of $10^{-5}$  and act as a dark matter. We then focus on constructing averaged metrics and present a generally covariant averaging process which decomposes the metric into Vielbeins and parallel transports them with a relativistic Wegner-Wilson operator to a single point where they can then be averaged. The Vielbeins are chosen in a Lagrangian formalism. The functionality of the process is demonstrated in specific examples in two and three space dimensions. This involves the numerical solution of partial differential equations by the aid of the simulation toolkit Gascoigne.

%% file: intro.tex
\chapter{Introduction}

The averaging problem is one of the greatest unresolved problems of general relativity. It concerns the question of how to define an average over tensor quantities in a physically sensible manner when they are defined on separate spacetime points. Such an average can only be easily defined for scalar quantities and most of the proposed averaging techniques thus far have failed to average higher-order tensors in a generally covariant way. Key to the problem is the construction of an averaged metric that describes the corresponding smoothed manifold. Such a metric would allow for the classification of manifolds, but an immediate focus are applications in astrophysics and cosmology. In particular, knowledge of the averaged metric gives us knowledge of the averaged causal structure of a spacetime. In a cosmological setting, it would give us the paths of light propagation through the averaged spacetime and as all our observations are necessarily of light this must be clearly understood. More fundamentally, the ability to take an averaged metric will enable us to construct a physically meaningful metric for a system of bodies. While obviously key, there is more to general relativity than the metric. Other important tensors are the stress-energy tensor and the Einstein tensor linking the dynamics of spacetime to the energy and momentum content of the universe. Neither of these can in general be averaged in a physically-meaningful way, but the average dynamics of a domain will be governed by Einstein's equation averaged across that domain. \\

This thesis focuses on these different aspects of the averaging problem, concentrating first on the averaging of the Einstein tensor in a simple, cosmological setting. In chapter \ref{problem} we give an introduction to cosmological issues that are connected to the ``dark energy'' problem and specifically the attempts to address this with cosmological backreaction. Modern cosmology is built on the Friedmann-Lema\^itre-Robertson-Walker (FLRW) metric, which is maximally-symmetric on hypersurfaces of constant time. The extreme isotropy of the Cosmic Microwave Background and the ``cosmological principle'' together imply that the universe is both homogeneous and isotropic to a high degree of precision. This makes the assumption of an FLRW background reasonable. We have by now overwhelming evidence that when the observations of distant supernovae (of type Ia) combined with those of CMB and LSS are interpreted in this model, the universe must be undergoing a phase of accelerated expansion. This was first detected in 1998 \cite{RiessEtAl98,PerlmutterEtAl98} and has been refined by different studies since, examples being \cite{RiessEtAl06,WoodVaseyEtAl07}. Combined with observations of the cosmic microwave background  \cite{WMAP,WMAP3} and  the large-scale structure of the universe \cite{Colless03,SDSS00,Cole05,Percival06} this has lead to the $\Lambda$CDM model of the universe which is in good agreement with all confirmed observational data. In this model the universe is described as FLRW on large scales, almost flat and composed of only around 4\% standard matter and about 20\% ``cold dark matter'' which interacts only through gravity. The acceleration can only be generated in this model if 76\% of the energy density in the universe is made of the so-called dark energy.\\

Probably the most important task of modern cosmology is to understand the nature of this dark energy, which may also have implications on particle physics. The most common source suggested for the ``dark energy'' is Einstein's cosmological constant $\Lambda$. This is usually explained with the energy density of the vacuum polarisation, but estimates show that it should then be many orders of magnitude larger. The most common alternatives are scalar fields minimally-coupled to gravity \cite{Peebles87,Wetterich88,Caldwell98} which in particular cases can act in the present universe to mimic a cosmological constant. While successful, it would perhaps be preferable to test models which do not require new physics.\\

One physical possibility to explain the origin of dark energy is based on the non-linearity of the Einstein tensor and the consequence that the averaged Einstein tensor is not the same as the Einstein tensor of the averaged metric. The difference between these two tensors in a cosmological setting must necessarily yield corrections to Einstein's equation which is known as ``cosmological backreaction''. This was first recognized in the 60's by Shirokov and Fisher \cite{Shirokov63}. After this the idea remained generally unconsidered until the 80's \cite{Ellis84} when Ellis brought many questions about averaging in relativity and cosmology to a wider audience. The discovery of dark energy has however created increased interest in this problem and recent efforts aim to explain dark energy as a pure backreaction effect. However, most of the works done in this field are based on averaging schemes that either directly or indirectly depend on the used coordinate system raising the question of whether or not they are physical. The most common formalism used at present is the so-called Buchert formalism introduced in 2000 \cite{Buchert99,Buchert01}. This separates cosmology into spacelike hypersurfaces and uses a three-dimensional average inspired by Newtonian gravity. Alternatives with improved averaging schemes include a generally-covariant scheme introduced by Zalaletdinov \cite{Zalaletdinov93,Zalaletdinov08}. Whether the backreaction can form the observed dark energy is still subject of debate \cite{Wetterich01,Rasanen04,Rasanen06,LiSchwarz07,LiSchwarz07-2,Rasanen08} but there are few quantitative predictions from realistic cosmological models.\\

In chapter \ref{paper} we address this issue. It has been recognised for some time that the backreaction from linear perturbations on an FLRW universe must be small \cite{Wetterich01,Rasanen04}. This has, however, never been quantified. We set up the Buchert formalism in ``Newtonian gauge'' to avoid any problems from gauge modes and then use a ``Boltzmann integrator'' to calculate the linear perturbations in particular cosmological models. Doing this for the $\Lambda$CDM model yields a backreaction that is far too small to drive the acceleration of the universe, and of the order of the rough predictions. For the alternative ``Einstein-de Sitter'' cosmology without dark energy the backreaction is larger but still negligible. The effective fluid also mimics a dark matter instead of a dark energy. The analysis can be extended to smaller scales using the ``halo model'' of structure formation. This is a phenomenological model based on numerical simulations and observations of large-scale structure. Although the approximation breaks down at small scales it provides an estimate for the corrections from ``quasilinear'' perturbations. We find that these are of the same order as the linear effect. The corrections also still act as a dark matter and not a dark energy.\\
 
Chapter \ref{process} forms the core of this thesis because it contains the formulation of a generally covariant averaging process which can be used to construct an averaged metric within the framework of general relativity. The chapter starts with an introduction to the constituents needed for this process and their modes of operation. First we present a Wegner-Wilson line operator reformulated for general relativity, which is used to parallel transport tensor quantities from one spacetime point to another along the geodesic connecting them. The metric can only be averaged if it is first separated across a tetrad field. We present a variational formalism to select a particular tetrad from among the several options determined by the metric. This is done with a proposed Lagrangian based on the covariant derivative of the tetrad. We refer to the field selected as the ``maximally smooth'' tetrad field. We end by presenting the actual averaging procedure itself along with an explanation of its functionality.\\

Any reasonable averaging process must leave a space of constant curvature invariant. These spaces are also the clearest aid to visualizing the process and have an immediate cosmological relevence as they form the basis of the FLRW metric. We consider the three possible constant curvature spaces in two dimensions in chapter \ref{twodimensions}. After employing stereographic projection onto a plane, the different constituents of the averaging process are computed in terms of the coordinates of the projection plane. It is then shown that these spaces are invariant under averaging, as should be expected.\\

In chapter \ref{twosphere} we turn to a more physically-interesting case. We take a two-sphere whose metric is slightly perturbed and calculate the constituents of the averaging process for a general perturbation function. This involves the solution of a geodesic equation that is parametrized by a parameter different from the arc length to account for the changes due to the perturbation. The Lagrangian formulation for the maximally smooth dyad field leads to a differential equation of Neumann type, which cannot be solved analytically. In principle the solution to this equation, which determines the dyads,  can be found numerically.\\

The analysis of chapter \ref{twosphere} is extended to the three-sphere in chapter \ref{threesphere}. It is shown that the averaging process leaves the smooth three-sphere invariant as should be expected. We then consider the case of a linearly-perturbed three-sphere which is directly relevent to cosmology and to closed linearly-perturbed FLRW models in particular. Varying the Lagrangian to find the maximally smooth triad results in differential equations analogous to those in chapter \ref{twosphere} but significantly more complicated. The equations are intractable both analytically and numerically. However, in more suitable coordinates these equations would be better controlled and they act as a proof of concept for the averaging process applied to immediately relevent models.\\

We solve the differential equation for the maximally smooth dyad field of a perturbed two-sphere in chapter \ref{gauss}, with the aid of the simulation toolkit Gascoigne. The particular model chosen is that of a gaussian perturbation. It is shown that the averaging process does not yield the expected result.\\

In chapter \ref{julelagrangian} we analyse the shortcomings of the averaging method and present a method to overcoming the problems with the formulation of an improved Lagrangian. The proposed Lagrangian specifies a tetrad that is rotated in a manner taking into account the curvature of the manifold characterised by the Ricci scalar. The strength of the average is therefore directly linked to the fluctuations in the manifold.\\

In chapter \ref{conclusions} we summarize the investigations and the open issues and give a brief outlook on future work.

%% file: problem.tex
\chapter{The Backreaction Problem}
\label{problem}

\section{The Standard Model of Cosmology}
\label{standardmodel}

\subsection{The Friedmann-Lema\^{i}tre-Robertson-Walker Model}

The cosmological principle states that the universe is homogeneous and isotropic. On account of this, Friedmann-Lema\^{i}tre-Robertson-Walker (FLRW) model has become the basis of the standard ``big bang'' model of cosmology. It is based on an exact solution to Einstein's equation (\ref{einsteindef}),
\be
G_{\mu\nu}=R_{\mu\nu}-\frac{1}{2}g_{\mu\nu}R=8\pi G T_{\mu\nu}+\Lambda g_{\mu\nu},
\label{einstein}
\ee
with a metric that is homogeneous and isotropic on constant time hypersurfaces,
\be
ds^2=dt^2-a^2(t)\left(\frac{dr^2}{1-kr^2}+r^2d\theta ^2+r^2\sin ^2\theta d\phi ^2\right).
\label{flrw}
\ee
By a suitable choice of units for $r$ the constant $k$ can be chosen to have the value +1, 0 or -1, indicating constant time hypersurfaces of positive, zero or negative curvature, respectively. The only dynamical variable in this metric is the time-dependent scale factor $a(t)$. The universe is assumed to be filled with a perfect fluid with the energy-momentum tensor
\be
T^{\mu\nu}=(\rho+p)u^{\mu}u^{\nu}-pg^{\mu\nu}.
\label{perfectfluid}
\ee
Here, $p$ denotes the pressure, $\rho$ the energy density, and $u^{\mu}$ the velocity of the fluid in the comoving frame. The terms ``pressure'' and ``density'' must be interpreted in a very general manner and correspond to the usual sense only in case of a perfect fluid. A negative value for the pressure, for example, is not unusual in cosmology.\\

Inserting the energy-momentum tensor (\ref{perfectfluid}) and the FLRW metric (\ref{flrw}) into Einstein's equations (\ref{einstein}) we find the Friedmann equations,
\begin{gather}
\left(\frac{\dot{a}}{a}\right)^2+\frac{k}{a^2}=\frac{8\pi G}{3}\rho+\frac{\Lambda}{3},\label{friedmann}\\
\frac{\ddot{a}}{a}=-\frac{4\pi G}{3}(\rho+3p)+\frac{\Lambda}{3}.\label{raychaudhuri}
\end{gather}
The latter is sometimes referred to as the Raychaudhuri equation and demonstrates the fundamentally attractive nature of gravity, and the repulsive effect of a positive cosmological constant $\Lambda$. We can also see that, if we interpret the cosmological constant as a uniformly distributed fluid, its energy density and pressure are defined as
\be
\rho_\Lambda\equiv\frac{\Lambda}{8\pi G}\qquad\mathrm{and}\qquad p_\Lambda\equiv-\frac{\Lambda}{8\pi G}.
\label{darkenergy}
\ee
While a universe containing matter and a cosmological constant fits the observations, its tiny energy scale has lead many authors to set $\Lambda=0$ and instead employ a time-varying fluid known as ``dark energy'' to mimic a cosmological constant. The density and pressure of the dark energy are still often denoted by $\rho_\Lambda(t)$ and $p_\Lambda(t)$. The most popular model is known as quintessence \cite{Wetterich88,Peebles87,Caldwell98} and consists of a minimally-coupled scalar field $\varphi$ with a potential $V(\varphi)$. The energy density and pressure for such a field are $\rho_\Lambda=\frac{1}{2}\dot{\varphi}^2+V(\varphi)$ and $p_\Lambda=\frac{1}{2}\dot{\varphi}^2-V(\varphi)$. For a sufficiently flat potential, $\dot{\varphi}\ll V(\varphi)$ we find $\rho_\Lambda\simeq-p_\Lambda$, mimicking a cosmological constant.\\

The Raychaudhuri equation (\ref{raychaudhuri}) can be derived from equation (\ref{friedmann}) and the energy-momentum conservation (\ref{conservationdef}),
\be
\dot{\rho}+3\frac{\dot{a}}{a}(\rho+p)=0.
\label{conservation}
\ee
Therefore, equations (\ref{friedmann}) and (\ref{conservation}) form an equivalent, commonly-used pair of dynamical equations to the Friedmann equations. We furthermore introduce the Hubble rate,
\be
H(t)\equiv\frac{\dot{a}}{a},
\label{hubble}
\ee
With this definition the usual set of dynamical equations for the universe can be expressed in the following way,
\be
H^2+\frac{k}{a^2}=\frac{8\pi G}{3}\rho+\frac{\Lambda}{3},\\
\dot{\rho}+3H(\rho+p)=0.
\ee
Furthermore, the Hubble rate allows us to define the critical density
\be
\rho_c(t)\equiv\frac{3H^2}{8\pi G},
\ee
for which the universe is exactly flat $k=0$. A larger energy density would imply a closed universe with $k=1$, and a smaller value would imply an open universe with $k=-1$. It is customary to express the matter density $\rho_m(t)$ (including baryonic matter and dark matter), the radiation density $\rho_\mathrm{rad}(t)$ (photons and neutrinos) and the effective density of dark energy $\rho_\Lambda(t)$ as fractions of the critical density:
\be
\Omega_m(t)\equiv\frac{\rho_m(t)}{\rho_c(t)},\qquad\Omega_\mathrm{rad}(t)\equiv\frac{\rho_\mathrm{rad}(t)}{\rho_c(t)},\quad\Omega_\Lambda(t)\equiv\frac{\rho_\Lambda(t)}{\rho_c(t)}.
\ee 
Sometimes, even the geometry term $k/a^2$ in the Friedmann equation (\ref{friedmann}) is assigned an energy density contribution, which is defined as
\be
\Omega_k(t)\equiv-\frac{k}{a^2H^2}.
\ee
Then, we find immediately that FLRW universes obey
\be
\Omega_m+\Omega_\mathrm{rad}+\Omega_\Lambda+\Omega_k=1.
\ee

The set of equations so far is not closed. To solve for $a(t)$ we must further specify an equation of state,
\be
p(t)=w(t)\rho(t),
\ee
for the dominant component of the energy density. For example, if the universe is flat $(k=0)$ and radiation dominated $(w=1/3)$ we find $a\propto t^{1/2}$. In a matter dominated flat universe, on the other hand, we can neglect the pressure, such that $w=0$. The scale factor then behaves as $a \propto t^{2/3}$. If the equation of state is $w<-1/3$, the universe accelerates, as can be seen from the Raychaudhuri equation. For a cosmological constant, $w=-1$ and we find $a\propto e^{Ht}$. \\

\begin{figure}[t]
\begin{center}
\includegraphics[width=0.9\textwidth]{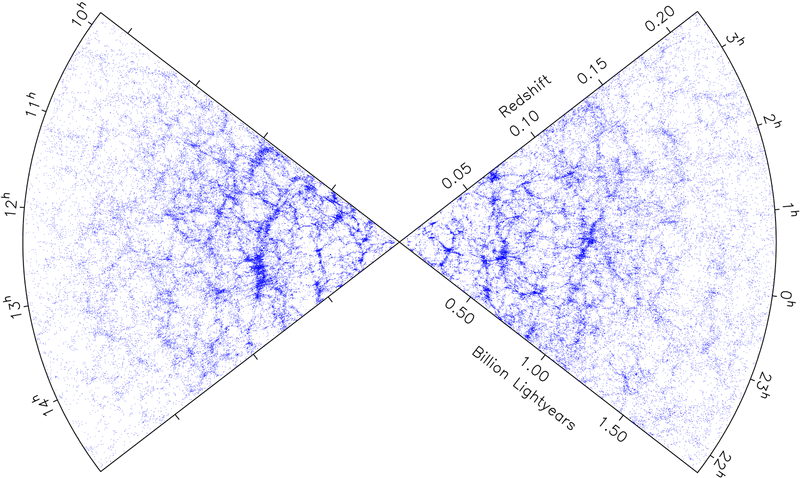}
\caption{Map of the galaxy distribution from the 2dF Galaxy Redshift Survey \cite{Colless03}}.
\label{cone}
\end{center}
\end{figure}

As the universe expands, the other galaxies are moving away from us and the light they emit gets redshifted on the way. This can be used to determine their distance from our galaxy. Assuming that the laws of atomic and molecular physics haven't changed since the emission of the light of wavelength $\lambda_\mathrm{emit}$, the redshift $z$ in the spectral lines is defined from the observed wavelength $\lambda_\mathrm{obs}$ as
\be
1+z=\frac{\lambda_\mathrm{obs}}{\lambda_\mathrm{emit}}.
\ee
In terms of the scale factor $a(t)$ the redshift is given by
\be
1+z=\frac{a_0}{a(t)},
\ee
where $a_0$ is value of the scale factor at present $t=t_0$. Redshift surveys such as the Sloan Digital Sky Survey (SDSS) \cite{SDSS00} and the 2dF Galaxy Redshift Survey (2dFGRS)~\cite{Colless03} measure the redshifts of hundreds of thousands of galaxies and produce maps of the galaxy distribution to determine the large-scale structure (LSS) of the universe. The 2dFGRS measured the spectra for 245,591 objects and probed the structure in the local universe out to $z\simeq 0.3$ (see Figure \ref{cone}). The SDSS is ongoing \cite{SDSSDR6} and currently covers 585,719 galaxies up to $z\simeq 1$ with a mean at $z\simeq 0.3$ and 103,647 quasars up to $z\simeq 3$ with a mean at $z\simeq 1.5$. For a better comparison to observations, cosmological parameters are usually expressed as functions of redshift~$z$ instead of the time coordinate $t$. In the local universe the expansion, commonly called the Hubble Flow, is smooth, and for small distances $d$ and small radial velocities $v$ of the galaxies the Hubble law
\be
v=H_0d
\ee
holds. This can be used to measure the Hubble constant $H_0$, the present value of the Hubble rate. Combining a number of different observation methods, the Hubble Telescope Space Key Project \cite{Freedman06} found 
\be
H_0=100~h~\mathrm{km~s^{-1}Mpc^{-1}}\qquad\mathrm{with}\qquad h=0.72\pm 0.08,
\ee
giving a value of $t_0=H_0^{-1}=13.7$ Gyr for the age of a flat universe with the fractional energy densities $\Omega^0_\Lambda=0.7$ and $\Omega^0_m=0.3$ at the present epoch. A precise measurement of $H_0$ is important for the determination of many cosmological parameters including in turn the current values of the fractional energy density contributions, $\Omega^0_m$, $\Omega^0_\mathrm{rad}$ and $\Omega^0_\Lambda$. The measurement of the current deceleration paramter,
\be
q_0\equiv-\frac{\ddot{a}_0}{a_0H_0^2},
\ee
is more involved and requires the comparison of the apparent magnitudes of high-redshift supernovae Ia with those of lower redshift. After calibration the supernovae of type Ia have a nearly uniform maximum luminosity and can therefore be used as ``standard candles'' for distance measurements.\\

Another important quantity is conformal time, the distance photons could have traveled since $t=0$:
\be
\tau=\int_0^t\frac{dt'}{a(t')}=\int_0^a\frac{da'}{a'}\frac{1}{a'H(a')}.
\ee
Just like $t$, $z$ and $a$, the conformal time $\tau$ serves as a reference parameter when discussing the evolution of the universe. Furthermore, particles which are separated by distances larger than $\tau$ have never been in causal contact with each other. Therefore, it is also called the comoving horizon, or more precisely the comoving particle horizon. For a flat matter-dominated universe with $a\propto t^{2/3}$ it is identical to the comoving Hubble radius or Hubble length,
\be
R_H\equiv\frac{1}{aH}.
\ee
Particles separated further than this are currently not in causal contact. Since the expansion of the real universe has changed over the time, the comoving horizon and the comoving Hubble length are actually not identical, but we can estimate the particle horizon to be of the order of the Hubble length today.

\subsection{Cosmological Perturbation Theory}
\label{cosperttheo}

As one can see from Figure \ref{cone}, the observed matter distribution in the universe is neither homogeneous nor isotropic on small scales. It is dominated by giant bubble-like voids separated by sheets and filaments of galaxies, with superclusters appearing as occasional relatively dense nodes. Quantitative estimates employing the fractal dimension of the data suggest that on scales above $\sim 70$Mpc$/h$ the large scale structure becomes homogeneous. See for example \cite{Thieberger08} for a recent analysis of the SDSS data, in broad agreement with older studies \cite{Amendola99,Yadav05}. However, it has been suggested that a large cold spot in the WMAP data is generated by the largest observed void with a radius of $\sim 140$ Mpc \cite{Rudnick07}. Regardless of this, we can say that structure beneath a scale of $70$Mpc is largely inhomogeneous. This structure must have grown from small initial fluctuations via gravitational instability. \\

Since the universe has been expanding from a much smaller state, the very early universe was made up of a hot plasma. This plasma was opaque since the photons were interacting constantly with the plasma through Thomson scattering. As the universe expanded the plasma cooled down adiabatically until the electrons and protons recombined to hydrogen atoms. The photons scattered off the neutral atoms and continued their path freely through the now transparent universe. These photons from the surface of last scattering ($z$=1100) can be observed today as cosmic microwave background (CMB). Due to the expansion of the universe their wavelength has been stretched and they now form a black-body spectrum of 2.726 K. The primeval plasma, tightly-coupled to the photons by Thomson scattering, underwent acoustic oscillations driven by gravity and radiation pressure which are imprinted on the CMB as temperature anisotropies in the form of acoustic peaks in the angular power spectrum. These are measured to high precision by CMB experiments such as the Wilkinson Microwave Anisotropy Probe (WMAP) \cite{WMAP,WMAP3,WMAP(2)} and ACBAR \cite{Acbar08} and contain much information about cosmology at the time of recombination. In particular, they tell us about the corresponding fluctuations in the matter density which were the seeds of the large scale structure we observe today. The inflationary scenario links these seeds in turn to quantum fluctuations in the very early universe and provides a relatively natural justification for the amplitude and near scale-independent power spectrum the observations imply for the primordial perturbations. However, the density fluctuations remain small enough to safely assume that the universe was very nearly of FLRW type in the early universe. In the early universe, therefore, we can apply linear perturbation theory (see \cite{Kodama85, Mukhanov92} for a review) to find the dynamical equations for the inhomogeneities. \\

The full metric describing the mildly inhomogeneous universe is split into the background metric plus perturbations,
\be
g_{\mu\nu}=g^{(0)}_{\mu\nu}+\delta g_{\mu\nu}.
\ee
The line element for the background FLRW metric in Cartesian coordinates is
\be
ds^2=g^{(0)}_{\mu\nu}(x)dx^\mu dx^\nu=a^2(\tau)(d\tau^2-\gamma_{ij}dx^idx^j),
\ee
where $d\tau=a^{-1}dt$ is the conformal time and 
\be
\gamma_{ij}=\delta_{ij}\left(1+\frac{1}{4}k(x^2+y^2+z^2)\right)^{-2}.
\ee
The metric perturbations may be decomposed into scalar, vector and tensor perturbations according to the way they transform under coordinate transformations on the hypersurfaces of constant conformal time. The decomposition theorem states that this decomposition is preserved by Einstein's equation to first order, and that each of these types of perturbations evolves independently. The most general form of the line element for the background and perturbations is
\be
ds^2&=a^2(\tau)\big((1+2\Phi)d\tau^2-2({\mathcal{D}}_iB+S_i)dx^id\tau\\
&-\left((1-2\Psi)\gamma_{ij}+2{\mathcal{D}}_i{\mathcal{D}}_jE+({\mathcal{D}}_jF_i+{\mathcal{D}}_iF_j)+h_{ij}\right)dx^idx^j\big).
\ee
The covariant derivative on the three-dimensional background hypersurface with respect to some coordinate $i$ is denoted by ${\mathcal{D}}_i$. Since vector perturbations $(S_i, F_i)$ decay kinematically in an expanding universe and tensor perturbations $(h_{ij})$ grow at a slower rate than scalar perturbations $(\phi,\psi,B,E)$ and represent gravitational waves, which do not couple to energy density and pressure inhomogeneities, we concentrate on scalar perturbations. Therefore, we start from the most general form of the scalar metric perturbations
\be
\delta g^{(s)}_{\mu\nu}=a^2(\tau)\begin{pmatrix}2\phi&-{\mathcal{D}}_iB\\-{\mathcal{D}}_iB&2(\psi\gamma_{ij}-{\mathcal{D}}_i{\mathcal{D}}_jE)\end{pmatrix}.
\label{cospertmetric}
\ee
There is a freedom in the choice of coordinate system with which to study the spacetime. The most common gauges are the synchronous gauge, where one sets $\phi=0$, $B=0$ and the conformal Newtonian, or longitudinal, gauge with $E=0$, $B=0$. The advantage of synchronous gauge is that it closely resembles Minkowski space with purely spatial perturbations, because we have chosen the hypersurfaces of freely falling observers. Moreover, there is an unambigious time parameter which eases numerical implementation. The drawback of synchronous gauge is that one freedom remains unfixed and this introduces unphysical artifiacts, or gauge modes, which must be dealt with carefully. (See \cite{PressVishniac80,Bardeen80} for more discussion on this.) A more convenient choice is the Newtonian gauge, where one chooses hypersurfaces of isotropic expansion rate perturbation. If we restrict our attention to spacetimes without anisotropic stress, which by the Einstein equations implies $\phi=\psi$, the line element
\be
ds^2=a^2(\eta)\big((1+2\phi)d\eta^2-(1-2\phi)\gamma_{ij}dx^idx^j\big),
\label{newtonian}
\ee
is the same as the Newtonian approximation to GR, whence $\phi$ may be interpreted as a generalization of the Newtonian potential.

The best way to get around the gauge problem is to express all quantities in a gauge-invariant way. This approach was pioneered by Bardeen \cite{Bardeen80}. Consider infinitesimal coordinate transformations
\be
x^\mu\rightarrow\tilde{x}^\mu=x^\mu+\xi^\mu(x),
\ee
with the four-vector $\xi^\mu=(\xi^0,\xi^i)$. Then one can show that only $\xi^0$ and the function $\xi$, which is defined as the solution to
\be
{\mathcal{D}}^i{\mathcal{D}}_i\xi={\mathcal{D}}_i\xi^i,
\ee
preserve the scalar nature of the metric fluctuations and therefore contribute to their transformation. The most general gauge transformation for scalar perturbations in terms of these functions is
\be
\tau\rightarrow\tilde{\tau}=\tau+\xi^0(x)\qquad\mathrm{and}\qquad x^i\rightarrow\tilde{x}^i=x^i+\gamma^{ij}{\mathcal{D}}_j\xi(x).
\ee
Denoting the derivative of a function $f$ with respect to the conformal time $\tau$ by $f'$, the change in the scalar perturbation functions $\phi$, $\psi$, $B$ and $E$ under the gauge transformation can be expressed as
\be
\tilde{\phi}=\phi-\frac{a'}{a}\xi^0-\xi^0,\quad\tilde{\psi}=\psi+\frac{a'}{a}\xi^0,\quad\tilde{B}=B+\xi^0-\xi',\quad\tilde{E}=E-\xi.
\ee
Although these functions are not gauge-invariant, certain combinations of them are, and the simplest gauge-invariant combinations are the Bardeen potentials
\be
\Phi=\phi+\frac{1}{a}((B-E')a)'\qquad\mathrm{and}\qquad\Psi-\frac{a'}{a}(B-E').
\ee
In Newtonian gauge (\ref{newtonian}) the Bardeen potentials $\Phi$ and $\Psi$ coincide with the scalar perturbations $\phi$ and $\psi$, respectively. \\

Depending on the choice of problem one wishes to study, the corresponding energy momentum tensor can now be decomposed in a similar way. At very high energies it is no longer reasonable to believe that a hydrodynamical description of matter will be valid. The matter will then be described in terms of fields as is for example done for inflationary cosmology. With a different motivation the same formalism is used in the quintessence model of dark energy. The theory of a scalar field minimally coupled to gravity is given by the action
\be
S=\int\left(\frac{1}{2}D^\alpha \varphi D_\alpha \varphi-V(\varphi)\right)\sqrt{-g}d^4x,
\ee
where $D_\mu$ denotes the covariant derivative (\ref{defcov}) and $V(\varphi)$ is the potential of the scalar field. The corresponding energy-momentum tensor is
\be
T^\mu_\nu=D^\mu\varphi D_\nu\varphi-\left(\frac{1}{2}D^\alpha\varphi D_\alpha\varphi-V(\varphi)\right)\delta^\mu_\nu.
\ee
To be consistent with the metric (\ref{cospertmetric}) the scalar field $\varphi(x)$ must also be approximately homogeneous and can be decomposed into
\be
\varphi(x)=\overline{\varphi}(t)+\delta\varphi(x)
\label{sffluctuation}
\ee
where an overbar represents the homogeneous component. The energy-momentum tensor can thus also be decomposed into background and perturbed parts,
\be
T^\mu_\nu=\overline{T}^\mu_\nu+\delta T^\mu_\nu ,
\ee
where $\delta T^\mu_\nu$ is linear in matter and metric perturbations $\delta\varphi$ and $\delta g_{\alpha\beta}$. Inserting this and the metric (\ref{cospertmetric}) into (\ref{sffluctuation}) gives the background energy-momentum tensor
\be
\overline{T}^0_0=\frac{1}{2a^2}\overline{\varphi}'^2+V(\overline{\varphi})=\overline{\rho}_\varphi,~\overline{T}^0_i=0,~ \overline{T}^i_j=\left(-\frac{1}{2a^2}\overline{\varphi}'^2+V(\overline{\varphi})\right)\delta^i_j=-p_\varphi\delta^i_j ,
\label{backgroundsfse}
\ee
and the first-order perturbation
\be
\delta T^0_0&=a^{-2}(-\overline{\varphi}'^2\phi+\overline{\varphi}'\delta\varphi'+V_{,\varphi} a^2\delta\varphi),\quad \delta T^0_i=a^{-2}\overline{\varphi}'\delta\varphi_{,i},\\
\delta T^i_j&=(\overline{\varphi}'^2\phi+\overline{\varphi}'\delta\varphi'+V_{,\varphi} a^2\delta\varphi)\delta^i_j .
\ee
Here, the comma with the space index means differentiation with respect to the corresponding coordinate and $V_{,\varphi}=dV/d\varphi$. This energy-momentum tensor can also be expressed in a completely gauge-invariant form if one wishes. From (\ref{backgroundsfse}) we can see that if the field is slowly-rolling so that $\overline{\varphi}^2\ll a^2V(\overline{\varphi})$, $p_\varphi=-\rho_\varphi$ and the scalar field mimics a cosmological constant. In the opposite limit $p_\varphi=\rho_\varphi$ and the scalar field acts as a ``super-stiff'' fluid.\\

For the hydrodynamical description of matter we can linearise the density by $\rho=\overline{\rho}(1+\delta)$ for a dimensionless density perturbation $\delta$, and write the barotropic pressure as $p=w\rho$. Then one can find the linearised energy-momentum tensor
\be
  T^0_0=\overline{\rho}(1+\delta), \; T^0_i=\overline{\rho}(1+w)v_i, \; T^i_j=\overline{\rho}(w+c_s^2\delta)\delta^i_j
\ee
with the background sound speed $c_s^2=\partial p/\partial\rho$.\\

Both the perturbed metric and energy-momentum tensor are then inserted into Einstein's equation and the resulting equations are combined in such a way, that they form a gauge-invariant set of equations for the Bardeen potentials. These equations can be split into equations for the various different forms of energy. The conservation equations for fluids interacting only through gravity, such as neutrinos and cold dark matter, can be then separated from the photon/baryon system since in the concordance model the only important interaction is Thomson scattering of photons by electrons. One then solves the Boltzmann equation for the photons coupled to a perfect baryon fluid. (See for example \cite{MaBertschinger95} for details.)\\

The perturbations are best described in terms of harmonic modes, characterized by the magnitude of the wavevector $k=\sqrt{k^ik_i}$. In flat space, these take the form of Fourier modes,
\be
A(\mathbf{x})=\int_{\mathbf{k}}A(\mathbf{k})e^{-i\mathbf{k}\cdot\mathbf{x}}\frac{d^3k}{(2\pi)^3}.
\ee
In linear perturbation theory, the different Fourier modes evolve uncorrelated from one-another. This implies that their initial statistical distribution is conserved, and one may link their power spectrum at late times directly to the initial power spectrum. A given mode initially has the power spectrum
\be
  \av{A^*(\mathbf{k})A(\mathbf{k}')}=\frac{2\pi^2}{k^3}\mathcal{P}_{\psi}(k)\delta(\mathbf{k-k}'),
\label{powerspectrum}
\ee
where $\mathcal{P}_\psi(k)$ is the primordial power spectrum of the metric fluctuations and the Dirac delta function enforces statistical homogeneity. According to the theory of inflation the power spectrum is nearly a scale-invariant Harrison-Zel'dovich-Peebles spectrum and follows a power law 
\be
\mathcal{P}_{\psi}(k)\propto k^{n_s-1},
\ee
where $n_s$ is called the spectral index. For adiabatic fluctuations $n_s\approx 1$.\\

Numerical CMB packages such cmbeasy \cite{Doran03}, CAMB \cite{CAMB} or cmbfast \cite{Seljak96} solve the Einstein equations, along with Boltzmann equations for the photons and neutrinos, and perfect fluid equations for the baryons and CDM. (For this reason they are often referred to as ``Boltzmann codes''.) The solutions to these equations then tell us how the amplitude of each wavemode changes between the earliest times and recombination and can be used to wrap (\ref{powerspectrum}) onto the CMB by integrating
\be
  C_l\propto\int\mathcal{P}_\psi(k)\left|\Delta_{Tl}(k)\right|^2d(\ln(k))
\ee
where $\Delta_{Tl}$ is the photon ``brightness function'' expanded across the spherical harmonics and $l$ is the multipole. Observations of the CMB combined with studies of the large-scale structure and Supernova type Ia data remain consistent with an initial, Gaussian distribution of adiabatic perturbations with a spectral index very close to Harrison-Zel'dovich-Peebles \cite{WMAP3}.\\

\begin{figure}[t]
\center{\includegraphics[width=0.6\textwidth]{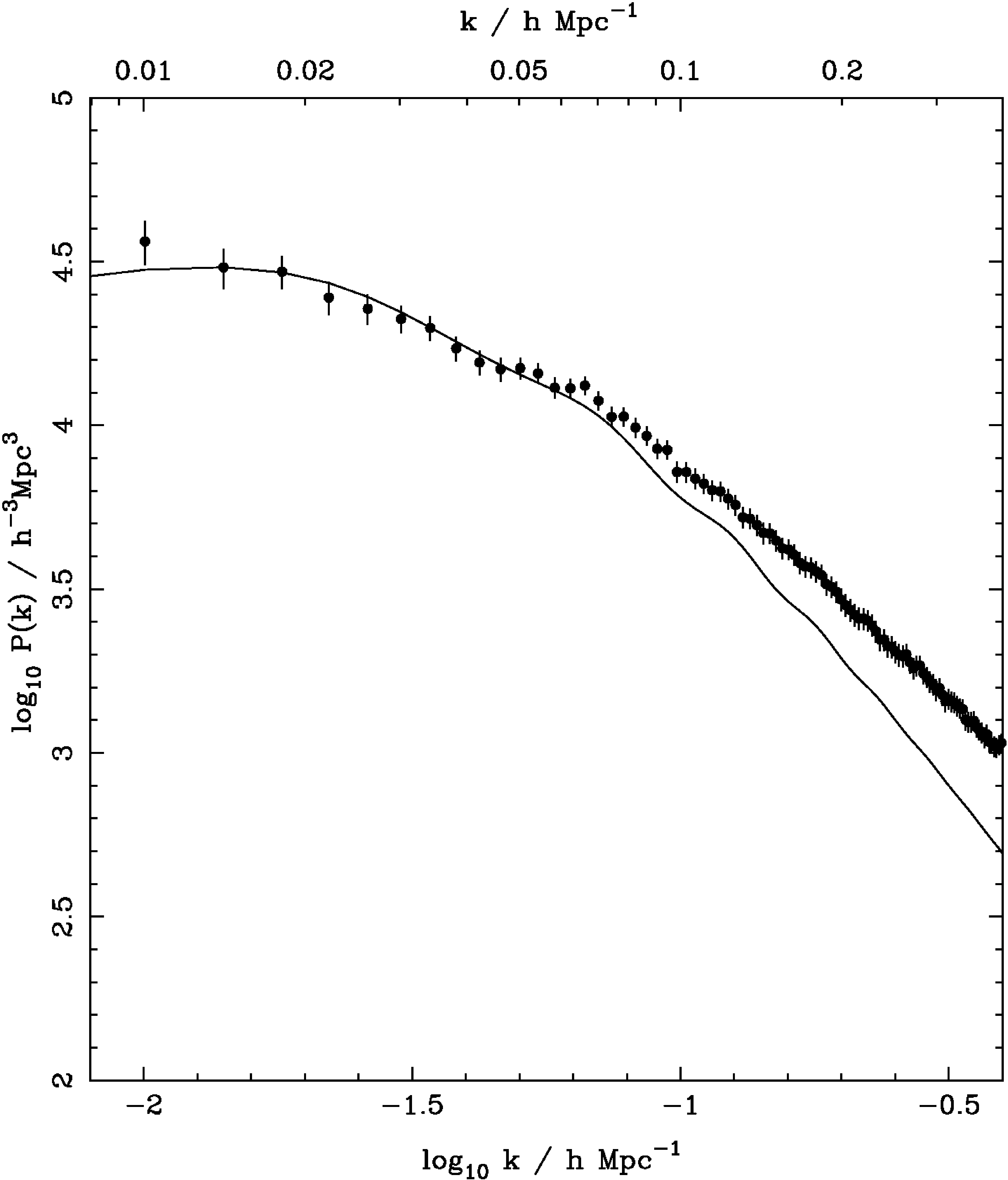}}
\caption{The observed matter power spectrum from the SDSS Luminous Red Galaxy survey (figure taken from \cite{Percival06}). The solid line is the $\Lambda$CDM prediction.}
\label{MatterPower}
\end{figure}

CAMB, cmbfast and cmbeasy also evolve the matter perturbations onward to the present day. The quantity $|\delta(k)|^2$ is known as the matter power spectrum $P(k)$ and quantifies the nature of the clustering of baryons and CDM (see Figure \ref{MatterPower}). The solid line in the figure is the prediction from the linear concordance $\Lambda$CDM model, while the data points are from the SDSS Luminous Red Galaxy survey \cite{Percival06}. For large scales and a Harrison-Zel'dovich primordial power spectrum, $P(k)\propto k$, while on small scales $P(k)\propto k^{-2}$. The turnover reflects the time at which the energy densities matter and radiation were approximately equivalent and can be seen to be at $z\approx 4500$. It is important to notice that modes outside of the Hubble horizon ($k<1/aH$) are gauge-dependent. While in the synchronous gauge they retain a power spectrum $P(k)\propto k$, in the Newtonian gauge they grow as one tends to larger scales. Within the horizon the two gauges agree, as they should for observable quantities.\\

Imprinted on the matter power spectrum are ripples entirely analogous to the oscillations in the CMB. These are the imprint of the oscillations the baryons underwent prior to decoupling from the photons and their wavelength provides a ``standard ruler'' with which to probe the expansion of the universe since the formation of the CMB (see for instance \cite{Percival07} and its references). Above $k^{-1}_{\mathrm{NL}}\approx 0.1~\mathrm{Mpc}/h$, the matter power spectrum is dominated by nonlinear effects and difficult to evaluate. This is clearly evidenced in Figure \ref{MatterPower} with the data points increasingly diverging from the linear prediction on smaller scales. It may be determined for specific models from $N$-body simulations \cite{Springel05}, evaluated from exact inhomogeneous solutions to Einstein's equations \cite{Krasinski}, or evaluated phenomenologically using the halo model of galaxy clustering as in, for example, the Halofit code \cite{Smith02}.

\subsection{The $\mathbf{\Lambda}$CDM model}

From observations of galactic rotation curves, velocity dispersions in galaxy clusters, and mass tracing via gravitational lensing, we know that there must be more mass in the universe than can be observed. Since this energy component of the universe is not interacting with photons it has to be non-baryonic and ``dark'', but it has to interact gravitationally and form structures. This implies that it moves non-relativistically and is ``cold''. Therefore, this energy component has been given the name ``cold dark matter'' (CDM).\\

The direct evidence for the existence of an additional dark energy component comes from distance measurements of type Ia supernovae \cite{PerlmutterEtAl98,RiessEtAl98,RiessEtAl06,WoodVaseyEtAl07}, which indicate that the expansion of the universe is currently accelerating. This fact can only be explained in terms of a FLRW model if it contains a dark energy component with a negative pressure. Recently, Melchiorri {\it et al} used combined data sets to constrain the redshift of transition from deceleration to acceleration, finding a transition at $z_{\mathrm{acc}}=0.76\pm0.10$ and equality between the dark energy and matter at $z_{\mathrm{eq}}=0.40\pm0.08$ \cite{Melchiorri07}. \\

The power-law $\Lambda$CDM or ``concordance'' model is the simplest known model that is in agreement with all astronomical observations, including the CMB, the cosmic expansion history determined from supernovae type Ia data, and the matter power spectrum, measured for example by Lyman $\alpha$ Forest, galaxy clustering and motions, gravitational lensing, cluster studies, and 21 cm topography. The model assumes the universe to be spatially flat, homogeneous and isotropic on large scales. It has a cosmological constant and is composed of ordinary matter, radiation, and dark matter. The primordial fluctuations in this model are adiabatic, nearly scale-invariant Gaussian random fluctuations. CMB anisotropy measurements combined with the LSS data and numerical studies (e.g. \cite{Springel05}) indicate that 76\% of the energy density in the universe is from dark energy, 20\% is from dark matter, and only 4\% is baryonic matter  \cite{2df,Tegmark06}. \\

One issue with the $\Lambda$CDM model concerns the CDM and structure formation on small scales. The model predicts a significantly higher number of dwarf galaxies in our local group than are found \cite{Moore99,Klypin99}. This is called the ``missing satellite'' or ``missing dwarf'' problem. Although improvements in the observation technology diminish this discrepancy \cite{Simon07}, the problem still remains unsolved. There has also been a long-standing discrepency between CMB predictions and LSS observations of $\sigma_8$, the normalisation of the matter power spectrum, with the CMB tended to suggest lower values than that measured by LSS probes. These are now converging (see for example \cite{Acbar08}) but some tension remains. Furthermore, the observed void of radius $\sim 140$ Mpc \cite{Rudnick07} is far outside the current expectations of the $\Lambda$CDM model.\\

Other than the problems on small scales the $\Lambda$CDM model has been phenomenologically very successful in fitting a wide range of cosmological data. However, it remains purely phenomenological and does not provide any explanation of the nature of dark energy and dark matter. If dark energy is explained by a cosmological constant a severe fine-tuning problem arises. Estimating the Casimir force of vacuum fluctuations in the universe gives a value of the order of $M_P^4$ with the Planck mass $M_P$, but the cosmological constant is of the order of $\Lambda\approx 10^{-124}M_P^4$. Why the cosmological constant is so extremely tiny, but not quite zero, is an unresolved problem known as the ``cosmological constant problem''. If we interpret the cosmological constant as a fluid with a dynamical equation of state, we have to explain why this component was negligible for most of the expansion history of the universe and became dominant only recently. This problem is known as the ``coincidence problem''.
\\


It is important to notice that models without a cosmological constant are consistent with the WMAP data alone, provided $H_0\leq 45~\mathrm{km~s^{-1}Mpc^{-1}}$ and the primordial power spectrum is slightly modified \cite{WMAP3,Sarkar07}. The flat, matter-dominated case is the so-called Einstein-DeSitter (EdS) model. Combined with large-scale structure observations (particularly the baryon acoustic oscillations), supernova data and measurements of local dynamics these models seem unlikely, but are nevertheless useful toy models to test the data and challenge the $\Lambda$CDM model.

\section{The Averaging Problem}

Considering that the expansion of the universe has changed from deceleration to acceleration in the recent past, when the formation of structure has grown increasingly non-linear on small scales, it seems sensible to suspect that these effects may be connected. If so, this would be an elegant way to solve the coincidence problem. Solving Einstein's equation on large scales for a general inhomogeneous universe raises the ``averaging problem'', which we will now discuss in detail.\\

The first important indication that the standard model of cosmology is based on an incorrect dynamical equation came from Shirokov and Fisher in 1963 (\cite{Shirokov63}, and reprinted as \cite{Shirokov98}). They realized that in the standard approach an averaged energy-momentum tensor, corresponding to a continuous matter distribution, is used, while the left-hand side of Einstein's equation (\ref{einstein}) remains unaveraged. Instead, the solution is assumed to be the Einstein tensor for an averaged metric,
\be
G_{\mu\nu}(\av{g_{\rho\lambda}})=8\pi G\av{T_{\mu\nu}}+\Lambda\av{g_{\mu\nu}}. 
\ee
This equation is clearly wrong, since from the nonlinear nature of the Einstein tensor we know
\be
\av{G_{\mu\nu}(g_{\rho\lambda})}\neq G_{\mu\nu}(\av{g_{\rho\lambda}}).
\ee
Shirokov and Fisher suggested introducing a correction term $8\pi G T^g_{\mu\nu}$ into Einstein's equation,
\be
G_{\mu\nu}(\av{g_{\rho\lambda}})=8\pi G\av{T_{\mu\nu}}+8\pi G T^g_{\mu\nu}+\Lambda\av{g_{\mu\nu}},
\ee
which depends on the fluctuations of the microscopic gravitational field and is of purely geometrical origin. They called it a ``polarization term'' and remarked that it might be interpreted formally as extra terms in the energy and pressure of the cosmic fluid. \\

The authors proposed the following averaging process 
\be
\av{g_{\mu\nu}(x)}=\frac{\int_{\mathcal{D}}g_{\mu\nu}(x+x')\sqrt{-g}~d^4x'}{\int_{\mathcal{D}}\sqrt{-g}~d^4x'},
\ee
where $g_{\mu\nu}(x)$ is the exact value of the metric tensor at the point $x$, and the domain of averaging ${\mathcal{D}}$, containing the points $x$, is assumed to be large compared to the fluctuations and small compared to the scale at which systematic variations of the mass or metric distribution are possible.\\

Shirokov and Fisher applied this averaging process to the Newtonian approximation and estimated the effect of small fluctuations. Their main result was that the ``polarization term'' mimics repulsive gravitation and prevents the Big Bang singularity, although they stated that this result was preliminary, since they extended their investigations beyond the applicability of perturbation theory.\\

The averaging problem remained generally unconsidered until Ellis emphasized its importance in great detail in 1984 \cite{Ellis84}. He discussed different scales of inhomogeneity and argued, that a suitable averaging process has to reproduce the same metric on the largest scales both when applied only once or when applied subsequently to the intermediate scales. Such a process has to be able to smooth out high-curvature phenomena such as cusps, caustics and singularities in a consistent way.  Furthermore, the averaged Einstein equation has to describe the same dynamical behaviour of the manifold on all scales when one additionally includes the correction term. Ellis also raised the ``fitting problem'', which he discussed in more detail in a subsequent paper with Stoeger \cite{Ellis87}. This concerns the problem of finding the FLRW model which is the best fit to the inhomogeneous real universe, in the same sense that an idealized perfect sphere is fitted to the real earth. This analogy is taken further by posing the question which coordinates in the manifolds of different averaging scale must be identified with one another, in order to be able to quantify the deviation from the best fit FLRW model and hence to indicate how good the fit actually is.\\

Noonan \cite{Noonan84,Noonan85} defined the macroscopic average $\av{Q'}$ of a microscopic quantity $Q'$ by
\be
\av{Q'}=\frac{\int Q'(x')\sqrt{-g(x')}d^4x'}{\int\sqrt{-g(x')}d^4x'}.
\ee
Applied to the Newtonian weak-field, slow-motion approximation, the author found that the contribution by the small-scale gravitational fields to the macroscopic density and energy-momentum tensor are the Newtonian gravitational energy density and energy-momentum tensor, respectively. However, $g(x')$ is assumed to be the determinant of the macroscopic metric and it is not explained how that metric is obtained.\\

Following pioneering work by Brill and Hartle \cite{Brill64}, Isaacson defined an effective energy-momentum tensor for gravitational waves \cite{Isaacson67} and showed its gauge-invariance after averaging for high-frequency waves. The author realized that tensors defined on different spacetime points cannot be compared directly, and therefore simple volume averaging destroys the tensor character of the averaged quantities. Therefore, he used the ``bivector of geodesic parallel displacement'', denoted by $V_\mu{}^{\nu'}(x,x')$. This bivector transforms as a vector with respect to coordinate transformations at either $x$ or $x'$ and, assuming that $x$ and $x'$ are sufficiently close together to ensure the existence of a unique geodesic of the metric $g_{\mu\nu}$ between them, given the vector $A_{\nu'}$ at $x'$ then $A_{\mu}=V_\mu{}^{\nu'}A_{\nu'}$ is the unique vector at $x$ which can be obtained by parallel-transporting $A_{\nu'}$ from $x'$ to $x$ along the geodesic. The average of a tensor $T_{\mu\rho}$ is then defined as
\be
\av{T_{\mu\rho}(x)}=\int V_\mu{}^{\nu'}(x,x')V_\rho{}^{\lambda'}(x,x')T_{\nu'\lambda'}(x')f(x,x')d^4x',
\label{isaacson}
\ee
where $f(x,x')$ is a weighting function which falls smoothly to zero when $x$ and $x'$ differ by the averaging radius, and which is furthermore normalized to unity,
\be
\int f(x,x')d^4x'=1.
\ee

Although Isaacson's procedure looks quite promising at first sight, Stoeger, Helmi and Torres \cite{StoegerEtAl99} realized that it cannot be used to average the metric: the parallel-transported metric at $x'$ is identical to the metric at $x$ since the covariant derivative of the metric vanishes everywhere,
\be
V_\mu{}^{\nu'}(x,x')V_\rho{}^{\lambda'}(x,x')g_{\nu'\lambda'}(x')=g_{\mu\rho}(x).
\ee
The metric is hence invariant under Isaacson's averaging procedure, $\av{g_{\mu\rho}}=g_{\mu\rho}$. Stoeger, Helmi and Torres proposed instead another averaging process,
\be
\av{Q(x)}=\frac{\int Q(x+x')\sqrt{-g(x+x')}~d^4x'}{\int\sqrt{-g(x+x')}~d^4x'}.
\ee
The authors applied this process to weak-field and perturbed FLRW models and demonstrate approximately tensor character of the averaged quantities to linear order. Unfortunately, this is not necessarily the case in general and the result is in any event only approximate.\\

Many of the early approaches to the Averaging Problem employed spatial averaging. This was applied to perturbed FLRW models (e.g. \cite{Boersma97,Futamase96,Kasai93,StoegerEtAl99}), after implementing spacetime-slicing (e.g. \cite{Futamase89}) or in the context of perturbed Newtonian cosmologies (e.g. \cite{BuchertEhlers95}). A review can be found in \cite{Krasinski}. The most promising approach is the Buchert formalism and we will therefore discuss it in detail.

\section{The Buchert Formalism}

Although a simple form of averaging cannot be used to construct an averaged metric or to average over other tensor quantities in a generally covariant way, it can still be used to find certain averages of scalars. Scalar fields are defined to be invariant under general coordinate transformations and their values on different spacetime points can therefore be compared directly. To make use of this fact, we need to decompose Einstein's equation into a set of dynamical equations for scalar quantities. This is usually done in the framework of the 3+1 formalism of general relativity, which we briefly review here. In line with most of the literature in this field, we use a metric of signature $(-,+,+,+)$ for the rest of this section.

\subsection{Einstein's Equation in 3+1 Form}
\label{ADM}

The 3+1 formalism, sometimes also called the ADM formalism, is a standard method in general relativity and an essential ingredient of numerical relativity. Details about this formalism can be found for example in \cite{Gourgoulhon07,York79,Wald,Buchert01}. Provided that the spacetime $(\mathcal{M},\bf{g})$ is globally hyperbolic, we may foliate $\mathcal{M}$ with a family of spacelike hypersurfaces $\Sigma$, such that each hypersurface is a level surface of a smooth scalar field $t$, which we will later identify with the time coordinate. Then the entire future and past history of the universe can be evaluated from the conditions on one of the ``Cauchy surfaces'' $\Sigma$. The normal unit vector $\bf{n}$ to the surface $\Sigma$ being timelike must satisfy
\be
  \mathbf{n}\cdot\mathbf{n}=-1 .
\ee
It is necessarily colinear to the metric dual $\boldsymbol{\nabla}t$ of the gradient 1-form $\mathbf{d}t$ and the proportionality factor is called the lapse function $\alpha$
\be
  \mathbf{n}=-\alpha\boldsymbol{\nabla}t ,
\ee
where $\alpha\equiv\left(-\boldsymbol{\nabla}t\cdot\boldsymbol{\nabla}t\right)^{-1/2}=-\left(\langle\mathbf{d}t,\boldsymbol{\nabla}t\rangle\right)^{-1/2}$. To ensure that $\mathbf{n}$ is future oriented, $\alpha$ must be larger than zero and in particular never vanishes for a regular foliation. Since $\langle\mathbf{d}t,\alpha\mathbf{n}\rangle=1$ the hypersurfaces are Lie dragged by the normal evolution vector $\alpha\mathbf{n}$.\\

Now introduce coordinate systems $x^i$ on each hypersurface $\Sigma$ that vary smoothly between neighbouring hypersurfaces, such that $x^\mu=(t,x^1,x^2,x^3)$ form well-behaved coordinate systems on $\mathcal{M}$. The 1-form $\mathbf{d}t$ is dual to the time vector $\boldsymbol{\partial}_t$. Hence, just like the normal evolution vector, the time vector Lie drags the hypersurfaces. In general these vectors do not coincide. The difference defines the shift vector $\boldsymbol{\beta}$
\be
  \boldsymbol{\partial}_t=\alpha\mathbf{n}+\boldsymbol{\beta} .
\label{shift}
\ee
The shift vector is purely spatial since $\langle\mathbf{d}t,\boldsymbol{\beta}\rangle=\langle\mathbf{d}t,\boldsymbol{\partial}_t\rangle-\langle\mathbf{d}t,\alpha\mathbf{n}\rangle=0$. In terms of the natural basis $\partial_\mu$ of the introduced coordinates $x^\mu$ the normal unit vector can then be expressed as
\be
  n^\mu=\frac{1}{\alpha}(1,-\beta^i) .
\ee 
With the projection operator into $\Sigma$
\be
  h_{\mu\nu}=g_{\mu\nu}+n_\mu n_\nu 
\ee
the components of the 3-metric $h_{ij}$ become
\be
  h_{ij}=g_{\mu\nu}h^\mu_i h^\nu_j .
\ee
From here one may find that the line element in these coordinates is
\be
  ds^2=(-\alpha^2+\beta_i\beta^i)dt^2+2\beta_idtdx^i+h_{ij}dx^idx^j ,
\ee
where $\beta_i=h_{ij}\beta^j$.\\

The projection of the gradient $\nabla_\mu$ of the unit normal vector defines the extrinsic curvature tensor
\be
  K_{ij}=-h_i^\mu\nabla_\mu n_j=-\nabla_i n_j .
\ee
Using Frobenius' theorem we find that the normal unit vector $n^\mu$ is hypersurface orthogonal if and only if 
\be
n_{[\mu}\nabla_\nu n_{\lambda]}=0,
\label{frobenius}
\ee
where square brackets denote antisymmetrization on the enclosed indices. Therefore, we can express the extrinsic curvature tensor as the Lie derivative (\ref{liedef})
\be
  K_{ij}=-\frac{1}{2}\mathcal{L}_nh_{ij}
\label{excurv}
\ee
and the evolution of the 3-metric $h_{ij}$ is given by the Lie derivative along the normal evolution vector $\alpha\mathbf{n}$. \\

We can use the freedom in the choice of coordinates to set the shift vector $\beta^i=0$. We further identify the normal unit vector $n^\mu$ with the fluid four-velocity $u^\mu$, thereby identifying the hypersurfaces as the rest-space of the fluid. The line element in these coordinates is given by
\be
ds^2=-\alpha^2 dt^2+h_{ij}dx^idx^j,
\ee
and the four-velocity of the observers becomes
\be
n^\mu=\left(\frac{1}{\alpha},\mathbf{0}\right),\qquad n_\mu=(-\alpha,\mathbf{0}).\\
\ee
Using (\ref{shift}) and (\ref{excurv}) the evolution equation then becomes
\be
 \alpha\mathcal{L}_nh_{ij}=\mathcal{L}_{\partial_t}h_{ij}-\mathcal{L}_\beta h_{ij}=\del_t{h}_{ij}=\alpha\dot{h}_{ij}=-2\alpha K_{ij} ,
\label{buchertmetricev}
\ee
where an overdot denotes a total derivative with respect to the proper time given by $d\tau=\alpha dt$, $\mathcal{D}_i$ is the covariant derivative on the 3-surface.\\

Now we perform a 3+1 decomposition of the stress-energy tensor
\be
  T_{\mu\nu}=\rho n_\mu n_\nu+ph_{\mu\nu} \iff
  \rho=T_{\mu\nu}n^\mu n^\nu, \quad ph_{ij}=T_{ij}
\ee
and project the Einstein equations onto the hypersurface and along its normal. This results in the Hamilton constraint equation
\be
  \mathcal{R}+K^2-K^i_jK^j_i=16\pi G\rho,
\label{rconstraint}
\ee
where $\mathcal{R}$ is the Ricci scalar on $\Sigma$ and $K=K^i_i$; the momentum constraint equation,
\be
  \mathcal{D}_j\left(K^{ij}-h^{ij}K\right)=0 ,
\ee
and, with the ``acceleration'', defined as
\be
\label{acceleration}
a^\mu\equiv n^\nu \nabla_\nu n^\mu=\dot{n}^\mu,\quad a_i=h_i^\mu a_\mu,
\ee
which describes the deviations from a geodesic flow, the evolution equation for the extrinsic curvature tensor can be expressed as
\be
\dot{K}_{ij}=\mathcal{R}_{ij}+KK_{ij}&-4\pi Gh_{ij}\left(\rho-p\right)-(\mathcal{D}_ja^i+a^ia_j).
\label{kevolution}
\ee

Now we can introduce the expansion tensor $\Theta_{ij}\equiv-K_{ij}$ and decompose it in terms of kinematical quantities and their scalar invariants,
\be
\Theta_{ij}\equiv-K_{ij}=\nabla_i n_j=\omega_{ij}+\sigma_{ij}+\frac{1}{3}h_{ij}\theta.
\ee 
These are the trace-free antisymmetric ``rotation tensor'', 
\be
\omega_{ij}\equiv \nabla_{[i}n_{j]},
\ee
the trace-free symmetric ``shear tensor''
\be
\sigma_{ij}\equiv \nabla_{(i}n_{j)}-\frac{1}{3}h_{ij}\nabla_k n^k,
\ee
where round brackets denote symmetrization on the enclosed indices, and the ``rate of expansion'', which is the trace of the expansion tensor, 
\be
\theta\equiv \nabla_k n^k.
\ee
Using once more Frobenius' theorem (\ref{frobenius}), we find that the manifold representing the universe is globally hyperbolic if and only if it is irrotational, and hence $\omega_{ij}=0$.\\

With the ``rate of shear'' $\sigma$, defined by
\be
\sigma^2\equiv \frac{1}{2}\sigma^i{}_j\sigma^j{}_i ,
\ee
we write down two out of the three scalar invariants of the expansion tensor
\be
-K\equiv\theta\qquad\mathrm{and}\qquad\frac{1}{2}\left(K^2-K^i{}_jK^j{}_i\right)=\frac{1}{3}\theta^2-\sigma^2.
\label{buchertscalars}
\ee
The four-divergence of the acceleration field is
\be
\mathcal{A}\equiv \nabla_\mu a^\mu=\mathcal{D}_ia^i+a_ia^i .
\ee
Inserting this and (\ref{rconstraint}) into the trace of (\ref{kevolution}) gives the Raychaudhuri equation,
\be
\dot{\theta}=-\frac{1}{3}\theta^2-2\sigma^2-4\pi G(\rho+3p)-\mathcal{A}.
\label{buchertray}
\ee

\subsection{The Buchert Equations}
\label{sectionbuchert}
Having decomposed Einstein's equations into a system of equations for scalar quantities, we now want to find their averaged form. Following Buchert \cite{Buchert99,Buchert01}, we define the volume of a domain $\mathcal{D}\subset\Sigma$ of the hypersurface by
\be
V=\int_{\mathcal{D}}\sqrt{h}d^3\mathbf{x}.
\ee
Then we define the average of a scalar field $A$ in this domain as
\be
\av{A}=\intdom{A}.
\label{AverageDef}
\ee
Since $\del_t$ and $d^3\mathbf{x}$ commute, we can furthermore introduce the effective Hubble rate by
\bdm
3\tilde{H}_\mathcal{D}=3\frac{\del_t\dom{a}}{\dom{a}}\equiv\frac{\del_t V}{V}=\frac{1}{V}\frac{\partial}{\partial t}\int_{\mathcal{D}}\sqrt{h}d^3\mathbf{x}
   =\intdom{\frac{1}{2}h^{ij}\alpha\dot{h}_{ij}} .
\edm
Using (\ref{buchertscalars}) and (\ref{buchertmetricev}) we thus have
\be
3\tilde{H}_\mathcal{D}=3\frac{\del_t\dom{a}}{\dom{a}}=-\av{\alpha K}=\av{\alpha\theta}=\av{\tilde{\theta}} .
\ee
Hence, the effective Hubble rate in a domain is given by the average of the scaled expansion $\tilde{\theta}\equiv\alpha\theta$. We can also quickly find the commutator between time and space derivatives for any scalar field $A$,
\be
  \av{\frac{\del A}{\del t}}-\frac{\partial}{\partial t}\av{A}=\av{A\tilde{\theta}}-\av{A}\av{\tilde{\theta}} .
\label{BuchertTimeAverage}
\ee
The averaged Raychaudhuri equation (\ref{buchertray}) for the scaled densities $\tilde{\rho}\equiv\alpha^2\rho$ and $\tilde{p}\equiv\alpha^2p$ is
\be
3\frac{\del_t^2\dom{a}}{\dom{a}}+4\pi G\av{\tilde{\rho}+3\tilde{p}}=\tilde{\mathcal{Q}}_\mathcal{D}+\tilde{\mathcal{P}}_\mathcal{D} ,
\label{bucherta}
\ee
and with the scaled spatial Ricci scalar $\tilde{\mathcal{R}}\equiv\alpha^2\mathcal{R}$ the averaged Hamilton constraint (\ref{rconstraint}) becomes
\be
6\tilde{H}_\mathcal{D}^2-16\pi G\av{\tilde{\rho}}=-\left(\tilde{\mathcal{Q}}_\mathcal{D}+\av{\tilde{\mathcal{R}}}\right) .
\label{buchertb}
\ee

Here, we have introduced the domain dependent ``backreaction terms'': the ``kinematical backreaction'',
\be
\tilde{\mathcal{Q}}_\mathcal{D}\equiv\frac{2}{3}\av{(\tilde{\theta}-\av{\tilde{\theta}})^2}-2\av{\tilde{\sigma}^2} ,
\label{buchertq}
\ee
with the scaled shear scalar $\tilde{\sigma}\equiv\alpha\sigma$, and the ``dynamical backreaction'',
\be
\tilde{\mathcal{P}}_\mathcal{D}\equiv\av{\tilde{\mathcal{A}}}+\av{\dot{\alpha}\tilde{\theta}} ,
\ee
with the scaled acceleration divergence $\tilde{\mathcal{A}}\equiv\alpha^2\mathcal{A}$. To ensure that (\ref{buchertb}) is an integral of (\ref{bucherta}), we additionally need the integrability condition
\be
\del_t\tilde{\mathcal{Q}}_\mathcal{D}\!+\!6\tilde{H}_\mathcal{D}\tilde{Q}_\mathcal{D}\!+\!\del_t\av{\tilde{\mathcal{R}}}\!+\!2\tilde{H}_\mathcal{D}\av{\tilde{\mathcal{R}}}\!+\!4\tilde{H}_\mathcal{D}\tilde{P}_\mathcal{D}\!-\!16\pi G\left(\del_t\av{\tilde{\rho}}\!+\!3\tilde{H}_\mathcal{D}\av{\tilde{\rho}\!+\!\tilde{p}}\right)=0 .
\label{buchertc}
\ee
Having now an averaged system of Einstein's equation in scalar form, we can fit it to a flat FLRW model as follows. Define the effective energy density and pressure as
\begin{gather}
\rho_{\mathrm{eff}}\equiv\av{\tilde{\rho}}-\frac{\tilde{\mathcal{Q}}_\mathcal{D}}{16\pi G}-\frac{\av{\tilde{\mathcal{R}}}}{16\pi G} ,\label{buchertrho}\\
p_\mathrm{eff}\equiv\av{\tilde{p}}-\frac{\tilde{\mathcal{Q}}_\mathcal{D}}{16\pi G}+\frac{\av{\tilde{\mathcal{R}}}}{48\pi G}-\frac{\tilde{\mathcal{P}}_\mathcal{D}}{12\pi G}.\label{buchertp}
\end{gather}
Then the averaged equations (\ref{bucherta}) and (\ref{buchertb}) can be expressed as
\begin{gather}
3\frac{\del_t\dom{a}}{\dom{a}}+4\pi G\left(\rho_\mathrm{eff}+3p_\mathrm{eff}\right)=0 ,\label{buchert1}\\
6\tilde{H}_\mathcal{D}^2-16\pi G\rho_\mathrm{eff}=0 \label{buchert2},
\end{gather}
and the integrability condition (\ref{buchertc}) has exactly the form of a conservation equation law
\be
\del_t\rho_\mathrm{eff}+3\tilde{H}_\mathcal{D}\left(\rho_\mathrm{eff}+p_\mathrm{eff}\right)=0 .
\label{buchert3}
\ee
This set of equations is known as ``Buchert equations''. They show that, despite the non-commutativity of time and space derivatives, the averaged variables obey the same equations as the variables in the flat FLRW model. Therefore, in the limit of vanishing domain of averaging the equations (\ref{buchert1}), (\ref{buchert2}) and (\ref{buchert3}) smoothly reduce to the equations (\ref{friedmann}), (\ref{raychaudhuri}) and (\ref{conservation}), respectively. However, as pointed out by Buchert (e.g. in \cite{Buchert071,Buchert072}), there are correction terms to the flat FLRW variables that counteract gravity. As we can see from the definition of the effective energy density (\ref{buchertrho}), the first term in the kinematical backreaction (\ref{buchertq}) is positive and hence reduces the value of the energy density. This stems from the fact that an average correlates the local fluctuations, which then act in the sense of a global ``kinematical pressure''. Therefore, one cannot exclude large-scale effects from averaging inhomogeneities with the argument that the perturbations are of small amplitude and act gravitationally.  \\

The Buchert equations reduce the solution of the averaging problem for scalars to finding an effective equation of state that relates the effective densities,
\be
w_{\mathrm{eff}}=\frac{p_{\mathrm{eff}}}{\rho_{\mathrm{eff}}}. 
\ee
If, however, we consider the backreaction terms independently as an effective fluid, it has an equation of state
\be
w_\mathcal{D}=-\frac{1}{3}\frac{\av{\tilde{\mathcal{R}}}-4\tilde{\mathcal{P}}_\mathcal{D}-3\tilde{\mathcal{Q}}_\mathcal{D}}{\av{\tilde{\mathcal{R}}}+\tilde{\mathcal{Q}}_\mathcal{D}} 
\ee
We can then see from this result that should these terms dominate the matter contribution
with $\tilde{\mathcal{Q}}_\mathcal{D}<-\tilde{\mathcal{P}}_\mathcal{D}$ and hence $w_\mathcal{D}<-\frac{1}{3}$, the modifications act to accelerate the averaged scale factor.\\

The statement that the kinematical backreaction term $\tilde{\mathcal{Q}}_\mathcal{D}$ in Buchert's equations can play the role of a cosmological constant has been confirmed by many authors (see e.g. \cite{Sicka99,Matarrese05,LiSchwarz07-2,Rasanen03,Rasanen06}, for a review, see \cite{Buchert071}). A promising approach has been suggested by Wiltshire \cite{Wiltshire07_2,Wiltshire07_3}. He introduced two scales, one that belongs to ``finite infinity'' regions, and one that describes the scale factor of negatively curved voids. Then the Buchert formalism is applied and the average spherically symmetric geometry is reconstructed in terms of a spatially averaged scale factor and time. According to the author, this ``fractal bubble'' model might simulateously resolve key anomalies relating to primordial lithium abundances, CMB ellipticity, the Hubble bubble feature and the expansion age. The key to the latter is the fact that within this model an observer in a galaxy measures a significantly older expansion age of the universe than that of a $\Lambda$CDM model, giving a volume average age of the universe of 18.6Gyr. \\

\section{The Lema\^itre-Tolmann-Bondi and Swiss Cheese Models}

The Lema\^itre-Tolmann-Bondi (LTB) solution is the most general spherically symmetric solution to Einstein's equations for dust, 
\be
G_{\mu\nu}=8\pi G\rho n_\mu n_\nu .
\ee
Because of the spherical symmetry the line element is expressed in spherical coordinates plus time,
\be
ds^2=-dt^2+\frac{R'(t,r)^2}{1+E(r)}dr^2+R(t,r)^2(d\theta^2+\sin^2\theta d\phi^2),
\ee
where the functions $R(t,r)$ and $E(r)$ are related to each other and to the energy density $\rho(r,t)$ as follows
\be
\dot{R}(t,r)^2&=8\pi G\frac{M(r)}{R(t,r)}+E(r),\\
\rho(t,r)&=\frac{M'(r)}{R(t,r)^2R'(t,r)} ,
\ee
where dots and primes denote derivatives with respect to $t$ and $r$, respectively. $E(r)$ can be interpreted as the total energy per unit mass and $M(r)$ as the mass within the sphere of comoving radial coordinate $r$.\\

This model is a useful toy for studying backreaction effects, because the symmetry allows quantitative studies without the use of approximations. Therefore, it is probably the most studied model in this context. In the earliest work \cite{Celerier00} the author used the LTB model as a toy model to obtain a reasonable fit to supernovae luminosity densities and showed that the acceleration implied by the supernovae data can be explained by a large scale inhomogeneity without the need for a cosmological constant. A later study \cite{Rasanen04} showed that backreaction slows down the expansion if measured in terms of the proper time, but speeds it up if measured in terms of the energy density or the scale factor. The authors of \cite{Nambu05} used a solution with both a region with positive spatial curvature and a region with negative spatial curvature. They found that after the region of positive spatial curvature begins to re-collapse, the averaged universe starts accelerated expansion. This implies a strong coupling between averaged scalar curvature and kinematical backreaction. It is important to notice, that such an effect cannot be studied in the LTB model with vanishing scalar curvature, since for this model $\tilde{\mathcal{Q}}_\mathcal{D}=0$ \cite{Paranjape06}. \\

However, although there are many studies about the averaged luminosity and angular diameter distances (see for example \cite{AlnesEtAl05,Enqvist07,Mattsson07,VanderveldEtAl06}) in this model, it remains merely a toy model which and unsuitable to describe the real universe, since it violates, for example, the cosmological principle. On the other hand, it is a suitable choice to model the observed voids in the universe and is as such implemented in the swiss-cheese model.\\

The swiss-cheese model is a more realistic model which consists of LTB patches embedded in a flat FLRW background containing only matter. In \cite{Marra071,Marra072} the authors did not analyze the averaged domain dynamics in this model but instead the propagation of photons. They found that the light-cone average of the density as a function of redshift is affected by inhomogeneities in a way that the phenomenological homogeneous model (which is identical to the EdS background) behaves as if it has a dark-energy component. The effect arises because, as the universe evolves, a photon spends increasingly more time in the large voids than in the thin high-density structures. The authors conclude further that within their toy model the voids must have a present size of $\sim 250$ Mpc to be able to mimic the $\Lambda$CDM model. In contrast, photon propagation in a swiss-cheese model has been independently studied \cite{Biswas07} and very small overall effects were found if the observer sits outside the void, while they are large if he sits inside.

\section{Super-Hubble Fluctuations}

In inflationary cosmology matter is treated as a scalar field $\varphi$ in the context of cosmological perturbation theory (which was introduced in section \ref{cosperttheo}). It was suggested that backreactions from perturbations of wave-lengths larger than the Hubble radius, generated during inflation, can explain the accelerated expansion of the universe without the need for a cosmological constant. In \cite{Barausse05} and \cite{KolbEtAl05} the authors computed the luminosity distance-redshift relation in a perturbed flat matter-dominated universe, taking into account the presence of cosmological inhomogeneities up to second order in perturbation theory in the adiabatic case. They found that the time evolution of the super-Hubble modes produce a large variance of the deceleration parameter, which could mimick dark energy. \\

A number of studies (e.g. \cite{Flanagan05,Geshnizjani05,HirataSeljak05,Rasanen05}) have criticised this result. In \cite{Geshnizjani05,Rasanen05} it was shown that the corrections only amount to a renormalization of local spatial curvature and hence that their magnitude is tightly constrained by observations. The authors of \cite{MartineauBrandenberger05} pointed out that this argument only excludes the approach of \cite{Barausse05,KolbEtAl05}, since in that work only the leading gradient terms in the energy-momentum tensor are considered. \\

The effects of long-wavelength scalar metric perturbations with the complete effective energy-momentum tensor were studied in \cite{MartineauBrandenberger05}. Their starting point was the metric without anisotropic stress in Newtonian gauge (\ref{newtonian}). Then, both the Einstein and energy-momentum tensor are expanded up to second order in metric $\phi$ and matter perturbations $\delta\varphi$. The linear equations are assumed to be satisfied, and the remnants are spatially averaged, providing the equation for a new background metric which takes into account the backreaction of linear fluctuations computed up to second order
\be
G_{\mu\nu}=8\pi G(T_{\mu\nu}+\tau_{\mu\nu}),
\ee 
where $\tau_{\mu\nu}$ consists of terms quadratic in metric and matter fluctuations and is called the effective energy-momentum tensor. Its explicit form was first calculated in \cite{Mukhanov97}:

\be
\tau_{00}=\frac{1}{8\pi G}\left(12H\av{\phi\dot{\phi}}-3\av{(\dot{\phi})^2}+9a^{-2}\av{(\nabla\phi)^2}\right)+\av{(\delta\dot{\varphi})^2}\\
+a^{-2}\av{(\nabla\delta\varphi)^2}+\frac{1}{2}V''(\overline{\varphi})\av{\delta\varphi^2}+2V'(\overline{\varphi})\av{\phi\delta\varphi} ,
\ee
and
\be
\tau_{ij}\!=\!a^2\delta_{ij}\bigg(\frac{1}{8\pi G}\left((24H^2\!+\!16\dot{H})\av{\phi^2}\!+\!24H\av{\dot{\phi}\phi}\!+\!\av{(\dot{\phi})^2}\!+\!4\av{\phi\ddot{\phi}}\!-\!\frac{4}{3}a^{-2}\av{(\nabla\phi)^2}\right)\\
+\!4\dot{\overline{\varphi}}^2\av{\phi^2}\!+\!\av{(\delta\dot{\varphi})^2}\!-\!a^{-2}\av{(\nabla\delta\varphi)^2}\!-\!4\dot{\overline{\varphi}}\av{\delta\dot{\varphi}\phi}\!-\!\frac{1}{2}V''(\overline{\varphi})\av{\delta\varphi^2}\!+\!2V'(\overline{\varphi})\av{\phi\delta\varphi}\bigg)
\ee

The role of backreaction of super-Hubble modes in those inflationary models in which inflation ends through the reheating dynamics of $\varphi$ was considered in \cite{MartineauBrandenberger05}. This work found that then the effective energy-momentum tensor acts as a tracker during the period of radiation domination, but redshifts less rapidly than matter in the matter era. Using standard values for the preheating temperature and the amplitude of the inflation following preheating , they found that this mechanism leads to a possible explanation of dark energy.

It should be commented that the issue of super-Hubble fluctuations is complicated by gauge-dependence. From the matter power spectrum it is quick to see that adiabatic fluctuations on such scales are highly-gauge dependent as in the Newtonian gauge the $P(k)$ grows for decreasing $k$ in contrast to the synchronous gauge where $P(k)$ remains well-behaved. It has been shown \cite{Geshnizjanibrandenberger02,GeshnizjaniBrandenberger03} that super-Hubble isocurvature modes are gauge-independent but it is still fair to say that the backreaction of super-Hubble fluctuations on the evolution of the universe is a controversial topic.

\section{Macroscopic Gravity}

Macroscopic gravity is a non-perturbative, geometrical approach to resolve the averaging problem by its reformulation as the problem of the macroscopic description of classical gravitation. This theory was suggested by Zalaletdinov \cite{Zalaletdinov92,Zalaletdinov93,Zalaletdinov08}. \\

According to this approach, the average value of a tensor field $Q^{\alpha\ldots}_{\beta\ldots}(x)$ on the manifold $\cal{M}$ over a region $\Sigma\subset\cal{M}$ at a supporting point $x\in\Sigma$ as
\be
\langle Q^{\alpha\ldots}_{\beta\ldots}(x)\rangle=\frac{1}{V_{\Sigma}}\int_{\Sigma}(\widehat{\cal V}^{-1})^\alpha{}_{\mu'}(x',x)\ldots\widehat{{\cal V}}_{\beta}{}^{\nu'}(x',x)\ldots Q^{\mu'\ldots}_{\nu'\ldots}(x')d\Omega'.
\label{zalaletdinoveq}
\ee
Here $V_{\Sigma}$ denotes the volume of the averaging region $\Sigma$ and $d\Omega$ denotes the invariant volume element. $\widehat{{\cal V}}_{\nu}{}^{\beta}(x',x)$ is the averaging bivector which is required to satisfy the two conditions,
\begin{align}
\mathrm{(i)}&\qquad \lim_{x'\rightarrow x}\widehat{{\cal V}}_{\beta}{}^{\nu'}(x',x)=\delta_{\beta}{}^{\nu'},\label{bivprop1}\\
\mathrm{(ii)}&\qquad\widehat{{\cal V}}_{\beta}{}^{\nu'}(x',x)=\delta_{\beta}{}^{\nu'}\quad\mathrm{for~vanishing~curvature.}\label{bivprop2}
\end{align}  
The first condition ensures the existence of the inverse bivector, $\widehat{{\cal V}}_{\alpha}{}^{\mu'}(\widehat{{\cal V}}^{-1})^\beta{}_{\mu'}=\delta_{\alpha}{}^{\beta}$, and the correct limit for $V_{\Sigma}\rightarrow 0$, the second condition guarantees a correct flat space limit.\\

Since averaging and covariant derivation of a tensor field do not commute in a curved space, a second bivector, the coordination bivector  $\widehat{{\cal W}}_{\mu}{}^{\alpha}(x',x)$, is introduced. This bivector Lie-drags the averaging region from $x'$ to $x$ along integral lines, such that a comparison of the averages at different points is possible. It also obeys the two properties (\ref{bivprop1}) and (\ref{bivprop2}), but is also required to satisfy
\begin{align}
\mathrm{(iii)}&\qquad\widehat{{\cal W}}_{\beta}{}^{\nu'}{}_{;\nu'}=0,\label{bivprop3} \\
\mathrm{(iv)}&\qquad\widehat{{\cal W}}_{[\alpha}{}^{\mu'}{}_{,\beta]}+\widehat{{\cal W}}_{[\alpha}{}^{\mu'}{}_{,\nu'}\widehat{{\cal W}}_{\beta]}{}^{\nu'}=0.\hspace{2.2cm}\label{bivprop4}
\end{align}
The first condition ensures volume preservation while Lie-dragging, and the second condition ensures analyticity of the averaged tensor field. It is reasonable to choose the averaging bivector to coincide with the ccordination one, $\widehat{{\cal V}}_{\beta}{}^{\nu'}=\widehat{{\cal W}}_{\beta}{}^{\nu'}$, and such a choice turns out to simplify the formalism greatly. \\

Next, the bilocal extension of the Christoffel symbols is defined as
\be
{\cal F}^\alpha{}_{\beta\gamma}=(\widehat{{\cal W}}^{-1})^\alpha{}_{\mu'}(\widehat{{\cal W}}_{\beta}{}^{\mu'}{}_{,\gamma}+\widehat{{\cal W}}_{\beta}{}^{\mu'}{}_{;\nu}\widehat{{\cal W}}_{\gamma}{}^{\nu'}).
\ee
When averaged by (\ref{zalaletdinoveq}), this serves as affine connection for the averaged manifold $\overline{{\cal F}}^{\alpha}{}_{\beta\gamma}$. From this the curvature tensor $M^\alpha{}_{\beta\gamma\delta}$ is calculated in the usual way. It is called the induction tensor. Contracted it gives $M_{\mu\nu}$.\\

Then the connection correlation tensor is introduced,
\be
Z^\alpha{}_{\beta\gamma}{}^{\mu}{}_{\nu\sigma}\equiv Z^\alpha{}_{\beta[\gamma}{}^{\mu}{}_{\nu\sigma]}=\av{{\cal{F}}^\alpha{}_{\beta[\gamma}{\cal{F}}^\mu{}_{\nu\sigma]}}-\overline{\cal{F}}^\alpha{}_{\beta[\gamma}\overline{\cal{F}}^\mu{}_{\nu\sigma]} ,
\ee 
and the quantity
\be
Z^\alpha{}_{\mu\nu\beta}=2Z^\alpha{}_{\mu[\rho}{}^{\rho}{}_{|\nu|\beta]},
\ee
where the index between bars $||$ does not participate in the antisymmetrization, and
the quantity $Q_{\mu\nu}=Q^\rho{}_{\mu\rho\nu}$ defined by
\be
Q^\alpha{}_{\beta\gamma\delta}=2Z^\mu{}_{\beta[\gamma}{}^{\alpha}{}_{|\mu|\delta]}
\ee
give the induction tensor $Z^\alpha{}_{\mu\nu\beta}-\frac{1}{2}\delta^\alpha_\beta Q_{\mu\nu}$. Eventually, the averaged Einstein equations can be expressed as the field equations of macroscopic gravity:
\be
\overline{g}^{\alpha\epsilon}M_{\epsilon\beta}-\frac{1}{2}\delta^\alpha_\beta\overline{g}^{\mu\nu}M_{\mu\nu}=-\kappa\av{{T^\alpha_\beta}^{\mathrm{(micro)}}}+\big(Z^\alpha{}_{\mu\nu\beta}-\frac{1}{2}\delta^\alpha_\beta Q_{\mu\nu}\big)\overline{g}^{\mu\nu} ,
\ee
where $\overline{g}^{\mu\nu}$ is the averaged metric.\\

If all correlation functions vanish and the macroscopic spacetime is highly symmetric, this equation reduces to the usual Einstein equations
\be
M_{\alpha\beta}-\frac{1}{2}\overline{g}_{\alpha\beta}\overline{g}^{\mu\nu}M_{\mu\nu}=-\kappa T_{\alpha\beta}^{(\mathrm{hydro})} .
\ee
It has been shown \cite{Zalaletdinov08} that, in the special case of the FLRW metric with an energy-momentum tensor of a perfect fluid, this macroscopic gravity equation takes the form of the Friedmann equations with additional terms $\rho_{\mathrm{grav}}$ and $p_{\mathrm{grav}}$ that obey the equation of state $p_{\mathrm{grav}}=-\frac{1}{3}\rho_{\mathrm{grav}}$, therefore mimicking dark energy.\\

In additionally, with reasonable cosmological assumptions Buchert's formalism can be realized as a consistent limit of the macroscopic gravity formalism with identical corrections to the Friedmann equations \cite{Paranjape07}.

\section{Conclusions}

We have presented in this chapter an overview of standard cosmology, some of the problems it faces and discussed in some detail one particular approach at solving these, concerning the averaging problem. The different approaches to explain the observed acceleration of the universe as a consequence of deviations from exact FLRW symmetry without invoking the presence of a cosmological constant or dark energy have attracted much attention in recent years. However, there is also much criticism and controversy about this ansatz. As pointed out by Ishibashi and Wald in \cite{IshibashiWald06}, an alternative model must be compatible with observations. They showed on a specific choice of time-slicing that a quantity representing the scale factor may accelerate even in Minkowski spacetime without any physical cause. Furthermore, they calculated Buchert's kinematical backreaction for two disconnected dust filled FLRW models, obtaining the requirement for an average acceleration although all observers see only deceleration. They concluded that acceleration, as defined by the Buchert formalism, can easily arise as a gauge artifact produced by the choice of slicing or the choice of averaging domain. This shows quite clearly the need for a generally covariant approach to the averaging problem which is not subjected to any gauge ambiguities. Macroscopic gravity is a systematic attempt to set up such an approach, but it has not been widely accepted.

%% file: paper.tex
\chapter{Cosmological Backreaction from Perturbations}
\label{paper}

\section{Introduction}

There are two main ways in which averaging can be seen to be a vital, if implicit, component in cosmology. On the one hand, cosmology is founded on the FLRW metric which is assumed to be the large-scale average of the true, inhomogeneous metric. This statement cannot be proven without a proper generally covariant averaging scheme capable of smoothing inhomogeneous metrics. While it is certainly logical that the metric of the universe on the largest scales should resemble an FLRW metric (and recent studies \cite{Lu07,McClure07} have considered this problem from an observational point of view) this is not a solid proof.\\

The other way in which averaging in cosmology has a vital impact concerns the non-linearity of the Einstein tensor. Even should we possess a covariantly averaged metric on large scales we cannot directly employ this in the Einstein equation, since the Einstein tensor of the average metric is not equivalent to the average of the inhomogeneous Einstein tensor. The corrections to the dynamics are known as the ``cosmological backreaction''. In the remainder of the thesis we will develop a generally covariant averaging procedure for metrics in general relativity, but in this chapter we consider a simple test-case to evaluate quantitatively the deviations from a standard FLRW evolution arising from linear and mildly nonlinear perturbations. Such a calculation complements the recent studies by \cite{VanderveldEtAl07}, in which the authors reconstruct the impact of backreaction effects from the observational data, \cite{KhosraviEtAl07} where the authors evaluate the size of the effective density of the backreaction as a function of redshift in a structured Robertson-Walker model and \cite{Rasanen08} which considers a statistical ``peak model'' for large scale structure in an Einstein-de Sitter universe. While the current literature generally concerns exact inhomogeneous rather than perturbative models these models are not constructed to accurately model the universe. In contrast, the deviations introduced by linear perturbations are expected to be small when averaged over a domain approaching the Hubble volume in size but the model is viable phenomenologically and the perturbations can be calculated numerically employing popular Boltzmann codes such as cmbeasy or cmbfast. Employing the Buchert formalism we will calculate the deviations in both the concordance, $\Lambda$CDM model and in a toy EdS universe with a low Hubble rate. We model mildly nonlinear (``quasilinear'') scales using a modified Halofit \cite{Smith02} code.\\

This chapter is based on work performed in collaboration with Iain Brown and Georg Robbers \cite{Behrend07} and was recently published in JCAP. In this chapter an overdot represents a derivative with respect to coordinate time $t$ rather than proper time $\tau$, and in line with much of the literature we again employ a signature $(-,+,+,+)$.

\section{The Buchert Formalism in Newtonian Gauge}

While much of the previous study into backreaction in perturbed FLRW spaces has been undertaken in synchronous gauge (see for example \cite{LiSchwarz07, LiSchwarz07-2}), this gauge contains unphysical modes and the metric perturbation can grow to be relatively large. While this is not necessarily an issue, Newtonian gauge is unambiguous and the variable $\phi$ remains small and well-defined across almost all scales. Moreover, it is easily incorporated into the cmbeasy Boltzmann code and should serve as a complementary probe to the previous studies, although it should be remembered that we cannot directly compare calculations performed in different gauges. It has been appreciated for some time \cite{Wetterich01,Rasanen03} that the impact on large-scale evolution from perturbations is not expected to be large, something clearly seen in Newtonian gauge; since the perturbations themselves are consistently small across almost all scales the impact from backreaction would na\"ively be expected to be at most of the order of $10^{-5}$.\\

As we intend to evaluate the perturbations with a multi-fluid Boltzmann code in Newtonian gauge some modifications to the formalism presented in sections \ref{ADM} and \ref{sectionbuchert} is necessary. In particular, we do not identify the normal vector $n^\mu$ to a hypersurface $\Sigma_t$ with the 4-velocity of any one fluid and will instead connect the lapse function to the Newtonian potential. As $u^\mu$ and $n^\mu$ are not equal, we cannot employ the acceleration as in equation (\ref{acceleration}) and our Raychaudhuri equation will look slightly different. Moreover, the split of the stress-energy tensor with respect to the foliation does not coincide with the definitions used in the Boltzmann codes and this will introduce a further modification in both the Friedmann and Raychaudhuri equations.\\

The Einstein equations separated with respect to this foliation can be written
\beas
  \mathcal{R}+K^2-K^i_jK^j_i=16\pi G\rho+2\Lambda, \quad D_j\left(K^{ij}-h^{ij}K\right)=8\pi Gj^i, \\
  \frac{1}{\alpha}\dot{K}_{ij}=\mathcal{R}_{ij}-2K^a_iK_{aj}+KK_{ij}-8\pi GS_{ij}+4\pi Gh_{ij}\left(S-\rho\right)-\Lambda h_{ij}
   -\frac{1}{\alpha}\mathcal{D}_i\mathcal{D}_j\alpha
\eeas
where $S_{ij}=T_{ij}$ and $\rho=T_{\mu\nu}n^\mu n^\nu$. We also have the normal vector and extrinsic curvature, which we write as
\be
n^\mu=\left(\frac{1}{\alpha},\mathbf{0}\right), \quad n_\mu=(-\alpha,\mathbf{0}), \quad K_{ij}=-\frac{1}{2\alpha}\dot{h}_{ij}, \\
\ee
recalling that $\dot{h}_{ij}=\del_th_{ij}$. We again define an average Hubble rate by
\be
  3\frac{\dom{\dot{a}}}{\dom{a}}=\frac{\dot{V}}{V}=-\av{\alpha K}.
\ee
The average of the Hamiltonian constraint can then be written as the average Friedmann equation,
\be
\label{AvFriedmann}
  \left(\frac{\dom{\dot{a}}}{\dom{a}}\right)^2=\frac{8\pi G}{3}\av{\alpha^2\rho}+\frac{\Lambda}{3}\av{\alpha^2}
   -\frac{1}{6}\left(\Qd+\Rd\right)
\ee
with the kinematical backreaction
\be
  \Qd=\av{\alpha^2\left(K^2-K^i_jK^j_i\right)}-\frac{2}{3}\av{\alpha K}^2
\ee
and
\be
  \Rd=\av{\alpha^2\mathcal{R}}
\ee
the correction arising from the spatial curvature. The trace of the evolution equation for the extrinsic curvature is
\be
  \alpha\dot{K}-\alpha K^i_j\alpha K^j_i+\alpha\mathcal{D}^i\mathcal{D}_i\alpha=4\pi G\alpha^2(\rho+S)-\alpha^2\Lambda .
\ee
Averaging this, employing the commutation relation and using the averaged Friedmann equation (\ref{AvFriedmann}) gives us the average Raychaudhuri equation,
\be
\label{AvRay}
  \frac{\dom{\ddot{a}}}{\dom{a}}=-\frac{4\pi G}{3}\av{\alpha^2(\rho+S)}+\frac{\Lambda}{3}\av{\alpha^2}+\frac{1}{3}\left(\Qd+\Pd\right)
\ee
with
\be
  \Pd=\av{\alpha D^iD_i\alpha}-\av{\dot{\alpha}K},
\ee
the dynamical backreaction.\\

Finally we have the integrability condition. Writing $\dom{H}=\av{\mathcal{H}}$ where $\mathcal{H}=-\alpha K$ is a local Hubble rate, and using local matter continuity we find that
\bea
\lefteqn{
  \frac{\partial_t\left(\dom{a}^6\dom{\mathcal{Q}}\right)}{\dom{a}^6}+\frac{\partial_t\left(\dom{a}^2\av{\alpha^2\mathcal{R}}\right)}{\dom{a}^2} 
   =12\av{\alpha^2\left(\frac{8\pi G}{3}\rho+\frac{\Lambda}{3}\right) \frac{\dot{\alpha}}{\alpha}} } \nonumber \\ &&+16\pi G\left(\av{\alpha^2\mathcal{S}}\av{\mathcal{H}}-\av{\alpha^2\mathcal{S}\mathcal{H}}\right)+6\Lambda\left(\av{\alpha^2\mathcal{H}}-\av{\alpha^2}\av{H}\right)-4\dom{H}\dom{\mathcal{P}}.
\eea
It is easy to see that in dust-filled synchronous gauge models, with $\alpha=1$ and $\mathcal{S}=0$, the source on the right-hand side vanishes.  For non-dust models, there is still an extra source term dependant on the local pressure and the local and averaged Hubble rates. In \cite{LiSchwarz07,LiSchwarz07-2} the authors make much use of the integrability condition, employing it iteratively to recover the backreaction at higher-orders in perturbation theory. In Newtonian gauge the source term is somewhat complicated. Since we will be evaluating the backreactions from perturbations calculated in a Boltzmann code, we will not make use of the integrability condition.

\section{The Connection with Linear Perturbation Theory}
\label{NewtGauge}

The Buchert equations are exact for any inhomogeneous model; for further progress we must specify this model. An increasingly common approach is to employ a model that averages out to be Robertson-Walker on large scales but is a relatively realistic approximation to the local universe on smaller scales. We choose instead to use a linearly-perturbed FLRW model, employing Newtonian gauge. We consider a universe filled with a cosmological constant and pressureless dust. This is a reasonable approximation to the current universe on the largest scales. Should the backreaction turn out to be significant the cosmological constant can be reduced or set to zero, while should the backreaction be insufficient to account for observations it can still remain phenomenologically viable. Working to first-order in the gravitational potential $\phi$, we consider only scalar perturbations and work with the line-element
\be
  ds^2=-(1+2\phi)dt^2+a^2(t)(1-2\phi)\delta_{ij}dx^idx^j
\ee
where the scale factor $a$ is distinct from the averaged scale factor $\dom{a}$. We neglect the tensor perturbations as sub-dominant to the scalars although a full approach should naturally take these into account. When performing our averages across scales approaching the Hubble horizon, we consider the quadratic order in linear perturbations and neglect the averages of pure first- and second-order perturbations. Were we to consider averaging across relatively small scales then these neglected terms should also be taken into account for a full evaluation. Likewise we can neglect the second-order vector and tensor perturbations that scalars inevitably source, but again these should be taken into account for a full treatment on small scales. These issues would require a generally-covariant averaging procedure. Since we are considering purely scalar perturbations, all vector quantities can be written as the gradient of a scalar quantity; in particular, $v^i=-\partial^i\psi$ for some velocity potential $\psi$.\\

By unambiguously identifying the 3+1 and Newtonian gauge coordinates and expanding quantities up to the second-order in perturbations (see also \cite{Mukhanov92}), we can quickly see that
\be
  \alpha^2=1+2\phi, \quad \alpha\approx 1+\phi-\frac{1}{2}\phi^2, \quad
  \alpha^{-1}\approx 1-\phi+\frac{3}{2}\phi^2, \quad \alpha^{-2}\approx 1-2\phi+4\phi^2 .
\ee
We also have
\be
  h_{ij}=a^2(t)(1-2\phi)\delta_{ij}, \quad h^{ij}\approx a^{-2}(t)(1+2\phi+4\phi^2)\delta^{ij} .
\ee
We can now proceed to evaluate the geometric quantities we need in the averaged equations. The extrinsic curvature is
\be
  \alpha K^i_j=-\left(\frac{\dot{a}}{a}-\dot{\phi}(1+2\phi)\right)\delta^i_j .
\ee
Using the Ricci scalar on the spatial hypersurface the curvature correction is
\be
  \Rd=\av{\alpha^2\mathcal{R}}=\frac{2}{a^2}\av{\left(2\nabla^2\phi+3(\nabla\phi)^2+12\phi\nabla^2\phi\right)}
\ee
where $\nabla=(\partial/\partial x,\partial/\partial y,\partial/\partial z)$ as in standard vector calculus.\\

From the extrinsic curvature we can immediately see that
\bdm
  3\frac{\dom{\dot{a}}}{\dom{a}}=-\av{\alpha K}=3\av{\frac{\dot{a}}{a}-\dot{\phi}(1+2\phi)}
\edm
implying
\be
  \frac{\dom{\dot{a}}}{\dom{a}}=\frac{\dot{a}}{a}-\av{\dot{\phi}(1+2\phi)}
\ee
giving the relationship between the physical averaged scale factor and the scale factor employed in the perturbative approximation. If $\dot{\phi}=0$ then these coincide; this occurs in an Einstein-de Sitter universe, or when one considers a domain sufficiently small that its time variation can be neglected but sufficiently large that linear perturbation theory may be employed. We can now rapidly find the kinematical backreaction,
\be
  \Qd=6\left(\av{\left(\frac{\dot{a}}{a}-\dot{\phi}(1+2\phi)\right)^2}-\av{\frac{\dot{a}}{a}-\dot{\phi}(1+2\phi)}^2\right)
    =6\left(\av{\dot{\phi}^2}-\av{\dot{\phi}}^2\right) .
\ee
Using the covariant derivative on the 3-surface and
\be
  \dot{\alpha}K=3\dot{\phi}^2-3\frac{\dot{a}}{a}\dot{\phi}(1-2\phi),
\ee
the dynamical backreaction is
\be
  \Pd=3\frac{\dot{a}}{a}\av{\dot{\phi}(1-2\phi)}-3\av{\dot{\phi}^2}+\frac{1}{a^2}\av{\nabla^2\phi+2\phi\nabla^2\phi-2(\nabla\phi)^2} .
\ee
This gives us the geometric information that we need.\\

Consider now the fluids. We separated the stress-energy tensor with respect to the normal vector
\be
  n^\mu=\left(1-\phi+\frac{3}{2}\phi^2,\mathbf{0}\right) .
\ee
However, the stress-energy tensor employed in the Boltzmann code is defined with respect to the 4-velocity and for the case of pressureless dust is typically taken to be
\be
  T_{\mu\nu}=\varepsilon u_\mu u_\nu
\ee
where $\varepsilon$ is the energy density as defined in this reference frame, which will be linearised to $\varepsilon=\bkr(1+\delta)$ for an average $\bkr$. The density $\rho$ and FLRW density $\varepsilon$ thus do not coincide unless the velocity is orthogonal to the normal, which is not in general the case. The 4-velocity, to second-order in first-order perturbations, can be seen to be
\be
  u^\mu=\left(1-\phi+\frac{1}{2}\left(a^2v^2+3\phi^2\right),v^i\right)
\ee
with $v_i=\delta_{ij}v^j$. Then the density is
\be
  \rho=T_{\mu\nu}n^\mu n^\nu=\varepsilon\left(u_\mu n^\mu\right)^2=\varepsilon\left(1+\frac{1}{2}a^2v^2\right)^2=\varepsilon\left(1+a^2v^2\right) .
\ee
Scaled by the lapse function, the quantity that enters the Buchert equations is then
\be
  \av{\alpha^2\rho}=\bkr\left(1+\delta+2\phi+a^2v^2+2\phi\delta\right)
\ee
and so we define a ``density correction''
\be
  \Td=\frac{8\pi G}{3}\bkr\av{\delta+2\phi+2\phi\delta+a^2v^2} .
\ee

The Friedmann equations can then be written as
\be
  \left(\frac{\dom{\dot{a}}}{\dom{a}}\right)^2=\frac{8\pi G}{3}\bkr_m+\frac{\Lambda}{3}+\Delta F, \quad
  \frac{\dom{\ddot{a}}}{\dom{a}}=-\frac{4\pi G}{3}\bkr_m+\frac{\Lambda}{3}+\Delta R
\ee
with corrections
\be
\label{DeltaFrieds}
  \Delta F=\Td-\frac{1}{6}\left(\Qd+\Rd\right), \quad
  \Delta R=-\frac{1}{2}\Td+\frac{1}{3}\left(\Qd+\Pd\right) .
\ee
The effective pressure and density of the corrections are then
\be
\label{EffectiveFluid}
  \frac{8\pi G}{3}\rho_{\mathrm{eff}}=\Td-\frac{1}{6}\left(\Qd+\Rd\right), \quad
  16\pi Gp_{\mathrm{eff}}=\frac{1}{3}\Rd-\Qd-\frac{4}{3}\Pd
\ee
and so the effective equation of state is
\be
\label{EoS}
  w_\mathcal{D}=-\frac{1}{3}\left(\frac{\Rd-4\Pd-3\Qd}{\Rd-6\Td+\Qd}\right) .
\ee

We take the domain to be large enough to allow us to neglect the averages of first order quantities on the background. Specifically, we take the domain size to be of the order of the comoving Hubble scale, across which first-order averages can consistently be neglected. While this implies that our analysis can only be taken to apply on the very largest scales, taking the domain to be so large also implies that we can invoke the ergodic principle to convert our spatial averages into ensemble averages; we exploit this in the next section. On such large scales, the modifications become
\bea
\label{Rd}
  \Rd&=&\frac{6}{a^2}\av{(\nabla\phi)^2+4\phi\nabla^2\phi}, \\
\label{Pd}
  \Pd&=&6\av{\dot{\phi}^2}, \\
\label{Qd}
  \Qd&=&-6\frac{\dot{a}}{a}\av{\dot{\phi}\phi}-3\av{\dot{\phi}^2}+\frac{2}{a^2}\av{\phi\nabla^2\phi-(\nabla\phi)^2}, \\
\label{Td}
  \Td&=&\frac{8\pi G}{3}\bkr\av{2\phi\delta+a^2v^2} .
\eea
We can see that, depending on the signs of $\av{\phi\nabla^2\phi}$ and $\av{\phi\delta}$, for an Einstein de-Sitter universe with $\dot{\phi}=0$ (as considered in \cite{TanakaEtAl06}), one might get either an enhancement or reduction of both the effective Hubble rate and of the effective acceleration.

\section{Ergodic Averaging}
\label{ErgodicAveraging}

The simplest attack on the problem is to employ a Boltzmann code such as cmbeasy \cite{Doran03} or cmbfast \cite{Seljak96}. The use of linear theory automatically implies that we are working on relatively large scales; we can render the system developed in the last section more tractable by taking our domain to approach the Hubble volume itself -- large enough, at least, that we can employ the ergodic theorem and convert the spatial averages into averages across a statistical ensemble. (Similar approaches were also employed in \cite{Wetterich01,Rasanen03}.)\\

Consider first the general average $\av{A(t,\mathbf{x})B(t,\mathbf{x})}$ where $\{A,B\}\in\{\phi,\delta,v,\dot{\phi}\}$. Then
\be
  \av{A(t,\mathbf{x})B(t,\mathbf{x})}=\int_{\mathbf{k}}\int_{\mathbf{k}'}\av{A(t,\mathbf{k})B^*(t,\mathbf{k}')}e^{-i(\mathbf{k-k}')\cdot\mathbf{x}}\frac{d^3k}{(2\pi)^3}\frac{d^3k'}{(2\pi)^3} .
\ee
At linear order and in the absence of decoherent sources, different wavemodes are decoupled from one another and the evolution equations depend only on the magnitude of $k$ and so we can write
\be
  A(t,\mathbf{k})=\alpha(\mathbf{k})A(t,k)
\ee
for any quantity. The initial correlation between two arbitrary quantities is
\be
  \av{\alpha(\mathbf{k})\beta^*(\mathbf{k}')}=\frac{2\pi^2}{k^3}\mathcal{P}_{\psi}(k)\delta(\mathbf{k-k}').
\ee
Inserting this into the correlation and integrating once yields
\be
  \av{A(t,\mathbf{x})B^*(t,\mathbf{x})}\approx\int_k\mathcal{P}_\psi(k)A(t,k)B^*(t,k)\frac{dk}{k}.
\ee

The averages involving gradients are only a little more complicated; for $\av{(\nabla\phi)^2}$ we have
\bea
  \av{(\nabla\phi)^2}&=&\int_{\mathbf{k}}\int_{\mathbf{k}'}(-i\mathbf{k})(i\mathbf{k}')\av{\phi(t,\mathbf{k})\phi^*(t,\mathbf{k}')}e^{-i(\mathbf{k-k}')\cdot\mathbf{x}}\frac{d^3k}{(2\pi)^3}\frac{d^3k'}{(2\pi)^3} \\
  &=&\int_kk^2\mathcal{P}_\psi(k)\left|\phi(t,k)\right|^2\frac{dk}{k} ,
\eea
which is positive-definite as it should be. The other curvature term is
\bea
  \av{\phi\nabla^2\phi}&=&-\int_{\mathbf{k}}\int_{\mathbf{k}'}\mathbf{k}'^2\av{\phi(t,\mathbf{k})\phi^*(t,\mathbf{k}')}e^{-i(\mathbf{k-k}')\cdot\mathbf{x}}\frac{d^3k}{(2\pi)^3}\frac{d^3k'}{(2\pi)^3} \\
  &=&-\int_kk^2\mathcal{P}_\psi(k)\left|\phi(t,k)\right|^2\frac{dk}{k} .
\eea
This term is then negative definite.\\

The kinematical, dynamical, curvature and density corrections (\ref{Rd}-\ref{Td}), then become
\bea
  \Rd&=&-\frac{18}{a^2}\int k^2\mathcal{P}_\psi(k)\left|\phi(t,k)\right|^2\frac{dk}{k}, \\
  \Qd&=&6\int\mathcal{P}_\psi(k)\left|\dot{\phi}(t,k)\right|^2\frac{dk}{k}, \\
  \Pd&=&-3\int\mathcal{P}_\psi(k)\left(\left|\dot{\phi}(t,k)\right|^2+\frac{\dot{a}}{a}\left(\phi(t,k)\dot{\phi}^{*}(t,k)+\phi^*(t,k)\dot{\phi}(t,k)\right)
\right. \\ && \left. \quad +\frac{4}{3}\frac{k^2}{a^2}\left|\phi(t,k)\right|^2\right)\frac{dk}{k}, \\
  \Td&=&\frac{8\pi G}{3}\bkr\left(\int k^2\mathcal{P}_\psi(k)a^2(t)\left|v(t,k)\right|^2\frac{dk}{k} \right. \\ && \left. \quad -\int\mathcal{P}_\psi(k)\left(\phi(t,k)\delta^*(t,k)+\phi^*(t,k)\delta(t,k)\right)\frac{dk}{k}\right) .
\eea
For a more realistic approach where we separate the baryons from the cold dark matter one can readily see that the contribution from the density can be expanded as
\bea
  \Td&=&\frac{8\pi G}{3}\int k^2\mathcal{P}_\psi(k)\sum_a\bkr_a\left|v_a(t,k)\right|^2\frac{dk}{k}
  \nonumber \\ && \quad
   -\int\mathcal{P}_\psi(k)\left(\phi(t,k)\sum_a\bkr_a\delta_a^*(t,k)+\phi^*(t,k)\sum_a\bkr_a\delta_a(t,k)\right)\frac{dk}{k} .
\eea
Since Boltzmann codes tend to be written in conformal time we must these into conformal time expressions using $d/dt=(1/a)d/d\eta$; denoting conformal time quantities with a tilde we thus have
\be
  v=\frac{\tilde{v}}{a}, \quad
  \dot{\phi}=\frac{\phi'}{a}, \quad
  \frac{\dot{a}}{a}=\frac{1}{a}\frac{a'}{a}
\ee
in the expressions above.

\section{Corrections to the FLRW Picture from Linear Perturbations}
\label{Backfast}

We incorporate the above formalism into the cmbeasy Boltzmann code and run consistency checks with the cmbfast code, modified to output conformal Newtonian quantities. We will evaluate the terms across all post-recombination redshifts and employ an infra-red cut-off at the comoving Hubble scale, $k_{\mathrm{min}}=1/\eta$, avoiding the unphysical gauge-dependent super-horizon contributions. The small-scale limit $k_{\mathrm{max}}$ of our domain, determined from the stability of the integration with respect to changing $k_{\mathrm{max}}$ is about $k_{\mathrm{max}}\approx 30$Mpc$^{-1}$ and we integrate to $k_\mathrm{max}=100$Mpc$^{-1}$. Naturally, we do not claim that our results at such small scales are complete, merely that we are evaluating the contribution of such large-scale modes on these scales.\\

At a linear level, there are scaling relations that hold to a high level of accuracy; it is obvious from Poisson's equation that
\be
  \phi\propto\delta/k^2 ;
\ee
from Euler's equation $\dot{v}+(\dot{a}/a)v\propto\nabla\phi$ we can also predict that
\be
  \left|v\right|\propto\delta/k .
\ee
From here, one may immediately state that on smaller scales where $1/k^4\ll 1/k^2$ and for an approximately scale-invariant primordial power spectrum,
\be
  \Qd(k)\propto\frac{\delta^2}{k^4}, \quad \Td(k)\propto\Pd(k)\propto\Rd(k)\propto\frac{\delta^2}{k^2}
\ee
where $\Qd(k)=6\mathcal{P}_\psi(k)|\dot{\phi}|^2$ is the integrand of $\Qd$, with similar definitions for $\Td(k)$, $\Rd(k)$ and $\Pd(k)$. $\Qd$ is thus generally subdominant to the other corrections except on very large scales. This also implies that each correction is approximately of the form
\be
  \mathcal{A}_D=\alpha\int\left|\delta\right|^2\frac{\mathcal{P}_\psi}{k^3}dk
\ee
for some constant $\alpha$, reminiscent of the approximation in equation (32) of \cite{Wetterich01}.\\

\begin{center}
\begin{figure}
\begin{center}\includegraphics[width=0.45\textwidth]{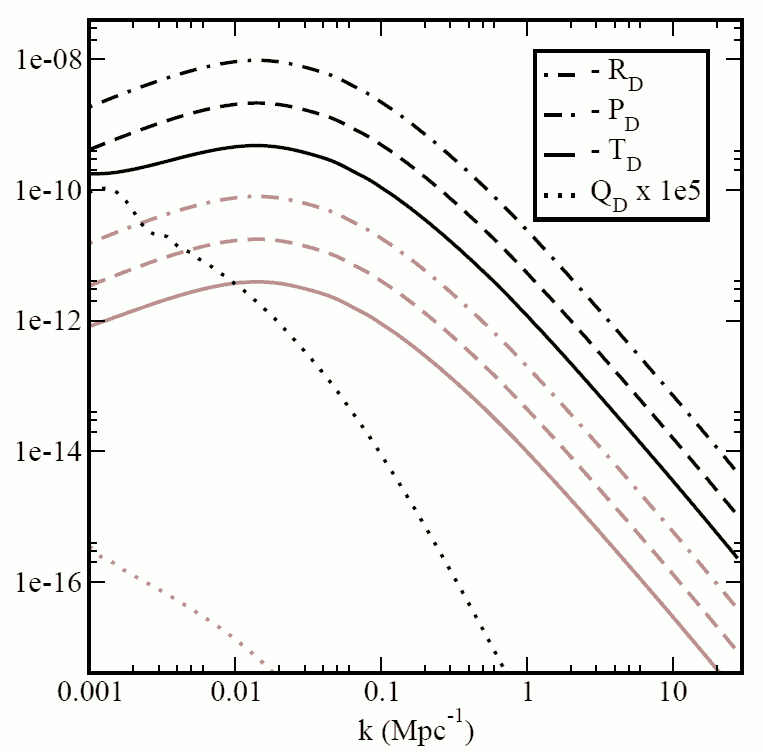}\quad\includegraphics[width=0.45\textwidth]{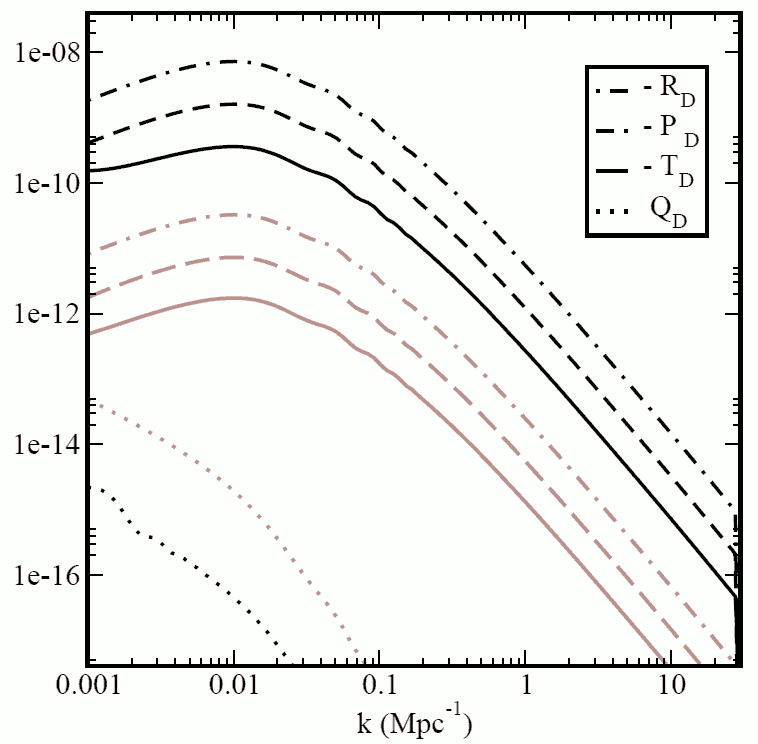}\end{center}
\caption{The correction terms $\Td$, $\Pd$, $\Rd$ and $\Qd$ as a function of $k$ at $z=10$ (black) and $z=0$ (brown) for (left) the sample Einstein de-Sitter model and (right) the WMAPIII concordance model.}
\label{Figure-EdSComparisonz10z0}
\end{figure}
\end{center}

\begin{center}
\begin{figure}
\begin{center}\includegraphics[width=0.45\textwidth]{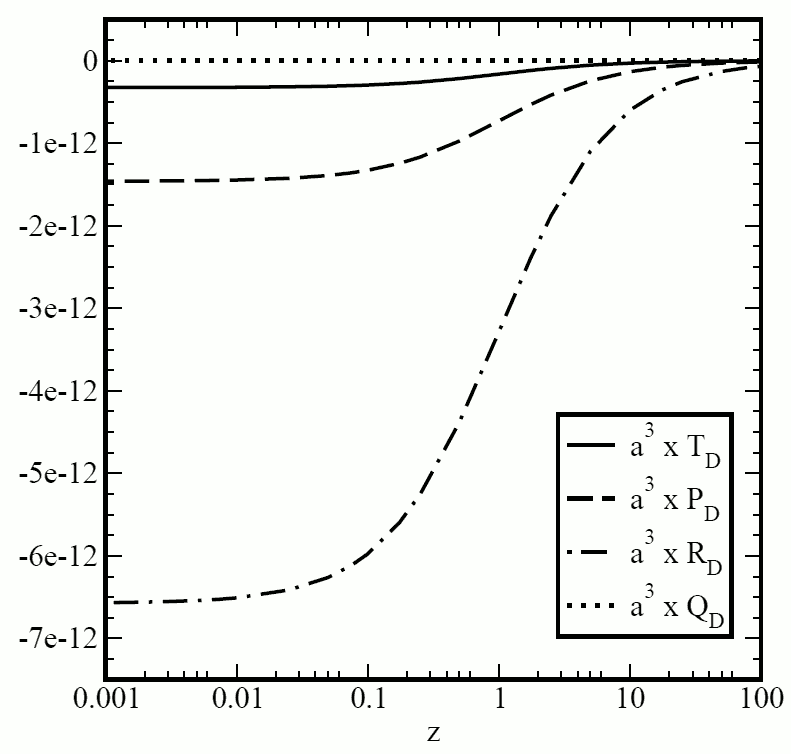}\quad\includegraphics[width=0.45\textwidth]{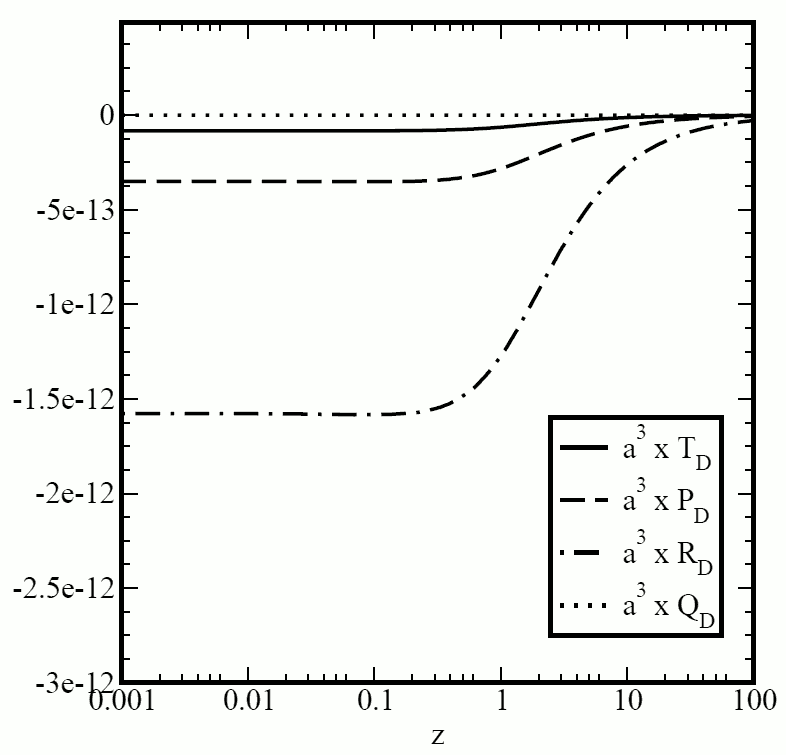}\end{center}
\caption{The evolution of the corrections for (left) the sample EdS model and (right) the WMAPIII concordance model.}
\label{Figure-EdSEvolution}
\label{Figure-LCDMEvolution}
\end{figure}
\end{center}

For our models we take a low-Hubble constant Einstein de-Sitter model with $\Omega_b=0.05$, $\Omega_m=1$ and $h=0.41$ and a WMAPIII $\Lambda$CDM concordance model. In both cases we consider only adiabatic initial conditions. The left panel in Figure \ref{Figure-EdSComparisonz10z0} shows the (integrands of the) four different correction terms at redshifts of $z=10$ and $z=0$ for the EdS case and the right panel the same for the $\Lambda$CDM case, and in Figure \ref{Figure-EdSEvolution} we present the correction terms as a function of $z$ for the EdS and $\Lambda$CDM cases; the premultiplication by $a^3(z)$ acts as a volume normalisation. The subdominance of $\Qd(k)$ and proportionality between $\Rd$, $\Td$ and $\Pd$ is very clear, with $\Rd$ the strongest correction.\\

\begin{center}
\begin{figure}
\begin{center}\includegraphics[width=0.45\textwidth]{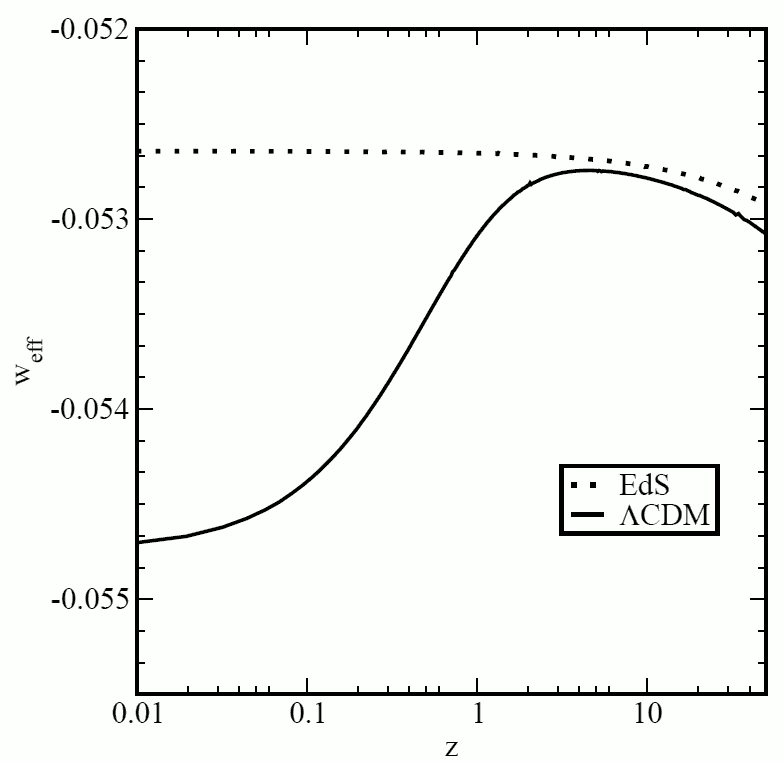}\quad\includegraphics[width=0.45\textwidth]{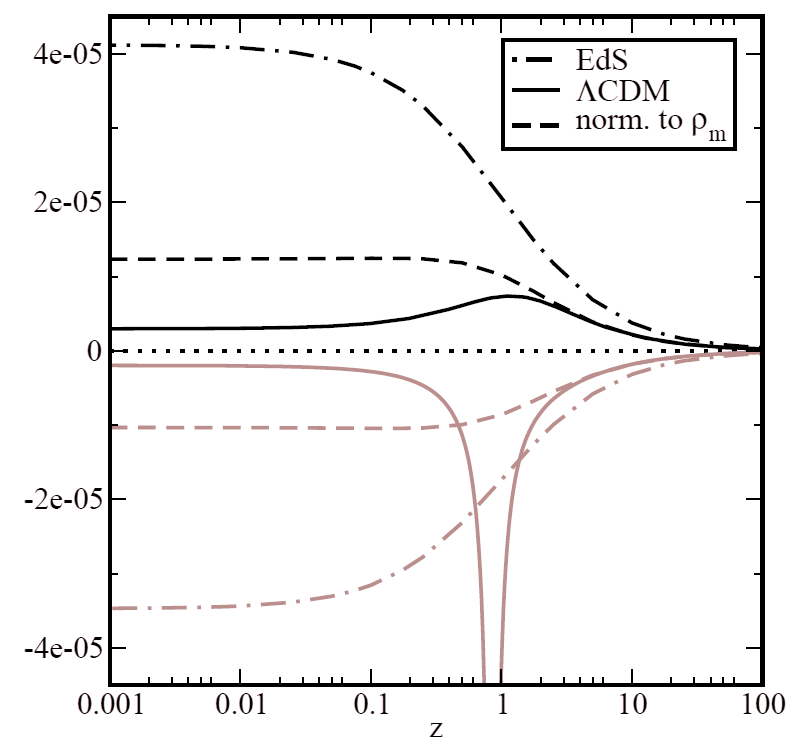}\end{center}
\caption{Left: The effective equation of state $\dom{w}$ as a function of redshift for a sample Einstein de-Sitter and WMAP concordance models. Right: The impact of the corrections onto the Friedmann equation (black) and the Raychaudhuri equation (brown), normalised in the $\Lambda$CDM case to the matter content (dashed) and to the standard equations (solid).}
\label{Figure-weff}
\label{Figure-Impacts}
\end{figure}
\end{center}

From the proportionality of the other corrections (which holds up to relatively large scales) we can write $\Td=t\Rd$ and $\Pd=p\Rd$ and express the equation of state (\ref{EoS}) as
\be
  \dom{w}\approx-\frac{1}{3}\left(\frac{1-4p}{1-6t}\right) .
\ee
which is simple to evaluate numerically by selecting some pivot scale on which to evaluate the ratios $p(z)$ and $t(z)$. Doing so at the current epoch and a pivot scale of $k=0.01$Mpc$^{-1}$, we find $p\approx 2/9$ and $t\approx 1/20$. Both of these are also only slowly evolving. This gives an estimate of the average equation of state as $\dom{w}\approx -1/19$. In the left panel of Figure \ref{Figure-weff} we plot the evolution of $\dom{w}$ for both the model EdS and $\Lambda$CDM cases evaluated directly from equation (\ref{EoS}). In both cases the effective equation of state from linear perturbations remains around the order $\dom{w}\approx -1/19$ and the correction terms to the usual FLRW as a whole thus act as a form of non-standard dark matter and \emph{not} as a dark energy. These results compare reasonably well with the estimate in \cite{Wetterich01} where the author found $\dom{w}\approx-1/27$\footnote{They also found $\dom{w}\approx -1/15$ for a clustering cosmon field but we have not considered such a component here.}.\\

To quantify the impact on the Friedmann and Raychaudhuri equations, define
\be
  F_m=\frac{8\pi G\bkr_{m0}}{3a^3}, \quad R_m=\frac{4\pi G\bkr_{m0}}{3a^3}
\ee
as the standard contribution from baryons and CDM. We can then consider $\Delta F/F_m$ and $\Delta R/R_m$ as a well-defined measure of the impact of the correction terms in both EdS and $\Lambda$CDM models. By equation (\ref{EffectiveFluid}), $\Delta F/F_m$ for an EdS model is just $\rho_\mathrm{eff}/\bkr_m$. For the sake of clarity we choose to focus on deviations from the standard behaviour but, naturally, we can re-express these quantities in terms of the effective energy density and pressure.\\

For the EdS case we can see in Figure \ref{Figure-Impacts} that the total impact from linear perturbations tends to $\sim 4\times 10^{-5}$ on the Friedmann equation and $\sim -3.2\times 10^{-5}$ on the Raychaudhuri equation; the corrections thus act with a positive effective density and with insufficiently negative pressure to accelerate the universe and instead act to decelerate it at a negligible level. The dashed curves in Figure \ref{Figure-Impacts} are for the $\Lambda$CDM case; they tend to impacts of $\sim 1.3\times 10^{-5}$ on the Friedmann equation and $\sim -1\times 10^{-5}$ on the Raychaudhuri equation. The behaviour, as might be expected, is qualitatively similar to that in the EdS case and the impact is significantly less (by a factor of roughly $10/3$ at the current epoch).\\

This compares reasonably well with the approximation in \cite{Wetterich01} which in our notation is
\be
  \frac{\rho_\mathrm{eff}}{\overline{\rho}_m}=\frac{9a^2}{4}(8\pi G\overline{\rho}_m)\int\frac{\mathcal{P}_\psi}{k^3}dk .
\ee
Evaluating this approximation for the EdS universe, we find that this underestimates the correction terms with respect to our more detailed study by a factor of about 30\%. In the same paper, the author found a ``cosmic virial theorem'',
\be
  \overline{p}_m=-p_\mathrm{eff}
\ee
where $\overline{p}_m$ is the correction to the matter pressure arising from gravitational interactions. Crudely modelling the matter pressure as $\overline{p}_m\approx (1/3)\overline{\rho}_m\av{v^2}$ we find that $\overline{p}_m$ underestimates $-p_\mathrm{eff}$, again by $\sim$30\%, but the proportionality between them holds remarkably well for a wide range of redshifts.\\

For the $\Lambda$CDM case we can consider an alternative normalisation that directly quantifies the impact on a universe with a cosmological constant,
\bdm
  \frac{\Delta F}{\dom{\dot{a}}/\dom{a}}=\frac{\Delta F}{F_m+\Lambda/3}, \quad
  \frac{\Delta R}{|\dom{\ddot{a}}/\dom{a}|}=\frac{\Delta R}{|R_m-\Lambda/3|} .
\edm
The benefit of doing so is that it demonstrates the declining contribution of the corrections with respect to the cosmological constant; however, it also introduces a singularity at a redshift of $z\approx 0.8$ when the reference FLRW universe undergoes a transition from deceleration to acceleration. The impact on both the Friedmann and Raychaudhuri equations tends to $\sim 10^{-6}$ and the maximum contribution is at a redshift of $z\approx 1.3$.

\section{Corrections to the FLRW Picture from Quasilinear Perturbations}
\label{Halofit}

In the previous section we evaluated the impact on the large-scale evolution of the universe from linear perturbations, demonstrating that it is small, as should be expected, but also that it acts as a dark matter and not a dark energy. In this section we employ the halo model to estimate the impact of perturbations on smaller, quasilinear scales which are much less understood.\\

The publicly-available Halofit code \cite{Smith02} converts a linear CDM power spectrum into the CDM power spectrum from the halo model. We employ a modified Halofit code that instead takes the linear matter power spectrum and estimates the nonlinear matter power spectrum. The square-root of this power spectrum can then be used to estimate the non-linear density contrast $\delta_{\mathrm{NL}}(k)$ and we can then recover the other relevant quantities from the scaling relations $v\propto\delta/k$, $\phi\propto\delta/k^2$. If we define
\be
  f(k)=\frac{\delta_{\mathrm{L}}(k)}{\delta_{\mathrm{NL}}(k)}
\ee
we can employ this as a scaling factor to recover estimates for the quasilinear behaviour of $\delta(k)$, $v(k)$, $\phi(k)$ and $\dot{\phi}(k)$. We can then employ the formalism we developed for the linear case to estimate the impact in the quasilinear case, retaining the same domain size. We should immediately note two problems with this technique. Firstly, while the scaling relation for the velocity is extremely precise for linear perturbations it is not for non-linear perturbations and so we have automatically introduced a source of error. Perhaps more importantly, we are employing a formalism developed for linear perturbations in which different wavemodes decouple from one another, allowing us to separate the statistical problem into transfer functions and a primordial power spectrum. This does not hold for a nonlinear problem. For both of these reasons, our results are only intended to be taken as good approximations on quasilinear scales until the velocity virialises, at which point a more detailed study is necessary.\\

\begin{center}
\begin{figure}
\begin{center}\includegraphics[width=0.45\textwidth]{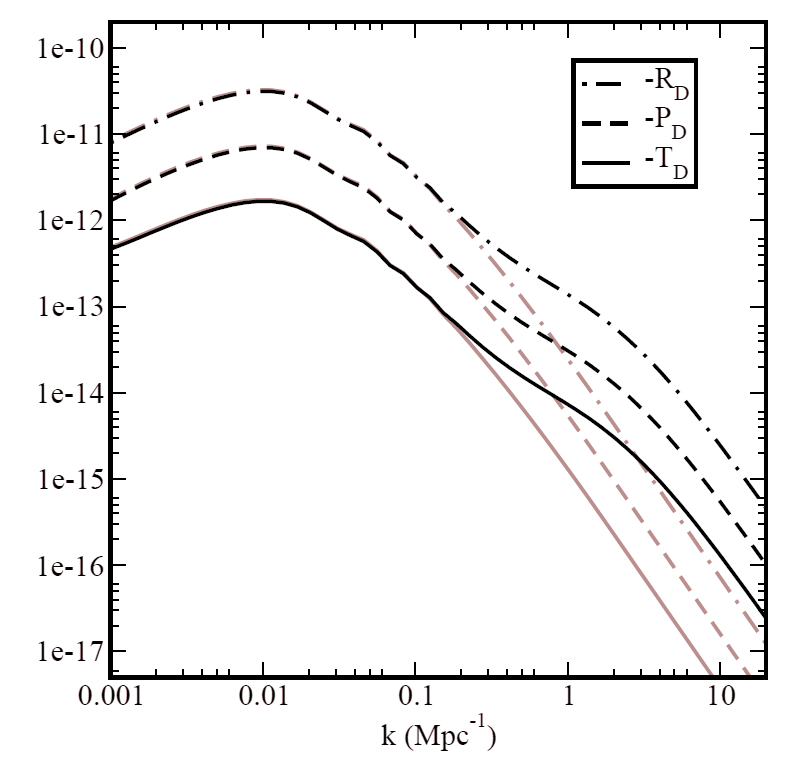}\quad\includegraphics[width=0.45\textwidth]{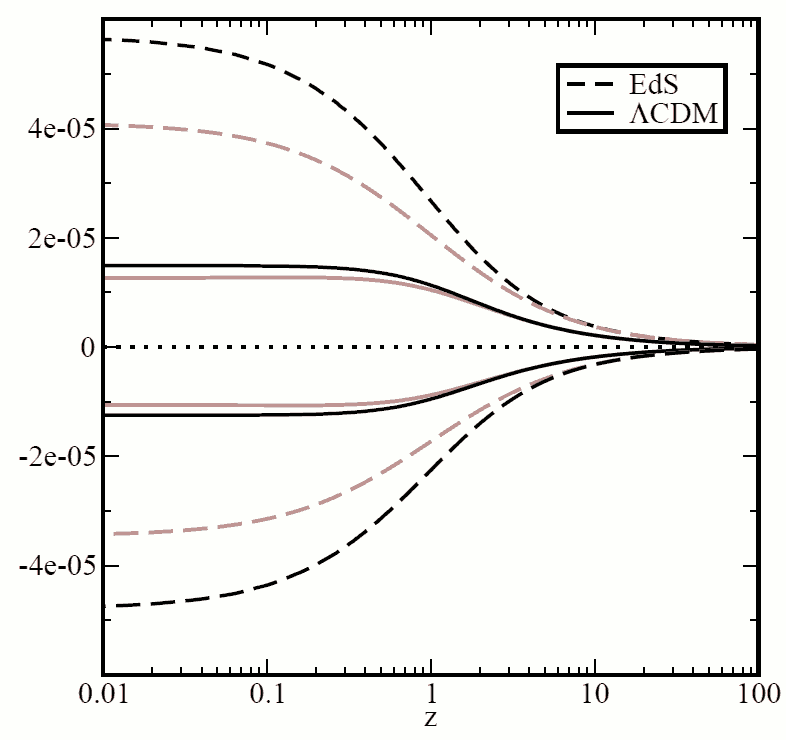}\end{center}
\caption{Left: The halo model approximations to the quasilinear corrections for the WMAPIII concordance model at $z=0$. Right: The halo model approximations to the quasilinear corrections to the Friedmann (positive) and Raychaudhuri (negative) equations; the linear prediction is in brown.}
\label{Figure-NLLCDM}
\label{Figure-NLImpact}
\end{figure}
\end{center}

In Figure \ref{Figure-NLLCDM} we present approximations for the corrections to the $\Lambda$CDM model at redshift $z=0$. As before, the kinematical backreaction is strictly negligible and as we have merely scaled the previous corrections by the same quantity, the effective equation of state remains unchanged. The total impacts on the Friedmann and Raychaudhuri equations for both are shown in Figure \ref{Figure-NLImpact}. We see that  although the halo model provides a boost in power on smaller scales, and even though $\Rd$ and $\Td$ in particular include factors of $k^2$ in the integral that increase the contribution from smaller scales, the impact is not significantly greater than from linear scales. More quantitively, for an EdS-universe the impact at the current epoch on the Friedmann equation is of order $5.6\times10^{-5}$ and on the Raychaudhuri equation is of order $-4.7\times10^{-5}$. Normalised to the matter content, the impact at the current epoch from $\Lambda$CDM perturbations is of order $1.6\times10^{-5}$  on the Friedmann equation and of order $-1.4\times10^{-5}$ on the Raychaudhuri equation. Normalised to the full evolution these become $4.4\times10^{-6}$ and $-2.7\times10^{-6}$, respectively.

Our results compare well with the observational estimate from Vanderveld \emph{et. al.} \cite{VanderveldEtAl07} and the structured FLRW estimate of Khosravi \emph{et. al.} \cite{KhosraviEtAl07}. It is also in broad agreement with the larger-scale estimates of Li and Schwarz \cite{LiSchwarz07,LiSchwarz07-2} although their method allows them to evaluate the backreaction on much smaller scales at which they recover significantly greater impacts, up to about $10^{-1}$ at scales of $50$Mpc. This should perhaps be treated with some wariness, however, since they work to fourth-order perturbation theory, at which level vector and tensor perturbations should be considered in a consistent approach. Recently, R\"as\"anen employed a statistical peak structure model at relatively small scales in an EdS universe \cite{Rasanen08} and recovered significantly larger modifications than our large-scale perturbative approach. Although the backreaction remained at the $10^{-2}$ level he found the curvature contribution to be significant. As with Li and Schwarz, R\"as\"anen's results are not contradictory to ours given the different scales considered in the two cases. Moreover he considered only EdS universes and approximate transfer functions.

\section{Discussion and Conclusions}

We have presented quantitive estimates of the corrections in Newtonian gauge to a standard perturbative FLRW model from an explicit averaging procedure, which can be separated into four distinct quantities: the kinematic backreaction, which remains strictly negligible across the scales of interest, the dynamic backreaction, the impact of spatial curvature and a correction to the FLRW energy density. In line with expectation, the impact from linear perturbations is insignificant and of the order of $10^{-5}$ for both $\Lambda$CDM and an Einstein de-Sitter model. Specifically, the effective energy density arising at the background level from inhomogeneities is $\rho_\mathrm{eff}\approx (4\times 10^{-5})\bkr_m$ and $\rho_\mathrm{eff}\approx(1.3\times 10^{-5})\bkr_m$ for the EdS and $\Lambda$CDM models respectively. Moreover, we have presented estimates arising from halo model corrections on quasilinear scales which are not much larger than those from the linear perturbations, the largest contribution arising from the quasilinear modes in an EdS universe remaining below $10^{-4}$; we find $\rho_\mathrm{eff}\approx(5.6\times 10^{-5})\bkr_m$ for the EdS model and $\rho_\mathrm{eff}\approx(1.6\times 10^{-5})\bkr_m$ for $\Lambda$CDM. This is in broad agreement with recent calculations \cite{VanderveldEtAl07,KhosraviEtAl07,LiSchwarz07-2,Rasanen08}. Additionally, the total effective equation of state arising from the different modifications does not act as a dark energy and instead as a dark matter with a slowly-varying equation of state $\dom{w}\approx -1/19$ for $z\in(0,10)$. The impact is to decelerate the universe and, contrary to other studies, the corrections impact positively on the Friedmann equation implying a positive effective pressure.\\

This does not, however, necessarily imply that there are no observational consequences arising from inhomogeneities, particularly on small scales, as our analysis is limited to the impacts in very large volumes and breaks down when mode-coupling and virialisation become significant. In particular, one can readily imagine situations in which the local Hubble rate is significantly larger than the global average (the so-called ``Hubble bubble''). The impact of local inhomogeneities directly on luminosity distances has also long been well studied and remains an active area of interest. One avenue for further research would thus be to employ the same formalism and consider more accurate models on smaller scales. This remains very much an active field and one possibility we wish to consider is to employ the fully non-linear approaches to cosmology of Vernizzi and Langlois and Enqvist \emph{et. al.} \cite{LangloisVernizzi05_1,LangloisVernizzi05_2,LangloisVernizzi06_1,EnqvistEtAl06}.\\

However, useful as this work will be, it will be necessarily limited until we have a usable, generally-covariant averaging procedure. On nonlinear scales one cannot neglect vector and tensor perturbations, which will be of a similar amplitude to the scalar perturbations. Without a generally-covariant averaging procedure, then, we cannot perform an average on relatively small scales without missing potentially significant contributions from the vector and tensor modes. Moreover, the Buchert averaging procedure relies on a vanishing vorticity, which holds true down to approximately galactic scales but breaks down if one wishes to consider, for example, the contribution from neutron stars. Perhaps more importantly, it relies on taking scalar projections of the Einstein equations and averaging these. This naturally loses much information about the system one is considering. It would be preferable to be able to directly average the Einstein and stress-energy tensors and so recover the large-scale behaviour directly. We will not address this point further in this thesis and instead refer to Zalaletdinov \cite{Zalaletdinov08} for encouraging recent work in this area. \\


To summarise, in this chapter we have demonstrated the existence of the cosmological backreaction arising from Newtonian gauge perturbations to a FLRW universe on the largest scales. In the domains we have chosen the averages of the perturbations themselves can be taken to vanish and leading contribution comes from averages of the squares of the linear perturbations. We found that the total modifications to the FLRW picture from such modes remains of the order of $10^{-5}\overline{\rho}_m$, both for the concordance model and for a low Hubble rate Einstein-de Sitter model, and that their equation of state is only marginally negative and certainly not sufficient to drive an accelerated expansion. We have also extended the analysis to smaller scales using the Halofit code and determined that the modifications to the FLRW behaviour remain negligible on smaller scales. In doing so, we have demonstrated that one can quantitatively study the backreaction in realistic cosmological models.

%% file: process.tex
\chapter{A Generally Covariant Averaging Process}
\label{process}

In this chapter we present a generally covariant averaging process in the framework of general relativity. It involves the decomposition of the metric into tetrads via a Lagrangian formulation and then parallel transport of the tetrads with a reformulated, general relativistic Wegner-Wilson line operator. In the first section we give an overview over the various constituents that we need to formulate the process in the subsequent section. Sections of this chapter are based on work undertaken for my diploma thesis \cite{Behrend03}.

\section{Constituents of the Averaging Process}

\subsection{Wegner-Wilson Lines in General Relativity}

In QCD, the  Wegner-Wilson line operator is used to transport quantities from one point to another at finite separation in a gauge-invariant manner. To build a covariant averaging process, we need the general relativistic analogue of this operator, one which parallel transports tensor quantities from one spacetime point to another in a generally covariant way. To define this operator, which we will call the ``connector'', we follow the calculations in \cite{Nachtmann96} and reformulate them in the framework of general relativity. \\

\begin{figure}[t]
\begin{center}\includegraphics[width=0.8\textwidth]{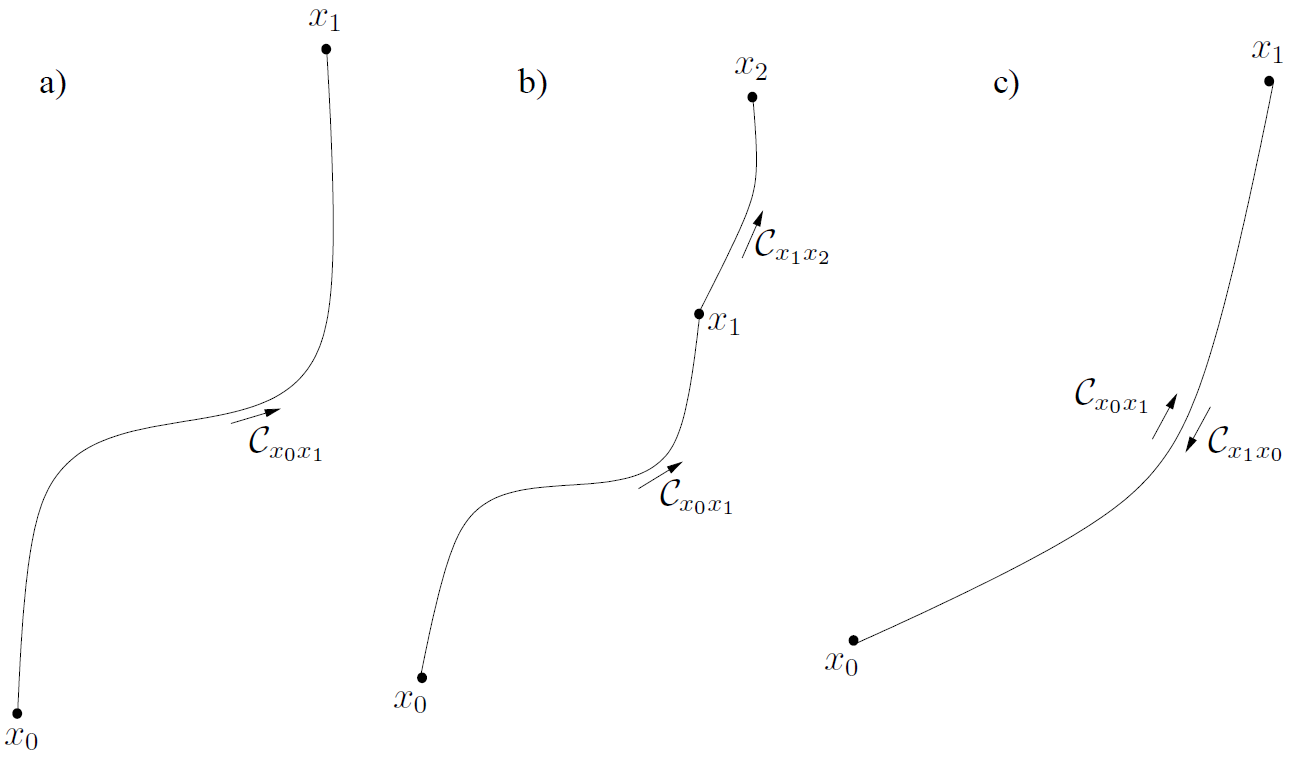}\end{center}
\caption{Different curves in spacetime: a) connecting two points, b) connecting three points and c) oriented in opposite directions.}
\label{wilsonlines}
\end{figure}

Let ${\bf\Gamma}_{\mu}(x)$ denote four $4\times4$ matrices with the components
\be
\left({\bf\Gamma}_{\mu}(x)\right)_{\nu}^{\lambda}=\Gamma_{\mu\nu}^{\lambda}(x),
\ee 
where $\Gamma_{\mu\nu}^{\lambda}(x)$ are the Christoffel symbols (\ref{christoffeldef}). We define the connector along a spacetime curve ${\cal C}_{x_0x_1}$, which connects the points $x_0$ and $x_1$ (Figure \ref{wilsonlines} a)), in the following way:
\be
V(x_1,x_0;{\cal C}_{x_0x_1})={\cal P}\exp\left(-\int_{{\cal C}_{x_0x_1}}dz^{\mu}~{\bf\Gamma}_{\mu}(z)\right).
\label{connector1}
\ee
${\cal P}$ denotes the path-ordering of the integral. To show what this means, we choose an explicit parametrisation for the curve, 
\be
{\cal C}_{x_0x_1}=\{(z^{\mu})|z^{\mu}=z^{\mu}(\tau),~\tau_0\leq\tau\leq\tau_1,~z^{\mu}(\tau_0)=x_0^{\mu},~z^{\mu}(\tau_1)=x_1^{\mu}\},
\label{parametrisation}
\ee
such that we can express the connector as
\be
V(\tau_1)={\cal P}\exp\left(-\int_{\tau_0}^{\tau_1}d\tau\frac{dz^{\mu}(\tau)}{d\tau}~{\bf\Gamma}_{\mu}(z(\tau))\right).
\label{connector2}
\ee
For $\tau_1=\tau_0$ the connector reduces to
\be
V(\tau_1=\tau_0)=\mathbbm{1}.
\label{eins}
\ee

Now we divide the interval $(\tau_1-\tau_0)$ into $N$ small pieces, denoted by $\Delta\rho_i=\rho_i-\rho_{i-1}$ for $i=1,\ldots,N$. For very fine granulation the exponential in the connector can be expanded in a power series, 
\be
V(\tau_1)=\lim_{N\to\infty}\left(\left(\mathbbm{1}-\Delta\rho_N\frac{dz^{\mu}(\rho_N)}{d\tau}{\bf\Gamma}_{\mu}(z(\rho_N))\right)\ldots\left(\mathbbm{1}-\Delta\rho_1\frac{dz^{\mu}(\rho_1)}{d\tau}{\bf\Gamma}_{\mu}(z(\rho_1))\right)\right).
\ee
Path-ordering means that matrices with higher values of $\rho$ are placed on the left, such that $\tau_0=\rho_0<\rho_1<\ldots<\rho_N=\tau_1$.\\

Due to the path-ordering, the connector has the property that for two adjoining curves ${\cal C}_{x_0x_1},{\cal C}_{x_1x_2}$ (Figure \ref{wilsonlines} b)) the connectors are multiplied
\be
V(x_2,x_0;{\cal C}_{x_0x_1}+{\cal C}_{x_1x_2})=V(x_2,x_1;{\cal C}_{x_1x_2})V(x_1,x_0;{\cal C}_{x_0x_1}).
\label{prop1}
\ee

Furthermore, let ${\cal C}_{x_1x_0}$ be the same curve as ${\cal C}_{x_0x_1}$, oriented in the opposite direction (Figure \ref{wilsonlines} c)). Then the connector obeys
\be
V(x_1,x_0;{\cal C}_{x_0x_1})V(x_0,x_1;{\cal C}_{x_1x_0})=\mathbbm{1}.
\label{prop2}
\ee
This can be seen by splitting and expanding the curve ${\cal C}_{x_1x_0}$ in the same way as ${\cal C}_{x_0x_1}$. Since $z^{\mu}(\tau)$ has then the reverse sign, we find
\be
V(\tau_0)=\lim_{N\to\infty}\left(\left(\mathbbm{1}+\Delta\rho_1\frac{dz^{\mu}(\rho_1)}{d\tau}{\bf\Gamma}_{\mu}(z(\rho_1))\right)\ldots\left(\mathbbm{1}+\Delta\rho_N\frac{dz^{\mu}(\rho_N)}{d\tau}{\bf\Gamma}_{\mu}(z(\rho_N))\right)\right) ,
\ee
such that in the product (\ref{prop2}) all $\Delta\rho_i$ terms cancel.

\subsection{Transformation Law for the Connector}

The connector (\ref{connector2}) is the solution to the differential equation
\be
\frac{dV^{\beta}{}_{\alpha}(\tau)}{d\tau}=-\frac{dz^{\mu}(\tau)}{d\tau}~\Gamma_{\mu\nu}^{\beta}(z(\tau))V^{\nu}{}_{\alpha}(\tau).
\label{diffeq}
\ee
with $\tau=\tau_1$ and the initial condition, from (\ref{eins}),
\be
V^{\beta}{}_{\alpha}(\tau_0)=\delta^{\beta}{}_{\alpha}.
\label{bound}
\ee

Under a coordinate transformation $z'(\tau)=f(z(\tau))$, contravariant and covariant vectors transform as
\be
A'^{\mu}=\frac{\del z'^{\mu}}{\del z^{\nu}}A^{\nu}\equiv U^{\mu}{}_{\nu}A^{\nu}\qquad\mathrm{and}\qquad A'_{\mu}=\frac{\del z^{\nu}}{\del z'^{\mu}}A_{\nu}\equiv (U^{-1})^{\nu}{}_{\mu}A_{\nu}.
\label{translaw}
\ee

Therefore, the transformation laws for $dz^{\rho}/d\tau$ and $\del/\del z^{\rho}$ are
\be
\frac{dz^{\rho}}{d\tau}=\frac{dz'^{\gamma}}{d\tau}(U^{-1})^{\rho}{}_{\gamma}\qquad\mathrm{and}\qquad\frac{\del}{\del z^{\rho}}=U^{\varepsilon}{}_{\rho}\frac{\del}{\del z'^{\varepsilon}}
\ee
and the Christoffel symbols transform as
\begin{align}
{\Gamma'}_{\gamma\delta}^{\beta}(z'(\tau))
&=U^{\beta}{}_{\mu}(U^{-1})^{\rho}{}_{\gamma}(U^{-1})^{\lambda}{}_{\delta}\Gamma_{\rho\lambda}^{\mu}(z(\tau))+U^{\beta}{}_{\mu}\frac{\del (U^{-1})^{\mu}{}_{\delta}}{\del z'^{\gamma}}
\end{align}

Furthermore, we find from $d(U^{\beta}{}_{\mu}(U^{-1})^{\mu}{}_{\delta})/d\tau=0$ that
\be
\frac{dz'^{\gamma}}{d\tau}(U^{-1})^{\mu}{}_{\delta}\frac{\del U^{\beta}{}_{\mu}}{\del z'^{\gamma}}=-\frac{dz'^{\gamma}}{d\tau}\frac{\del(U^{-1})^{\mu}{}_{\delta}}{\del z'^{\gamma}}U^{\beta}{}_{\mu}.
\ee

Consider now the following quantity:
\be
\widetilde{V}^{\beta}{}_{\alpha}(\tau)=U^{\beta}{}_{\mu}(\tau)V^{\mu}{}_{\nu}(\tau)(U^{-1}(\tau_0))^{\nu}{}_{\alpha}.
\label{connectortrafo}
\ee

Differentiation of this with respect to $\tau$ gives 
\be
\frac{d}{d\tau}\widetilde{V}^{\beta}{}_{\alpha}&(\tau)
=\frac{dU^{\beta}{}_{\mu}(\tau)}{d\tau}V^{\mu}{}_{\nu}(\tau)(U^{-1}(\tau_0))^{\nu}{}_{\alpha}+U^{\beta}{}_{\mu}(\tau)\frac{dV^{\mu}{}_{\nu}(\tau)}{d\tau}(U^{-1}(\tau_0))^{\nu}{}_{\alpha}\\
&=\frac{dU^{\beta}{}_{\mu}}{d\tau}\left((U^{-1})^{\mu}{}_{\delta}U^{\delta}{}_{\sigma}\right)V^{\sigma}{}_{\nu}(U^{-1}(\tau_0))^{\nu}{}_{\alpha}-U^{\beta}{}_{\mu}\left(\frac{dz^{\rho}}{d\tau}\Gamma_{\rho\lambda}^{\mu}V^{\lambda}{}_{\nu}\right)(U^{-1}(\tau_0))^{\nu}{}_{\alpha}\\
&=\frac{dz^{\rho}}{d\tau}\frac{\del U^{\beta}{}_{\mu}}{\del z^{\rho}}(U^{-1})^{\mu}{}_{\delta}\widetilde{V}^{\delta}{}_{\alpha}-U^{\beta}{}_{\mu}\frac{dz^{\rho}}{d\tau}\Gamma_{\rho\lambda}^{\mu}(U^{-1})^{\lambda}{}_{\delta}U^{\delta}{}_{\sigma}V^{\sigma}{}_{\nu}(U^{-1}(\tau_0))^{\nu}{}_{\alpha}\\
&=-\frac{dz'^{\gamma}}{d\tau}~\left(U^{\beta}{}_{\mu}(U^{-1})^{\rho}{}_{\gamma}(U^{-1})^{\lambda}{}_{\delta}\Gamma_{\rho\lambda}^{\mu}-(U^{-1})^{\mu}{}_{\delta}\frac{\del U^{\beta}{}_{\mu}}{\del z'^{\gamma}}\right)\widetilde{V}^{\delta}{}_{\alpha}\\
&=-\frac{dz'^{\gamma}}{d\tau}~{\Gamma'}_{\gamma\delta}^{\beta}~\widetilde{V}^{\delta}{}_{\alpha}.
\ee
This equation is identical to that for the connector (\ref{diffeq}) in new coordinates, and because of the existence and uniqueness theorem, which states that there exists only one solution to a first order differential equation which satisfies a given initial condition, we can identify $\widetilde{V}^{\beta}{}_{\alpha}(\tau)$ with $V'^{\beta}{}_{\alpha}(\tau)$. Thus, (\ref{connectortrafo}) is the transformation law for the connector.\\ 

This transformation law shows that the first index of the connector transforms as a contravariant vector at $z(\tau)$, while the second index transforms as a covariant vector at $z(\tau_0)$. In this sense the connector connects different spacetime points. Hence, the indices of $V^{\beta}{}_{\alpha}(\tau)$ are lowered and raised by applying the metric on the corresponding point,
\be
\widehat{V}_{\alpha}{}^{\beta}(\tau)=g_{\alpha\mu}(z(\tau))V^{\mu}{}_{\nu}(\tau)g^{\nu\beta}(z(\tau_0)).
\label{connectorhat}
\ee

Differentiation with respect to $\tau$ gives
\be
&\frac{d\widehat{V}_{\alpha}{}^{\beta}(\tau)}{d\tau}=\frac{d}{d\tau}\left(g_{\alpha\mu}(z(\tau))V^{\mu}{}_{\nu}(\tau)g^{\nu\beta}(z(\tau_0))\right)\\
&\qquad=\frac{dg_{\alpha\mu}(z(\tau))}{d\tau}V^{\mu}{}_{\nu}(\tau)g^{\nu\beta}(z(\tau_0))+g_{\alpha\mu}(z(\tau))\frac{dV^{\mu}{}_{\nu}(\tau)}{d\tau}g^{\nu\beta}(z(\tau_0))\\
&\qquad=\frac{dg_{\alpha\mu}(z(\tau))}{d\tau}g^{\mu\sigma}(z(\tau))g_{\sigma\rho}(z(\tau))V^{\rho}{}_{\nu}(\tau)g^{\nu\beta}(z(\tau_0))\\
&\qquad\quad-g_{\alpha\mu}(z(\tau))\frac{dz^{\lambda}(\tau)}{d\tau}\Gamma_{\lambda\rho}^{\mu}(z(\tau))V^{\rho}{}_{\nu}(\tau)g^{\nu\beta}(z(\tau_0))\\
&\qquad=\frac{dz^{\lambda}(\tau)}{d\tau}\left(\frac{\del g_{\alpha\mu}(z(\tau))}{\del z^{\lambda}}g^{\mu\sigma}(z(\tau))-g_{\alpha\mu}(z(\tau))\Gamma_{\lambda\rho}^{\mu}(z(\tau))g^{\rho\sigma}(z(\tau))\right)~\widehat{V}_{\sigma}{}^{\beta}(\tau).
\ee
Inserting the explicit expression for the Christoffel symbols (\ref{christoffeldef}) yields the differential equation
\be
\frac{d\widehat{V}_{\alpha}{}^{\beta}(\tau)}{d\tau}=\frac{dz^{\lambda}(\tau)}{d\tau}~\Gamma_{\lambda\alpha}^{\sigma}(z(\tau))\widehat{V}_{\sigma}{}^{\beta}(\tau), 
\label{invdiffeq}
\ee
and thus (\ref{connectorhat}) is the solution to this differential equation with the initial condition
\be
\widehat{V}_{\alpha}{}^{\beta}(\tau_0)=\delta_{\alpha}{}^{\beta}.
\label{zwei}
\ee
We refer to this as the ``hatted connector''.

\subsection{Effect on Tensor Quantities}\label{effect}

We define the application of the connector (\ref{connector2}) to a contravariant vector $A^{\alpha}(z(\tau_0))$ at the point $z(\tau_0)$ as
\be
\tilde{A}^{\beta}(z(\tau))\equiv V^{\beta}{}_{\alpha}(\tau)A^{\alpha}(z(\tau_0)).
\ee
Differentiation with respect to $\tau$ gives
\be
\frac{d}{d\tau}\tilde{A}^{\beta}(z(\tau))&=\frac{d}{d\tau}\left(V^{\beta}{}_{\alpha}(\tau)A^{\alpha}(z(\tau_0))\right)\\
&=-\frac{dz^{\mu}(\tau)}{d\tau}\Gamma_{\mu\nu}^{\beta}(z(\tau))V^{\nu}{}_{\alpha}(\tau)A^{\alpha}(z(\tau_0))\\
&=-\frac{dz^{\mu}(\tau)}{d\tau}\Gamma_{\mu\nu}^{\beta}(z(\tau))\tilde{A}^{\nu}(z(\tau)),
\ee
where we used (\ref{diffeq}). Therefore, the covariant derivative of $\tilde{A}^{\beta}(z(\tau))$ along the curve $z(\tau)$ vanishes,
\be
\frac{D\tilde{A}^{\beta}(z(\tau))}{D\tau}\equiv\frac{d}{d\tau}\tilde{A}^{\beta}(z(\tau))+\frac{dz^{\mu}(\tau)}{d\tau}\Gamma_{\mu\nu}^{\beta}(z(\tau))\tilde{A}^{\nu}(z(\tau))=0 .
\label{paratrans1}
\ee
With (\ref{eins}) we find that $\tilde{A}^{\beta}(z(\tau_0))=A^{\beta}(z(\tau_0))$ and conclude that the application of the connector to a contravariant vector $A^{\alpha}(z(\tau_0))$ at $z(\tau_0)$ yields the unique parallel transported vector $\tilde{A}^{\beta}(z(\tau))$ at $z(\tau)$. \\

For a covariant vector $\tilde{A}_{\gamma}(z(\tau))$ at $z(\tau)$ we find
\be
\tilde{A}_{\gamma}(z(\tau))&=g_{\gamma\beta}(z(\tau))\tilde{A}^\beta(z(\tau))=g_{\gamma\beta}(z(\tau))V^{\beta}{}_{\alpha}(\tau)g^{\alpha\delta}(z(\tau_0))A_{\delta}(z(\tau_0))\\
&=\widehat{V}_{\gamma}{}^{\delta}(\tau)A_{\delta}(z(\tau_0)),
\ee
where we used (\ref{connectorhat}). Differentiation with respect to $\tau$ and employing (\ref{invdiffeq}) gives
\be
\frac{d}{d\tau}\tilde{A}_{\gamma}(\tau)&=\frac{d}{d\tau}\left(\widehat{V}_{\gamma}{}^{\delta}(\tau)A_{\delta}(z(\tau_0))\right)\\
&=\frac{dz^{\mu}(\tau)}{d\tau}\Gamma_{\mu\gamma}^{\nu}(z(\tau))\widehat{V}_{\nu}{}^{\delta}(\tau)A_{\delta}(z(\tau_0))\\
&=\frac{dz^{\mu}(\tau)}{d\tau}\Gamma_{\mu\gamma}^{\nu}(z(\tau))\tilde{A}_{\nu}(z(\tau)).
\ee
Hence the covariant derivative of $\tilde{A}_{\gamma}(z(\tau_0))$ along the curve $z(\tau)$ vanishes,
\be
\frac{D\tilde{A}_{\gamma}(\tau)}{D\tau}\equiv\frac{d}{d\tau}\tilde{A}_{\gamma}(z(\tau))-\frac{dz^{\mu}(\tau)}{d\tau}\Gamma_{\mu\gamma}^{\nu}(z(\tau))\tilde{A}_{\nu}(z(\tau))=0,
\label{paratrans2}
\ee
and with (\ref{zwei}) we find $\tilde{A}_{\gamma}(\tau_0)=A_{\gamma}(\tau_0)$ and thus the application of the hatted connector (\ref{connectorhat}) to a covariant vector $A_{\delta}(z(\tau_0))$ at $z(\tau_0)$ yields the unique parallel transported vector $\tilde{A}_{\gamma}(\tau)$ at $z(\tau)$.\\

In a similar way, one can show that general tensors are parallel transported by applying the connector to each contravariant and the hatted connector to each covariant index. We will now turn to two special cases, namely, the tangent vector and the metric.\\

A geodesic is defined as a curve $z(\tau)$ whose tangent vector is always self-parallel. Therefore, such a curve has to obey
\be
\frac{dz^{\beta}}{d\tau}(\tau)=V^{\beta}{}_{\alpha}(\tau)\frac{dz^{\alpha}}{d\tau}(\tau_0).
\ee
Differentiation with respect to $\tau$ gives
\be
\frac{d}{d\tau}\frac{dz^{\beta}}{d\tau}(\tau)
&=\frac{d}{d\tau}V^{\beta}{}_{\alpha}(\tau)\frac{dz^{\alpha}}{d\tau}(\tau_0)=-\frac{dz^{\mu}(\tau)}{d\tau}~\Gamma_{\mu\nu}^{\beta}(z(\tau))V^{\nu}{}_{\alpha}(\tau)\frac{dz^{\alpha}}{d\tau}(\tau_0).
\ee
Thus, we find the following differential equation for the curve, identical to the geodesic equation (\ref{geodef}):
\be
\frac{d^2z^{\beta}}{d\tau^2}(\tau)+\Gamma_{\mu\nu}^{\beta}(z(\tau))\frac{dz^{\mu}}{d\tau}(\tau)\frac{dz^{\nu}}{d\tau}(\tau)=0.
\label{geodesic}
\ee

The parallel transported metric $\tilde{g}_{\mu\nu}(\tau)$ obeys the differential equation
\be
\frac{d}{d\tau}\tilde{g}_{\mu\nu}(z(\tau))
&=\frac{d}{d\tau}\left(\widehat{V}_{\mu}{}^{\alpha}(\tau)\widehat{V}_{\nu}{}^{\beta}(\tau)g_{\alpha\beta}(z(\tau_0))\right)\\
&=\frac{d\widehat{V}_{\mu}{}^{\alpha}(\tau)}{d\tau}\widehat{V}_{\nu}{}^{\beta}(\tau)g_{\alpha\beta}(z(\tau_0))+\widehat{V}_{\mu}{}^{\alpha}(\tau)\frac{d\widehat{V}_{\nu}{}^{\beta}(\tau)}{d\tau}g_{\alpha\beta}(z(\tau_0))\\
&=\frac{dz^{\lambda}(\tau)}{d\tau}\Gamma_{\lambda\mu}^{\rho}(z(\tau))\widehat{V}_{\rho}{}^{\alpha}(\tau)\widehat{V}_{\nu}{}^{\beta}(\tau)g_{\alpha\beta}(z(\tau_0))\\
&\quad+\widehat{V}_{\mu}{}^{\alpha}(\tau)\frac{dz^{\lambda}(\tau)}{d\tau}\Gamma_{\lambda\nu}^{\rho}(z(\tau))\widehat{V}_{\rho}{}^{\beta}(\tau)g_{\alpha\beta}(z(\tau_0))\\
&=\frac{dz^{\lambda}(\tau)}{d\tau}\left(\Gamma_{\lambda\mu}^{\rho}(z(\tau))\tilde{g}_{\rho\nu}(z(\tau))+\Gamma_{\lambda\nu}^{\rho}(z(\tau))\tilde{g}_{\rho\mu}(z(\tau))\right)=\frac{d}{d\tau}g_{\mu\nu}(z(\tau))
\label{parallelmetric}
\ee
with the initial condition
\be
\tilde{g}_{\mu\nu}(\tau_0)=g_{\mu\nu}(\tau_0).
\ee
From the existence and uniqueness theorem we conclude that $\tilde{g}_{\mu\nu}(z(\tau))=g_{\mu\nu}(z(\tau))$. This is consistent with the fact that the covariant derivative of the metric always vanishes (\ref{defcovdermetric}).

As a consequence, we find an important relation between $\widehat{V}(x_1,x_0;{\cal C}_{x_0x_1}))$ and $V(x_1,x_0;{\cal C}_{x_0x_1})$. From
\be
\delta^{\lambda}{}_{\nu}=g^{\lambda\mu}(z(\tau))g_{\mu\nu}(z(\tau))=g^{\lambda\mu}(\tau)\widehat{V}_{\mu}{}^{\alpha}(\tau)\widehat{V}_{\nu}{}^{\beta}(\tau)g_{\alpha\beta}(\tau_0)=V^{\lambda}{}_{\beta}(\tau)\widehat{V}_{\nu}{}^{\beta}(\tau)
\ee
and (\ref{prop2}), we conclude that in general
\be
\widehat{V}(x_1,x_0;{\cal C}_{x_0x_1})=(V^T)(x_0,x_1;{\cal C}_{x_1x_0}).
\label{connectorrelation}
\ee

\subsection{The Tetrad Formalism}

As we saw in the last subsection, the parallel transported metric $\tilde{g}_{\mu\nu}(z(\tau))$ is identical to $g_{\mu\nu}(z(\tau))$. This is the reason why a covariant averaging process such as the one suggested by Isaacson (\ref{isaacson}) cannot be used to construct an averaged metric. To overcome this problem, we decompose the metric into a set of vectors with the help of the tetrad formalism.\\

Consider a set of four continuous, pointwise orthonormal vector fields $E_a{}^\mu(x)$ with $a=0,1,2,3$. An orthonormal set must be linearly independent, and so $E_a{}^\mu(x)$ forms a vector space basis for the tangent Minkowski space at each point $x$ of the manifold. Three of the vectors of this basis are necessarily spacelike and one is timelike. We shall always so label the vectors that $E_0{}^\mu(x)$ is timelike. Then the condition of orthonormality may be written as
\be
E_a{}^{\mu}(x)E_b{}^{\nu}(x)g_{\mu\nu}(x)=\eta_{ab},
\label{tetradeq}
\ee
where $\eta_{ab}=\eta^{ab}=\mathrm{diag}(1,-1,-1,-1)$ is the Minkowski metric. We can now interpret the index $a$ as a Lorentz vector index which is raised and lowered by the Minkowski metric. Then we find
\be
E^a{}_{\mu}(x)E_b{}^{\mu}(x)=\delta^a_b.
\ee
Provided the manifold is time and spacetime orientable, we choose $E_0{}^\mu(x)$ to be future-pointing and $E_i{}^\mu(x)$ to be a right-handed triad of vectors. In this case $E_a{}^\mu(x)$ is called a right-handed orthochronous Minkowski tetrad.\\

The components of a given tensor field $T^{\mu\nu}{}_{\rho\sigma}(x)$ in the orthonormal basis are
\be
T^{ab}{}_{cd}(x)=E^a{}_\mu(x)E^b{}_\nu(x)E_c{}^\rho(x)E_d{}^\sigma(x)T^{\mu\nu}{}_{\rho\sigma}(x).
\ee

If the Lorentz vector fields are subject to a spacetime-dependent Lorentz transformation,
\be
\tilde{E}^b{}_{\mu}(x)=\Lambda^b{}_a(x)E^a{}_{\mu}(x),
\ee
the metric transforms as
\be
\tilde{g}_{\mu\nu}(x)&=\tilde{E}^c{}_{\mu}(x)\tilde{E}^d{}_{\nu}(x)\eta_{cd}(x)\\
&=\Lambda^c{}_a(x)E^a{}_{\mu}(x)\Lambda^d{}_bE^b{}_{\nu}(x)\eta_{cd}(x)\\
&=E^a{}_{\mu}(x)E^b{}_{\nu}(x)\eta_{ab}(x)\\
&=g_{\mu\nu}(x).
\label{invarianttetrad}
\ee
The metric is therefore gauge-invariant under local Lorentz transformations and only a certain class of local tetrads is determined by the metric.\\

Finally, we remark that under general coordinate transformations the tetrads still transform as coordinate vectors,
\be
\tilde{E}^a{}_{\nu}(x')=\frac{\del x^{\mu}}{\del x'^{\nu}}E^a{}_{\mu}(x).
\ee

\section{The Averaging Process}\label{ouraveragingprocess}

\begin{figure}
\begin{center}\includegraphics[width=0.6\textwidth]{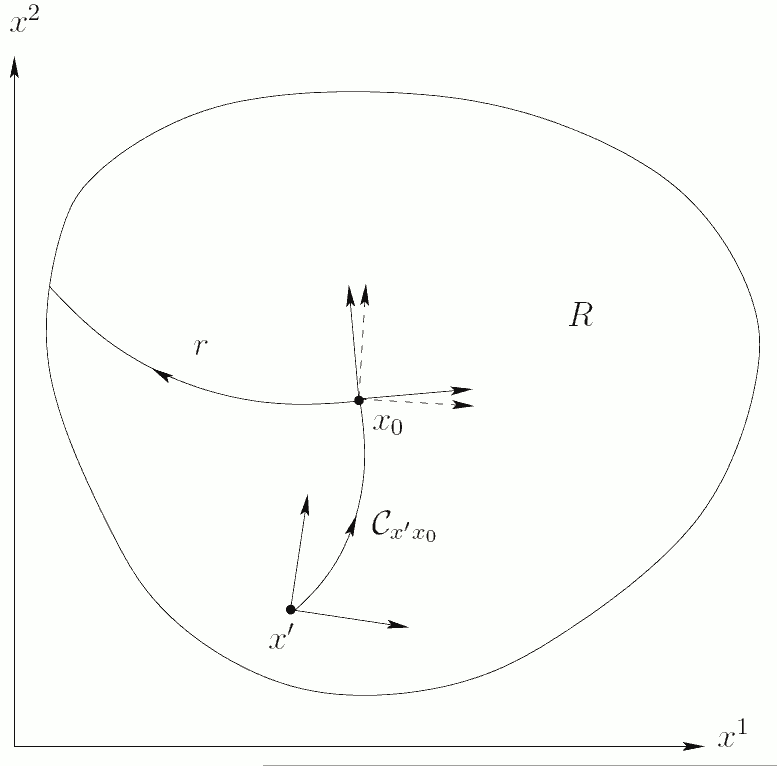}\end{center}
\caption{Parallel transport of all tetrads in a spacelike geodesic $R$ region to the center.}
\label{tetradfig}
\end{figure}

To perform any sort of averaging we first need to define the domain of averaging. Let $x_0$ be a particular point of the manifold. Provided the domain is a sufficiently small neighborhood around $x_0$, it contains no conjugate points and hence there is a unique geodesic between each point of the domain and the center $x_0$. We take the  geodesic congruence generated by all spacelike vectors of the tangent space at $x_0$. We then parametrize the geodesics by their length $s$ and cut them off at the geodesic distance $s=r=$ const. from the origin. We call the resulting region the spacelike ``geodesic region'' $R$ of radius $r$ around $x_0$. This region is bounded by the congruence of null geodesics. \\

The metric inside the geodesic region is then decomposed into a right-handed orthochronous Minkowski tetrad,
\be
g_{\mu\nu}(x)=E^a{}_{\mu}(x)E^b{}_{\nu}(x)\eta_{ab}.
\label{ourtetradfield}
\ee
Since the result of the averaging process will depend on the choice of $E^a{}_{\mu}(x)$, we need to choose a particular gauge. This is done by emplying a Lagrangian formulation. We define the ``maximally smooth tetrad field'' $(E_\mathrm{MS})^a{}_\mu(x)$ to be the tetrad field that minimizes 
\be
S=\int_R \mathcal{L}_\mathrm{MS}\sqrt{-g(x)}d^4x,
\label{ouraction}
\ee
within the geodesic region $R$ with the Lagrangian 
\be
 \mathcal{L}_\mathrm{MS}=\left(D_\mu E^a{}_\rho(x)\right)\left(D_\nu E^b{}_\sigma(x)\right)g^{\mu\nu}(x)g^{\rho\sigma}(x)\eta_{ab}.
\label{mslagrangian}
\ee
This Lagrangian describes the variation of the covariant derivative of the tetrads. It should be noted that (\ref{ouraction}) has to be minimized within a certain class of tetrads $E^a{}_\mu(x)$ that obey (\ref{ourtetradfield}), namely, the group of proper, orthochronous Lorentz transformations. Furthermore, since we want to explicitly break gauge-invariance, the covariant derivative $D_\mu$ does not include the spin connection.\\

Recall the expression for the hatted connector $\widehat{V}_{\nu}{}^{\mu}(x_0,x';{\cal C}_{x'x_0})$, defined by (\ref{connectorrelation}) and (\ref{connector1}),
\be
\widehat{V}(x_0,x';{\cal C}_{x'x_0})={\cal P}\exp\left(-\int_{{\cal C}_{x'x_0}}dz^{\mu}~{\bf\Gamma}_{\mu}(z)\right).
\label{inverseconnector}
\ee

The hatted connector allows for the parallel transport of the maximally smooth tetrads $(E_\mathrm{MS})^a{}_\mu(x)$ from all points $x'$ in the geodesic region $R$ to the center $x_0$ (see Figure \ref{tetradfig}), where we perform an average defined by
\be
\av{(E_\mathrm{MS})^a{}_{\nu}(x_0)}=\int_R f(x_0,x';{\cal C}_{x'x_0})\widehat{V}_{\nu}{}^{\mu}(x_0,x';{\cal C}_{x'x_0})(E_\mathrm{MS})^a{}_{\mu}(x')\sqrt{-g(x')}~d^4x'.
\label{ourprocess}
\ee
Here, $f(x_0,x';{\cal C}_{x'x_0})$ is a positive weighting function along the geodesics ${\cal C}_{x'x_0}$, which smoothly falls off to zero at the geodesic distance~$r$ and is furthermore normalised to unity over the geodesic region $R$,
\be
\int_R f(x_0,x';{\cal C}_{x'x_0})\sqrt{-g(x')}~d^4x'=1.
\label{fnormalisation}
\ee
This procedure has to be repeated for all points $x$ of the manifold. We can then recompose the averaged metric from the averaged tetrads by
\be
\av{g_{\mu\nu}(x)}=\av{(E_\mathrm{MS})^a{}_{\mu}(x)}\av{(E_\mathrm{MS})^b{}_{\nu}(x)}\eta_{ab}.
\label{ouraveragedtetrad}
\ee 

\section{Summary}

In this chapter we have defined a covariant averaging process which can be used to smooth metrics within the framework of general relativity. We presented a Lagrangian approach to determining a maximally smooth tetrad field used to decompose the metric at each point in the averaging domain. We defined a general relativistic analogue of the Wegner-Wilson operator to parallel transport all tetrads in the averaging domain to the centre and we defined an average at this point. The averaged tetrads can then be employed to rebuild the averaged metric at the centre point. Repeating the procedure for all points of the manifold finally yields the averaged metric. To demonstrate the functionality of this process and to visualize its effects, we apply it to various toy models in the rest of this thesis.

%% file: twodimensions.tex
\chapter{Averaging Constant Curvature Spaces in Two Dimensions}
\label{twodimensions}

As we have seen in section \ref{standardmodel}, standard cosmology is based on the assumption of a homogeneous and isotropic universe. The only spacetime metric consistent with this assumption is the FLRW metric (\ref{flrw}), which is homogeneous and isotropic on constant time hypersurfaces. This provides us with a strong physical motivation for studying such geometries. A more immediate motivation concerns the averaging process itself. Surfaces of constant curvature should be invariant under any reasonable averaging process and testing this should be an immediate aim. To study the geometry of constant curvature spaces, it is easiest to consider the case with two spatial dimensions. In this case, the space of constant curvature is either a positively curved two-sphere $(S^2)$, a flat plane $(\mathbbm{R}^2)$, or a negatively curved hyperbolic plane $(H^2)$. In the next section we will apply the averaging process to the two-sphere $(S^2)$ and investigate its effects in detail. The analysis of the flat plane $(\mathbbm{R}^2)$ and the hyperbolic plane $(H^2)$ will be the subject of the subsequent sections.

\section{Averaging the Sphere}

The two-sphere of radius $a$ can be embedded in a three-dimensional Euclidean space with coordinates $(\xi^1,\xi^2,\xi^3)$. This enables us to visualize it and to easily read off the line element. In spherical coordinates
\be
\xi^1&=a\sin\theta\cos\phi,\\
\xi^2&=a\sin\theta\sin\phi,\\
\xi^3&=a\cos\theta,
\label{firsttrans}
\ee
where $0\leq\theta\leq\pi$ and $0\leq\phi\leq2\pi$, the line element is given by $ds^2=a^2d\theta^2+a^2\sin^2\theta d\phi^2$. There is an obvious similarity between this line element and the FLRW line element (\ref{flrw}). We could equally have used coordinates $(r,\theta)$ in which the line element takes the form $ds^2=a^2(1-r^2)^{-1}dr^2+a^2r^2d\theta^2$. However, both line elements lead to metrics whose corresponding geodesics have a complicated parameter representation. To circumvent this problem, we take the line element  $ds^2=a^2d\theta^2+a^2\sin^2\theta d\phi^2$ and furthermore apply a stereographic projection into the tangent plane of the south pole. The geodesics through the origin of the projection plane are then given by straight lines. This simplifies the calculation of the connector and hence the calculation of the averaged metric considerably.

\subsection{The Metric from Stereographic Projection}

\begin{figure}
\begin{center}\includegraphics[width=0.6\textwidth]{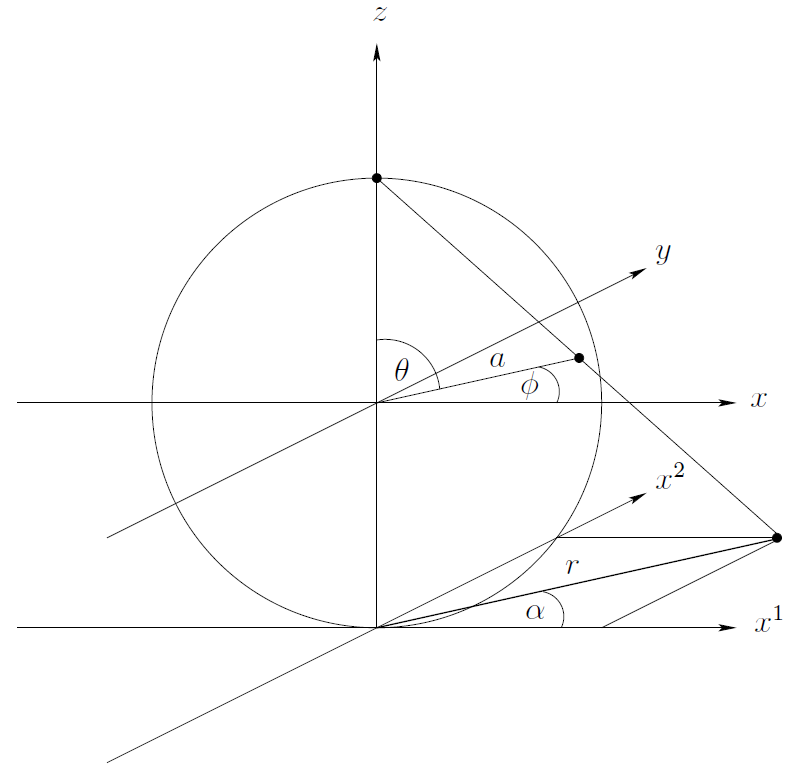}\end{center}
\caption{Stereographic projection of the two-sphere into the $(x^1x^2)$-plane}
\label{projection}
\end{figure}

The stereographic projection of the two-sphere into the tangent plane of the south pole is illustrated in Figure \ref{projection}. The corresponding coordinate transformation is given by
\be
\alpha&=\phi,\\
r&=2a\tan(\pi/2-\theta/2). 
\label{secondtrans}
\ee

In terms of the new coordinates, the line element becomes 
\be
ds^2=\frac{16a^4}{(4a^2+r^2)^2}~dr^2+a^2\sin^2\left(\pi-2\arctan\left(\frac{r}{2a}\right)\right)~d\alpha^2.
\ee
Using the addition theorems this can be simplified to the following form: 
\be
ds^2=\frac{16a^4}{(4a^2+r^2)^2}\left(dr^2+d\alpha^2\right).
\ee
This reduction of the line element to a simple form comes at a cost: The angle $\alpha$ is undetermined at $r = 0$, causing severe problems. Therefore, we change to Cartesian coordinates on the projection plane,
\be
x^1&=r\cos\alpha,\\
x^2&=r\sin\alpha.
\label{thirdtrans}
\ee
If we additionally introduce the function
\be
\D L^2(x)=4a^2+(x^1)^2+(x^2)^2,
\ee
we can express the line element in the following compact form:
\be 
ds^2=\frac{16a^4}{L^4}\left((dx^1)^2+(dx^2)^2\right).
\ee
The associated metric and its inverse are
\be
g_{ij}=\left(\frac{2a}{L(x)}\right)^4\delta_{ij}\qquad\mathrm{and}\qquad g^{ij}=\left(\frac{L(x)}{2a}\right)^4\delta^{ij},
\label{metric}
\ee
where the indices $i,j\in\{1,2\}$.

\subsection{The Geodesics and the Connector}

The two-dimensional analogue of the four-dimensional tetrad is the ``dyad''. To find the connector in the $(x^1x^2)$-coordinates, which parallel transports dyads from any point of the geodesic region to the origin, we first need to determine the geodesic congruence generated by all tangent vectors of the tangent space at the origin. The partial derivatives of the metric (\ref{metric})
\be
\frac{\del g_{ij}}{\del x^k}=-\frac{4x_k}{L(x)^ 2}g_{ij}
\ee
yield the Christoffel symbols
\be
\Gamma^i_{jk}(x)=-\frac{2}{L(x)^2}\left(x_j\delta^i_k+x_k\delta^i_j-x^i\delta_{jk}\right).
\label{christoffel}
\ee
Inserting these into the geodesic equation (\ref{geodef}) gives the differential equation for the components $z^i(\tau)$ of the geodesics,
\be
L(z(\tau))^2~\frac{d^2z^i}{d\tau^2}-4z_j\frac{dz^j}{d\tau}\frac{dz^i}{d\tau}+2z^ig_{jk}\frac{dz^j}{d\tau}\frac{dz^k}{d\tau}=0.
\ee
For the initial conditions $z^i(0)=0$ and $dz^i/d\tau(0)=\mathrm{const.}$ the solution to this equation is
\be
z^i(\tau)=2a\tan\left(\frac{\tau}{2a}\right)\frac{dz^i}{d\tau}(0).
\label{geo}
\ee
With 
\be
L(z(\tau))^2=4a^2\cos^{-2}\left(\frac{\tau}{2a}\right)\left(\left(\frac{dz^1}{d\tau'}(0)\right)^2+\left(\frac{dz^1}{d\tau'}(0)\right)^2\right)
\ee
we find for the arc length
\be
s&=\int_0^\tau d\tau'\sqrt{g_{ij}(z(\tau'))\frac{dz^i}{d\tau'}\frac{dz^j}{d\tau'}}\\
&=\tau\left(\left(\frac{dz^1}{d\tau'}(0)\right)^2+\left(\frac{dz^1}{d\tau'}(0)\right)^2\right)^{-1/2}.
\ee
Thus, if we choose $\tau$ to be the arc length, $\tau=s$, we find
\be
\left(\frac{dz^1}{d\tau'}(0)\right)^2+\left(\frac{dz^2}{d\tau'}(0)\right)^2=1.
\label{idinitialcond}
\ee
This property makes it possible to introduce an angle $\gamma$ with $0\leq\gamma<2\pi$, such that the initial conditions for the geodesics are given by
\be
\frac{dz^1}{d\tau}(0)=\cos\gamma\qquad\mathrm{and}\qquad\frac{dz^2}{d\tau}(0)=\sin\gamma.
\label{initialcond}
\ee
Recall the general form of the connector (\ref{connector1}),
\be
V(x_1,x_0;{\cal C}_{x_0x_1})={\cal P}\exp\left(-\int_{{\cal C}_{x_0x_1}}dz^i~{\bf\Gamma}_i(z)\right).
\label{connectordef}
\ee
We parametrize the connector by $\tau$ and insert the Christoffel symbols (\ref{christoffel}) and the geodesic (\ref{geo}) into this expression. Considering the parallel transport from the origin parametrized by $\tau=0$ to the point parametrized by $\tau$ gives 
\be
V(\tau,0;{\cal C}_{0\tau})&={\cal P}\exp\left(-\int_0^{\tau}\left(\frac{dz^1}{d\tau}(0)\frac{1}{\cos^2(\frac{\tau'}{2a})}{\bf\Gamma}_1(z(\tau'))
+\frac{dz^2}{d\tau}(0)\frac{1}{\cos^2(\frac{\tau'}{2a})}{\bf\Gamma}_2(z(\tau'))\right)d\tau'\right)\\
&={\cal P}\exp\left(\left(\frac{1}{a}\int_0^{\tau}\tan\left(\frac{\tau'}{2a}\right)~d\tau'\right)\mathbbm{1}\right)\\
&={\cal P}\exp\left(-2\ln\left(\cos\left(\frac{\tau'}{2a}\right)\right)\bigg|_0^{\tau}\mathbbm{1}\right)\\
&=\cos^{-2}(\frac{\tau}{2a})\mathbbm{1}.
\ee
Applying the inverse of this,
\be
V(0,\tau;{\cal C}_{\tau 0})=V^{-1}(\tau,0;{\cal C}_{0\tau})=\cos^2\left(\frac{\tau}{2a}\right)\mathbbm{1},
\ee 
to a contravariant vector at $z(\tau)$ yields the parallel transported vector at $z(0)=0$. The parallel transport of contravariant vectors is achieved through the application of the hatted connector (\ref{connectorhat}),
\be
\widehat{V}_j{}^i(0,\tau;{\cal C}_{\tau 0})=g_{jk}(0)V^k{}_l(0,\tau;{\cal C}_{\tau 0})g^{li}(\tau)={(V^T)}_j{}^i(\tau,0;{\cal C}_{0\tau}).
\ee
The hatted connector appears in the averaging process and according to the calculation above its explicit form in this case is given by
\be
\widehat{V}_j{}^i(0,\tau;{\cal C}_{\tau 0})=\cos^{-2}\left(\frac{\tau}{2a}\right)\delta^i_j.
\label{hattedconnector}
\ee

\subsection{The Maximally Smooth Dyad Field}
\label{smoothdyad}

In this section we want to find the maximally smooth dyad field. We start by writing the metric in terms of dyads,
\be
g_{ij}(x)=E^a{}_i(x) E^b{}_j(x)~\delta_{ab}.
\label{ortho}
\ee
In two spatial dimensions the indices $a,b,\dots$ are raised and lowered by $\delta_{ab}$ and the indices $i,j,\dots$ by $g_{ij}$. Let a reference field be
\be
\widetilde{E}^a{}_i(x)=\left(\frac{2a}{L(x)}\right)^2\delta^a_i.
\label{startingfield}
\ee 
The maximally smooth dyad field has to be a rotation of this dyad field by a position dependent angle $\phi(x^1,x^2)$:
\be
E^a{}_i=U^a{}_b~\widetilde{E}^b{}_i\quad\mathrm{with}\quad U^a{}_b=\begin{pmatrix}\cos\phi(x^1,x^2)&\sin\phi(x^1,x^2)\\-\sin\phi(x^1,x^2)&\cos\phi(x^1,x^2)\end{pmatrix}. 
\ee  
The partial derivative of the reference field is
\be
\del_i\widetilde{E}^a{}_j=-\frac{4a^2}{L^4}2x_i\delta^a{}_j=-\frac{2x_i}{L^2}\widetilde{E}^a{}_j.
\ee
Hence, the covariant derivative can be expressed as
\be
D_i\widetilde{E}^a{}_j=\del_i\widetilde{E}^a{}_j-\Gamma^k_{ij}\widetilde{E}^a{}_k=\frac{2}{L^2}\left(x_j\delta_i^k-x^kg_{ij}\right)\widetilde{E}^a{}_k,
\label{kov}
\ee
The partial derivative of the rotation matrix $U_{ab}$ is
\be
\del_iU^a{}_b=\epsilon^a{}_cU^c{}_b(\del_i\phi),
\ee
where $\epsilon^a{}_b$ is the totally antisymmetric tensor density,
\be
\epsilon^a{}_b=\begin{pmatrix}0&1\\-1&0\end{pmatrix},
\ee
and so we find for the covariant derivative of the maximally smooth dyad field: 
\be
D_iE^a{}_j=(\del_iU^a{}_b)\widetilde{E}^b{}_j+U^a{}_bD_i\widetilde{E}^b{}_j=\epsilon^a{}_cU^c{}_b(\del_i\phi)\widetilde{E}^b{}_j+\frac{2}{L^2}U^a{}_b\left(x_j\delta_i^k-x^kg_{ij}\right)\widetilde{E}^a{}_k.
\ee
We will furthermore need the following identities:
\begin{align}
\delta_{cf}U^e{}_cU^f{}_d&=\delta_{cd}\label{id1}\\
U^a{}_c\delta_{ab}\epsilon^b{}_fU^f{}_d&=\epsilon_{cd}\label{id2}\\
\epsilon^a{}_c\epsilon^b{}_f\delta_{ab}&=\delta_{cf}\label{id3}
\end{align}
We now find the maximally smooth dyad field by varying the Lagrangian (\ref{mslagrangian}). For the chosen reference field, this Lagrangian is 
\be
\mathcal{L}_\mathrm{MS}&=\left(D_iE^a{}_j\right)\left(D_kE^b{}_l\right)g^{ik}g^{jl}\delta_{ab}\\
&=\left(\epsilon^a{}_eU^e{}_c(\del_i\phi)\widetilde{E}^c{}_j+U^a{}_cD_i\widetilde{E}^c{}_j\right)\left(\epsilon^b{}_fU^f{}_d(\del_k\phi)\widetilde{E}^d{}_l+U^b{}_dD_k\widetilde{E}^d{}_l\right)g^{ik}g^{jl}\delta_{ab}\\
&=\epsilon^a{}_eU^e{}_c\widetilde{E}^c{}_j\epsilon^b{}_fU^f{}_d(\del_i\phi)(\del_k\phi)\widetilde{E}^d{}_lg^{ik}g^{jl}\delta_{ab}+U^a{}_c(D_i\widetilde{E}^c{}_j)\epsilon^b{}_fU^f{}_d(\del_k\phi)\widetilde{E}^d{}_lg^{ik}g^{jl}\delta_{ab}\\
&\quad+U^b{}_d(D_k\widetilde{E}^d{}_l)\epsilon^a{}_eU^e{}_c(\del_i\phi)\widetilde{E}^c{}_jg^{ik}g^{jl}\delta_{ab}+U^a{}_cU^b{}_d(D_i\widetilde{E}^c{}_j)(D_k\widetilde{E}^d{}_l)g^{ik}g^{jl}\delta_{ab}.
\label{smoothlagrangian}
\ee
This can be simplified by applying the identities (\ref{id1}-\ref{id3}) and inserting the expression (\ref{kov}) for the covariant derivative:
\be
\mathcal{L}_\mathrm{MS}=\delta_{cd}\widetilde{E}^c{}_j&\widetilde{E}^d{}_lg^{jl}g^{ik}(\del_i\phi)(\del_k\phi)+2\epsilon_{cd}\frac{2}{L^2}(x_j\delta^m_i-x^mg_{ij})\widetilde{E}^c{}_m\widetilde{E}^d{}_lg^{jl}g^{ik}(\del_k\phi)\\
&+\delta_{cd}\widetilde{E}^c{}_m\widetilde{E}^d{}_n\frac{4}{L^4}x^mx^n.
\ee
With (\ref{ortho}) and (\ref{startingfield}) and this expression gives
\be
&\mathcal{L}_\mathrm{MS}=2g^{ik}(\del_i\phi)(\del_k\phi)+\epsilon_{cd}\widetilde{E}^c{}_m\widetilde{E}^d{}_l\frac{4}{L^2}\left(g^{mk}x^l-x^mg^{lk}\right)(\del_k\phi)+\frac{8}{L^4}g_{mn}x^mx^n.
\ee
Introducing the vector
\be
u^k\equiv\epsilon_{cd}\widetilde{E}^c{}_m\widetilde{E}^d{}_l\frac{4}{L^2}\left(g^{mk}x^l-x^mg^{lk}\right),
\label{vectoru}
\ee
the Lagrangian becomes
\be
\mathcal{L}_\mathrm{MS}=2g^{ik}(\del_i\phi)(\del_k\phi)+u^k(\del_k\phi)+\frac{4}{L^4}g_{mn}x^mx^n.
\ee

If we now vary the corresponding generally covariant action 
\be
S=\int_R d^2x\sqrt{g}~\mathcal{L}_\mathrm{MS}=\int_R d^2x\sqrt{g}~\left(2(\del_i\phi)(\del_k\phi)g^{ik}+u^k(\del_k\phi)+\frac{4}{L^4}g_{mn}x^mx^n\right)
\ee
with respect to $\phi$, we find
\be
\delta S&=\int_Rd^2x\sqrt{g}\left(4g^{ik}(\partial_i \phi)(\partial_k\delta\phi)+2u^k(\partial_k\delta\phi)\right)\\
&=\int_Rd^2x\partial_k\left(4\sqrt{g}~g^{ik}(\partial_i \phi)\delta\phi+2\sqrt{g}~u^k\delta\phi\right)\\
&+\int_Rd^2x\left(-4\partial_k(\sqrt{g}~g^{ik}(\partial_i \phi))-2\partial_k(\sqrt{g}~u^k)\right)\delta\phi=0.
\label{firstaction}
\ee
The fact that the integration area $R$ is finite leads to two conditions that ought to be fulfilled by the dyad field. The first term in (\ref{firstaction}) can be converted to a line integral according to Gauss' law,
\be
\int_{\partial R}ds~\sqrt{g}~n_k\left(4g^{ik}(\partial_i\phi)+2u^k\right)\delta\phi=0.
\ee
For arbitrary $\delta\phi$ this leads to the boundary condition
\be
\frac{\partial\phi}{\partial n}=-\frac{1}{2}n_ku^k.
\label{dgl1}
\ee
on $\partial R$. With the general identities
\be
g^{ij}\partial_kg_{ij}&=\frac{1}{g}\partial_kg\\
\partial_k\sqrt{g}&=\frac{1}{2}(\sqrt{g})^{-1}\partial_kg
\ee
the covariant derivative $D_k$ of any vector $A^k$ can be expressed as
\begin{equation}
D_kA^k=\partial_kA^k+\frac{1}{2}g^{km}(\partial_lg_{km})A^l=\partial_kA^k+\frac{1}{2g}(\partial_lg)A^l=(\sqrt{g})^{-1}\partial_k(\sqrt{g}~A^k).
\label{anyvector}
\end{equation}
Hence, for arbitrary $\delta\phi$ the second term in (\ref{firstaction}) leads to the differential equation on $R$
\be
g^{ik}\left(\partial_i\partial_k\phi\right)&=-\frac{1}{2}D_ku^k.
\label{dgl2}
\ee
In this case the covariant derivative of $u^k$ vanishes, $D_ku^k=0$, as can be seen from (\ref{vectoru}). \\

The geodesics through the origin (\ref{geo}) are parametrized by $\tau$ and $\gamma$, where $\gamma$ is the initial direction of the geodesic, as can be seen from ($\ref{initialcond}$). We will call these coordinates the ``geodesic coordinates''. They are parametrized as follows:
\be
z^1(\tau,\gamma)&=2a\tan\left(\frac{\tau}{2a}\right)\cos\gamma\\
z^2(\tau,\gamma)&=2a\tan\left(\frac{\tau}{2a}\right)\sin\gamma
\label{geocoord}
\ee
where $0\leq\gamma<2\pi$ and $0\leq\tau<\infty$. The contour line $\del$R of the geodesic region $R$ is defined as all points of geodesic distance $s=r$ from the origin. It therefore has the parameter representation  
\be
\alpha^1(\gamma)&=2a\tan\left(\frac{r}{2a}\right)\cos\gamma,\\
\alpha^2(\gamma)&=2a\tan\left(\frac{r}{2a}\right)\sin\gamma.
\ee

The coordinate invariant definition of the normal vector to $\del R$ is
\be
\nu_k=\sqrt{g(r,\gamma)}~\epsilon_{kl}\frac{\del\alpha^l}{\del\gamma},
\ee
and the normal unit vector is
\be
n_k=\left(g^{ij}(r,\gamma)\nu_i\nu_j\right)^{-1/2}\nu_k.
\ee

Explicit calculation shows that $n_ku^k=0$ and we find the differential equation
\be
(\del_i\del_k\phi)\delta^{ik}=0\quad\mathrm{in\ the\ area\ } R\qquad\mathrm{and}\qquad\frac{\del\phi}{\del n}=0 \quad\mathrm{on\ }\del R.
\label{neumann}
\ee
The solution to this equation is $\phi=0$. We have thus seen that the dyad field (\ref{startingfield}) was already the maximally smooth dyad field, and so
\be
(E_\mathrm{MS})^a{}_i(x)=\left(\frac{2a}{L(x)}\right)^2\delta_{ai}.
\label{dyadfield}
\ee

\subsection{The Result of the Averaging Process}

In the last sections we have determined all the constituents that we need for the averaging process (\ref{ourprocess}). The general form of the average is
\be
\av{(E_\mathrm{MS})^a{}_j(x)}=\int f(x,x';{\cal C}_{x'x})\widehat{V}_j{}^i(x,x';{\cal C}_{x'x})(E_\mathrm{MS})^a{}_i(x')\sqrt{g(x')}~d^2x'.
\label{avprocess}
\ee
Since we calculated the connector (\ref{hattedconnector}) in terms of the affine parameter $\tau$ it is useful to carry out the integral along the geodesics through the origin (\ref{geocoord}), which were parametrized by $\gamma$ and $\tau$. Expressing the maximally smooth dyad field (\ref{dyadfield}) and the connector (\ref{hattedconnector}) in terms of these coordinates, we see that 
\be
(E_\mathrm{MS})^a{}_i(\gamma,\tau)&=\cos^2\left(\frac{\tau}{2a}\right)\delta^a{}_i,\\
\widehat{V}_j{}^i(z(\gamma,\tau))&=\cos^{-2}\left(\frac{\tau}{2a}\right)~\delta_j^i,
\ee
and insert these into the averaging process (\ref{avprocess}). In terms of the coordinates (\ref{geocoord}) the square root of the determinant of the metric is
\be
\sqrt{g(\gamma,\tau)}=2a\sin\left(\frac{\tau}{2a}\right)\cos\left(\frac{\tau}{2a}\right).
\ee
Expressed completely in terms of $\gamma$ and $\tau$, the averaged dyad field at the origin becomes:
\be
\av{(E_\mathrm{MS})^a{}_j(0,0)}&=\int_0^r\int_0^{2\pi} f(z(\gamma,\tau))\widehat{V}_j{}^i(z(\gamma,\tau))(E_\mathrm{MS})^a{}_i(\gamma,\tau)\sqrt{g(\gamma,\tau)}~d\tau d\gamma\\
&=\left(\int_0^r\int_0^{2\pi} f(z(\gamma,\tau))\sqrt{g(\gamma,\tau)}~d\tau d\gamma\right)\delta^a{}_j=\delta^a{}_j.
\ee
The term in brackets equals one since the averaging function $f(z(\gamma,\tau))$ is taken to be normalized to unity over all space. Hence, the averaging process returns the averaged metric at the origin
\be
\av{{g}_{ij}(0,0)}=\av{(E_\mathrm{MS})^a{}_i(0,0)}\av{(E_\mathrm{MS})^b{}_j(0,0)}\delta_{ab}=\delta_{ij}.
\ee
To get the full averaged metric, we need to repeat this procedure for each point of the sphere. Doing so, however, leaves the results for each point expressed in different coordinates. To express them all in the same coordinate system we need to know the transformation law which we now turn to

\subsection{The Coordinate Transformation}

\begin{figure}
\begin{center}
\includegraphics[width=0.6\textwidth]{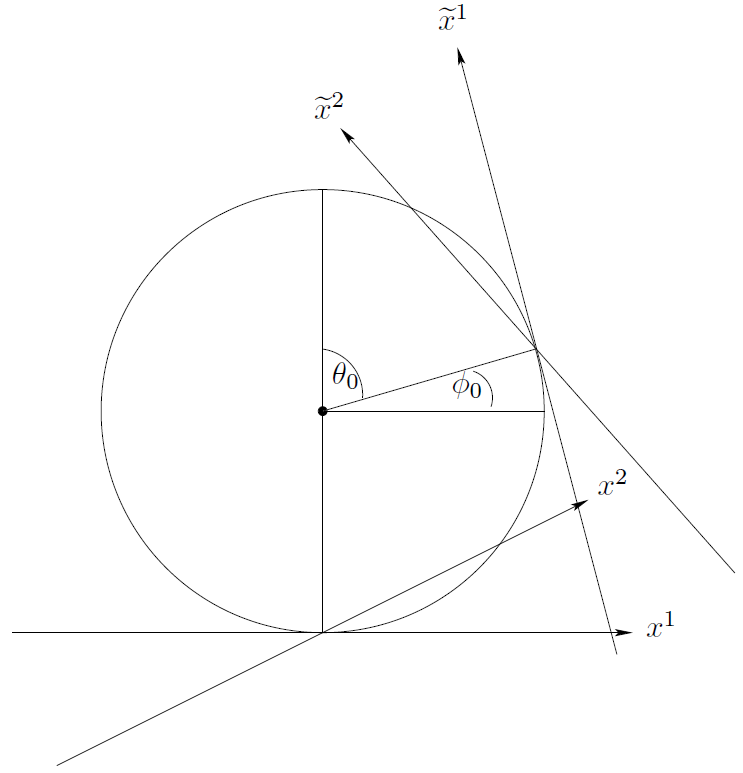}
\caption{Coordinate transformation from the $(x^1x^2)$-plane to the $(\widetilde{x}^1\widetilde{x}^2)$-plane and back}
\label{transformation}
\end{center}
\end{figure}

Let the coordinates of the sphere be ${\bf x}=(x,y,z)$ and those of the stereographic projection plane at the south pole ${\bf x}_S=(0,0,-a)$ be $(x^1,x^2)$. We want the coordinate transformation to the coordinates $(\widetilde{x}^1,\widetilde{x}^2)$ of the stereographic projection plane at another point ${\bf x}_0=(x_0,y_0,z_0)$ $=(a\sin\theta_0\cos\phi_0,a\sin\theta_0\sin\phi_0,a\cos\theta_0)$ of the sphere (see Figure \ref{transformation}). This coordinate transformation can be decomposed into three steps: the inverse stereographic projection of $(x^1,x^2)$ onto the sphere ${\bf x}=(x,y,z)$, the rotation of ${\bf x}_0$ to the south pole ${\bf x}_S$ of the sphere and the associated transformation to the coordinates $\widetilde{{\bf x}}=(\widetilde{x},\widetilde{y},\widetilde{z})$, and finally the stereographic projection into the tangent plane of the new south pole $(\widetilde{x}^1,\widetilde{x}^2)$.\\

The inverse stereographic projection from $(x^1,x^2)$ to the coordinates of sphere ${\bf x}=(x,y,z)$ is a composition of the coordinate transformations (\ref{firsttrans}), (\ref{secondtrans}), and (\ref{thirdtrans}):
\be
x=\frac{4 a^2x^1}{L^2},\qquad y=\frac{4 a^2x^2}{L^2},\qquad z=a\left(1-\frac{8 a^2}{L^2}\right).
\label{stereo2}
\ee

The rotation $D$ of the sphere consists of three rotation matrices $D_z$, $D_y$ and $D_z^{-1}$. First $D_z$ rotates the sphere around the $z$-axis until the point ${\bf x}_0$ intersects with the $(xz)$-plane then $D_y$ rotates the sphere around the $y$-axis until the point $D_z^{-1}{\bf x}_0$ is aligned with the south pole, and finally $D_z^{-1}$ rotates the sphere back to the original orientation. The rotation $D=D_zD_yD_z^{-1}$ is of the following form:
\be
D=\begin{pmatrix}\cos\phi_0&-\sin\phi_0&0\\\sin\phi_0&\cos\phi_0&0\\0&0&1\end{pmatrix}\begin{pmatrix}-\cos\theta_0&0&-\sin\theta_0\\0&1&0\\\sin\theta_0&0&-\cos\theta_0\end{pmatrix}\begin{pmatrix}\cos\phi_0&\sin\phi_0&0\\-\sin\phi_0&\cos\phi_0&0\\0&0&1\end{pmatrix}.
\ee

The stereographic projection of the rotated sphere with the coordinates $\widetilde{{\bf x}}=(\widetilde{x},\widetilde{y},\widetilde{z})$ into the tangent plane of the new south pole with the coordinates $(\widetilde{x}^1,\widetilde{x}^2)$ is given by
\be
\widetilde{x}^1=\frac{2a\widetilde{x}}{(a-\widetilde{z})},\qquad\widetilde{x^2}=\frac{2a\widetilde{y}}{(a-\widetilde{z})}.
\ee

The explicit transformation from the coordinates $(x^1,x^2)$ to the coordinates $(\widetilde{x}^1,\widetilde{x}^2)$ and vice versa are given in appendix \ref{sphereappend}.

\subsection{The M\"obius Transformation}

If we identify the $(x^1x^2)$- and $(\widetilde{x}^1\widetilde{x}^2)$-projection planes with complex planes $\mathbbm{C}$, we can express the coordinate transformation (\ref{coordtrans}) as a M\"obius transformation. The general form of a M\"obius transformation is 
\be
z\longrightarrow \widetilde{z}=\frac{az+b}{cz+d},
\ee
where $a,b,c,d\in\mathbbm{C}$ and $ad-bc\ne0$. If we use extended complex planes $\mathbbm{C}\cup\{\infty\}$, we can furthermore include the north pole of the sphere in the transformation. In that case the point $z=-d/c$ is mapped to the complex infinity $\widetilde{z}=\infty$ and the point $z=\infty$ is mapped to $\widetilde{z}=a/c$.\\

One method of finding the M\"obius transformation that describes the coordinate transformation (\ref{coordtrans}) is to insert the complex numbers $a=a_1+ia_2$, $b=b_1+ib_2$, $c=c_1+ic_2$, and $d=d_1+id_2$ and to compute
\be
\widetilde{z}&=\widetilde{x}_1+i\widetilde{x}_2=\frac{(a_1+ia_2)(x^1+ix^2)+(b_1+ib_2)}{(c_1+ic_2)(x^1+ix^2)+(d_1+id_2)}\\
&=\frac{(a_1x^1\!-\!a_2x^2\!+\!b_1)+i(a_2x^1\!+\!a_1x^2\!+\!b_2)}{(c_1x^1\!-\!c_2x^2\!+\!d_1)+i(c_2x^1\!+\!c_1x^2\!+\!d_2)}\cdot\frac{(c_1x^1\!-\!c_2x^2\!+\!d_1)-i(c_2x^1\!+\!c_1x^2\!+\!d_2)}{(c_1x^1\!-\!c_2x^2\!+\!d_1)-i(c_2x^1\!+\!c_1x^2\!+\!d_2)}
\label{longway}
\ee
explicitly. Then $a,b,c$ and $d$ follow by comparison of coefficients.\\

A more elegant approach originates from M\"obius transformations preserving the cross-ratio
\be
\frac{z-z_1}{z-z_3}\cdot\frac{z_2-z_3}{z_2-z_1}=\frac{\widetilde{z}-\widetilde{z}_1}{\widetilde{z}-\widetilde{z}_3}\cdot\frac{\widetilde{z}_2-\widetilde{z}_3}{\widetilde{z}_2-\widetilde{z}_3}.
\label{crossratio}
\ee
From here one can compute the M\"obius transformation if at least three points and their images $z_1\longrightarrow\widetilde{z_1}$, $z_2\longrightarrow\widetilde{z_2}$, and $ z_3\longrightarrow\widetilde{z_3}$ are provided.\\

To construct the M\"obius transformation in this way, we list here some particular points of the $(x^1x^2)$-plane and their associated images in the $(\widetilde{x}^1\widetilde{x}^2)$-plane. First of all, there is the origin of the $(x^1x^2)$-projection plane and its image in the $(\widetilde{x}^1\widetilde{x}^2)$-plane, which we name $\widetilde{x}_0$. Secondly, there is the origin of the $(\widetilde{x}^1\widetilde{x}^2)$-plane and its preimage in the $(x^1x^2)$-plane, which we call $x_0$. Finally, there are the two fixed points $x_{F1}$ and $x_{F2}$ of the transformation, which are the stereographic projections of the antipodal intersection points of the sphere with the rotations axis into both projection planes:
\be
(0,0)&\longleftrightarrow\widetilde{x}_0=\left(-\frac{2a\cos\theta_0\cos\phi_0}{1-\cos\theta_0},-\frac{2a\sin\theta_0\cos\phi_0}{1-\cos\theta_0}\right),\\
x_0=\left(\frac{2a\sin\theta_0\cos\phi_0}{1-\cos\theta_0},\frac{2a\sin\theta_0\sin\phi_0}{1-\cos\theta_0}\right)&\longleftrightarrow(0,0),\\
x_{F1}=(2a\sin\phi_0,-2a\cos\phi_0)&\longleftrightarrow\widetilde{x}_{F1}=(2a\sin\phi_0,-2a\cos\phi_0),\\
x_{F2}=(-2a\sin\phi_0,2a\cos\phi_0)&\longleftrightarrow\widetilde{x}_{F2}=(-2a\sin\phi_0,2a\cos\phi_0).\notag
\label{list}
\ee

 We choose the two fixed points $z_1=\widetilde{z}_1=(x^1_{F1}+ix^2_{F2})$ and $z_2=\widetilde{z}_2=(x^1_{F2}+ix^2_{F2})$, as well as the origin of the $(x^1x^2)$-plane $z_3=0$ and its image $\widetilde{z}_3=(2a\sin\theta_0\cos\phi_0/(1+\cos\theta_0)+i2a\sin\theta_0\sin\phi_0/(1+\cos\theta_0))$. This leads to the following representation of the M\"obius transformation:
\be
\widetilde{z}=\frac{z-2a\cot(\theta_0/2)(\cos\phi_0+i\sin\phi_0)}{(1/2a)\cot(\theta_0/2)(\cos\phi_0-i\sin\phi_0)z+1}.
\ee
 
\subsection{The Metric Transformation}

Having derived the coordinate transformation from the stereographic projection plane of the south pole to an arbitrary projection plane and back, we now turn our attention to the way the metric behaves under such transformations. In general, the metric transforms as a covariant tensor field of rank two,
\be
\widetilde{g}_{kl}(\widetilde{x}^1,\widetilde{x}^2)=\frac{\del x^i}{\del\widetilde{x}^k}\frac{\del x^j}{\del\widetilde{x}^l}~g_{ij}(x^1,x^2).
\label{generalmetrictrans}
\ee
For the coordinate transformation (\ref{coordtrans}) the Jacobi matrix elements are given by (\ref{coordder}). Recalling the metric in terms of the $(x^1x^2)$-coordinates (\ref{metric}), which were the coordinates of the stereographic projection plane around the south pole,
\be
g_{ij}(x^1,x^2)=\left(\frac{2a}{L}\right)^4\delta_{ij},
\ee
we can now use (\ref{generalmetrictrans}) to compute the transformed metric in the projection plane specified by $\theta_0$ and $\phi_0$:
\be
\widetilde{g}_{kl}(\widetilde{x}^1,\widetilde{x}^2)=\frac{\del x^i}{\del\widetilde{x}^k}\frac{\del x^j}{\del\widetilde{x}^l}~\left(\frac{2a}{L}\right)^4\delta_{ij}=\left(\frac{2a}{\widetilde{L}}\right)^4\delta_{kl}.
\ee
This result is not unexpected, since the plain two-sphere is invariant under spatial rotations. Hence, applying the averaging process to the origin of the transformed coordinate system is analogous to averaging in the original coordinate system and yields the averaged metric
\be
\av{\widetilde{g}_{kl}(0,0)}=\delta_{kl}. 
\label{averagedtransmetric}
\ee
For the transformation of the averaged metric back to the original coordinate system we need to assume that it transforms as a tensor, which is justifiable considering the general covariance of the averaging process. Keeping in mind that in terms of the $x^1x^2$-coordinates the origin of the transformed coordinate system $(\widetilde{x}^1,\widetilde{x}^2)=(0,0)$ is given by $(x^1,x^2)=(x_0^1,x_0^2)$, the transformation law can be expressed as
\be
\av{g_{ij}(x_0^1,x_0^2)}=\frac{\del\widetilde{x}^k}{\del x^i}\bigg|_{(x^1,x^2)=(x_0^1,x_0^2)}\frac{\del\widetilde{x}^l}{\del x^j}\bigg|_{(x^1,x^2)=(x_0^1,x_0^2)}\av{\widetilde{g}_{kl}(0,0)}.
\label{metricbacktrans}
\ee
The required Jacobi matrix elements are obtained from the back transformation (\ref{coordtrans2}) by partial derivation. The explicit results are given in (\ref{coordder2}). Inserting them along with the averaged metric at the origin of the transformed coordinate system (\ref{averagedtransmetric}) into  the back transformation formula (\ref{metricbacktrans}) returns the averaged metric at the point $(x_0^1,x_0^2)=(2a\sin\theta_0\cos\phi_0/(1-\cos\theta_0),-2a\sin\theta_0\sin\phi_0/(1-\cos\theta_0))$,
\be
\av{g_{ij}(x_0^1,x_0^2)}&=\frac{64a^4}{(L^2-(8a^2-L^2)\cos\theta_0+4a\sin\theta_0(x^1\cos\phi_0+x^2\sin\phi_0))^2}\bigg|_{(x^1,x^2)=(x_0^1,x_0^2)}~\delta_{ij}\notag\\
&=\frac{1}{4}(1-\cos\theta_0)^2\delta_{ij}.
\ee
Doing the same for all points of the sphere leads to the assembled averaged metric in the $(x^1x^2)$-plane,
\be
\av{g_{ij}(x^1,x^2)}=\frac{1}{4}(1-\cos\theta)^2\delta_{ij}.
\label{nearlythere}
\ee

This result can be best interpreted when expressing $\theta$ in terms of $x^1$ and $x^2$. From (\ref{firsttrans}) we get
\be
\theta=\arctan\frac{\sqrt{x^2+y^2}}{z},
\ee
and from the stereographic projection formula (\ref{stereo2}) we know that
\be
z=a\left(1-\frac{8a^2}{L^2}\right).
\ee
Combining these coordinate transformations and inserting them into (\ref{nearlythere}) leads to the complete averaged metric in the stereographic projection plane of the south pole,
\be
\av{g_{ij}(x^1,x^2)}=\left(\frac{2a}{L}\right)^4\delta_{ij}.
\ee

Comparing this to the original unaveraged metric (\ref{metric}) leads to the conclusion that the averaging process preserves the plain two-sphere. Considering that this is the two-dimensional space of constant positive curvature, this is in complete agreement with our requirements on a reasonable averaging process which would be expected to leave smooth manifolds invariant.\\

An obvious question to ask is whether this result also holds for the space of constant negative curvature, namely the hyperbolic plane $(H^2)$, and the space of constant zero curvature, the flat plane $(\mathbbm{R}^2)$. The investigation of this will be the topic of the next sections.

\section{Averaging the Hyperbolic Plane}

\subsection{The Stereographic Projection of the Hyperbolic Plane}

In contrast to the two-sphere, the two-dimensional space of constant negative curvature cannot be embedded in a three-dimensional Euclidean space. While this space can be embedded in a five-dimensional Euclidean space this does not help its visualization. On the other hand, it can be embedded in a three-dimensional Minkowski space with the line element $ds^2=dx^2+dy^2-dz^2$, where it becomes a hyperboloid of two sheets. The parameter representation of the embedded hyperbolic plane is given by 
\be
x&=a\sinh\theta\cos\phi,\\
y&=a\sinh\theta\sin\phi,\\
z&=a\cosh\theta,
\label{newfirsttrans}
\ee

\begin{figure}[t]
\begin{center}
\includegraphics[width=0.6\textwidth]{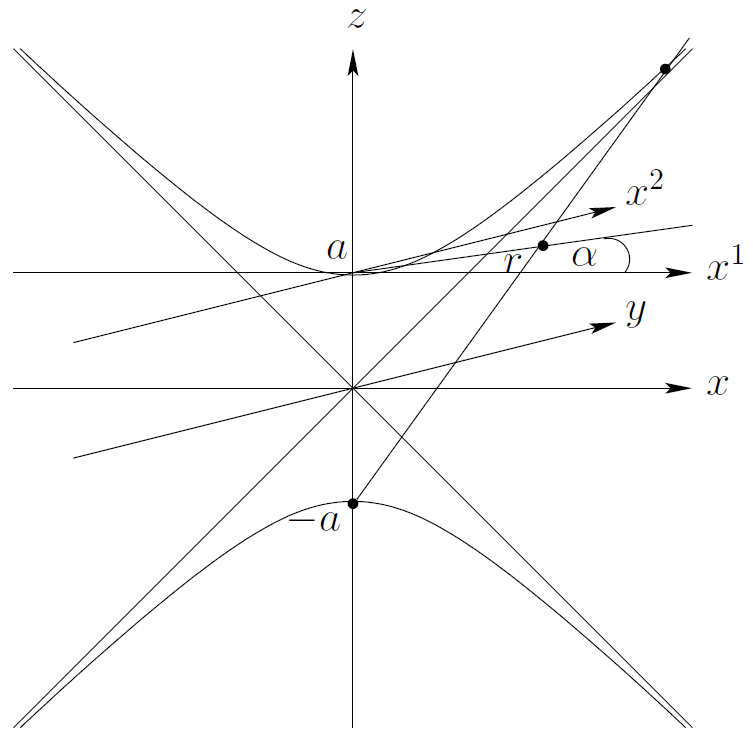}
\caption{Stereographic projection of the hyperbolic plane into the $(x^1x^2)$-plane}
\label{projection2}
\end{center}
\end{figure}

where $a=$const. The associated line element has the form
\be
ds^2=a^2d\theta^2+a^2\sinh^2\theta d\phi^2.
\ee
We will apply the analogue of the stereographic projection we used in the case of the two-sphere, illustrated in Figure \ref{projection}. Concentrating on the upper sheet of the hyperboloid, we project it from the maximum point of the lower sheet onto the plane which intersects the minimum point of the upper sheet. This is illustrated in Figure \ref{projection2}. The corresponding coordinate transformation is
\be
\alpha&=\phi,\\
r&=2a\tanh\left(\frac{\theta}{2}\right),
\label{newsecondtrans}
\ee
and, in terms of the new coordinates, the line element is
\be
ds^2=\frac{16a^4}{(4a^2-r^2)^2}(dr^2+r^2d\alpha^2).
\ee
Since the angle $\alpha$ is undetermined for $r=0$, we again convert to Cartesian coordinates,
\be
x^1&=r\cos\alpha,\\
x^2&=r\sin\alpha,
\label{newthirdtrans}
\ee
which yields the line element
\be
ds^2=\frac{16a^4}{(4a^2-(x^1)^2-(x^2)^2)^2}((dx^1)^2+(dx^2)^2).
\ee
The metric and its inverse are then the same as for the two-sphere (\ref{metric}),
\be
g_{ij}=\left(\frac{2a}{L}\right)^4\delta_{ij}\qquad\mathrm{and}\qquad g^{ij}=\left(\frac{L}{2a}\right)^4\delta^{ij},
\label{newmetric}
\ee
except that here $L^2$ is defined as
\be
L^2=4a^2-(x^1)^2-(x^2)^2.
\label{newL}
\ee
The Christoffel symbols associated with the metric (\ref{newmetric}) are then those for the sphere (\ref{christoffel}) apart from the new definition of $L^2$ and a minus sign,
\be
\Gamma^i_{jk}=\frac{2}{L^2}\left(x_j\delta^i_k+x_k\delta^i_j-x^ig_{jk}\right),
\label{newchristoffel}
\ee
and the geodesic equation is hence of the form
\be
L^2~\frac{d^2z^i}{d\tau^2}+4z_j\frac{dz^j}{d\tau}\frac{dz^i}{d\tau}-2z^ig_{jk}\frac{dz^j}{d\tau}\frac{dz^k}{d\tau}=0.
\ee
For the initial conditions $z^i(0)=0$ and $dz^i/d\tau(0)=\mathrm{const.}$ the solution to this equation is given by
\be
z^i(\tau)=2a\tanh\left(\frac{\tau}{2a}\right)\frac{dz^i}{d\tau}(0).
\label{newgeo}
\ee
Again, we find that if $\tau$ is equal the arc length, the initial conditions have to fulfill the identity
\be
\left(\frac{dz^1}{d\tau'}(0)\right)^2+\left(\frac{dz^1}{d\tau'}(0)\right)^2=1,
\ee
and we introduce an angle $\gamma$ such that the initial conditions for the geodesics are given by
\be
\frac{dz^1}{d\tau}(0)=\cos\gamma\qquad\mathrm{and}\qquad\frac{dz^2}{d\tau}(0)=\sin\gamma.
\label{newinitialcond}
\ee
Thus, the geodesic coordinates, which we need in order to carry out the integral in the averaging process (\ref{avprocess}), are given by
\be
z^1(\tau,\gamma)&=2a\tanh\left(\frac{\tau}{2a}\right)\cos\gamma\\
z^2(\tau,\gamma)&=2a\tanh\left(\frac{\tau}{2a}\right)\sin\gamma,
\label{newgeocoord}
\ee
where $0\leq\gamma<2\pi$ and $0\leq\tau<\infty$.

\subsection{The Result of the Averaging Process}

To apply the averaging process (\ref{avprocess}) we still need to determine the connector  (\ref{connectordef}) and the maximally smooth dyad field. In terms of the geodesic coordinates (\ref{newgeocoord}) the hatted connector for the projected hyperbolic plane is
\be
\widehat{V}_j{}^i(0,\tau;{\cal C}_{\tau 0})=\cosh^{-2}\left(\frac{\tau}{2a}\right)\delta^i_j.
\label{newhattedconnector}
\ee
To calculate the maximally smooth dyad field we make the same ansatz as for the two-sphere (\ref{startingfield}) except with the new definition of $L^2$. Varying the action gives again the differential equation
\be
\left(\del_i\del_k\phi\right)\delta^{ik}=0\quad\mathrm{in\ the\  area\ }R\qquad\mathrm{and}\qquad\frac{\del\phi}{\del n}=0\quad\mathrm{on\ }\del R,
\label{newneumann}
\ee
where $\del_R$ is the geodesic region of radius $r$,
\be
\alpha^1(\gamma)&=2a\tanh\left(\frac{r}{2a}\right)\cos\gamma,\\
\alpha^2(\gamma)&=2a\tanh\left(\frac{r}{2a}\right)\sin\gamma.
\ee
The equation (\ref{newneumann}) is again solved by $\phi(x^1,x^2)=0$. In terms of the geodesic coordinates (\ref{newgeocoord}) the maximally smooth dyad field is thus given by
\be
E^a{}_i(\gamma,\tau)=\cosh^2\left(\frac{\tau}{2a}\right)\delta^a{}_i.
\ee
Inserting this and the hatted connector (\ref{newhattedconnector}) into the averaging process (\ref{avprocess}), and  expressing the other quantities in terms of $\gamma$ and $\tau$ as well, the averaged dyad field at the origin becomes:
\be
\av{E^a{}_j(0,0)}=\int_0^r\int_0^{2\pi} f(z(\gamma,\tau))\widehat{V}_j{}^i(z(\gamma,\tau))E^a{}_i(\gamma,\tau)\sqrt{g(\gamma,\tau)}~d\tau d\gamma=\delta^a_j,
\ee
where we used again the fact that the averaging function is normalized to unity over all space. Hence, the averaging process returns the following averaged metric at the origin:
\be
\av{{g}_{ij}(0,0)}=\av{E^a{}_i(0,0)}\av{E^b{}_j(0,0)}\delta_{ab}=\delta_{ij}.
\ee
As before we need the coordinate transformation to the stereographic projection plane of any other point of the embedded hyperbolic plane and back, as well as the metric transformation law.\\

Rotations that map the hyperbolic plane into itself are different from the rotations that map the two-sphere into itself. Having embedded the hyperbolic plane into a three-dimensional Minkowski space, we have to account for the minus sign in the line element $ds^2=dx^2+dy^2-dz^2$. Therefore, rotations around the z-axis are the same as in the case of the Euclidean space, but rotations around the x-axis and y-axis have to be replaced by hyperbolic rotations. The rotation matrix $D$, which rotates the point ${\bf x}_0=(x_0,y_0,z_0)=(a\sinh\theta_0\cos\phi_0,a\sinh\theta_0\sin\phi_0,a\cosh\theta_0)$ to the minimum point of the upper sheet of the embedded hyperbolic plane $(0,0,a)$, is again constructed from the three rotation matrices $D_zD_yD_z^{-1}$. The first matrix, $D_z$, rotates the hyperbolic plane around the $z$-axis until the point ${\bf x}_0$ intersects with the $(xz)$-plane. The second, $D_y$, rotates the hyperbolic plane around the $y$-axis until the point $D_z^{-1}{\bf x}_0$ aligns with the minimum point, and the third, $D_z^{-1}$, rotates the hyperbolic plane back to the original orientation. Hence, the rotation $D=D_zD_yD_z^{-1}$ is
\begin{gather}
D=\begin{pmatrix}\cos\phi_0&-\sin\phi_0&0\\\sin\phi_0&\cos\phi_0&0\\0&0&1\end{pmatrix}\begin{pmatrix}\cosh\theta_0&0&\sinh\theta_0\\0&1&0\\\sinh\theta_0&0&\cosh\theta_0\end{pmatrix}\begin{pmatrix}\cos\phi_0&\sin\phi_0&0\\-\sin\phi_0&\cos\phi_0&0\\0&0&1\end{pmatrix}.
\end{gather}
In this case, the inverse stereographic projection from $(x^1,x^2)$ to the coordinates of the hyperbolic plane ${\bf x}=(x,y,z)$ is given by
\be
x=\frac{4 a^2x^1}{L^2},\qquad y=\frac{4 a^2x^2}{L^2},\qquad z=a\left(\frac{8 a^2}{L^2}-1\right),
\ee
and the stereographic projection of the rotated hyperbolic plane with the coordinates $(\widetilde{x},\widetilde{y},\widetilde{z})$ into the tangent plane of the new minimum point with the coordinates $(\widetilde{x}^1,\widetilde{x}^2)$ is
\be
\widetilde{x}^1=\frac{2a\widetilde{x}}{(a+\widetilde{z})},\qquad\widetilde{x}^2=\frac{2a\widetilde{y}}{(a+\widetilde{z})}.
\ee
The complete coordinate transformation from the coordinates $(x^1,x^2)$ to the coordinates $(\widetilde{x}^1,\widetilde{x}^2)$ and vice versa is given in appendix \ref{hyperappend}.\\

In terms of the coordinates $(x^1,x^2)$ the origin of the $(\widetilde{x}^1\widetilde{x}^2)$-plane is
\be
(x_0^1,x_0^2)=\left(2a\tanh\left(\frac{\theta_0}{2}\right)\cos\phi_0,2a\tanh\left(\frac{\theta_0}{2}\right)\sin\phi_0\right).
\label{neworigin}
\ee
Since the rotation axis associated with $D$ is in the $xy$-plane and therefore does not intersect with the embedded hyperbolic plane, the fixed points of the coordinate transformations (\ref{newcoordtrans}) and (\ref{newcoordtrans2}) are not as easy to find as in the case of the sphere. We calculate the M\"obius transformation by computing (\ref{longway}) and comparing the coefficients. In this way we find the the M\"obius transformation
\be
\widetilde{z}=\frac{z-2a\tanh(\theta_0/2)(\cos\phi_0+i\sin\phi_0)}{(1/2a)\tanh(\theta_0/2)(-\cos\phi_0+i\sin\phi_0)z+1}.
\ee
For the transformation of the metric we use the Jacobi matrix elements of the coordinate transformation (\ref{newcoordtrans2}), which are given in (\ref{newcoordder}). Inserting them along with the metric (\ref{newmetric}) into the transformation law, we find the metric in terms of the coordinates $(\widetilde{x}^1,\widetilde{x}^2)$,
\be
\widetilde{g}_{kl}(\widetilde{x}^1,\widetilde{x}^2)=\frac{\del x^i}{\del\widetilde{x}^k}\frac{\del x^j}{\del\widetilde{x}^l}\left(\frac{2a}{L}\right)^4\delta_{ij}=\left(\frac{2a}{\widetilde{L}}\right)^4\delta_{kl}.
\ee
As for the two-sphere the metric is invariant under the coordinate transformation and we already know the result of the averaging process in the rotated coordinate system:
\be
\av{\widetilde{g}_{kl}(0,0)}=\delta_{kl}. 
\label{newaveragedtransmetric}
\ee
For the inverse transformation of the averaged metric into the original coordinates $(x^1,x^2)$ we use the Jacobi matrix elements of the coordinate transformation (\ref{newcoordtrans}), which are given in (\ref{newcoordder2}). Inserting them along with the averaged metric at the origin of the transformed coordinate system into (\ref{metricbacktrans}) returns the averaged metric at the point $(x_0^1,x_0^2)$, which is given by (\ref{neworigin}):
\be
\av{g_{ij}(x_0^1,x_0^2)}&=\frac{64a^4}{(L^2\!+\!(8a^2\!-\!L^2)\cosh\theta_0\!-\!4a\sinh\theta_0(x^1\cos\phi_0\!+\!x^2\sin\phi_0))^2}\bigg|_{(x^1,x^2)=(x_0^1,x_0^2)}\delta_{ij}\\
&=\frac{1}{4}(1+\cosh\theta_0)^2\delta_{ij}.
\ee
Therefore, the complete averaged metric in the $(x^1x^2)$-plane is
\be
\av{g_{ij}(x^1,x^2)}=\frac{1}{4}(1+\cosh\theta)^2\delta_{ij}.
\label{newnearlythere}
\ee
For the correct interpretation of this result we express $\theta$ again in terms of $x^1$ and $x^2$ by first transforming it to $x,y$ and $z$, which follows from (\ref{newfirsttrans}) ,
\be
\theta=\arctanh\frac{\sqrt{x^2+y^2}}{z},
\ee
and then applying the $z$-component of the stereographic projection
\be
z=a\left(\frac{8a^2}{L^2}-1\right).
\ee
The result is 
\be
\av{g_{ij}(x^1,x^2)}=\left(\frac{2a}{L}\right)^4\delta_{ij}.
\ee
This equals the unaveraged metric (\ref{newmetric}). We conclude that the averaging process preserves not only the plain two-sphere but also the hyperbolic plane.

\section{Averaging the Flat Plane}

For the sake of completeness we present the case of the flat plane ($\mathbbm{R}^2$) explicitly, although it follows from the two previous cases for $a\rightarrow\infty$. Due to its simplicity the flat plane does not need to be embedded in any three-dimensional space. The line element is 
\be
ds^2=(dx^1)^2+(dx^2)^2,
\ee
and the metric and its inverse are
\be
g_{ij}=\delta_{ij}\qquad\mathrm{and}\qquad g^{ij}=\delta^{ij},
\label{flatmetric}
\ee
where $i,j=1,2$. All Christoffel symbols vanish and therefore the geodesic equation is
\be
\frac{d^2z^i}{d\tau^2}=0.
\ee
For the initial conditions $z^i(0)=0$ and $dz^i/d\tau(0)=0$ the solution to the geodesic equation is the bundle of straight lines through the origin,
\be
z^i(\tau)=\frac{dz^i}{d\tau}(0)~\tau.
\ee
The parameter $\tau$ is equals the arc length if
\be
\left(\frac{dz^1}{d\tau}(0)\right)^2+\left(\frac{dz^2}{d\tau}(0)\right)^2=1,
\ee
and we again introduce an angle $\gamma$ such that the initial conditions for the geodesics are given by
\be
\frac{dz^1}{d\tau}(0)=\cos\gamma\qquad\mathrm{and}\qquad\frac{dz^2}{d\tau}(0)=\sin\gamma.
\ee
The geodesic coordinates are given by
\be
z^1(\tau,\gamma)&=\tau\cos\gamma\\
z^2(\tau,\gamma)&=\tau\sin\gamma,
\label{flatgeocoord2}
\ee
where $0\leq\gamma<2\pi$ and $0\leq\tau<\infty$. In terms of these coordinates the maximally smooth dyad field and the hatted connector are 
\be
E^a{}_i(\gamma,\tau)=\delta^a{}_i\qquad\mathrm{and}\qquad\widehat{V}_j{}^i(z(\gamma,\tau))=\delta^i_j.
\ee
Inserting them into the averaging process results in the averaged dyad field,
\be
\av{E^a{}_j(0,0)}=\delta^a{}_j.
\ee
Therefore, the averaging process returns the metric
\be
\av{{g}_{ij}(0,0)}=\av{E^a{}_i(0,0)}\av{E^b{}_j(0,0)}\delta_{ab}=\delta_{ij}
\ee
at the origin. The coordinate transformation is a pure translation from the plane with the coordinates $(x^1,x^2)$ to the plane with the coordinates $(\widetilde{x}^1,\widetilde{x}^2)$, and the reference point $(x_0^1,x_0^2)$ in the untransformed coordinates becomes the origin $(0,0)$ of the transformed coordinates. Therefore, the coordinate transformations are
\be
\widetilde{x}^1&=x^1+x_0^1,\qquad\qquad\qquad x^1=\widetilde{x}^1-x_0^1,\\
\widetilde{x}^2&=x^2+x_0^2,\qquad\qquad\qquad x^2=\widetilde{x}^2-x_0^2.
\label{flatcoordtrans}
\ee
This can be expressed as a M\"obius transformation for $z=x_0^1+ix_0^2$ and $\widetilde{z}=\widetilde{x}_0^1+i\widetilde{x}_0^2$,
\be
\widetilde{z}=\frac{1\cdot z\pm (x_0^1+ix_0^2)}{0\cdot z+1}.
\ee
The Jacobi matrix elements for the two transformations (\ref{flatcoordtrans}) are given by
\be
\frac{\del\widetilde{x}^1}{\del x^1}=1,\qquad\frac{\del\widetilde{x}^1}{\del x^2}=0,\qquad\qquad\qquad\frac{\del x^1}{\del \widetilde{x}^1}=1,\qquad\frac{\del x^1}{\del \widetilde{x}^2}=0,\\
\frac{\del\widetilde{x}^2}{\del x^1}=0,\qquad\frac{\del\widetilde{x}^2}{\del x^2}=1,\qquad\qquad\qquad\frac{\del x^2}{\del \widetilde{x}^1}=0,\qquad\frac{\del x^2}{\del \widetilde{x}^2}=1.
\ee
From the metric transformation law (\ref{generalmetrictrans}) we find the metric in the $\widetilde{x}^1\widetilde{x}^2$-coordinates, 
\be
\widetilde{g}_{kl}(\widetilde{x}^1,\widetilde{x}^2)=\frac{\del x^i}{\del\widetilde{x}^k}\frac{\del x^j}{\del\widetilde{x}^l}g_{kl}(x^1,x^2)=\delta_{kl},
\ee
and from the back transformation (\ref{metricbacktrans}) we get the averaged metric at the point $(x_0^1,x_0^2)$:
\be
\av{g_{ij}(x_0^1,x_0^2)}=\frac{\del\widetilde{x}^k}{\del x^i}\bigg|_{(x^1,x^2)=(x_0^1,x_0^2)}\frac{\del\widetilde{x}^l}{\del x^j}\bigg|_{(x^1,x^2)=(x_0^1,x_0^2)}\av{\widetilde{g}_{kl}(0,0)}=\delta_{ij}.
\ee
The complete averaged metric in the coordinates $(x^1,x^2)$ equals the unaveraged metric (\ref{flatmetric}),
\be
\av{g_{ij}(x^1,x^2)}=\delta_{ij}.
\ee\\

In summary, we have shown in this chapter that the averaging process preserves all two-dimensional spaces of constant curvature. The obvious question to ask now is what effect the averaging process would have on a space of contant curvature, say a two-sphere, if we added a small perturbation. This will the topic of the next chapter.

%% file: twosphere.tex
\chapter{The Perturbed Two-Sphere}
\label{twosphere}

In the previous chapter we tested our averaging formulation on the conceptually trivial examples of two-dimensional spaces of constant curvature and verified that it leaves such spaces invariant. In this chapter we extend our discussion of the averaging process in two dimensions to the more interesting case of a sphere whose metric is slightly perturbed and derive the expressions for a general perturbation function. A specific example is discussed in chapter \ref{gauss}.  

\section{The Perturbed Metric}

To begin, we recall that embedding the unperturbed sphere into a three-dimensional Euclidean space enabled us to specify the stereographic projection from the Euclidean coordinates of the sphere ${\bf x}=(x,y,z)$ to the coordinates of the projection plane $(x^1,x^2)$. The associated coordinate transformation (\ref{stereo2}) was 
\be
x&=4a^2x^1/L^2,\\
y&=4a^2x^2/L^2,\\
z&=a(1-8a^2/L^2),
\ee
where $a$ denotes the radius of the sphere and $L^2$ is
\be
L^2=(x^1)^2+(x^2)^2+4a^2.
\ee
We now apply a small perturbation defined by
\be
x_p&=(1+\eta f)x=(1+\eta f)4a^2x^1/L^2,\\
y_p&=(1+\eta f)y=(1+\eta f)4a^2x^2/L^2,\\
z_p&=(1+\eta f)z=(1+\eta f)a(1-8a^2/L^2),
\ee
with a space-dependent perturbation function $f=f(x^1,x^2)$ and a small parameter $\eta>0$. Up to first order in $\eta$ the line element takes the form
\be
ds^2=dx_p^2+dy_p^2+dz_p^2=\left(\frac{16a^4}{L^4}+\eta\frac{32a^4f}{L^4}\right)\left(dx^1\right)^2+\left(\frac{16a^4}{L^4}+\eta\frac{32a^4f}{L^4}\right)\left(dx^2\right)^2,
\ee
and hence the metric and its inverse are given by
\be
(g_{P})_{ij}=(1+2\eta f)\left(\frac{2a}{L}\right)^4\delta_{ij}\qquad\mathrm{and}\qquad (g_P)^{ij}=(1-2\eta f)\left(\frac{L}{2a}\right)^4\delta^{ij},
\label{pertmetric}
\ee                                                                                                                        
where the indices take the values $i,j=1,2$. 

\section{The Perturbed Geodesics}

\begin{figure}
\begin{center}\includegraphics[width=0.6\textwidth]{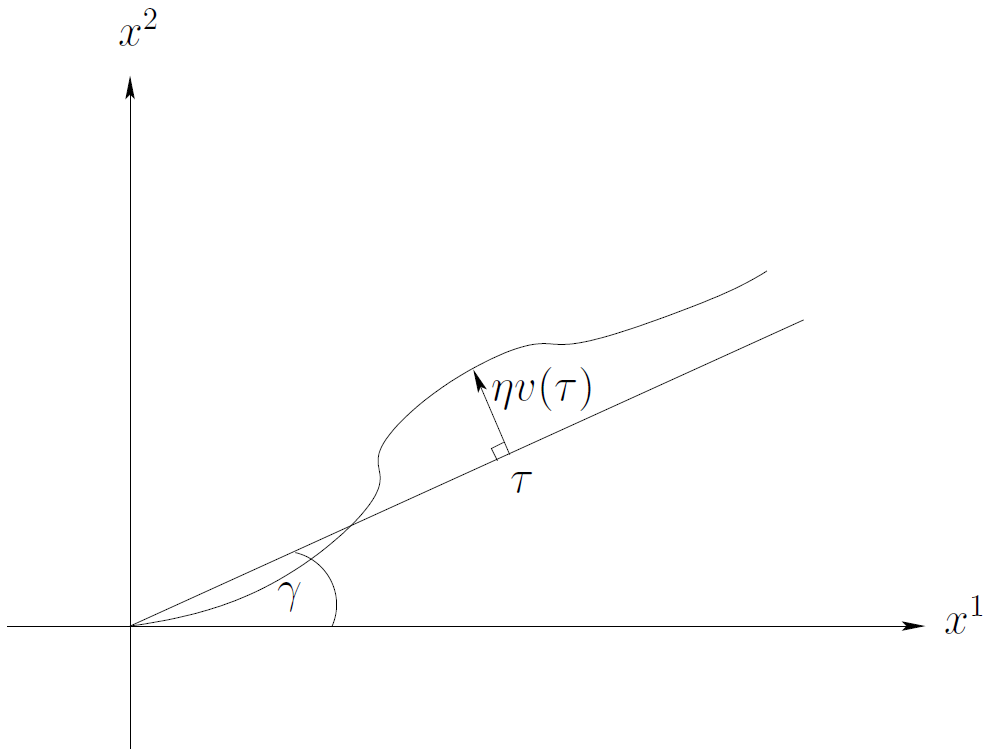}\end{center}
\caption{Parametrization of the perturbed geodesic.}
\label{transversal}
\end{figure}

In the case of the unperturbed two sphere the solution to the geodesic equation for the initial conditions $z^i(0)=0$ and $dz^i/d\tau(0)=$ const. (\ref{geo}) was given by
\be
z^i(\tau)=2a \tan\left(\frac{\tau}{2a}\right)\frac{dz^i}{d\tau}(0).
\label{unpertgeo} 
\ee
Since the imposed perturbation accounts for a change in the arc length of the geodesic, we are forced to use the general geodesic equation (\ref{pertgeoeq}), where the parametrization $\tau$ of the geodesic is not necessarily the arc~length~$s$. This equation is invariant under a reparametrization of the form $\tau'=\tau+\eta w(\tau)$ with a continuously differentiable function $w(\tau)$. That allows us to make the following ansatz for the parametrization of the perturbed geodesic:
\be
z_P^i(\tau)=z^i(\tau)+\eta v(\tau)\epsilon_{ij}\frac{dz^j}{d\tau}(\tau).
\label{pertgeo}
\ee
Here $v(\tau)$ is a twice continuous differentiable function which describes the longitudinal deviation of the geodesic caused by the perturbation. This parametrization is also visualized in Figure \ref{transversal}.
The arc length of the perturbed geodesic up to first order in $\eta$ is given by
\be
s(\tau)=\int_0^\tau d\tau'\sqrt{(g_P)_{ij}(z(\tau'))\frac{dz_P^i}{d\tau'}\frac{dz_P^j}{d\tau'}}=\tau+\eta \int_0^\tau f(\tau')d\tau',
\label{arclength}
\ee
with the inverse function
\be
\tau(s)=s-\eta\int_0^s f(s')ds'.
\label{invarclength}
\ee
Inserting the arc length (\ref{arclength}) and the derivative of (\ref{pertgeo}),
\be
\frac{dz_P^i}{d\tau}(\tau)=\frac{dz^i}{d\tau}(\tau)+\eta\frac{dv}{d\tau}(\tau)\epsilon^i{}_j\frac{dz^j}{d\tau}(\tau)+\eta v(\tau)\epsilon^i{}_j\frac{d^2z^j}{d\tau^2}(\tau),
\ee
into the equation for the perturbed geodesic (\ref{pertgeoeq}) yields the differential equation of a driven harmonic oscillator for the function $v(\tau)$,
\be
\frac{d^2v(\tau)}{d\tau^2}+\frac{v(\tau)}{a^2}=\frac{h(\tau)}{\cos^2(\frac{\tau}{2a})}.
\label{pertgeov}
\ee
Here, the driving force $h(\tau)$ has the dependence on the perturbation function $f(\tau)$
\be
h(\tau)=\frac{\del f}{\del x^2}\bigg|_{(x^1,x^2)=(z^1(\tau),z^2(\tau))}\frac{\del z_P^1}{d\tau}(0)-\frac{\del f}{\del x^1}\bigg|_{(x^1,x^2)=(z^1(\tau),z^2(\tau))}\frac{\del z_P^2}{d\tau}(0).
\label{hfunction}
\ee

The initial conditions are chosen such that the perturbed and unperturbed geodesics and their derivatives coincide at the origin, 
\be
z^i_P(0)=z^i(0)\qquad\mathrm{and}\qquad\frac{dz_P^i}{d\tau}(0)=\frac{dz^i}{d\tau}(0).
\ee
Therefore we are looking for solutions to the differential equation (\ref{pertgeov}) for the initial conditions
\be
 v(0)=0\qquad\mathrm{and}\qquad\frac{dv}{d\tau}(0)=0.
\ee
The general solution of the differential equation (\ref{pertgeov}) is composed of the solution of the homogeneous equation and a particular solution of the inhomogeneous equation, which can be found by variation of constants. It is given by \be
v(\tau)&=A\cos\left(\frac{\tau}{a}\right)+B\sin\left(\frac{\tau}{a}\right)\\
&-\frac{a}{2}\cos\left(\frac{\tau}{a}\right)\int_0^\tau\frac{1}{\sin(\frac{\tau'}{a})}\frac{h(\tau')}{\cos^2(\frac{\tau'}{2a})}d\tau'+\frac{a}{2}\sin\left(\frac{\tau}{a}\right)\int_0^\tau\frac{1}{\cos(\frac{\tau'}{a})}\frac{h(\tau')}{\cos^2(\frac{\tau'}{2a})}d\tau',
\ee
with $A,B=$ const. For the initial conditions $v(0)=0$ and $dv/d\tau(0)=0$ the constants $A$ and $B$ vanish. Hence, the solution is 
\be
v(\tau)=-\frac{a}{2}\cos\left(\frac{\tau}{a}\right)\int_0^\tau\frac{1}{\sin(\frac{\tau'}{a})}\frac{h(\tau')}{\cos^2(\frac{\tau'}{2a})}d\tau'+\frac{a}{2}\sin\left(\frac{\tau}{a}\right)\int_0^\tau\frac{1}{\cos(\frac{\tau'}{a})}\frac{h(\tau')}{\cos^2(\frac{\tau'}{2a})}d\tau'.
\ee  

\section{The Maximally Smooth Dyad Field}

Here we perform a calculation analogous to that in section \ref{smoothdyad}. In this case, we will derive the differential equation needed to determine the maximally smooth dyad field for a general metric and then specialise to the case of the perturbed sphere.\\

We decompose the metric into dyads according to
\begin{equation}
g_{ij}(x)=E^a{}_i(x)E^b{}_j(x)\delta_{ab}.
\label{orthonormality}
\end{equation}
We choose a reference dyad field $\widetilde{E}^b{}_i(x)$ and express the class of tetrads that obey (\ref{orthonormality}) as a rotation by a position dependent angle $\phi(x^1,x^2)$,
\begin{equation}
E^a{}_i(x)=U^a{}_b(x)\widetilde{E}^b{}_i(x)\qquad\mathrm{where}\qquad U^a{}_b(x)=\begin{pmatrix}\cos\phi(x)&\sin\phi(x)\\-\sin\phi(x)&\cos\phi(x)\end{pmatrix}.
\end{equation}
The partial derivative of the rotation matrix $U_{ab}$ is again
\begin{equation}
\partial_i U^a{}_b(x)=\epsilon^a{}_cU^c{}_b(x)(\partial_i \phi(x)),
\end{equation}
The generally covariant derivative of the dyad field then becomes
\be
D_i E^a{}_j&=(\partial_iU^a{}_b)\widetilde{E}^b{}_j+U^a{}_b(D_i\widetilde{E}^b{}_j)=\epsilon^a{}_cU^c{}_b(\partial_i \phi)\widetilde{E}^b{}_j+U^a{}_b(\partial_i\widetilde{E}^b{}_j)-U^a{}_b\Gamma^k_{ij}\widetilde{E}^b{}_k.
\ee
The Lagrangian for the maximally smooth dyad field (\ref{smoothlagrangian}) is
\be
\mathcal{L}_\mathrm{MS}&=(D_iE^a{}_j)(D_kE^b{}_l)g^{ik}g^{jl}\delta_{ab}=\epsilon^a{}_eU^e{}_c\widetilde{E}^c{}_j\epsilon^b{}_fU^f{}_d(\del_i\phi)(\del_k\phi)\widetilde{E}^d{}_lg^{ik}g^{jl}\delta_{ab}\\
&+2U^a{}_c(D_i\widetilde{E}^c{}_j)\epsilon^b{}_fU^f{}_d(\del_k\phi)\widetilde{E}^d{}_lg^{ik}g^{jl}\delta_{ab}+U^a{}_cU^b{}_d(D_i\widetilde{E}^c{}_j)(D_k\widetilde{E}^d{}_l)g^{ik}g^{jl}\delta_{ab}.
\ee
This can be simplified further by applying the identities (\ref{id1}-\ref{id3}) and (\ref{orthonormality}), which leaves
\be
\mathcal{L}_\mathrm{MS}=2(\partial_i \phi)(\partial_k\phi)g^{ik}+2(\partial_k \phi)(D_i\widetilde{E}^c{}_j)\widetilde{E}^d{}_l\epsilon_{cd}g^{ik}g^{jl}+(D_i\widetilde{E}^c{}_j)(D_k\widetilde{E}^d{}_l)\delta_{cd}g^{ik}g^{jl}.
\ee
We define the vector field
\be
u^k=(D_i\widetilde{E}^c{}_j)\widetilde{E}^d{}_l\epsilon_{cd}g^{ik}g^{jl}
\label{vectorfield}
\ee
and set the variation of $S=\int d^2x\sqrt{g}~\mathcal{L}_\mathrm{MS}$ with respect to $\phi$ equal to zero. As shown in section \ref{smoothdyad}, the result is a differential equation on $R$,
\be
g^{ik}\left(\partial_i\partial_k\phi\right)&=-\frac{1}{2}D_ku^k,
\label{DGL2}
\ee
with boundary conditions of Neumann type on $\partial R$,
\be
\frac{\partial\phi}{\partial n}=-\frac{1}{2}n_ku^k.
\label{DGL1}
\ee

To investigate whether the solution $\phi(x^1,x^2)$ to this differential equation depends on the choice of reference dyad field $\widetilde{E}^b{}_i(x)$, we choose a different reference field $\widehat{E}^c{}_i(x)$ and find the differential equation
\be
g^{ik}\left(\partial_i\partial_k\hat{\phi}\right)&=-\frac{1}{2}D_k\hat {u}^k,
\label{hatDGL1}
\ee
with the boundary condition 
\be
\frac{\partial\hat{\phi}}{\partial n}=-\frac{1}{2}n_k\hat{u}^k.
\label{hatDGL2}
\ee
Since the reference field $\widehat{E}^c{}_i(x)$ can be expressed as a rotation of the reference field $\widetilde{E}^b{}_i(x)$ by a position dependent angle $\alpha(x^1,x^2)$,
\be
\widehat{E}^c{}_i(x)=\widehat{U}^c{}_b\left(\alpha(x)\right)\widetilde{E}^b{}_i(x),
\ee
the vector field $\hat{u}^k$ becomes
\be
\hat{u}^k&=(D_i\widehat{E}^c{}_j)\widehat{E}^d{}_l\epsilon_{cd}(g_P)^{ik}(g_P)^{jl}\\
&=\left(\epsilon^c{}_e\widehat{U}^e{}_b(\del_i\alpha)\widetilde{E}^b{}_j+\widehat{U}^c{}_b(D_i\widetilde{E}^b{}_j)\right)\widehat{U}^d{}_f\widetilde{E}^f{}_l\epsilon_{cd}(g_P)^{ik}(g_P)^{jl}\\
&=2(g_P)^{ik}(\del_i\alpha)+(D_i\widetilde{E}^b{}_j)\widetilde{E}^f{}_l\epsilon_{bf}(g_P)^{ik}(g_P)^{jl}\\
&=2(g_P)^{ik}(\del_i\alpha)+u^k.
\ee
When inserting this in the differential equation (\ref{hatDGL1}) and the boundary condition (\ref{hatDGL2}), we can add the $\alpha$-dependent term to the left-hand side,
\be
g^{ik}\left(\partial_i\partial_k(\hat{\phi}+\alpha)\right)&=-\frac{1}{2}D_ku^k,
\ee
and
\be
\frac{\partial(\hat{\phi}+\alpha)}{\partial n}=-\frac{1}{2}n_ku^k.
\ee
The maximally smooth dyad field is a rotation of the reference field $\widehat{E}^c{}_i(x)$ by the solution $\hat{\phi}(x^1,x^2)$ to this differential equation. Since $\widehat{E}^c{}_i(x)$ is a rotation of the dyad $\widetilde{E}^b{}_i(x)$ by the angle $\alpha(x^1,x^2)$, we can also express the maximally smooth dyad field as a rotation of $\widetilde{E}^b{}_i(x)$ by the angle $\phi=(\hat{\phi}+\alpha)$. The differential equation is then identical to (\ref{DGL2}) and (\ref{DGL1}) and we conclude that the definition of the maximally smooth dyad field is independent of the choice of reference dyad field.\\

To find the maximally smooth dyad field for the perturbed sphere with the metric (\ref{pertmetric}) we first need to calculate the vector field (\ref{vectorfield})
\be
u^k=(D_i\widetilde{E}^c{}_j)\widetilde{E}^d{}_l\epsilon_{cd}(g_P)^{ik}(g_P)^{jl}.
\ee
In the two dimensional case with the general metric 
\be
g_{ij}=\begin{pmatrix}g_{11}&g_{12}\\g_{21}&g_{22}\end{pmatrix},
\ee
we can always choose the refernece field
\be
\widetilde{E}^1{}_i=\begin{pmatrix}\sqrt{g_{11}-(g_{12}^2/g_{22})}\\0\end{pmatrix}\qquad\mathrm{and}\qquad\widetilde{E}^2{}_i=\begin{pmatrix}g_{12}/\sqrt{g_{22}}\\\sqrt{g_{22}}\end{pmatrix}.
\ee
For $g_{11},g_{22}>0$ and $g=\det(g_{ij})>0$ this dyad field is well-defined and obeys
\be
\widetilde{E}^a{}_i\widetilde{E}^b{}_j\delta_{ab}=g_{ij}.
\ee
For the perturbed sphere we therefore choose the reference field to be
\be
\widetilde{E}^1{}_i=\begin{pmatrix}(4a^2/L^2)+\eta(4a^2f/L^2)\\0\end{pmatrix}\qquad\mathrm{and}\qquad\widetilde{E}^2{}_i=\begin{pmatrix}0\\(4a^2/L^2)+\eta(4a^2f/L^2)\end{pmatrix}.
\label{startingdyad}
\ee
Then the vector field (\ref{vectorfield}) becomes
\be
u^1=\frac{L^2x^2}{4a^4}-\eta\left(\frac{L^2fx^2}{2a^4}\!+\!\frac{L^4}{8a^4}\frac{\del f}{\del x^2}\right)\quad\mathrm{and}\quad u^2=-\frac{L^2x^1}{4a^4}+\eta\left(\frac{L^2fx^1}{2a^4}\!+\!\frac{L^4}{8a^4}\frac{\del f}{\del x^1}\right).
\label{ufield}
\ee
The covariant derivative of this vector field vanishes, $D_ku^k=0$, and thus the differential equation (\ref{DGL2}) reduces to
\be
(g_P)^{ik}\left(\partial_i\del_k\phi\right)=\frac{L^4}{16a^4}\left(1-2\eta f\right)\left(\frac{\del^2\phi}{(\del x^1)^2}+\frac{\del^2\phi}{(\del x^2)^2}\right)=0.
\ee
Thus, the differential equation for $\phi(x^1,x^2)$ on $R$ is
\be
\left(\frac{\del^2}{(\del x^1)^2}+\frac{\del^2}{(\del x^2)^2}\right)\phi(x^1,x^2)=\Delta\phi(x^1,x^2)=0.
\label{laplace}
\ee

Recalling the initial conditions for the perturbed geodesics $v(0)=0$ and $dv/d\tau(0)=0$, we can introduce an angle $\gamma$ with $0\leq\gamma<2\pi$ in the same way as for the unperturbed sphere (\ref{newinitialcond}),
\be
\frac{dz_P^1}{d\tau}(0)=\cos\gamma\qquad\mathrm{and}\qquad\frac{dz_P^2}{d\tau}(0)=\sin\gamma.
\label{geowinkel}
\ee
That means in particular that all functions which depend on the initial conditions acquire a dependence on $\gamma$ which must be included in the argument. We will therefore from now on denote $v(\tau)$ and $h(\tau)$ by $v(\tau,\gamma)$ and $h(\tau,\gamma)$. The geodesic coordinates are then given by
\be
z_P^1(\tau,\gamma)&=2a\tan\left(\frac{\tau}{2a}\right)\cos\gamma+\eta\frac{v(\tau,\gamma)}{\cos^2(\frac{\tau}{2a})}\sin\gamma\\
z_P^2(\tau,\gamma)&=2a\tan\left(\frac{\tau}{2a}\right)\sin\gamma-\eta\frac{v(\tau,\gamma)}{\cos^2(\frac{\tau}{2a})}\cos\gamma.
\label{geodesiccoordinates}
\ee

The geodesic region $R$ is defined as the region inside the entirety of points at geodesic distance $r$ from the origin that form $\del R$. The geodesic distance is the  arc length (\ref{arclength}) along the geodesics. In parameter representation the boundary $\del R$ is
\be
\alpha^1(\gamma)&=2a\tan\left(\frac{r}{2a}\right)\cos\gamma-\eta\frac{\cos\gamma}{\cos^2(\frac{r}{2a})}\int_0^rf(s',\gamma)ds'+\eta\frac{v(r,\gamma)}{\cos^2(\frac{r}{2a})}\sin\gamma\\
\alpha^2(\gamma)&=2a\tan\left(\frac{r}{2a}\right)\sin\gamma-\eta\frac{\sin\gamma}{\cos^2(\frac{r}{2a})}\int_0^rf(s',\gamma)ds'-\eta\frac{v(r,\gamma)}{\cos^2(\frac{r}{2a})}\cos\gamma.
\label{areaR}
\ee
An important quantity for the calculation of the Neumann boundary condition is $L^2$. Its value at the boundary $\del R$ is
\be
L^2\big|_{\del R}=\frac{4a^2}{\cos^2(\frac{r}{2a})}-\eta\frac{4a\tan(\frac{r}{2a})}{\cos^2(\frac{r}{2a})}\int_0^rf(s',\gamma)~ds',
\ee
and hence the square root of the determinant of the metric at the boundary is
\be
\sqrt{g_P}\big|_{\del R}=\cos^4\left(\frac{r}{2a}\right)\!+\!\eta 2f(r,\gamma)\cos^4\left(\frac{r}{2a}\right)\!+\!\eta\frac{2}{a}\cos^4\left(\frac{r}{2a}\right)\tan\left(\frac{r}{2a}\right)\int_0^r\!\!f(s',\gamma)ds'.
\ee
From this we can compute the generally covariant normal vector $\nu_k$ to $\del R$,
\be
\nu_k=\sqrt{g_P}\big|_{\del R}~\epsilon_{kl}\frac{\del\alpha^l}{\del\gamma},
\ee
the normal unit vector $n_k$ to $\del R$, 
\be
n_k=\left((g_P)^{ij}(r,\gamma)\nu_i\nu_j\right)^{-1/2}\nu_k,
\ee
and the value of the vector field $u^k$, given in (\ref{ufield}), at the boundary.
The Neumann boundary condition on $\del R$ can then be calculated from (\ref{DGL2}), giving
\be
\frac{\del\phi}{\del n}=\eta~\frac{h(r,\gamma)}{\cos^2\left(\frac{r}{2a}\right)}+\eta~\frac{1}{2a^2}\frac{1}{\cos^2\left(\frac{r}{2a}\right)}~\del_\gamma\left(\int_0^rf(s',\gamma)ds'\right).
\label{randbed}
\ee\\

In summary, the maximally smooth dyad field for the perturbed sphere is given by the solution to the differential equation (\ref{laplace}) on $R$ with the Neumann boundary condition (\ref{randbed}) on $\del R$, where $\del R$ is given by (\ref{areaR}) .

\section{The Perturbed Connector}

Using that
\be
\left(\frac{2a}{L}\right)^4~\bigg|_{(x^1,x^2)=(z^1(\tau,\gamma),z^2(\tau,\gamma))}=\cos^2\left(\frac{\tau}{2a}\right).
\ee
and the perturbed geodesic (\ref{pertgeo}) we find the Christoffel symbols
\be
\Gamma^i_{jk}\left(z_P(\tau,\gamma)\right)&=\frac{1}{a}\sin\left(\frac{\tau}{2a}\right)\cos\left(\frac{\tau}{2a}\right)\left(\frac{dz_P^i}{d\tau}(0)\delta_{jk}-\frac{dz_P^k}{d\tau}(0)\delta_{ij}-\frac{dz_P^j}{d\tau}(0)\delta_{ik}\right)\\
&\quad +\eta~\frac{1}{2a^2}v(\tau)\left(\epsilon_{il}\frac{dz_P^l}{d\tau}(0)\delta_{jk}-\epsilon_{kl}\frac{dz_P^l}{d\tau}(0)\delta_{ij}-\epsilon_{jl}\frac{dz_P^l}{d\tau}(0)\delta_{ik}\right)\\
&\quad +\eta~\left(\frac{\del f}{\del x^k}\delta_{ij}+\frac{\del f}{\del x^j}\delta_{ik}-\frac{\del f}{\del x^i}\delta_{jk}\right)\bigg|_{(x^1,x^2)=(z^1(\tau,\gamma),z^2(\tau,\gamma))}.
\ee
The connector of parallel transport along the geodesic $z_P^i(\tau,\gamma)$ from the origin to the point with parameter~$\tau$ is then
\be
V(\tau,0;{\cal C}_{0\tau})&={\cal P}\exp\left(-\int_0^\tau\frac{dz_P^i}{d\tau}(\tau',\gamma)~{\bf\Gamma}_i\left(z_P(\tau',\gamma)\right)d\tau'\right)\\
&={\cal P}\exp\bigg(\left(\frac{1}{a}\int_0^\tau \tan\left(\frac{\tau'}{2a}\right)d\tau'-\eta\int_0^\tau\frac{\del f}{\del\tau}(\tau',\gamma)~d\tau'\right)\begin{pmatrix}1&0\\0&1\end{pmatrix}\\
&\qquad+\eta~\bigg(-\frac{1}{2a^2}\int_0^\tau \frac{v(\tau',\gamma)}{\cos^2(\frac{\tau'}{2a})}d\tau'+\frac{1}{a^2}\int_0^\tau \tan^2\left(\frac{\tau'}{2a}\right)v(\tau',\gamma)d\tau'\\
&\qquad+\frac{1}{a}\int_0^\tau\tan\left(\frac{\tau'}{2a}\right)\frac{dv}{d\tau}(\tau',\gamma)d\tau'+\int_0^\tau\frac{h(\tau',\gamma)}{\cos^2(\frac{\tau'}{2a})}d\tau'\bigg)\begin{pmatrix}0&1\\-1&0\end{pmatrix}\bigg).
\label{connectorwithoutexpansion}
\ee
The integrand in (\ref{connectorwithoutexpansion}) is of the form $M_0(\tau')+\eta M_1(\tau')$ with two matrices $M_0(\tau')$ and $M_1(\tau')$. Defining the matrix $U_0$ to be
\be
U_0(\tau)={\cal P}\exp\left(\int^\tau d\tau'M_0(\tau')\right),
\ee
we can expand the general identity
\be
{\cal P}\exp\left(\int^\tau\!\!\! d\tau'\left(M_0(\tau')+\eta M_1(\tau')\right)\right)=U_0(\tau)~{\cal P}\exp\left(\eta\!\int^\tau\!\!\!d\tau'U_0^{-1}(\tau')M_1(\tau')U_0(\tau')\right)
\ee
to first order in $\eta$ and find
\be
{\cal P}\exp\left(\int^\tau\!\!\! d\tau'\left(M_0(\tau')+\eta M_1(\tau')\right)\right)=U_0(\tau)~\left({\mathbbm 1}+\eta\int^\tau\!\!\! d\tau'U_0^{-1}(\tau')M_1(\tau')U_0(\tau')\right).
\ee
For the connector (\ref{connectorwithoutexpansion}) the matrix $U_0(\tau)$ is given by
\be
U_0(\tau)&={\cal P}\exp\left(\int^\tau\!\!\! d\tau'M_0(\tau')\right)={\cal P}\exp\left(\left(\frac{1}{a}\int_0^{\tau}\tan\left(\frac{\tau'}{2a}\right)d\tau'\right)\mathbbm{1}\right)\\
&={\cal P}\exp\left(\frac{1}{a}\left(-2a\ln\left(\cos\left(\frac{\tau'}{2a}\right)\right)\right)\bigg|_0^{\tau}\mathbbm{1}\right)=\cos^{-2}\left(\frac{\tau}{2a}\right)\mathbbm{1},
\ee
and therefore the first order expansion of the connector becomes 
\be
V(\tau,0;{\cal C}_{0\tau})&=\cos^{-2}\left(\frac{\tau}{2a}\right)\left(1-\eta f(\tau,\gamma)+\eta f(0,\gamma)\right)\begin{pmatrix}1&0\\0&1\end{pmatrix}\\
&+\eta\cos^{-2}\left(\frac{\tau}{2a}\right)\bigg(\!\!-\frac{1}{2a^2}\int_0^\tau\!\!\frac{v(\tau',\gamma)}{\cos^2\left(\frac{\tau'}{2a}\right)}d\tau'+\frac{1}{a^2}\int_0^\tau\!\!\tan^2\left(\frac{\tau'}{2a}\right)v(\tau',\gamma)d\tau'\\
&\qquad\qquad+\frac{1}{a}\int_0^\tau\!\!\tan\left(\frac{\tau'}{2a}\right)\frac{dv}{d\tau}(\tau',\gamma)d\tau'+\int_0^\tau\!\!\frac{h(\tau',\gamma)}{\cos^2(\frac{\tau'}{2a})}d\tau'\bigg)\begin{pmatrix}0&1\\-1&0\end{pmatrix}.
\ee
Using (\ref{pertgeov}) and partial integration methods this expression can be further simplified for the boundary conditions $v(0,\gamma)=0$ and $dv/d\tau(0,\gamma)=0$:
\be
V(\tau,0;{\cal C}_{0\tau})&=\cos^{-2}\left(\frac{\tau}{2a}\right)\bigg(\left(1-\eta f(\tau,\gamma)+\eta f(0,\gamma)\right)\begin{pmatrix}1&0\\0&1\end{pmatrix}\\
&\qquad\quad+\eta\left(v'(\tau,\gamma)+\frac{1}{a}\tan\left(\frac{\tau}{2a}\right)v(\tau,\gamma)\right)\begin{pmatrix}0&1\\-1&0\end{pmatrix}\bigg).
\label{juleconnector}
\ee
For the averaging process we need the connector from the point with parameter $\tau$ to the origin and hence we need the inverse of this expression to first order in $\eta$, which is
\be
&V(0,\tau;{\cal C}_{\tau 0})=V^{-1}(\tau,0;{\cal C}_{0\tau})\\
&=\left(\cos^4\left(\frac{\tau}{2a}\right)\!\left(1\!+\!2\eta f(\tau)\!-\!2\eta f(0)\right)\right)\!\!\left(\cos^{-2}\left(\frac{\tau}{2a}\right)\!\left(1\!-\!\eta f(\tau)\!+\!\eta f(0)\right)\right)\!\!\begin{pmatrix}1&0\\0&1\end{pmatrix}\\
&-\left(\cos^4\left(\frac{\tau}{2a}\right)\!\left(1\!+\!2\eta f(\tau)\!-\!2\eta f(0)\right)\right)\!\!\left(\eta\cos^{-2}\left(\frac{\tau}{2a}\right)\!\left(v'(\tau)\!+\!\frac{1}{a}\tan\left(\frac{\tau}{2a}\right)v(\tau)\right)\right)\!\!\begin{pmatrix}0&1\\-1&0\end{pmatrix}.\notag
\ee
In components the result can be written as
\be
V^k{}_l(0,\tau;{\cal C}_{\tau 0})\!=\!\cos^2\left(\frac{\tau}{2a}\right)\left(\left(1\!+\!\eta f(\tau)\!-\!\eta f(0)\right)\delta^k{}_l-\eta\left(v'(\tau)+\frac{1}{a}\tan\left(\frac{\tau}{2a}\right)v(\tau)\right)\epsilon^k{}_l\right).\notag
\ee
For the parallel transport of dyads in the averaging process the hatted connector is needed and we find it to be of the form
\be
&\widehat{V}_j{}^i(0,\tau;{\cal C}_{\tau 0})=g_{jk}(0,\gamma)V^k{}_l(0,\tau;{\cal C}_{\tau 0})g^{li}(\tau,\gamma)\\
&=\cos^{-2}\left(\frac{\tau}{2a}\right)\!\!\left(\!\left(1\!+\!\eta f(0,\gamma)\!-\!\eta f(\tau,\gamma)\right)\delta^i_j\!-\!\eta\left(v'(\tau,\gamma)+\frac{1}{a}\tan\left(\frac{\tau}{2a}\right)v(\tau,\gamma)\right)\epsilon^i{}_j\!\right).
\label{sphereconnector}
\ee

\section{Alternative Computation of the Parallel Transport}

When the vector $E^a{}_i$ is parallel transported along the geodesic $z^i(s,\gamma)$, the angle between the vector and the tangent vector of the geodesic is preserved. We can use this to compute the parallel transported dyad at the origin in a direct way without the connector. \\

Recall that we expressed the maximally smooth dyad field $E^a{}_i$ in terms of the reference dyad field $\widetilde{E}^b{}_i$, which was given by (\ref{startingdyad}). In terms of the arc length (\ref{invarclength}) this reference field is
\be
\widetilde{E}^a{}_i(s,\gamma)=\delta^a{}_i~\cos^2\left(\frac{s}{2a}\right)\left(1+\eta f(s,\gamma)+\eta\frac{1}{a}\tan\left(\frac{s}{2a}\right)\int_0^sf(s',\gamma)ds'\right).
\label{arcdyad}
\ee

\begin{figure}
\begin{center}\includegraphics[width=0.6\textwidth]{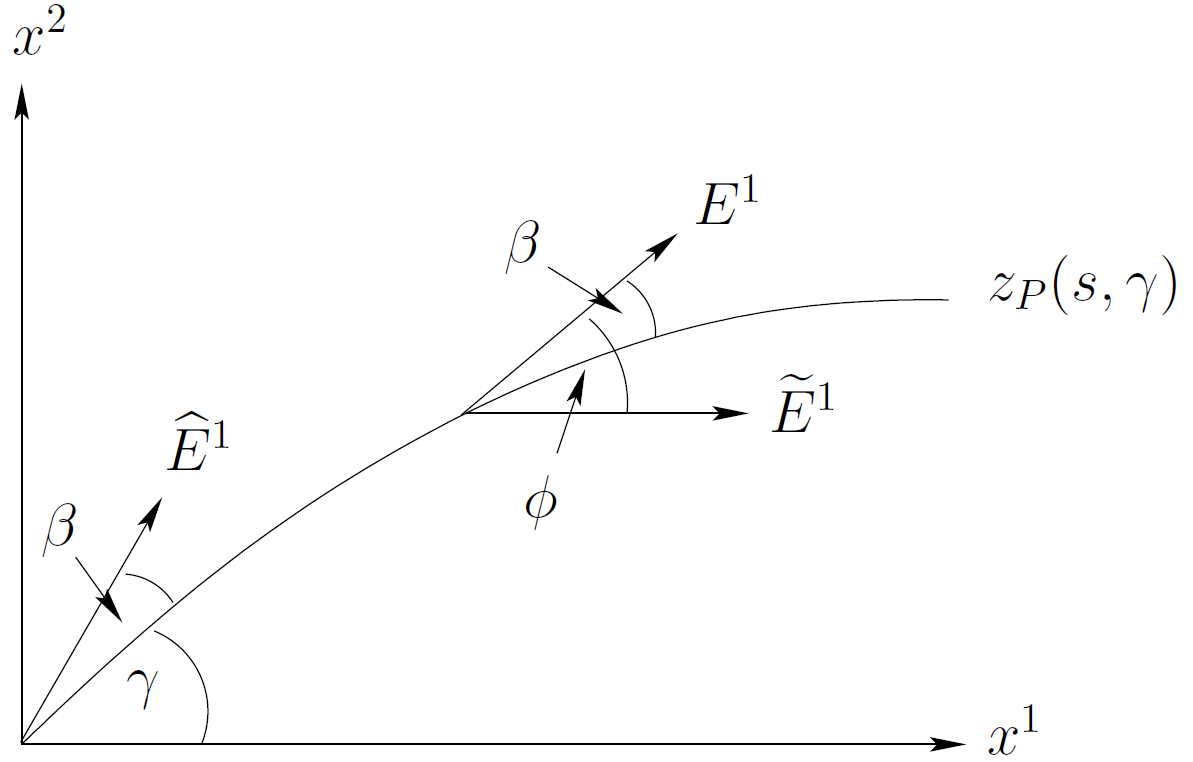}\end{center}
\caption{Definition of angles in the parallel transport of the dyads.}
\label{paratrans}
\end{figure}

The maximally smooth dyad field is a rotation of this reference dyad field by a position dependent angle $\phi(x^1,x^2)$, 
\be
E^a{}_i=U^a{}_b\widetilde{E}^b{}_i\qquad\mathrm{where}\qquad U^a{}_b=\begin{pmatrix}\cos\phi&\sin\phi\\-\sin\phi&\cos\phi\end{pmatrix}.
\label{maxsmooth}
\ee

Let the angle between $E^a{}_i(s,\gamma)$ and the tangent vector of the geodesic $z_P(s,\gamma)$ be denoted by $\beta$, as shown in Figure \ref{paratrans}. After the parallel transport the angle between the transported dyad $\widehat{E}^a{}_i(0,\gamma)$ at the origin and the geodesic $z_P(0,\gamma)$ is still $\beta$. Since the angle of the geodesic at the origin was, according to (\ref{geowinkel}), given by $\gamma$, we can express the parallel transported dyad as a rotation of the reference dyad field $\widetilde{E}^b{}_i$ by the angle 
\be
\widetilde{\phi}=\phi-\beta+\gamma.
\ee
Since 
\be
\gamma=\arccos\left(\frac{dz_P^1}{ds}(0,\gamma)\right)\qquad\mathrm{and}\qquad\beta=\arccos\left(\frac{dz_P^1}{ds}(s,\gamma)\right),
\ee
we need the perturbed geodesic (\ref{pertgeo}) in terms of the arc length (\ref{invarclength}), which is
\be
z_P^1(s,\gamma)&=2a\tan\left(\frac{s}{2a}\right)\cos\gamma-\eta\frac{\cos\gamma}{\cos^2(\frac{s}{2a})}\int_0^sf(s',\gamma)ds'+\eta v(s,\gamma)\frac{\sin\gamma}{\cos^2(\frac{s}{2a})}\\
z_P^2(s,\gamma)&=2a\tan\left(\frac{s}{2a}\right)\sin\gamma-\eta\frac{\sin\gamma}{\cos^2(\frac{s}{2a})}\int_0^sf(s',\gamma)ds'-\eta v(s,\gamma)\frac{\cos\gamma}{\cos^2(\frac{s}{2a})}.
\label{arclengthgeo}
\ee
From the derivative of this and the identity
\be
\left(\frac{dz_P^1}{ds}(s,\gamma)\right)^2+\left(\frac{dz_P^2}{ds}(s,\gamma)\right)^2=1,
\ee
we find 
\be
\cos\widetilde{\phi}&=\cos(\phi\!-\!\beta\!+\!\gamma)=\cos\phi\big(\cos\gamma\cos\beta+\sin\gamma\sin\beta\big)-\sin\phi\big(\sin\gamma\cos\beta-\sin\beta\cos\gamma\big)\\
&=\cos\phi\!\left(\frac{dz_P^1}{ds}(0)\frac{dz_P^1}{ds}(s)\!+\!\frac{dz_P^2}{ds}(0)\frac{dz_P^2}{ds}(s)\right)\!-\!\sin\phi\!\left(\frac{dz_P^2}{ds}(0)\frac{dz_P^1}{ds}(s)\!-\!\frac{dz_P^2}{ds}(s)\frac{dz_P^1}{ds}(0)\right)\\
&=\cos\phi\left(\frac{1}{\cos^2(\frac{s}{2a})}-\eta\frac{f(0,\gamma)}{\cos^2(\frac{s}{2a})}-\eta\frac{f(s,\gamma)}{\cos^2(\frac{s}{2a})}-\eta\frac{1}{a}\frac{\sin(\frac{s}{2a})}{\cos^3(\frac{s}{2a})}\int_0^sf(s',\gamma)ds'\right)\\
&-\sin\phi\left(\eta\frac{1}{\cos^2(\frac{s}{2a})}\frac{dv}{ds}(s,\gamma)+\eta\frac{v(s,\gamma)}{a}\frac{\sin(\frac{s}{2a})}{\cos^2(\frac{s}{2a})}\right)
\ee
and
\be
\sin\widetilde{\phi}&=\sin(\phi\!-\!\beta\!+\!\gamma)=\sin\phi\big(\cos\gamma\cos\beta+\sin\gamma\sin\beta\big)+\cos\phi\big(\sin\gamma\cos\beta-\sin\beta\cos\gamma\big)\\
&=\sin\phi\left(\frac{1}{\cos^2(\frac{s}{2a})}-\eta\frac{f(0,\gamma)}{\cos^2(\frac{s}{2a})}-\eta\frac{f(s,\gamma)}{\cos^2(\frac{s}{2a})}-\eta\frac{1}{a}\frac{\sin(\frac{s}{2a})}{\cos^3(\frac{s}{2a})}\int_0^sf(s',\gamma)ds'\right)\\
&+\cos\phi\left(\eta\frac{1}{\cos^2(\frac{s}{2a})}\frac{dv}{ds}(s,\gamma)+\eta\frac{v(s,\gamma)}{a}\frac{\sin(\frac{s}{2a})}{\cos^2(\frac{s}{2a})}\right).
\ee
The rotation $U(\widetilde{\phi})_j{}^i\widetilde{E}^a{}_i(s,\gamma)$ of the reference dyad field, (\ref{arcdyad}), does not yet give the correct result. The reason is that the length of the vector $E^a{}_i$ is changed when it is parallel transported. We can calculate the rate of change from a quantity which is conserved during the parallel transport, and this is the scalar product of the transported vector and the tangent vector of the geodesic,
\be
E^a{}_i(s,\gamma)\frac{dz_P^i}{ds}(s,\gamma)=\widehat{E}^a{}_i(0,\gamma)\frac{dz_P^i}{ds}(0,\gamma).
\ee
From this we find that to first order in $\eta$ the factor of change is given by
\be
C=1+2\eta f(0,\gamma).
\ee
Altogether, the parallel transported maximally smooth dyad at the origin is given by
\be
\widehat{E}^a{}_j(0,\gamma)=CU_j{}^i(\widetilde{\phi})\widetilde{E}^a{}_i(s,\gamma)
\ee
with the components
\be
\widehat{E}^1{}_1(0,\gamma)=\widehat{E}^2{}_2(0)=\cos\phi\left(1+\eta f(0)\right)-\sin\phi\left(\eta\frac{dv}{ds}(s)+\eta\frac{v(s)}{a}\tan\left(\frac{s}{2a}\right)\right),\\
\widehat{E}^1{}_2(0,\gamma)=-\widehat{E}^2{}_1(0,\gamma)=\sin\phi\left(1+\eta f(0)\right)+\cos\phi\left(\eta\frac{dv}{ds}(s)+\eta\frac{v(s)}{a}\tan\left(\frac{s}{2a}\right)\right).
\ee
This result is identical to the connector result, 
\be
\widehat{E}^a{}_j(0,\gamma)=\widehat{V}_j{}^i(0,s;{\cal C}_{s0})E^a{}_i(s,\gamma),
\ee
with the connector (\ref{sphereconnector}) and the maximally smooth dyad field (\ref{maxsmooth}) expressed in terms of the arc length $s$.

\section{The Averaging Function}

The weighting function $f(x_0,x';{\cal C}_{x'x_0})$ in the averaging process is required to be positive and normalised to unity,
\be
\int_R f(x_0,x';{\cal C}_{x'x_0})\sqrt{-g(x')}~d^4x'=1.
\ee
The simplest choice is a step function which drops to zero at the boundary $\del R$. With such a choice all tetrads of the geodesic region contribute to the average with equal weight and furthermore there are discontinuities at the boundary of the region. It is therefore preferable to choose a weighting function that has a maximum at the origin and falls off to zero smoothly at the averaging radius $r$. \\

The Beta function is defined as
\be
B(x,y)=\int_0^1t^{(x-1)}(1-t)^{(y-1)}dt=\frac{\Gamma(x)\Gamma(y)}{\Gamma(x+y)}.
\ee
For positive integers $m$ and $n$ it takes the value
\be
B(m,n)=\frac{(n-1)!(m-1)!}{(n+m-1)!}.
\ee
We can extend the interval $[0,1]$ to $[-r,r]$ if we set
\be
t=\frac{1}{2}\left(1+\frac{\tau^2}{r^2}\right)\qquad\mathrm{and}\qquad (1-t)=\frac{1}{2}\left(1-\frac{\tau^2}{r^2}\right).
\ee

\begin{figure}[h]
\center{\includegraphics[width=0.6\textwidth]{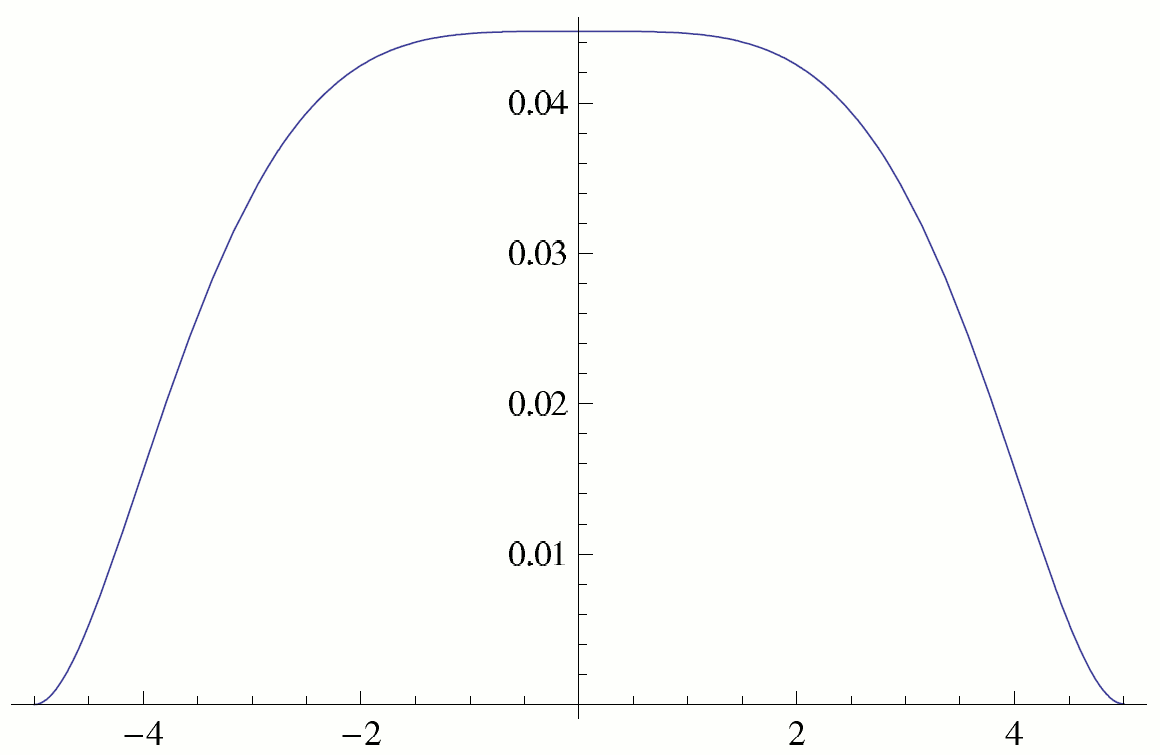}}
\caption{Normalised averaging function $f_2(\tau)$ for the averaging radius r=5 in the flat two-dimensional case $g_{ij}=\delta_{ij}$.}
\label{function}
\end{figure}

Motivated by the Beta function, reasonable choices for the averaging function are thus the normalised functions
\be
f_n(\tau)=Nt^n(1-t)^n,
\ee
where $N$ is the normalisation over the geodesic region. We choose the function $f_2(\tau)$, which is shown in Figure \ref{function} for the averaging radius $r=5$,
\be
f_2(\tau)=\frac{N}{16}\left(1-\frac{\tau^4}{r^4}\right)^2,
\ee

For the perturbed two-sphere we find
\be
N=\left(\int_0^{2\pi}\int_0^rf_2(\tau)\sqrt{g_P(\tau,\gamma)}d\gamma d\tau\right)^{-1}=I_0^{-1}-\eta~I_0^{-2}~I_f,
\label{norm}
\ee
where we defined
\be
I_0(\tau,\gamma)&=\int_0^{2\pi}\int_0^rf_2(\tau)\cos^4\left(\frac{\tau}{2a}\right)~d\gamma d\tau
\ee
and
\be
I_f(\tau,\gamma)&=\int_0^{2\pi}\int_0^r2f_2(\tau)f(\tau,\gamma)\cos^4\left(\frac{\tau}{2a}\right)~d\gamma d\tau.
\ee

With the normalised averaging function $f_2(\tau)$ we have now all constituents at hand to apply the averaging process to the perturbed two-sphere. This will be done in the next section.

\section{The Average Metric}

The general form of the averaging process, (\ref{ourprocess}), is
\be
\av{ E^a{}_j(x)}=\int f_2(x,x';{\cal C}_{x'x})\widehat{V}_j{}^i(x,x';{\cal C}_{x'x})E^a{}_i(x')\sqrt{g(x')}~d^2x'.
\ee
In terms of the geodesic coordinates (\ref{geodesiccoordinates}) the averaging process becomes
\begin{align}
\av{ E^a{}_j(0,\gamma)}&=\int_0^{2\pi}\int_0^r f_2(\tau)\widehat{V}_j{}^i(z(\tau,\gamma))E^a{}_i(\tau,\gamma)\sqrt{g_P(\tau,\gamma)}~d\gamma d\tau.
\end{align}
We insert the connector (\ref{sphereconnector}) and the maximally smooth dyad field, expressed as the rotation of the dyad field (\ref{startingdyad}), where the angle of rotation $\phi(\tau,\gamma)$ is the solution to the differential equation (\ref{laplace}) on the area $R$ with the Neumann boundary condition (\ref{randbed}) on the boundary (\ref{areaR}),
\be
E^a{}_j(\tau,\gamma)=\begin{pmatrix}\cos\phi(\tau,\gamma)&\sin\phi(\tau,\gamma)\\-\sin\phi(\tau,\gamma)&\cos\phi(\tau,\gamma)\end{pmatrix}_{ij}\delta^{ai}~(1+\eta f(\tau,\gamma))\cos^2\left(\frac{\tau}{2a}\right).
\ee

Furthermore, we define
\be
I_s(\tau,\gamma)&=\int_0^{2\pi}\int_0^rf_2(\tau)\sin(\phi(\tau,\gamma))\cos^4\left(\frac{\tau}{2a}\right)~d\gamma d\tau,\\
I_c(\tau,\gamma)&=\int_0^{2\pi}\int_0^rf_2(\tau)\cos(\phi(\tau,\gamma))\cos^4\left(\frac{\tau}{2a}\right)~d\gamma d\tau,
\ee
and the scalar fields
\be
A(\tau,\gamma)&=\cos^4\left(\frac{\tau}{2a}\right)\left(\frac{1}{a}\tan(\frac{\tau}{2a})v(\tau,\gamma)+\frac{d}{d\tau}v(\tau,\gamma)\right),\\
B(\tau,\gamma)&=\cos^4\left(\frac{\tau}{2a}\right)\left(f(0,\gamma)+2 f(\tau,\gamma)\right).
\label{aandb}
\ee

The components $\av{ E^a{}_j(0)}$ of the averaged dyad at the origin are then
\be
\av{E^1{}_1(0,\gamma)}&=\av{ E^2{}_2(0,\gamma)}=I_0^{-1}~I_c-\eta~I_0^{-2}~I_f~I_c\\
&+\eta~I_0^{-1}\int_0^{2\pi}\int_0^r\left(\cos(\phi(\tau,\gamma))B(\tau,\gamma)-\sin(\phi(\tau,\gamma))A(\tau,\gamma)d\gamma d\tau\right),\\
\av{ E^1{}_2(0,\gamma)}&=-\av{ E^2{}_1(0,\gamma)}=I_0^{-1}~I_s-\eta~I_0^{-2}~I_f~I_s\\
&+\eta~I_0^{-1}\int_0^{2\pi}\int_0^r\left(\cos(\phi(\tau,\gamma))A(\tau,\gamma)+\sin(\phi(\tau,\gamma))B(\tau,\gamma)d\gamma d\tau\right).
\label{avdyad}
\ee

The averaged metric at the origin is again diagonal and the metric elements are recomposed by the averaged dyad at the origin,
\be
\av{ (g_P)_{11}(0,\gamma)}=\av{ (g_P)_{22}(0,\gamma)}&=\left(\av{ E^1{}_1(0,\gamma)}\right)^2+\left(\av{ E^1{}_2(0,\gamma)}\right)^2
\ee
and
\be
\av{ (g_P)_{12}(0,\gamma)}=\av{ (g_P)_{21}(0,\gamma)}=0.
\ee

\section{The Coordinate Transformation}

Starting from the coordinate transformations which we used for the unperturbed sphere, (\ref{coordtrans}) and (\ref{coordtrans2}), we have to make two modifications. First, we notice that for a numerical calculation of the averaging process on a grid it is much easier to express the reference point ${\bf x}_0=(x_0,y_0,z_0)=(a\sin\theta_0\cos\phi_0,a\sin\theta_0\sin\phi_0,a\cos\theta_0)$ in terms of the coordinates of the stereographic projection plane. It is then given by
\be
x_0^1&=\frac{2a\sin\theta_0\cos\phi_0}{1-\cos\theta_0},\\
x_0^2&=\frac{2a\sin\theta_0\sin\phi_0}{1-\cos\theta_0}.
\ee

Second, we have to take the perturbation into account. The coordinate transformation is composed of three parts: the stereographic projection from the $(x^1x^2)$-plane onto the sphere, a rotation of the sphere, and a stereographic projection into the tangent $(\widetilde{x}^1\widetilde{x}^2)$-plane of the reference point ${\bf x}_0$. The perturbation enters the coordinate transformation only through the two stereographic projections, but it can be shown that the first order contributions cancel. We thus use the unperturbed coordinate transformation (\ref{coordtrans}) and (\ref{coordtrans2}).\\

To find the perturbed metric in the coordinate system of the reference point $(x_0^1,x_0^2)$ we recall that the transformation law of the metric is given by (\ref{generalmetrictrans}):
\be
(\widetilde{g}_P)_{kl}(\widetilde{x}^1,\widetilde{x}^2)=\frac{\del x^i}{\del \widetilde{x}^k}\frac{\del x^j}{\del \widetilde{x}^l}~(g_P)_{ij}(x^1,x^2).
\ee
Inserting the coordinate transformation (\ref{coordtrans2}) and the perturbed metric (\ref{pertmetric}) yields
\be
(\widetilde{g}_P)_{kl}(\widetilde{x}^1,\widetilde{x}^2)=\frac{16a^4}{\widetilde{L}^4}(1+2\eta f(\widetilde{x}^1,\widetilde{x}^2)).
\ee
The change to the coordinate system $(\widetilde{x}^1,\widetilde{x}^2)$ affects thus only the perturbation function $f(\widetilde{x}^1,\widetilde{x}^2)$, which has to be transformed to the new coordinates.\\

Applying the back transformation law (\ref{metricbacktrans}) to the averaged metric at the origin of the coordinate system $(\widetilde{x}^1,\widetilde{x}^2)$ we find the averaged metric at the reference point $(x_0^1,x_0^2)$ in terms of the original $(x^1x^2)$-coordinates to be 
\be
\av{ (g_P)_{ij}(x_0^1,x_0^2)}=\frac{\del\widetilde{x}^k}{\del x^i}\bigg|_{(x^1,x^2)=(x_0^1,x_0^2)}\frac{\del\widetilde{x}^l}{\del x^j}\bigg|_{(x^1,x^2)=(x_0^1,x_0^2)}\av{(\widetilde{g}_P)_{kl}(0,0)}.
\ee
Since the averaged metric is diagonal we only need the transformation law for the diagonal metric elements. By inserting (\ref{coordtrans}) we find 
\be
\av{(g_P)_{11}(x_0^1,x_0^2)}&=T\av{(\widetilde{g}_P)_{11}(0,0)}
\ee
and
\be
\av{(g_P)_{22}(x_0^1,x_0^2)}&=T\av{(\widetilde{g}_P)_{22}(0,0)},
\ee
where
\be
T=\frac{16a^4}{(4a^2+(x_0^1)^2+(x_0^2)^2)^2}.
\ee

The result for the averaged metric is therefore
\be
\av{ (g_P)_{12}(0,\gamma)}=\av{ (g_P)_{21}(0,\gamma)}=0
\ee
and
\be
\av{(g_P)_{11}(0,\gamma)}&=\av{ (g_P)_{22}(0,\gamma)}=\left(\av{ E^1{}_1(0,\gamma)}\right)^2+\left(\av{ E^1{}_2(0,\gamma)}\right)^2\\
&=\frac{16a^4}{(4a^2+x_0^2+y_0^2)^2}\bigg(\left(I_0^{-2}~I_c^2+I_0^{-2}~I_s^2\right)\\
&+2\eta\bigg(-I_0^{-3}~I_c^2\times I_f-I_0^{-3}~I_s^2~I_f\\
&+I_0^{-2}~I_c~\int_0^{2\pi}\int_0^rf_2(\tau)\left(\cos\phi~B-\sin\phi~A\right)d\gamma d\tau\\
&+I_0^{-2}~I_s~\int_0^{2\pi}\int_0^rf_2(\tau)\left(\cos\phi~A+\sin\phi~B\right)d\gamma d\tau\bigg)\bigg).
\label{ergebnis}
\ee\\

In this chapter we investigated the averaging process for a general imposed perturbation function $f(x^1,x^2)$ to the two-sphere. We will use the results of this chapter to compute an explicit example of such a function in chapter \ref{gauss}.

%% file: threesphere.tex
\chapter{Averaging the Three-Sphere}
\label{threesphere}

In the previous chapters we have considered the smooth and the linearly-perturbed two-sphere in some detail. However, despite the successes in those metrics we should demonstrate its validity in a higher dimension. To this aim we present a proof-of-concept for a three-sphere, first applying the averaging process to the smooth case and then to the linearly perturbed case. Such spaces form the basis of closed FLRW models.

\section{The Metric from Stereographic Projection}

As in the three dimensional case we start with a stereographic projection of the three-sphere into the tangent plane at the south pole. Let $(\xi^1,\xi^2,\xi^3,\xi^4)$ denote the Cartesian coordinates in four dimensions and $(x^1,x^2,x^3)$ the ones of the projection plane. The radius of the four sphere is denoted by $a$. Solving the equation
\be
\begin{pmatrix}0\\0\\0\\2a\end{pmatrix}+\lambda\left[\begin{pmatrix}\xi^1\\\xi^2\\\xi^3\\\xi^4\end{pmatrix}-\begin{pmatrix}0\\0\\0\\2a\end{pmatrix}\right]=\begin{pmatrix}x^1\\x^2\\x^3\\0\end{pmatrix}
\ee
for $\lambda$ returns the relations
\be
\begin{aligned}
x^i&=\frac{2a}{2a-\xi^4}~\xi^i,
\end{aligned}
\ee
where the index $i\in\{1,2,3\}$. To express $\xi^4$ in the coordinates of the projection plane we make use of the fact that the surface of the sphere is given by
\be
(\xi^1)^2+(\xi^2)^2+(\xi^3)^2+(\xi^4-a)^2=a^2.
\ee
With the quantity
\be
L^2=4a^2+(x^1)^2+(x^2)^2+(x^3)^2
\ee
we can express the coordinate $\xi^4$ as
\be
\xi^4=\frac{2a\left((x^1)^2+(x^2)^2+(x^3)^2\right)}{L^2}.
\ee
From here we can calculate the line element in terms of the coordinates of the projection plane,
\be
ds^2=(d\xi^1)^2+(d\xi^2)^2+(d\xi^3)^2+(d\xi^4)^2=\frac{16a^4}{L^4}\left((dx^1)^2+(dx^2)^2+(dx^3)^2\right).
\ee
The metric and its inverse are therefore
\be
g_{ij}=\left(\frac{2a}{L}\right)^4\delta_{ij}\qquad\mathrm{and}\qquad g^{ij}=\left(\frac{L}{2a}\right)^4\delta^{ij},
\label{threemetric}
\ee
with the indices $i,j \in\{1,2,3\}$.

\section{The Geodesics and the Connector}

As  for the two-sphere the Christoffel symbols of this metric are given by
\be
\Gamma^i_{jk}=\frac{2}{L^2}\left(x_j\delta^i_k+x_k\delta^i_j-x^i\delta_{jk}\right),
\ee
and the geodesic is
\be
L^2\frac{d^2z^i}{d\tau^2}-4z_j\frac{dz^j}{d\tau}\frac{dz^i}{d\tau}+2z^ig_{jk}\frac{dz^j}{d\tau}\frac{dz^k}{d\tau}=0.
\ee
For the initial conditions $z^i(0)=0$ and $dz^i/d\tau(0)=$ const. the solution to this equation is again 
\be
z^i(\tau)=2a\tan\left(\frac{\tau}{2a}\right)~\frac{dz^i}{d\tau}(0),
\ee
with the additional requirement 
\be
\left(\frac{dz^1}{d\tau}(0)\right)^2+\left(\frac{dz^2}{d\tau}(0)\right)^2+\left(\frac{dz^3}{d\tau}(0)\right)^2=1.
\label{threerequirement}
\ee
The latter identity ensures furthermore that $\tau$ is the arc length
\be
\tau=\int_0^\tau d\tau'\sqrt{g_{ij}(z(\tau'))\frac{dz^i}{d\tau'}\frac{dz^j}{d\tau}}.
\ee
The connector in this metric is 
\be
V(\tau,0;{\cal C}_{0\tau})={\cal P}\exp\left(-\int_0^\tau \cos^{-1}\left(\frac{\tau}{2a}\right)\frac{dz^i}{d\tau}(0)~{\bf \Gamma}_i(z(\tau'))~d\tau'\right)=\cos^2\left(\frac{\tau}{2a}\right)~\mathbbm{1},
\ee
and hence the hatted connector that appears in the averaging process is
\be
\widehat{V}_j{}^i(0,\tau;{\cal C}_{\tau0})=\cos^{-2}\left(\frac{\tau}{2a}\right)~\delta^i_j.
\ee

\section{The Maximally Smooth Triad Field}

\begin{figure}
\begin{center}\includegraphics[width=0.5\textwidth]{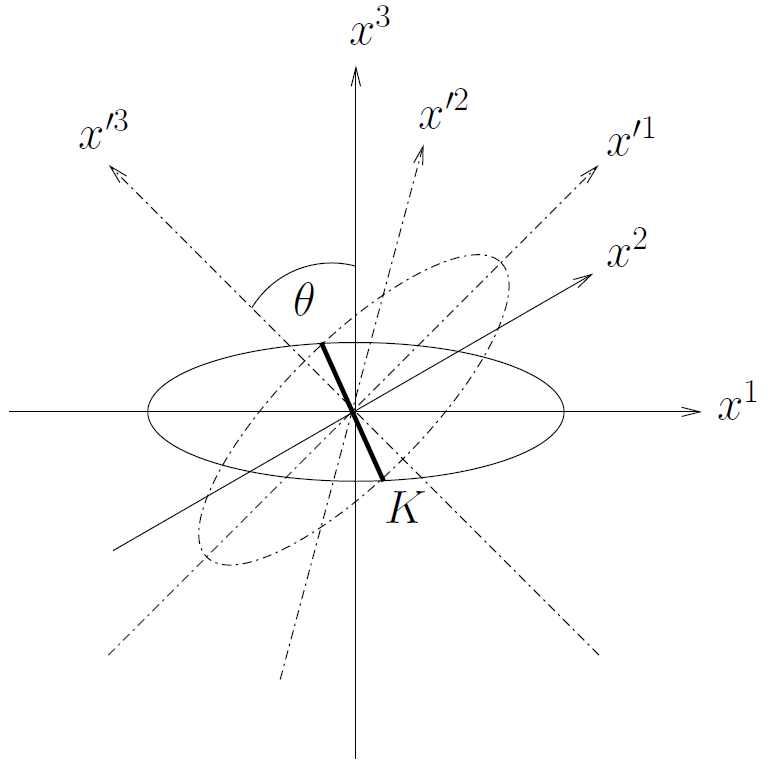}\end{center}
\caption{Definition of Euler angles}
\label{euler}
\end{figure}

The rotation group $SO(3)$ is the group of homogeneous linear transformations
\be
x'^a=R^a{}_bx^b
\ee
that preserve lengths and orientation. A general element of the group can be characterized by three angles. A common choice of angles are the Euler angles. Let the Cartesian coordinates of the unrotated frame be $(x^1,x^2,x^3)$ and the coordinates of the rotated frame be $(x'^1,x'^2,x'^3)$. Let further the intersection of the $(x^1x^2)$-plane and the $(x'^1x'^2)$-plane, namely the nodal line, be labelled $K$, as shown in Figure \ref{euler}. Then the Euler angles $(\phi,\psi,\theta)$ are defined as follows: $\phi$ is the angle between the $x^1$-axis and $K$, $\psi$ is the angle between $K$ and the $x'^1$-axis, and $\theta$ is the angle between the $x^3$-axis and the $x'^3$-axis. Any rotation can then be represented as a composition of a rotation about the $x^3$-axis by $\phi$ $(0\leq\phi<2\pi)$  and a rotation about the $x^1$-axis by $\theta$  $(0\leq\theta<2\pi)$ with a rotation about the $x^3$-axis by $\psi$ $(0\leq\psi<2\pi)$. In terms of the Euler angles the parametrization of a general rotation is

\be
&{\bf R}(\phi,\theta,\psi)={\bf R}(\psi)\cdot{\bf R}(\theta)\cdot{\bf R}(\phi)\\
&=\begin{pmatrix}\cos\psi&\sin\psi&0\\-\sin\psi&\cos\psi&0\\0&0&1\end{pmatrix}\begin{pmatrix}1&0&0\\0&\cos\theta&\sin\theta\\0&-\sin\theta&\cos\theta\end{pmatrix}\begin{pmatrix}\cos\phi&\sin\phi&0\\-\sin\phi&\cos\phi&0\\0&0&1\end{pmatrix}
\ee
\be
\!=\begin{pmatrix}\cos\psi\cos\phi\!-\!\sin\psi\cos\theta\sin\phi&\cos\psi\sin\phi\!+\!\sin\psi\cos\theta\cos\phi&\sin\theta\sin\psi\\-\!\sin\psi\cos\phi\!-\!\cos\psi\cos\theta\sin\phi&-\!\sin\psi\sin\phi\!+\!\cos\psi\cos\theta\cos\phi&\cos\psi\sin\theta\\\sin\theta\sin\phi&-\cos\phi\sin\theta&\cos\theta\end{pmatrix}.
\ee

Now we express the maximally smooth triad field $E^a{}_i$ as a rotation of the reference triad field $\widetilde{E}^b{}_i$. In terms of the Euler angles this is
\be
E^a{}_i=(R^T)^a{}_b(\phi,\theta,\psi)~\widetilde{E}^b{}_i=U^a{}_b(\phi,\theta,\psi)~\widetilde{E}^b{}_i.
\ee
To determine the differential equations for the Euler angles we insert this into the Lagrangian
\be
\mathcal{L}_\mathrm{MS}=(D_iE^a{}_j)(D_kE^b{}_l)g^{ik}g^{jl}\delta_{ab}.
\ee
With the substitution
\be
(D_i\widetilde{E}^a{}_j)\widetilde{E}^b{}_l~g^{ik}g^{jl}=(Q^{ab})^k
\label{threevectors}
\ee
we find
\be
\begin{aligned}
&\mathcal{L}_\mathrm{MS}=2(\del_i\psi)(\del_k\psi)g^{ik}\!+\!2(\del_i\theta)(\del_k\theta)g^{ik}\!+\!2(\del_i\phi)(\del_k)g^{ik}\!+\!4\cos\theta(\del_i\psi)(\del_k\phi)g^{ik}\\
&-\!2\left((\del_k\psi)\!+\!\cos\theta(\del_k\phi)\right)(Q^{12})^k\!+\!2\left((\del_k\psi)\!+\!\cos\theta(\del_k\phi)\right)(Q^{21})^k\\
&-\!2\left(\sin\psi(\del_k\theta)\!-\!\sin\theta\cos\psi(\del_k\phi)\right)(Q^{13})^k\!+\!2\left(\sin\psi(\del_k\theta)\!-\!\sin\theta\cos\psi(\del_k\phi)\right)(Q^{31})^k\\
&-\!2\left(\cos\psi(\del_k\theta)\!+\!\sin\theta\sin\psi(\del_k\phi)\right)(Q^{23})^k\!+\!2\left(\cos\psi(\del_k\theta)\!+\!\sin\theta\sin\psi(\del_k\phi)\right)(Q^{32})^k\\
&+\!(D_i\widetilde{E}^c{}_j)(D_k\widetilde{E}^d{}_l)g^{ik}g^{jl}\delta_{cd}.
\end{aligned}
\ee
Applying the principle of least action on a finite integration volume $R$ with boundary $\del R$ yields three partial differential equations with Neumann boundary equations. The differential equation for $\phi$ on $R$ is
\be
\begin{aligned}
&2(\sqrt{g})^{-1}\del_k\left(\sqrt{g}~g^{ik}(\del_i\phi)\right)+2\cos\theta(\sqrt{g})^{-1}\del_k\left(\sqrt{g}~g^{ik}(\del_i\psi)\right)-2\sin\theta g^{ik}(\del_i\psi)(\del_k\theta)\\
&=\cos\theta D_k(Q^{12})^k-\cos\theta D_k(Q^{21})^k-\sin\theta(Q^{12})^k(\del_k\theta)+\sin\theta(Q^{21})^k(\del_k\theta)\\
&-\sin\theta\cos\psi D_k(Q^{13})^k+\sin\theta\cos\psi D_k(Q^{31})^k-\cos\theta\cos\psi(Q^{13})^k(\del_k\theta)\\
&+\cos\theta\cos\psi(\del_k\theta)(Q^{31})^k+\sin\theta\sin\psi(Q^{13})^k-\sin\theta\sin\psi(Q^{31})^k(\del_k\psi)\\
&+\sin\theta\sin\psi D_k(Q^{23})^k-\sin\theta\sin\psi D_k(Q^{32})^k+\cos\theta\sin\psi(Q^{23})^k(\del_k\theta)\\
&-\cos\theta\sin\psi(Q^{32})^k(\del_k\theta)+\sin\theta\cos\psi(Q^{23})^k(\del_k\psi)-\sin\theta\cos\psi(Q^{32})^k(\del_k\psi)
\end{aligned}
\ee
with the following boundary condition on $\del R$:
\be
\begin{aligned}
2\frac{\del\phi}{\del n}+2\cos\theta\frac{\del\psi}{\del n}&=\cos\theta n_k(Q^{12})^k-\cos\theta n_k(Q^{21})^k-\sin\theta\cos\psi n_k(Q^{13})^k\\
&+\sin\theta\cos\psi n_k(Q^{31})^k-\sin\theta\sin\psi n_k(Q^{23})^k-\sin\theta\sin\psi n_k(Q^{32})^k.
\end{aligned}
\ee
Here, $n_k$ denotes the normal unit vector to $\del R$. The differential equation for $\theta$ on $R$ is
\be
\begin{aligned}
&2(\sqrt{g})^{-1}\del_k\left(\sqrt{g}~g^{ik}(\del_i\theta)\right)+2\sin\theta g^{ik}(\del_i\psi)(\del_k\phi)\\
&=\sin\psi D_k(Q^{13})^k-\sin\psi D_k(Q^{31})^k+\cos\theta\cos\psi(\del_k\phi)(Q^{13})^k+\cos\psi(Q^{13})^k(\del_k\psi)\\
&-\cos\theta\cos\psi(\del_k\phi)(Q^{31})^k-\cos\psi(Q^{31})^k(\del_k\psi)+\cos\psi D_k(Q^{23})^k-\cos\psi D_k(Q^{32})^k\\
&+\cos\theta\sin\psi(\del_k\phi)(Q^{23})^k-\sin\psi(\del_k\psi)(Q^{23})^k+\cos\theta\sin\psi(\del_k\phi)(Q^{32})^k\\
&+\sin\psi(\del_k\psi)(Q^{32})^k+\sin\theta(Q^{12})^k(\del_k\phi)-\sin\theta(Q^{21})^k(\del_k\phi)
\end{aligned}
\ee
with the boundary condition on $\del R$
\be
2\frac{\del\theta}{\del n}=\sin\psi n_k(Q^{13})^k-\sin\psi n_k(Q^{31})^k+\cos\psi n_k(Q^{23})^k-\cos\psi n_k(Q^{32})^k.
\ee
And finally the differential equation for $\psi$ on $R$ is
\be
\begin{aligned}
&2(\sqrt{g})^{-1}\del_k\left(\sqrt{g}~g^{ik}(\del_i\psi)\right)+2\cos\theta(\sqrt{g})^{-1}\del_k\left(\sqrt{g}~g^{ik}(\del_i\phi)\right)-2\sin\theta g^{ik}(\del_i\phi)(\del_k\theta)\\
&=D_k(Q^{12})^k-D_k(Q^{21})^k-\cos\psi(Q^{13})^k(\del_k\theta)+\cos\psi(Q^{31})^k(\del_k\theta)\\
&-\sin\theta\sin\psi(\del_k\phi)(Q^{13})^k+\sin\theta\sin\psi(\del_k\phi)(Q^{31})^k+\sin\psi(Q^{23})^k(\del_k\theta)\\
&-\sin\psi(Q^{32})^k(\del_k\theta)-\sin\theta\cos\psi(\del_k\phi)(Q^{23})^k+\sin\theta\cos\psi(\del_k\phi)(Q^{32})^k
\end{aligned}
\ee
with the boundary condition on $\del R$
\be
2\frac{\del\psi}{\del n}+2\cos\theta\frac{\del\phi}{\del n}=n_k(Q^{12})^k-n_k(Q^{21})^k.
\ee

\section{The Result of the Averaging Process}

To solve the differential equations for the maximally smooth triad field we first need to decompose the metric of the smooth three-sphere (\ref{threemetric}) into triads and choose among the options one reference triad field to start from. The easiest choice is
\be
\widetilde{E}^a{}_i=\left(\frac{2a}{L}\right)^2~\delta^a{}_i.
\label{smoothtriad}
\ee
The vector fields (\ref{threevectors}) then become
\be
\left(Q^{ij}\right)^k=\frac{L^2}{8a^4}(\delta^{ik}\delta^j{}_n-\delta^{jk}\delta^i{}_n)x^n.
\ee
Furthermore, we need the normal unit vector to the boundary of the geodesic region $\del R$. The requirement (\ref{threerequirement}) allows us to introduce the angles $\gamma_1\in[0,\pi)$ and $\gamma_2\in[0,2\pi)$, such that
\be
\frac{dz^1}{d\tau}(0)=\sin\gamma_1\cos\gamma_2,\quad
\frac{dz^2}{d\tau}(0)=\sin\gamma_1\sin\gamma_2,\quad
\frac{dz^3}{d\tau}(0)=\cos\gamma_1.
\label{threeangles}
\ee
Then we can express the boundary of the geodesic region $R$ of radius $r$ in terms of these angles,
\be
\begin{aligned}
\alpha^1(\gamma_1,\gamma_2)&=2a\tan\left(\frac{r}{2a}\right)\sin\gamma_1\cos\gamma_2,\\
\alpha^2(\gamma_1,\gamma_2)&=2a\tan\left(\frac{r}{2a}\right)\sin\gamma_1\sin\gamma_2,\\
\alpha^3(\gamma_1,\gamma_2)&=2a\tan\left(\frac{r}{2a}\right)\cos\gamma_1.
\end{aligned}
\ee
The coordinate invariant definition of the normal vector to $\del R$ is
\be
\nu_k=\sqrt{g(r,\gamma_1,\gamma_2)}~\epsilon_{klm}\frac{\del\alpha^l}{\del\gamma_1}\frac{\del\alpha^m}{\del\gamma_2},
\ee
and the normal unit vector is
\be
n_k=\left(g^{ij}(r,\gamma_1,\gamma_2)\nu_i\nu_j\right)^{-1/2}\nu_k.
\ee
In this case we find
\be
n_1=\cos^2\left(\frac{r}{2a}\right)\sin\gamma_1\cos\gamma_2,~~n_2=\cos^2\left(\frac{r}{2a}\right)\sin\gamma_1\sin\gamma_2,~~n_3=\cos^2\left(\frac{r}{2a}\right)\cos\gamma_1,
\ee
and the differential equations for the maximally smooth triad field reduce to
\be
\begin{aligned}
&\Delta\phi+\cos\theta\Delta\psi-\sin\theta~\nabla\theta\cdot\nabla\psi=\frac{2}{L^2}\big[-\sin\theta(x^2\del_1\theta-x^1\del_2\theta)-\cos\theta\cos\psi(x^3\del_1\theta-x^1\del_3\theta)\\
&+\sin\theta\sin\psi(x^3\del_1\psi-x^1\del_3\psi)+\cos\theta\sin\psi(x^3\del_2\theta-x^2\del_3\theta)+\sin\theta\cos\psi(x^3\del_2\psi-x^2\del_3\psi)\big]
\end{aligned}
\ee
for $\phi$,
\be
\begin{aligned}
&\Delta\theta-\sin\theta~\nabla\psi\cdot\nabla\phi=\frac{2}{L^2}\big[\cos\psi(x^3\del_1\psi-x^1\del_3\psi)-\sin\psi(x^3\del_2\psi-x^2\del_3\psi)\\
&-\sin\theta(x^2\del_1\phi-x^1\del_2\phi)+\cos\theta\cos\psi(x^3\del_1\phi-x^1\del_3\phi)-\cos\theta\sin\psi(x^3\del_2\phi-x^2\del_1\phi)\big]
\end{aligned}
\ee
for $\theta$, and
\be
\begin{aligned}
&\Delta\psi+\cos\theta\Delta\phi-\sin\theta~\nabla\theta\cdot\nabla\phi=\frac{2}{L^2}\big[-\cos\psi(x^3\del_1\theta-x^1\del_3\theta)\\
&-\sin\theta\sin\phi(x^3\del_1\phi-x^1\del_3\phi)+\sin\psi(x^3\del_2\theta-x^2\del_3\theta)-\sin\theta\cos\psi(x^3\del_2\phi-x^2\del_3\phi)\big]
\end{aligned}
\ee
for $\psi$. The boundary conditions reduce to
\be
\frac{\del\phi}{\del n}+\cos\theta~\frac{\del\psi}{\del n}=0,\qquad
\frac{\del\theta}{\del n}=0,\qquad
\frac{\del\psi}{\del n}+\cos\theta~\frac{\del\phi}{\del n}=0.
\ee
For this type of boundary condition the solution to the differential equation is  $\phi=\theta=\psi=$ const. up to a global rotation and we can set all the constants to zero. Hence, with the reference triad field (\ref{smoothtriad}) we have already found the maximally smooth triad field. The averaging process then yields
\be
\begin{aligned}
\av{ E^a{}_j(0)}&=\int_0^\infty\!\!\!\int_0^\pi\!\!\!\int_0^{2\pi}\!f_M(z(\tau,\gamma_1,\gamma_2))\widehat{Q}_j{}^i(z(\tau,\gamma_1,\gamma_2))E^a{}_i(\tau,\gamma_1,\gamma_2)\sqrt{g(\tau,\gamma_1,\gamma_2)}d\tau d\gamma_1 d\gamma_2\\
&=\delta^a_j\int_0^\infty\!\!\!\int_0^\pi\!\!\!\int_0^{2\pi}\!f_M(z(\tau,\gamma_1,\gamma_2))\sqrt{g(\tau,\gamma_1,\gamma_2)}d\tau d\gamma_1 d\gamma_2=\delta_{aj},
\end{aligned}
\ee
since the averaging function is normalized to unity. \\

Therefore, the averaged metric is identical to the original metric, 
\be
\av{g_{ij}(0)}=\delta_{ij}=g_{ij}(0),
\ee
and we conclude that the smooth three-sphere is not affected by the averaging process. 

\section{Averaging the Perturbed Three-Sphere}

In this section we investigate the effect of the averaging process when the surface of the three-sphere is subjected to arbitrary linear fluctuations. While we do not present a closed result, leaving such to future study, we demonstrate that our formalism is capable of dealing with such spaces. The obvious immediate application lies in cosmology, for which the perturbed three-sphere forms the basis of the closed FLRW model

\subsection{The Metric from Stereographic Projection}

Following the same procedure as for the two-sphere, we perturb the three-sphere slightly and apply the stereographic projection. This yields the coordinate transformation
\be\begin{aligned}
\xi_P^i&=(1+\eta f({\bf x}))\frac{4a^2}{L^2}x^i,\\
\xi_P^4&=(1+\eta f({\bf x}))\frac{2a\left((x^1)^2+(x^2)^2+(x^3)^2\right)}{L^2},
\end{aligned}\ee

where $i\in\{1,2,3\}$. The perturbation is encoded in the perturbation fuction $f({\bf x})$ and $\eta>0$ is a small constant. The line element of the perturbed sphere after the projection is to first order in $\eta$ given by
\be
ds^2=(d\xi_P^1)^2+(d\xi_P^2)^2+(d\xi_P^3)^2+(d\xi_P^4)^2=(1+2\eta f)\left(\frac{2a}{L}\right)^4\!\!\left((dx^1)^2+(dx^2)^2+(dx^3)^2\right)
\ee
and therefore the metric and its inverse are 
\be
g_{ij}=(1+2\eta f({\bf x}))\left(\frac{2a}{L}\right)^4\delta_{ij}\qquad\qquad\mathrm{and}\qquad\qquad g^{ij}=(1-2\eta f({\bf x}))\left(\frac{L}{2a}\right)^4\delta^{ij}.
\label{threepertmetric}
\ee

\subsection{The Geodesics and the Connector}

The Christoffel symbols of the metric (\ref{threepertmetric}) are
\be
\Gamma^i_{jk}=\frac{2}{L^2}\left(-x_k\delta^i{}_j-x_j\delta^i{}_k+x^i\delta_{jk}\right)+\eta\left(\frac{\del f}{\del x^k}\delta^i{}_j+\frac{\del f}{\del x^j}\delta^i{}_k-\delta^{im}\frac{\del f}{\del x^m}\delta_{jk}\right).
\ee
For the parametrization of the perturbed geodesic we need three parameters indicating the deviation from the unperturbed geodesic in each of the space dimensions. We therefore make the ansatz for the geodesic
\be
z_P^i(\tau)=z^i(\tau)+\eta~\delta^{ij}\epsilon_{jkl}v^k(\tau)\frac{dz^l}{d\tau}(\tau).
\ee
An appropriate choice for the direction of the $v^i(\tau)$ is perpendicular to the tangent vector of the unperturbed geodesic and hence we demand
\be
(g_P)_{ij}v^i(\tau)\frac{dz_P^j}{d\tau}(\tau)=0.
\label{perpendicular}
\ee
Making use once more of the requirement (\ref{threerequirement}) we find the arc length of the perturbed geodesic
\be
s(\tau)=\tau+\eta\int_0^\tau f(\tau')d\tau'. 
\ee
Since the arc length is changed through the perturbation we have to use the general geodesic equation (\ref{pertgeoeq}). With (\ref{perpendicular}) and the identity
\be
\left(L(z_P)\right)^2=4a^2\cos^{-2}\left(\frac{\tau}{2a}\right)
\ee
which holds to first order in $\eta$, we find 
\be
\begin{aligned}
&\left(\epsilon_{ijk}\frac{d^2v^j}{d\tau^2}+\epsilon_{ijk}\frac{v^j}{a^2}+\cos^{-2}\left(\frac{\tau}{2a}\right)\frac{dz^j}{d\tau}(0)\!\left(\delta_{ij}\frac{\del f}{\del x^k}\biggl|_{x=z}\!-\delta_{jk}\frac{\del f}{dx^i}\biggl|_{x=z}\right)\right)\frac{dz^k}{d\tau}(0)=0.
\end{aligned}
\ee
For non-vanishing initial values $dz^k/d\tau(0)\neq 0$ this can be further simplified to 
\be
\frac{d^2v^i}{d\tau^2}+\frac{v^i}{a^2}=\cos^{-2}\left(\frac{\tau}{2a}\right)\epsilon_{ijk}\frac{dz^j}{d\tau}(0)~\delta^{kl}\frac{\del f}{\del x^l}\biggl|_{x=z}.
\ee
These are three independent driven harmonic oscillator equations for the functions $v^i(\tau)$. We require the perturbed and unperturbed geodesics and their first derivatives to coincide at the origin,
\be
z_P^i(0)=z^i(0)\quad\mathrm{and}\quad\frac{dz_P^i}{d\tau}(0)=\frac{dz^i}{d\tau}(0),
\ee
and therefore we impose the initial conditions
\be
v^i(0)=0\quad\mathrm{and}\quad\frac{dv^i}{d\tau}(0)=0.
\ee
Now we switch to geodesic coordinates, namely the parameter along the geodesics, $\tau$, and the angles which describe the direction of the geodesic in the origin, $\gamma_1$ and $\gamma_2$. The latter were introduced in (\ref{threeangles}). In terms of these coordinates the first order expansion of the connector becomes
\be
\begin{aligned}
V(\tau,0;{\cal C}_{0\tau})=&\cos^{-2}\left(\frac{\tau}{2a}\right)\left(1-\eta f(\tau,\gamma_1,\gamma_2)+\eta f(0,\gamma_1,\gamma_2)\right)~{\mathbbm 1}\\
&+\eta \cos^{-2}\left(\frac{\tau}{2a}\right)\left(\del_\tau v_i(\tau,\gamma_1,\gamma_2)+\frac{1}{a}\tan\left(\frac{\tau}{2a}\right)v_i(\tau,\gamma_1,\gamma_2)\right)\delta^{ij}\Lambda_j\\
\label{firstconnector}
\end{aligned}
\ee
where the $\Lambda_j$ indicate the matrices
\be
\Lambda_1=\begin{pmatrix}0&0&0\\0&0&-1\\0&1&0\end{pmatrix},\qquad\Lambda_2=\begin{pmatrix}0&0&1\\0&0&0\\-1&0&0\end{pmatrix},\qquad\Lambda_3=\begin{pmatrix}0&-1&0\\1&0&0\\0&0&0\end{pmatrix}.
\ee
These are the generators of the rotation group. They form a Lie algebra with the commutator relations
\be
\left[\Lambda_i,\Lambda_j\right]=\epsilon_{ijk}\Lambda_k.
\ee
The $jk$-element of $\Lambda_i$ is given by
\be
\left(\Lambda_i\right)_{jk}=\epsilon_{ijk},
\ee
with which we can express the components of the inverse of the connector (\ref{firstconnector}) as
\be
\begin{aligned}
V^k{}_l(0,\tau;{\cal C}_{\tau 0})=&(V^{-1})^k{}_l(\tau,0;{\cal C}_{0\tau})\\
=&\cos^{-2}\left(\frac{\tau}{2a}\right)\left(1-\eta f(\tau,\gamma_1,\gamma_2)+\eta f(0,\gamma_1,\gamma_2)\right)\delta^k{}_l\\
&+\eta \cos^{-2}\left(\frac{\tau}{2a}\right)\left(\del_\tau v_i(\tau,\gamma_1,\gamma_2)+\frac{1}{a}\tan\left(\frac{\tau}{2a}\right)v_i(\tau,\gamma_1,\gamma_2)\right)\delta^{ij}\epsilon_{jkl}.
\end{aligned}
\ee
Finally, we need the hatted connector for the parallel transport of the triads in the averaging process, which is defined as
\be
\begin{aligned}
\widehat{V}_j{}^i(0,\tau;{\cal C}_{\tau 0})=&g_{jk}(0)V^k{}_l(\tau,0;{\cal C}_{0\tau})g^{li}(0)\\
=&\cos^{-2}\left(\frac{\tau}{2a}\right)\left(1+\eta f(\tau,\gamma_1,\gamma_2)-\eta f(0,\gamma_1,\gamma_2)\right)\delta_j{}^i\\
&+\eta \cos^{-2}\left(\frac{\tau}{2a}\right)\left(\del_\tau v_k(\tau,\gamma_1,\gamma_2)+\frac{1}{a}\tan\left(\frac{\tau}{2a}\right)v_k(\tau,\gamma_1,\gamma_2)\right)\epsilon_{kjl}\delta^{il}.
\end{aligned}
\ee 

\subsection{The Maximally Smooth Triad Field}

Since the metric (\ref{threepertmetric}) is proportional to unity, the best choice of reference field in geodesic coordinates is 
\be
\widetilde{E}^a{}_i=\cos^2\left(\frac{\tau}{2a}\right)\left(1+\eta f(\tau,\gamma_1,\gamma_2)\right)~\delta^a{}_i.
\ee
The vector fields (\ref{threevectors}) are
\be
(Q^{ij})^k=\epsilon^{ijm}\delta_{mn}\epsilon^{nkl}\left((1-2\eta f)\frac{L^2}{8a^4}\delta_{ls}x^s-\eta\frac{L^4}{16a^4}\frac{\del f}{\del x^l}\right).
\ee
For the differential equations that determine the maximally smooth triad field we need the value of these vector fields on the surface of constant geodesic distance $r=$, which is the boundary of the area $R$. With the function
\be
F(r,\gamma_1,\gamma_2)=\int_0^rf(\tau',\gamma_1,\gamma_2)d\tau'
\ee
the parameter representation of the surface $\del R$ can be expressed as
\be
\alpha^i(\gamma_1,\gamma_2)=2a\tan\left(\frac{r}{2a}\right)\frac{\del z^i}{\del\tau}(0)-\eta\frac{F(r,\gamma_1,\gamma_2)}{\cos^2\left(\frac{r}{2a}\right)}\frac{\del z^i}{\del\tau}(0)+\eta~\delta^{ij}\epsilon_{jkl}\frac{v^k(r,\gamma_1,\gamma_2)}{\cos^2\left(\frac{r}{2a}\right)}\frac{\del z^l}{\del\tau}(0).
\ee
The value of $L^2$ on the boundary $\del R$ is
\be
L^2\left(z(r)\right)=\frac{4a^2}{\cos^2\left(\frac{r}{2a}\right)}-\eta ~4aF(r,\gamma_1,\gamma_2)\frac{\sin\left(\frac{r}{2a}\right)}{\cos^3\left(\frac{r}{2a}\right)},
\ee
and the square root of the determinant of the metric on $\del R$ is
\be
\sqrt{g_P\left(z(r)\right)}=\left(1+\eta ~2f(r,\gamma_1,\gamma_2)+\eta~\frac{2}{a}F(r,\gamma_1,\gamma_2)\tan\left(\frac{r}{2a}\right)\right)\cos^4\left(\frac{r}{2a}\right).
\ee
Thus, the values of the non vanishing vector fields (\ref{threevectors}) on the boundary are
\be
&(Q^{12})^1\left(z(r)\right)=-(Q^{21})^1\left(z(r)\right)=-(Q^{23})^3\left(z(r)\right)=(Q^{32})^3\left(z(r)\right)\\
=&\left(\frac{1}{a}\frac{\sin\left(\frac{r}{2a}\right)}{\cos^3\left(\frac{r}{2a}\right)}-\eta\frac{F(r)}{2a^2}\frac{1}{\cos^4\left(\frac{r}{2a}\right)}-\eta\frac{F(r)}{a^2}\frac{\sin^2\left(\frac{r}{2a}\right)}{\cos^4\left(\frac{r}{2a}\right)}-\eta\frac{2f(r)}{a}\frac{\sin\left(\frac{r}{2a}\right)}{\cos^3\left(\frac{r}{2a}\right)}\right)\frac{\del z^2}{\del\tau}(0)\\
&+\eta\frac{1}{2a^2}\frac{1}{\cos^4\left(\frac{r}{2a}\right)}\left(v^3(r)\frac{\del z^1}{\del\tau}(0)-v^1(r)\frac{\del z^3}{\del\tau}(0)\right)-\eta\frac{1}{\cos^4\left(\frac{r}{2a}\right)}\frac{\del f}{\del x^2}\bigg|_{x=z(r)},
\ee
and
\be
\begin{aligned}
&(Q^{12})^2\left(z(r)\right)=-(Q^{21})^2\left(z(r)\right)=-(Q^{13})^3\left(z(r)\right)=(Q^{31})^3\left(z(r)\right)\\
=&\left(-\frac{1}{a}\frac{\sin\left(\frac{r}{2a}\right)}{\cos^3\left(\frac{r}{2a}\right)}+\eta\frac{F(r)}{2a^2}\frac{1}{\cos^4\left(\frac{r}{2a}\right)}+\eta\frac{F(r)}{a^2}\frac{\sin^2\left(\frac{r}{2a}\right)}{\cos^4\left(\frac{r}{2a}\right)}+\eta\frac{2f(r)}{a}\frac{\sin\left(\frac{r}{2a}\right)}{\cos^3\left(\frac{r}{2a}\right)}\right)\frac{\del z^1}{\del\tau}(0)\\
&-\eta\frac{1}{2a^2}\frac{1}{\cos^4\left(\frac{r}{2a}\right)}\left(v^2(r)\frac{\del z^3}{\del\tau}(0)-v^3(r)\frac{\del z^2}{\del\tau}(0)\right)+\eta\frac{1}{\cos^4\left(\frac{r}{2a}\right)}\frac{\del f}{\del x^1}\bigg|_{x=z(r)},
\end{aligned}
\ee
and
\be
\begin{aligned}
&(Q^{13})^1\left(z(r)\right)=-(Q^{31})^1\left(z(r)\right)=-(Q^{23})^2\left(z(r)\right)=(Q^{32})^2\left(z(r)\right)\\
=&\left(\frac{1}{a}\frac{\sin\left(\frac{r}{2a}\right)}{\cos^3\left(\frac{r}{2a}\right)}-\eta\frac{F(r)}{2a^2}\frac{1}{\cos^4\left(\frac{r}{2a}\right)}-\eta\frac{F(r)}{a^2}\frac{\sin^2\left(\frac{r}{2a}\right)}{\cos^4\left(\frac{r}{2a}\right)}-\eta\frac{2f(r)}{a}\frac{\sin\left(\frac{r}{2a}\right)}{\cos^3\left(\frac{r}{2a}\right)}\right)\frac{\del z^3}{\del\tau}(0)\\
&+\eta\frac{1}{2a^2}\frac{1}{\cos^4\left(\frac{r}{2a}\right)}\left(v^1(r)\frac{\del z^2}{\del\tau}(0))-v^2(r)\frac{\del z^1}{\del\tau}(0)\right)-\eta\frac{1}{\cos^4\left(\frac{r}{2a}\right)}\frac{\del f}{\del x^3}\bigg|_{x=z(r)}.
\end{aligned}
\ee
The normal unit vector to the surface of constant geodesic distance $\del R$ has the components
\be
\begin{aligned}
n_1&=\left(\cos^2\left(\frac{r}{2a}\right)+\eta f\cos^2\left(\frac{r}{2a}\right)+\eta\frac{F}{a}\cot\left(\frac{r}{2a}\right)-\eta \frac{3F}{a}\frac{\cos^3\left(\frac{r}{2a}\right)}{\sin\left(\frac{r}{2a}\right)}\right)\sin\gamma_1\cos\gamma_2\\
+&\eta\frac{1}{2a}\cot\left(\frac{r}{2a}\right)\left(\frac{\del F}{\del \gamma_1}\cos\gamma_1\cos\gamma_2-\frac{\del F}{\del \gamma_2}~\frac{\sin\gamma_2}{\sin\gamma_1}+v^2\cos\gamma_1-v^3\sin\gamma_1\sin\gamma_2\right),
\end{aligned}
\ee
and
\be
\begin{aligned}
n_2=&\left(\cos^2\left(\frac{r}{2a}\right)+\eta f\cos^2\left(\frac{r}{2a}\right)+\eta\frac{F}{a}\cot\left(\frac{r}{2a}\right)-\eta \frac{3F}{a}\frac{\cos^3\left(\frac{r}{2a}\right)}{\sin\left(\frac{r}{2a}\right)}\right)\sin\gamma_1\sin\gamma_2\\
+&\eta\frac{1}{2a}\cot\left(\frac{r}{2a}\right)\left(\frac{\del F}{\del \gamma_1}\cos\gamma_1\sin\gamma_2+\frac{\del F}{\del \gamma_2}~\frac{\cos\gamma_2}{\sin\gamma_1}-v^1\cos\gamma_1+v^3\sin\gamma_1\cos\gamma_2\right),
\end{aligned}
\ee
and
\be
\begin{aligned}
n_3=&\left(\cos^2\left(\frac{r}{2a}\right)+\eta f\cos^2\left(\frac{r}{2a}\right)+\eta\frac{F}{a}\cot\left(\frac{r}{2a}\right)-\eta \frac{3F}{a}\frac{\cos^3\left(\frac{r}{2a}\right)}{\sin\left(\frac{r}{2a}\right)}\right)\cos\gamma_1\\
&+\eta\frac{1}{2a}\cot\left(\frac{r}{2a}\right)\left(-\frac{\del F}{\del \gamma_1}\sin\gamma_1-v^2\sin\gamma_1\cos\gamma_2+v^1\sin\gamma_1\sin\gamma_2\right).
\end{aligned}
\ee
For simplicity we introduce the three scalar fields
\be
A_i(x^1,x^2,x^3)=\frac{2}{L^2}\delta_{ij}x^j-\eta\frac{\del f}{\del x^i}.
\ee
We can then express the coupled differential equations for the maximally smooth triad field as
\be
\begin{aligned}
\Delta\phi+\cos\theta~\Delta\psi-\sin\theta~\nabla\theta\cdot\nabla\psi&=-\sin\theta(A_2\del_1\theta\!-\!A_1\del_2\theta)-\cos\theta\cos\psi(A_3\del_1\theta\!-\!A_1\del_3\theta)\\
&+\sin\theta\sin\psi(A_3\del_1\psi\!-\!A_1\del_3\psi)+\cos\theta\sin\psi(A_3\del_2\theta\!-\!A_2\del_3\theta)\\
&+\sin\theta\cos\psi(A_3\del_2\psi\!-\!A_2\del_3\psi)\\
\Delta\theta+\sin\theta~\nabla\psi\cdot\nabla\phi&=\cos\psi(A_3\del_1\psi\!-\!A_1\del_3\psi)-\sin\psi(A_3\del_2\psi\!-\!A_2\del_3\psi)\\
&+\sin\theta(A_2\del_1\phi\!-\!A_1\del_2\phi)+\cos\theta\cos\psi(A_3\del_1\phi\!-\!A_1\del_3\phi)\\
&-\cos\theta\sin\psi(A_3\del_2\phi\!-\!A_2\del_3\phi),\\
\Delta\psi+\cos\theta~\Delta\phi-\sin\theta~\nabla\theta\cdot\nabla\phi&=-\cos\theta(A_3\del_1\theta\!-\!A_1\del_3\theta)-\sin\theta\sin\psi(A_3\del_1\phi\!-\!A_1\del_3\phi)\\
&+\sin\psi(A_3\del_2\theta\!-\!A_2\del_3\theta)-\sin\theta\cos\psi(A_3\del_2\phi\!-\!A_2\del_3\phi).\\
\end{aligned}
\ee
For the boundary conditions we introduce furthermore the quantities
\be
h_i(\tau,\gamma_1,\gamma_2)=\epsilon_{ijk}\frac{\del f}{\del x^j}\bigg|_{x=z}\frac{\del z_P^k}{\del\tau}(0).
\ee
Then the boundary conditions become
\be
\begin{aligned}
\frac{\del\phi}{\del n}+\cos\theta~\frac{\del\psi}{\del n}&=\eta \frac{\cos\theta}{\cos^2\left(\frac{r}{2a}\right)}\left(-\frac{1}{2a^2}\frac{\del F}{\del\gamma_2}+h_3(r,\gamma_1,\gamma_2)\right)\\
&+\eta\frac{\sin\theta\cos\psi}{\cos^2\left(\frac{r}{2a}\right)}\left(-\frac{1}{2a^2}\frac{\del F}{\del\gamma_1}\cos\gamma_2+\frac{1}{2a^2}\frac{\del F}{\del\gamma_2}\cot\gamma_1\sin\gamma_2+h_2(r,\gamma_1,\gamma_2)\right)\\
&+\eta\frac{\sin\theta\sin\psi}{\cos^2\left(\frac{r}{2a}\right)}\left(\frac{1}{2a^2}\frac{\del F}{\del\gamma_1}\sin\gamma_2+\frac{1}{2a^2}\frac{\del F}{\del\gamma_2}\cot\gamma_1\cos\gamma_2+h_1(r,\gamma_1,\gamma_2)\right),\\
\frac{\del\theta}{\del n}&=\eta\frac{\sin\theta}{\cos^2\left(\frac{r}{2a}\right)}\left(\frac{1}{2a^2}\frac{\del F}{\del \gamma_1}\cos\gamma_2-h_2(r,\gamma_1,\gamma_2)\right)\\
&+\eta\frac{\cos\psi}{\cos^2\left(\frac{r}{2a}\right)}\left(\frac{1}{2a^2}\frac{\del F}{\del\gamma_1}+\frac{1}{2a^2}\cot\gamma_1\cos\gamma_2+h_1(r,\gamma_1,\gamma_2)\right),\\
\frac{\del\psi}{\del n}+\cos\theta~\frac{\del\phi}{\del n}&=\eta\frac{1}{\cos^2\left(\frac{r}{2a}\right)}\left(\frac{1}{2a^2}\frac{\del F}{\del\gamma_2}+h_3(r,\gamma_1,\gamma_2)\right).
\end{aligned}
\ee\\

These coupled differential equations implicitly define the angles $\phi$, $\theta$ and $\psi$ and therefore the maximally smooth triad. Although the solution of these equations is very involved and therefore cannot present any specific example, the analysis shows that no principle difficulties arise when the method is applied to more than two space dimensions.

%% file: gauss.tex
\chapter{The Gaussian shaped Perturbation}
\label{gauss}

In this chapter we analyse numerically the results of chapter \ref{twosphere}, in which we derived the equations governing the average of a two-sphere perturbed by a function $f(x^1,x^2)$, for the specific example of a Gaussian perturbation. The differential equation for the maximally smooth dyad field cannot be solved analytically and we make use of the numerical toolkit Gascoigne. After briefly introducing the method of finite element solutions we apply it to the problem. This chapter is based on work performed in collaboration with Thomas Richter.

\section{Finite Element Solutions}

Consider Poisson's equation,
\be
-\Delta\phi =0
\label{starequation}
\ee
in the domain $\Omega$ with a Neumann type boundary condition on $\del \Omega$,
\be
\frac{\del\phi}{\del n}=\eta\frac{df}{d\tau}(\tau,\gamma)+\eta\frac{1}{2a^2}\frac{1}{\cos(\frac{r}{2a})} \del_\gamma\left(\int_0^rf(s',\gamma)ds'\right)\equiv g_n,
\label{starboundary}
\ee  
which we will solve numerically. A detailed discussion of the use of the finite element approximation is given in \cite{Ciarlet78}, while we present a brief review of the essential points.\\

The ``weak formulation'' of the boundary value problem is obtained by the multiplying (\ref{starequation}) with a test function $\xi$ and integrating over the domain $\Omega$ to give
\be
-\int_\Omega \Delta\phi\cdot\xi dx = 0.
\ee
Integration by parts gives
\be
\int_\Omega \nabla\phi\cdot\nabla\xi dx = \int_{\del \Omega}
\frac{\del\phi}{\del n}\cdot \xi do.
\ee
 The weak formulation of the partial differential equation is given by $\phi\in V$ (where $V$ is an appropriate function space) with
\be
\int_\Omega \nabla\phi\cdot\nabla\xi dx = \int_{\del \Omega} g_n\cdot \xi do,\quad\forall\xi\in V.
\label{weakstar}
\ee
If $\phi$ is solution of this weak equation and is sufficiently regular - the Laplacian of $\phi$ has to exist - then we can reverse the integration by parts and see that $\phi$ is also the solution to
\be
-\int_\Omega \Delta\phi\cdot\xi dx=0, \text{ with }
\int_{\del \Omega} \frac{\del\phi}{\del  n}\cdot\xi do =
\int_{\del \Omega} g_n\cdot\xi do,\quad\forall\xi\in V.
\ee
One can then show that $\phi$ is a solution to equation (\ref{starequation}) with the boundary condition (\ref{starboundary}) for every point of the domain $\Omega$. \\

In the finite element method one solves the weak formulation (\ref{weakstar}) by introducing discrete subspaces $V_h\subset V$ and searching for solutions $u_h\in V_h$. This requires the discretization of the spatial domain $\Omega$ with a regular triangulation $\Omega_h$, a mesh of ``finite elements'', and the introduction of local basis functions of $V_h$ on this mesh. The finite element triangulation $\Omega_h$ of the domain $\Omega$ is a set of open elements $K\in\Omega_h$ with the following properties:
\begin{enumerate}
  \item Every element $K\in\Omega_h$ is a quadrilateral and all
    interior angles $\alpha$ of $K$ are bound 
    uniformly $\alpha\le \alpha_0 < \pi$  for all elements of $\Omega_h$. 
  \item The ratio of the diameter of the element $d_K$ and the radius of the
    largest inscribed circle $\rho_K$ is uniformly bounded in the grid
    $d_K/\rho_K\le \rho_0$. 
  \item The closure of the elements covers the domain:
    \[
    \Omega = \bigcup \bar K.
    \]
  \item For every two elements $K^1$ and $K^2$ in $\Omega_h$ with
    $K^1\neq K^2$ the intersection $K^1\cap K^2$ is empty and the
    intersection of the closures $\bar K^2\cap \bar K^2$ is either
    empty, a common edge or a common point.
\end{enumerate}
On this finite element mesh $\Omega_h$ we define the space of piece-wise polynomial functions
\be
V_h = \{ v\in V : v_{|K}\in \text{span}\{1,x,y,xy,x^2,\dots,x^r
  y^r\}\}. 
\label{starspace}
\ee
Using the grid points $x_i$, the finite element basis $\{\xi_i\}$ of this function space is usually chosen as
\be
\xi_i(x_j) = \delta_{ij}.
\ee
Thus, every basis function $\xi_i$ differs from zero only in the elements adjacent to the node $x_i$. In terms of the basis the function $\phi_h\in V_h$ can be expressed as
\be
\phi_h = \sum_{i=1}^N \phi^i\xi_i,\text{ with }\phi^i\in
\mathds{R}. 
\ee

The numerical solution of (\ref{weakstar}) in the finite element discretization is then the solution $\phi_h\in V_h$ to
\be
\sum_{i=1}^N \phi^i\int_{\Omega} \nabla\xi_i\cdot\nabla\xi_j dx =
\sum_{i=1}^N \phi^i\int_{\Omega} g_n \cdot\xi_j dx,\;\forall
j=1,\dots,N. 
\label{discstar}
\ee
Since the basis functions $\xi_i$ have a small support, the integrals cover only small areas of the domain and every function $\xi_i$ couples only to a small number of other basis functions~$\xi_j$. For a two dimensional computation with piece-wise quadratic finite elements every basis function couples usually to no more than 25 neighboring basis functions. Equation (\ref{discstar}) gives a system of $N$ linear equations with $N$ unknowns. The resulting matrix is very sparse; every row has of the order of 25 non-zero entries. This system can be solved efficiently with a multi-grid algorithm \cite{Hackbusch}. 

\section{Approximation of curved boundaries}
 
\begin{figure}
\begin{center}
\includegraphics[width=0.8\textwidth]{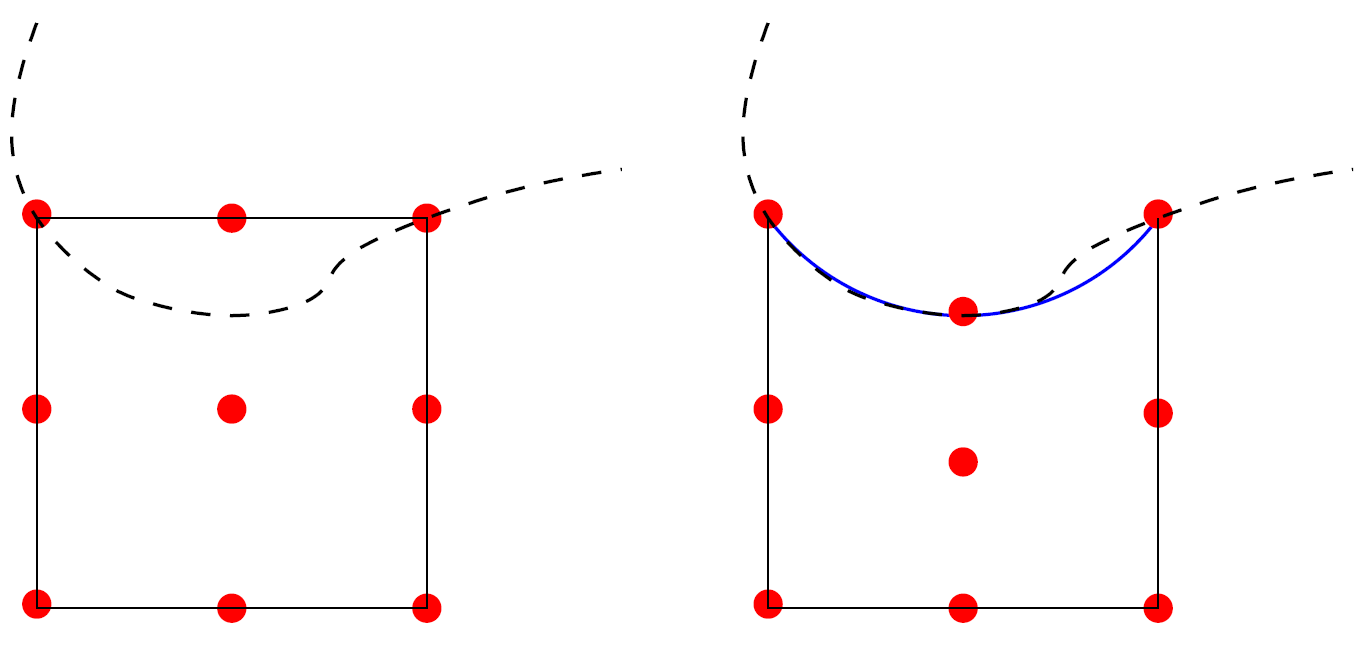}
\caption{\label{starfig}Approximation of a curved boundary with a linear and iso parametric (here quadratic) transformation $T_K$ to the computational element.}
\end{center}
\end{figure}

The discretized domain is not identical to the original continuous domain and the approximation by quadrilaterals leads to a significant error when approximating the curved boundary. This must be carefully considered. To gain a better approximation of the boundary we use so called iso-parametric finite elements. Instead of defining the finite element space directly on the computational grid $\Omega_h$, as in (\ref{starspace}), we define a reference function space on the unit square $\hat K=[0,1]^2$ 
\be
{\cal Q}^r = \{\hat v\in \text{span}\{1,x,y,xy,x^2,\dots,x^r y^r\}\},
\ee
and use a nonlinear mapping onto the computational elements $T_K:\hat K\to K\in \Omega_h$. This mapping $T_K$ is an element of the same reference space ${\cal Q}^r$. The finite element space $V_h$ is then given by
\be
V_h = \{v\in V : \exists \hat v\in {\cal Q}^r, v_{|K}=T_K(\hat v)\}.
\ee
To improve the definition of the mapping $T_K$ on an element $K$ at the curved boundary, we add further mesh points. Figure (\ref{starfig}) shows an example in which this iso-parametric approach results in a much finer approximation of the boundary.\\

This finite element method with iso-parametric elements is implemented in the software library \textsl{Gascoigne 3D} \cite{Gascoigne} developed by M. Braack, D. Meidner, R. Becker, T. Richter, B. Vexler at the universities of Heidelberg, Linz, Kiel and Pau, which we will employ in the remainder of this chapter.

\section{The Perturbed Two-Sphere Revisited}

\begin{figure}[]
\center{\includegraphics[width=0.45\textwidth]{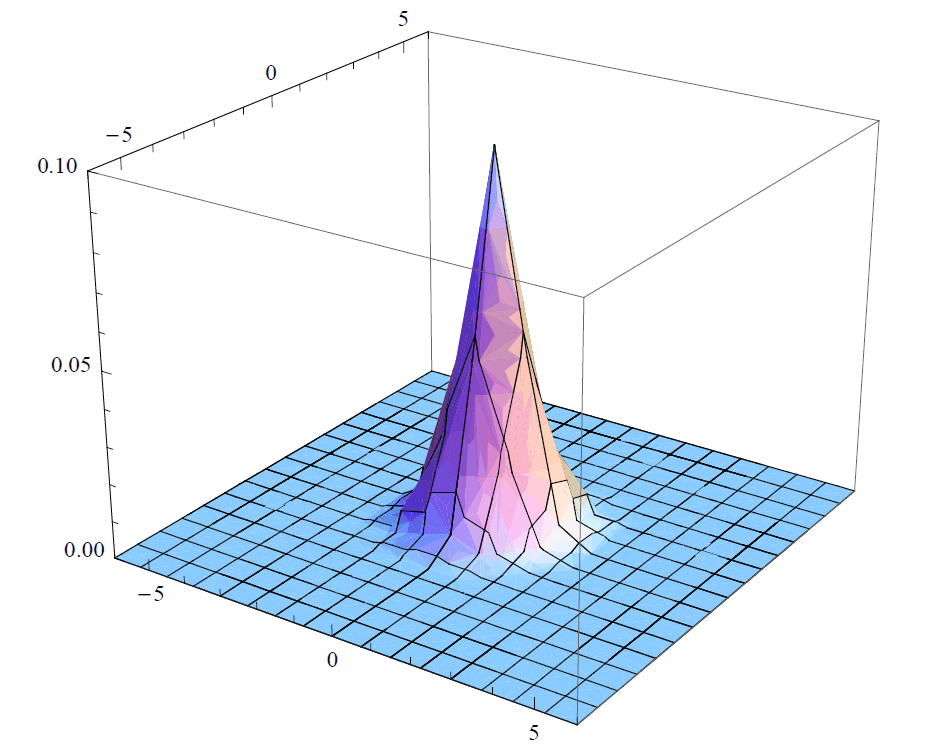}}
\caption{The Gaussian shaped perturbation function.}
\label{bild1}
\end{figure}

We now apply the formalism outlined above to the analysis of the perturbed sphere presented in chapter~\ref{twosphere}. As a concrete example, we choose a sphere of radius $a=10$ and a perturbation function of Gaussian form. In general, the Gaussian function is of the form
\be
f(x^1,x^2)=D \exp\left(-\frac{1}{\sigma^2}\left((x^1)^2+(x^2)^2\right)\right).
\ee
To keep the perturbation small, we choose the amplitude $D=0.1$ and standard deviation $\sigma=1$. The resulting function is shown in Figure~\ref{bild1}.\\

We would expect a reasonable averaging process, when applied to the perturbed two-sphere, to significantly diminish and broaden the perturbation function., which should furthermore tend to disappear for increasingly large averaging radii. At the same time the radius of the background sphere might be slightly altered. \\

To take the average we must first transform the perturbation to the coordinate system of the reference point $(x_0^1,x_0^2)$ using (\ref{coordtrans2}). Doing so, we see that the transformed Gaussian function is 
\be
f(\widetilde{x}^1&,\widetilde{x}^2)=D\exp\left(-\frac{1}{\sigma^2}~\left((x^1(\widetilde{x}^1))^2+x^2(\widetilde{x}^2)\right)^2)\right)\\
&=D\exp\left(-\frac{1}{\sigma^2}~\frac{16a^4\left((x_0^1)^2+(x_0^2)^2+2(x_0^1\widetilde{x}^1+x_0^2\widetilde{x}^2)+(\widetilde{x}^1)^2+(\widetilde{x}^2)^2\right)}{16a^4-8a^2(x_0^1\widetilde{x}^1+x_0^2\widetilde{x}^2)+((x_0^1)^2+(x_0^2)^2)((\widetilde{x}^1)^2+(\widetilde{x}^2)^2)}\right).
\ee
\begin{figure}[t]
\center{\includegraphics[width=0.4\textwidth]{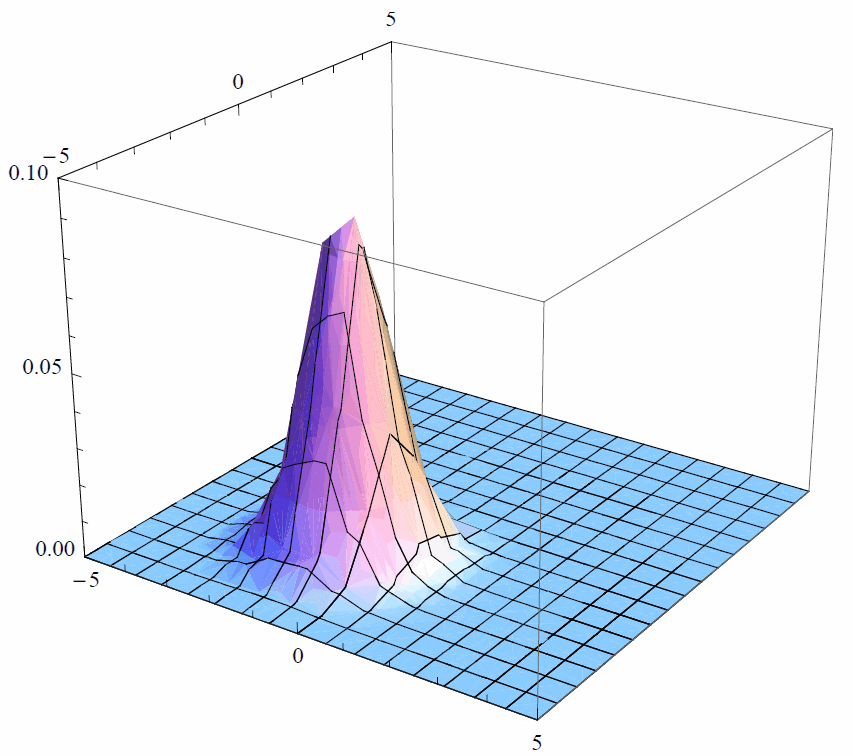}\hspace{1.cm}\includegraphics[width=0.4\textwidth]{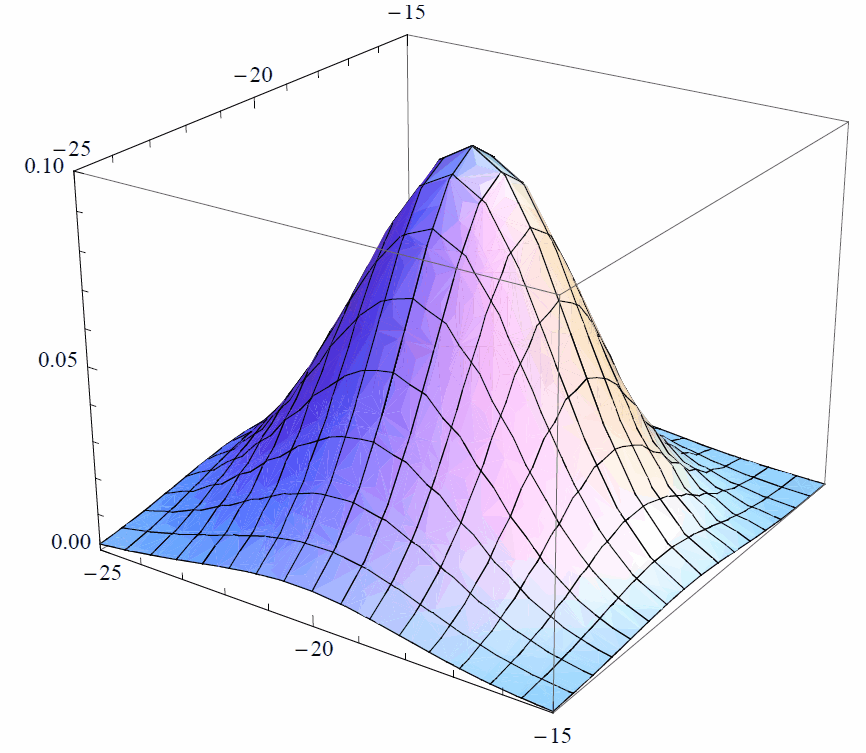}}
\caption{The Gaussian shaped perturbation function in the coordinate systems of the reference points $(x_0^1,x_0^2)=(1,2)$ and $(x_0^1,x_0^2)=(20,20)$.}
\label{bild2}
\end{figure}
From this we can see that the transformed function keeps the same amplitude but grows broader for increasing $(x_0^1,x_0^2)$. In Figure~\ref{bild2} we show the transformed function for the reference points $(x_0^1,x_0^2)=(1,2)$ and $(x_0^1,x_0^2)=(20,20)$. This clearly shows this broadening at distant reference points, which originates from the larger angle of rotation at such distances, and from the subsequent stereographic projection.\\

\begin{figure}[h]
\center{\includegraphics[width=0.5\textwidth]{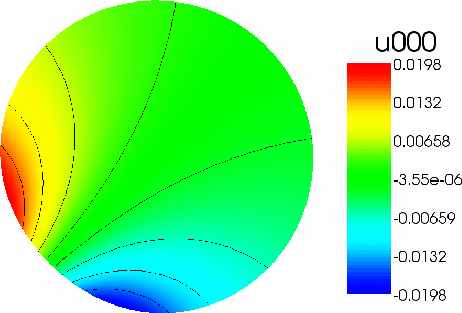}}
\caption{The result for $\phi(x^1,x^2)$ for the references point $(x_0^1,x_0^2)=(1,1)$ and $r=3$.}
\label{bild3}
\end{figure}

\begin{figure}[h]
\center{\includegraphics[width=0.5\textwidth]{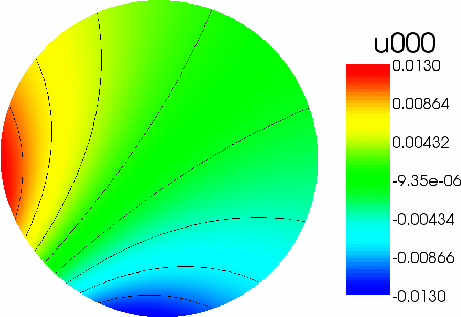}}
\caption{The result for $\phi(x^1,x^2)$ for the references point $(x_0^1,x_0^2)=(1,1)$ and $r=5$.}
\label{bild4}
\end{figure}

In terms of the geodesic coordinates the function $f(\tau,\gamma)$ is 
\be
f(\tau,\gamma)\!=\!D\!\exp\!\left(\!-\frac{1}{\sigma^2}~\frac{4a^2\big((x_0^1)^2\!+\!(x_0^2)^2\!+\!4a\tan(\frac{\tau}{2a})(x_0^1\cos\gamma\!+\!x_0^2\sin\gamma)\!+\!4a^2\tan^2(\frac{\tau}{2a})\big)}{4a^2\!-\!4a\tan(\frac{\tau}{2a})\big(x_0^1\cos\gamma\!+\!x_0^2\sin\gamma\big)\!+\!\big((x_0^1)^2\!+\!(x_0^2)^2\big)\tan^2(\frac{\tau}{2a})}\right)\notag
\ee
and the function $h(\tau,\gamma)$, defined in (\ref{hfunction}) is
\be
h(\tau,\gamma)=\frac{8a^2\big(4a^2+(x_0^1)^2+(x_0^2)^2\big)\big(x_0^1\sin\gamma-x_0^2\cos\gamma\big)~f(\tau,\gamma)}{\sigma^2\cos^2(\frac{\tau}{2a})\left(4a^2\!-\!4a\tan(\frac{\tau}{2a})\big(x_0^1\cos\gamma\!+\!x_0^2\sin\gamma\big)\!+\!\big((x_0^1)^2\!+\!(x_0^2)^2\big)\tan^2(\frac{\tau}{2a})\right)^2}.\notag
\ee
This function defines the longitudinal deviation $v(\tau,\gamma)$ of the perturbed geodesic via the differential equation (\ref{pertgeov}). With $v(\tau,\gamma)$ we can compute the connector (\ref{sphereconnector}) and the boundary of the geodesic region, $\del R$, which in parameter representation is given by (\ref{areaR}). The differential equation for the scalar field $\phi(x^1,x^2)$, which defines the maximally smooth dyad field, is the Laplace equation (\ref{laplace}) with the boundary condition (\ref{randbed}) on $\del R$. The numerical solution to this equation determined with Gascoigne for the reference point $(x_0^1,x_0^2)=(1,1)$ is shown in Figure \ref{bild3} for the averaging radius $r=3$ and in Figure \ref{bild4} for the averaging radius $r=5$. Since the perturbation function enters the differential equation only through the boundary condition, $\phi(x^1,x^2)$ is essentially zero in the region far away from the perturbation and also just across the perturbation. Furthermore, the maximum value is larger for the smaller averaging radius.

\begin{figure}[t]
\center{\includegraphics[width=0.4\textwidth]{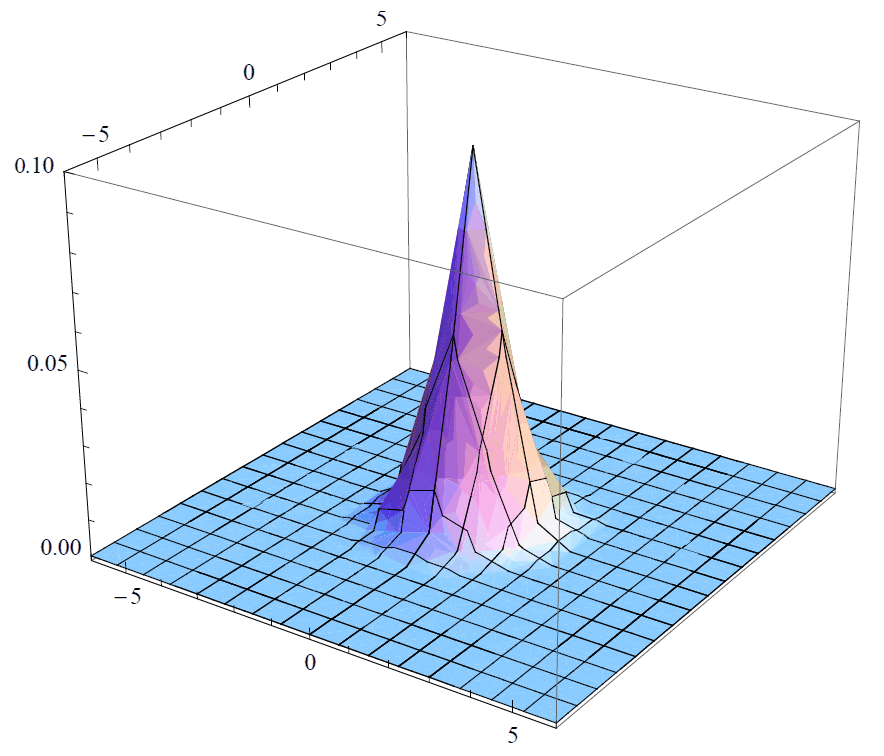}\hspace{1.cm}\includegraphics[width=0.4\textwidth]{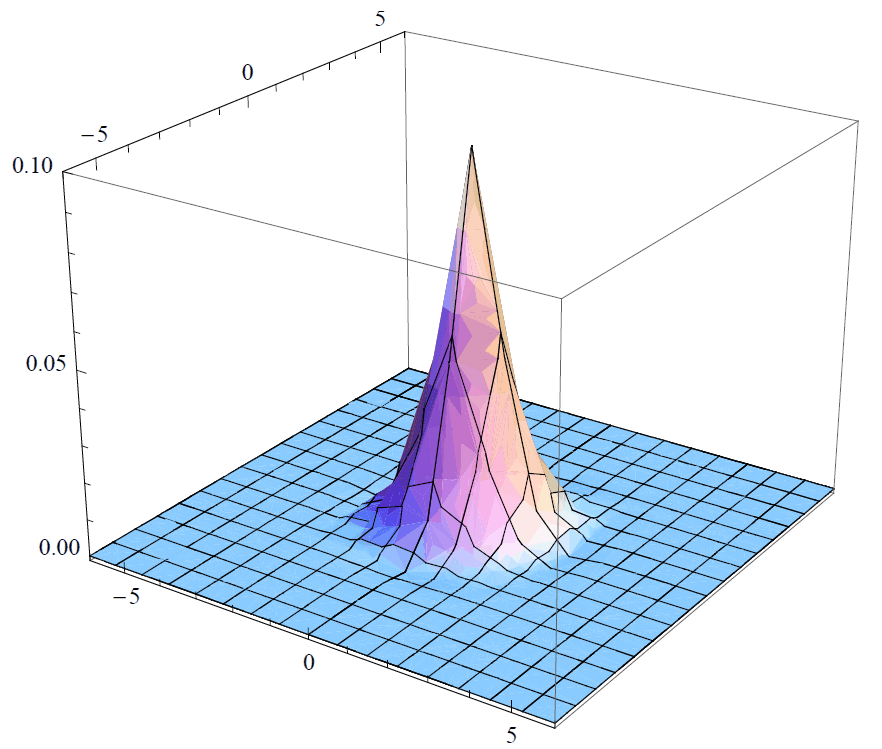}}
\caption{The perturbation before and after averaging over a region of radius r=3.}
\label{bild5}
\end{figure}

With the solution for $\phi(x^1,x^2)$ for all reference points in a domain around the perturbation function $f(x^1,x^2)$, we can construct the averaged metric from (\ref{ergebnis}). The initial perturbation and the averaged perturbation function for the averaging radius $r=3$ are shown in Figure~\ref{bild5}. It is clear that the perturbation function is not much altered by the averaging process and that the effect of averaging is much smaller than expected. A reasonable averaging process should diminish the perturbation significantly and therefore we must conclude that, in its current form, the averaging process is of limited use. We will therefore elaborate on potential problems in the process, and discuss possible improvements, briefly here and in greater depth in the next chapter.\\

\begin{figure}[h]
\center{\includegraphics[width=0.45\textwidth]{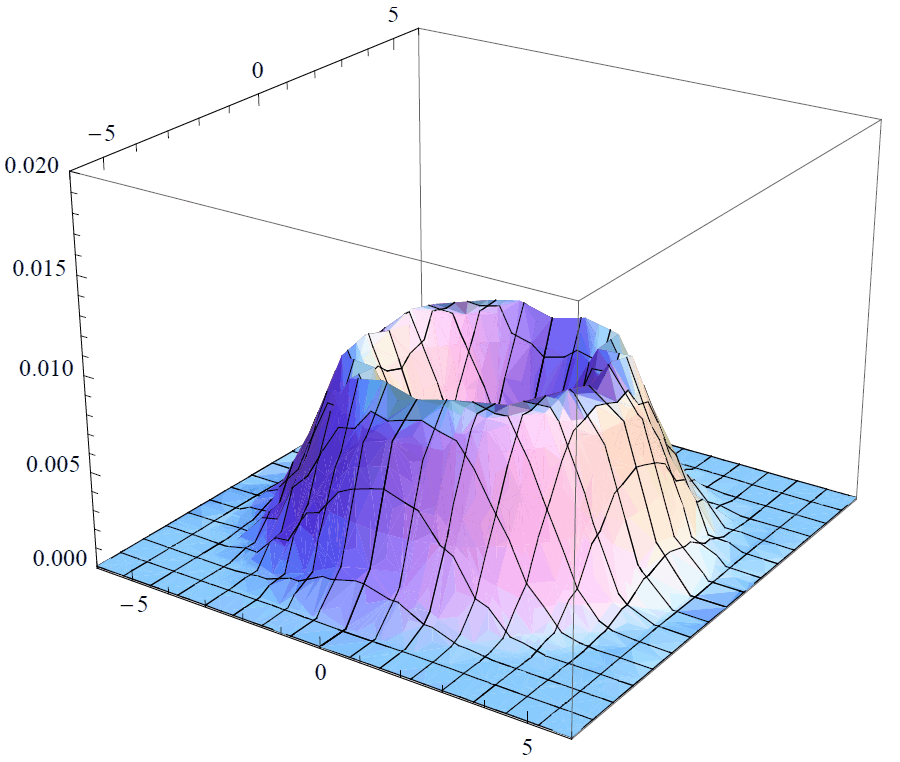}}
\caption{Variance for the averaging radius $r=3$.}
\label{bild6}
\end{figure}

The problem might arise from the extended spread of the bundle of parallel transported dyads at each reference point, causing the averaged dyads to be shortened when compared to the maximally smooth dyad. A first ansatz to improve the averaging process is therefore to renormalize the averaged dyad at each reference point witht the variance $\av{(\phi-\av{\phi})^2}=\av{\phi^2}-\av{\phi}^2$, but this can be shown to be untenable. For the perturbed two-sphere we find
\be
\av{\phi^2}-\av{\phi}^2&=I_0^{-1}~\int_0^{2\pi}\int_0^rf_2(\tau)\phi^2(\tau,\gamma)\cos^4\left(\frac{\tau}{2a}\right)d\gamma d\tau\\
&-I_0^{-2}~\left(\int_0^{2\pi}\int_0^rf_2(\tau)\phi(\tau,\gamma)\cos^4\left(\frac{\tau}{2a}\right)d\gamma d\tau\right)^2.
\ee
The solution for an averaging radius $r=3$ and for all reference points is shown in Figure \ref{bild6}. Due to the symmetry of the perturbation function at the reference point $(x_0^1,x_0^2)=(0,0)$, the boundary condition vanishes and therefore $\phi(x^1,x^2)=0$. The variance thus also vanishes for this reference point which can be seen from the central dip in Figure \ref{bild6}. The variance can therefore not be used to change the length of the averaged dyad at this point.\\

A more promising approach that we are currently working on is the renormalization of the averaged dyads with the renormalization factor 
\be
\alpha_\mathrm{ren}\equiv\left(1+\mu\frac{|\Delta R|}{|\Delta R|_\mathrm{max}}\right)^{-1},
\ee
where $\mu$ is a constant that must be defined according to the problem under consideration, $R$ is the Ricci scalar and $|\Delta R|_\mathrm{max}$ is the maximum value of $\Delta R$ in the averaging domain. This renormalization factor appears only in the integrals in (\ref{avdyad}) and not in the normalization (\ref{norm}). For the perturbed two-sphere the factor becomes
\be
\alpha_\mathrm{ren}=1-\eta\mu\left(\frac{2L^2}{a^4}\Delta f+\frac{L^2}{2a^4}\left(x\del_x(\Delta f)+y\del_y(\Delta f)\right)+\frac{L^4}{8a^4}\Delta(\Delta f)\right),
\ee
and can completely be included in the scalar field $B(\tau,\gamma)$, defined in (\ref{aandb}), to give
\be
B_\mathrm{ren}(\tau,\gamma)=B(\tau,\gamma)-\frac{|\Delta R|}{|\Delta R|_\mathrm{max}}.
\ee

This approach must alter the length of the averaged dyads and results are expected in the near future.\\

In this chapter we have applied the averaging process to the metric of a perturbed two-sphere for the specific example of a Gaussian perturbation. Solving the differential equation for the maximally smooth dyad field numerically with the toolkit Gascoigne has led to an averaged metric with a perturbation that is essentially unaltered by the averaging process. We elaborated on the reason for the small effect of the averaging process and presented an ansatz which might improve the process. Another approach which aims at the core of the problem will be presented in the next chapter.

%% file: julelagrangian.tex
\chapter{The Curvature Lagrangian}
\label{julelagrangian}

In this chapter we argue that the maximally smooth Vielbein field employed in the averaging process thus far is too close to the specific Vielbein field that is invariant under the process. We suggest a different Vielbein field which will improve the averaging effect by taking the local curvature of the manifold into account. This Vielbein field will be determined from a modification of the Lagrangian which determines the invariant Vielbein.

\section{Drawbacks of the Averaging Process}

In the previous section we applied the averaging process defined in chapter \ref{process} to a two-sphere perturbed with a Gaussian perturbation. A reasonable averaging process would be expected to diminish this function and tend to remove it for larger averaging radii. Instead, our procedure left the perturbation basically invariant and therefore the process needs to be modified to overcome this problem. We presented in the previous chapter a renormalization factor that might address the shortening of the Vielbeins under parallel transport. In this chapter we instead consider a redefinition of the Lagrangian used to determine the Vielbeins in the first place.\\

There is one particular Vielbein field that is invariant under the averaging process. This Vielbein field can be constructed by parallel transporting the reference Vielbein from the origin $\widetilde{E}^a{}_\mu(0)$ to each point of the geodesic region $R$ and therefore we will refer to it as the geodetic induced parallel field. Clearly, this Vielbein field is only determined up to global rotations. The averaging process parallel transports all Vielbeins back to the origin and the averaging returns the initial Vielbein $\widetilde{E}^a{}_i(0)$. The Lagrangian which defines the geodetic induced parallel field is
\be
\mathcal{L}_\mathrm{inv}=\left(\pounds_tE^a{}_\mu\right)\left(\pounds_tE^b{}_\nu\right)\delta_{ab}t^\mu t^\nu,
\label{juleinv}
\ee
where $\pounds_t$ denotes the Lie derivative with respect to the tangent vector $t^\mu$ to the geodesic bundle through the origin $z^\mu$. \\

The Lagrangian we chose to determine the Vielbeins was that for the maximally smooth Vielbeins (\ref{mslagrangian})
\be
 \mathcal{L}_\mathrm{MS}=\left(D_\mu E^a{}_\rho(x)\right)\left(D_\nu E^b{}_\sigma(x)\right)g^{\mu\nu}(x)g^{\rho\sigma}(x)\eta_{ab}.
\ee

This Lagrangian leads to a differential equation whose solution $\phi_\mathrm{MS}$ is very similar to the solution $\phi_\mathrm{inv}$ for the differential equation defined by (\ref{juleinv}). The average of this Vielbein field is therefore only mildly altered.\\ 

Knowing the Lagrangian that defines the invariant scalar field $\phi_\mathrm{inv}(x^1,x^2)$ allows us to modify this Lagrangian in order to find a scalar field that is definitely not invariant under the averaging process. The modification is chosen to be proportional to the Ricci scalar~$R$. Since the modifier should be dimensionless and the dimension of the Ricci scalar is $(\mathrm{length})^{-2}$ we must balance this with a factor of the dimension $(\mathrm{length})^2$. A reasonable choice is the length of the geodesic $s^2$. We therefore define the curvature Lagrangian
\be
\mathcal{L}_R=\left(\pounds_tE^a{}_\mu\right)\left(\pounds_tE^b{}_\nu\right)\delta_{ab}\left(t^\mu t^\nu+g^{\mu\nu}Rs^2\right).
\ee
For the flat case with $R=0$ the modification vanishes and the solution merges into the invariant scalar field. The concept of this Lagrangian is to construct the geodetic induced parallel field and to then rotate the Vielbeins according to the curvature at their location. The rotated Vielbeins are then parallel transported to a common point and averaged over to return the averaged Vielbein and therefore the averaged metric as before.\\

To further explain our approach, we return to consider the perturbed two-sphere, starting with a detailed summary of the important quantities and equations.

\section{Summary of the Perturbed Two-Sphere}

In this section we review the important quantities and equations for the perturbed two-sphere, which were derived in chapter \ref{twosphere}. We also compute the tangent vector that appears in the curvature Lagrangian in terms of the coordinates of the stereographic projection plane. \\

The metric for the perturbed two-sphere after stereographic projection is
\be
(g_P)_{ij}=(1+2\eta f(x^1,x^2))\left(\frac{2a}{L}\right)^4\delta_{ij}\quad\mathrm{and}\quad (g_P)^{ij}=(1-2\eta f(x^1,x^2))\left(\frac{L}{2a}\right)^4\delta^{ij},
\ee
where $L^2\equiv 4a^2+(x^1)^2+(x^2)^2$. The square root of the determinant is then
\be
\sqrt{g_P}=(1+2\eta f)\left(\frac{2a}{L}\right)^4
\ee
and a suitable choice for the reference dyad field is
\be
\widetilde{E}^a{}_i=(1+\eta f)\left(\frac{2a}{L}\right)^2\delta^a_i.
\label{julerefdyad}
\ee
The geodesic bundle through the origin is 
\be
z^1(s,\gamma)&=2a\tan\left(\frac{s}{2a}\right)\cos\gamma-\eta\frac{\cos\gamma}{\cos^2(\frac{s}{2a})}\int_0^sf(s',\gamma)ds'+\eta\frac{v(s,\gamma)}{\cos^2(\frac{s}{2a})}\sin\gamma,\\
z^2(s,\gamma)&=2a\tan\left(\frac{s}{2a}\right)\cos\gamma-\eta\frac{\cos\gamma}{\cos^2(\frac{s}{2a})}\int_0^sf(s',\gamma)ds'+\eta\frac{v(s,\gamma)}{\cos^2(\frac{s}{2a})}\sin\gamma,
\label{julegeodesics}
\ee
where $s$ is the arc length of the geodesics,
\be
s=\tau+\eta\int_0^\tau f(\tau',\gamma)d\tau'\qquad\mathrm{and}\qquad \tau=s-\eta\int_0^s f(s',\gamma)ds',
\ee
and $v(\tau,\gamma)$ is the solution to the differential equation
\be
\frac{d^2v(\tau)}{d\tau^2}+\frac{v(\tau)}{a^2}=\frac{h(\tau)}{\cos^2(\frac{\tau}{2a})}.
\label{julevequation}
\ee
The function $h(\tau,\gamma)$ is
\be
h(\tau)=\frac{\del f}{\del x^2}\bigg|_{(x^1,x^2)=(z^1(\tau),z^2(\tau))}\frac{\del z_P^1}{d\tau}(0)-\frac{\del f}{\del x^1}\bigg|_{(x^1,x^2)=(z^1(\tau),z^2(\tau))}\frac{\del z_P^2}{d\tau}(0).
\ee
The tangent vectors to the geodesics (\ref{julegeodesics}) are defined as
\be
t^1(s,\gamma)=\frac{dz^1}{ds}&=\frac{\cos\gamma}{\cos^2(\frac{s}{2a})}-\eta\frac{1}{a}\frac{\sin(\frac{s}{2a})}{\cos^3(\frac{s}{2a})}\cos\gamma\int_0^sf(s',\gamma)ds'-\eta\frac{\cos\gamma}{\cos^2(\frac{s}{2a})}f(s,\gamma)\\
&+\eta\frac{v(s,\gamma)}{a}\frac{\sin(\frac{s}{2a})}{\cos^3(\frac{s}{2a})}\sin\gamma+\eta\frac{\sin\gamma}{\cos^2(\frac{s}{2a})}\frac{dv}{ds}(s,\gamma),\\
t^2(s,\gamma)=\frac{dz^2}{ds}&=\frac{\sin\gamma}{\cos^2(\frac{s}{2a})}-\eta\frac{1}{a}\frac{\sin(\frac{s}{2a})}{\cos^3(\frac{s}{2a})}\sin\gamma\int_0^sf(s',\gamma)ds'-\eta\frac{\sin\gamma}{\cos^2(\frac{s}{2a})}f(s,\gamma)\\
&-\eta\frac{v(s,\gamma)}{a}\frac{\sin(\frac{s}{2a})}{\cos^3(\frac{s}{2a})}\cos\gamma-\eta\frac{\cos\gamma}{\cos^2(\frac{s}{2a})}\frac{dv}{ds}(s,\gamma).
\label{juletangent1}
\ee
In terms of geodesic coordinates $s$ and $\gamma$ we thus find that
\be
L^2(s,\gamma)=\frac{4a^2}{\cos^2(\frac{s}{2a})}-4\eta a\frac{\sin\left(\frac{s}{2a}\right)}{\cos^3\left(\frac{s}{2a}\right)}\int_0^sf(s',\gamma)ds'
\ee
and therefore that the metric is
\be
(g_P)_{ij}=\left(1+2\eta f(s,\gamma)+2\eta a\tan\left(\frac{s}{2a}\right)\int_0^sf(s',\gamma)ds'\right)\cos^4\left(\frac{s}{2a}\right)\delta_{ij}.
\ee
For constant arc length $s=r$, where $r$ is the averaging radius, the geodesics (\ref{julegeodesics}) define the boundary $\del R$ of the geodesic region $R$. The square root of the determinant at the boundary $\del R$ is then
\be
\sqrt{g_P}\big|_{\del R}=\left(1+2\eta f(r,\gamma)+\eta \frac{1}{a}\tan\left(\frac{r}{2a}\right)\int_0^rf(s',\gamma)ds'\right)\cos^4\left(\frac{r}{2a}\right)
\ee
and the reference dyad field becomes
\be
\widetilde{E}^a{}_i\big|_{\del R}=\left(1+\eta f(r,\gamma)+\eta\frac{1}{a}\tan\left(\frac{r}{2a}\right)\int_0^rf(s',\gamma)ds'\right)\cos^2\left(\frac{r}{2a}\right)\delta^a_i.
\ee
The normal vector to the boundary $\del R$ is
\be
n_1(\gamma)&=(1+\eta f)\cos^2\left(\frac{r}{2a}\right)\cos\gamma+\eta\frac{1}{a}\tan\left(\frac{r}{2a}\right)\cos^2\left(\frac{r}{2a}\right)\cos\gamma\int_0^rf(s',\gamma)ds'\\
&-\eta\frac{1}{2a}\frac{\cos\left(\frac{r}{2a}\right)}{\sin\left(\frac{r}{2a}\right)}\sin\gamma\frac{d}{d\gamma}\int_0^rf(s',\gamma)ds'+\eta\frac{v(s,\gamma)}{2a}\frac{\cos\left(\frac{r}{2a}\right)}{\sin\left(\frac{r}{2a}\right)}\sin\gamma,\\
n_2(\gamma)&=(1+\eta f)\cos^2\left(\frac{r}{2a}\right)\sin\gamma+\eta\frac{1}{a}\tan\left(\frac{r}{2a}\right)\cos^2\left(\frac{r}{2a}\right)\sin\gamma\int_0^rf(s',\gamma)ds'\\
&+\eta\frac{1}{2a}\frac{\cos\left(\frac{r}{2a}\right)}{\sin\left(\frac{r}{2a}\right)}\cos\gamma\frac{d}{d\gamma}\int_0^rf(s',\gamma)ds'-\eta\frac{v(s,\gamma)}{2a}\frac{\cos\left(\frac{r}{2a}\right)}{\sin\left(\frac{r}{2a}\right)}\cos\gamma,
\ee
and so
\be
t^i n_i\big|_{\del R}=1.
\ee\\

For the coordinate transformation from the geodesic coordinates to the coordinates of the stereographic projection plane, we need to invert (\ref{julegeodesics}). With
\be
F(s,\gamma)=\int_0^s f(s',\gamma)ds'
\ee
we find
\be
s&=2a\arctan\left(\frac{\sqrt{(x^1)^2+(x^2)^2}}{2a}\right)+\eta F(x^1,x^2),\\
\gamma&=\arctan\left(\frac{x^2}{x^1}\right)+\eta\frac{L^2}{4a^2}\frac{v(x^1,x^2)}{\sqrt{(x^1)^2+(x^2)^2}}.
\ee
The tangent vectors (\ref{juletangent1}) to the geodesics in terms of the coordinates $x^1$ and $x^2$ are therefore
\be
t^1(x^1,&x^2)=\frac{L^2}{4a^2}\frac{x^1}{\sqrt{(x^1)^2+(x^2)^2}}-\eta\frac{L^2}{4a^2}\frac{x^1 f(x^1,x^2)}{\sqrt{(x^1)^2+(x^2)^2}}-\eta\frac{L^4}{16a^4}\frac{x^2v(x^1,x^2)}{(x^1)^2+(x^2)^2}\\
&+\eta\frac{L^2x^2v(x^1,x^2)}{8a^4}+\eta\frac{L^4}{16a^4}\frac{x^1x^2}{((x^1)^2+(x^2)^2)}\frac{\del v}{\del x^1}+\eta\frac{L^4}{16a^4}\frac{(x^2)^2}{((x^1)^2+(x^2)^2)}\frac{\del v}{\del x^2},\\
t^2(x^1,&x^2)=\frac{L^2}{4a^2}\frac{x^2}{\sqrt{(x^1)^2+(x^2)^2}}-\eta\frac{L^2}{4a^2}\frac{x^2 f(x^1,x^2)}{\sqrt{(x^1)^2+(x^2)^2}}+\eta\frac{L^4}{16a^4}\frac{x^1v(x^1,x^2)}{(x^1)^2+(x^2)^2}\\
&-\eta\frac{L^2x^1v(x^1,x^2)}{8a^4}-\eta\frac{L^4}{16a^4}\frac{(x^1)^2}{((x^1)^2+(x^2)^2)}\frac{\del v}{\del x^1}-\eta\frac{L^4}{16a^4}\frac{(x^2)^2}{(x^1x^2+(x^2)^2)}\frac{\del v}{\del x^2}.
\label{juletangent2}
\ee
Furthermore, we find
\be
\eta\frac{d^2v}{ds^2}&=\eta\frac{L^2}{8a^4}\left(x^1\frac{\del v}{\del x^1}+x^2\frac{\del v}{\del x^2}\right)\\
&+\eta\frac{L^4}{16a^4((x^1)^2+(x^2)^2)}\left((x^1)^2\frac{\del^2v}{(\del x^1)^2}+2x^1x^2\frac{\del v}{\del x^1\del x^2}+(x^2)^2\frac{\del^2v}{(\del x^2)^2}\right).
\label{secondderivative}
\ee

We will need the tangent vectors (\ref{juletangent2}) for the Lagrangian (\ref{juleinv}) and equation (\ref{secondderivative}) for the proof that this Lagrangian defines the geodetic induced parallel field in the case of the perturbed two-sphere. This will be shown in the next section.

\section{The Geodetic Induced Parallel Field}
 
For the perturbed two-sphere we can compute the scalar field $\phi_\mathrm{inv}(x^1,x^2)$ that defines the geodetic induced parallel field with the connector, which was defined in (\ref{juleconnector}),
\be
V^k{}_l(\tau,0;{\cal C}_{0\tau})=\cos^{-2}\left(\frac{\tau}{2a}\right)&\bigg(\left(1-\eta f(\tau,\gamma)+\eta f(0,\gamma)\right)\delta^k_l\\
&+\eta\left(v'(\tau,\gamma)+\frac{1}{a}\tan\left(\frac{\tau}{2a}\right)v(\tau,\gamma)\right)\epsilon^k{}_l\bigg).
\ee
From this we find the hatted connector 
\be
&\widehat{V}_i{}^j(\tau,0;{\cal C}_{0\tau})=g_{ik}(\tau,\gamma)V^k{}_l(\tau,0;{\cal C}_{0\tau})g^{lj}(0,\gamma)\\
&=\cos\left(\frac{\tau}{2a}\right)\!\!\bigg(\left(1+\eta f(\tau,\gamma)-\eta f(0,\gamma)\right)\delta_i^j+\eta\left(v'(\tau,\gamma)+\frac{1}{a}\tan\left(\frac{\tau}{2a}\right)v(\tau,\gamma)\right)\epsilon_i{}^j\!\bigg),
\ee
where the prime denotes the partial derivative with respect to $\tau$. This connector can be used to parallel transport the reference dyad from the origin,
\be 
\widetilde{E}^a{}_i(0)=(1+\eta f(0))\delta^a_i
\ee
to all points of the manifold. The local rotation of the resulting dyad field as compared to the reference dyad field then defines the scalar field $\varphi_\mathrm{inv}(x^1,x^2)$,
\be
U_a{}_b(\varphi_\mathrm{inv}(x^1,x^2))\widetilde{E}^b{}_i(\tau,\gamma)=\widehat{V}_i{}^j(\tau,0;{\cal C}_{0\tau})\widetilde{E}^a{}_j(0,\gamma),
\ee
and the result for the scalar field is
\be
\varphi_\mathrm{inv}(x^1,x^2)=-\frac{v}{2a^2}\sqrt{(x^1)^2+(x^2)^2}-\frac{L^2}{4a^2}\frac{1}{\sqrt{(x^1)^2+(x^2)^2}}\left(x^1\frac{\del v}{\del x^1}+x^2\frac{\del v}{\del x^2}\right).
\label{julephi}
\ee
In geodesic coordinates this result is
\be
\varphi_\mathrm{inv}(s,\gamma)=-\frac{v}{a}\tan\left(\frac{s}{2a}\right)-\frac{\del v}{ds}.
\ee

The Lagrangian that describes the geodetic induced parallel field is
\be
\mathcal{L}_\mathrm{inv}=\left(\pounds_tE^a{}_i\right)\left(\pounds_tE^b{}_j\right)\delta_{ab}t^it^j.
\ee
With the definition of the Lie derivative (\ref{liedef}) this Lagrangian leads to the differential equation
\be
D_i\left(t^it^j(\del_j\phi_\mathrm{inv})\right)=-D_i\left(P^i+Q^i\right)
\label{julediff}
\ee
on the geodesic region $R$ with the boundary condition
\be
t^j\left(\del_j\phi_\mathrm{inv}\right)=-n_i\left(P^i+Q^i\right)
\label{julebound}
\ee
on $\del R$, where
\be
P^i&=Pt^i\equiv t^jt^k\epsilon_{df}\widetilde{E}^d{}_l\widetilde{E}^f{}_j(D_kt^l)t^i,
\ee
and
\be
Q^i&=Qt^i\equiv t^jt^k\epsilon_{df}(D_l\widetilde{E}^d{}_j)\widetilde{E}^f{}_kt^lt^i.
\ee
With (\ref{julerefdyad}) and (\ref{juletangent2}) we can see that
\be
Q+P&=\eta\frac{L^2v}{8a^4}+\eta\frac{L^2}{4a^4}\left(x^1\frac{\del v}{\del x^1}+\frac{\del v}{\del x^2}\right)\\
&+\eta\frac{L^4}{16a^4((x^1)^2+(x^2)^2)}\left((x^1)^2\frac{\del^2 v}{(\del x^1)^2}+2x^1x^2\frac{\del^2 v}{\del x^1\del x^2}+(x^2)^2\frac{\del^2 v}{(\del x^2)^2}\right).
\ee
An explicit calculation using (\ref{julephi}) shows that
\be
t^j(\del_j\varphi_\mathrm{inv})=-(P+Q)
\ee
and thus proves that $\phi_\mathrm{inv}(x^1,x^2)=\varphi_\mathrm{inv}(x^1,x^2)$ is actually the solution to the differential equation (\ref{julediff}) with the boundary condition (\ref{julebound}), and hence is the geodetic induced parallel field needed in the curvature Lagrangian, which we now turn to.

\section{The Curvature Lagrangian}

For the perturbed two-sphere the curvature Lagrangian is
\be
\mathcal{L}_R=\left(\pounds_tE^a{}_i\right)\left(\pounds_tE^b{}_j\right)\delta_{ab}\left(t^it^j+g^{ij}Rs^2\right).
\ee
This Lagrangian yields the differential equation
\be
D_i\left(t^it^j(1+2Rs^2)(\del_j\phi)\right)=-D_i\left(P^i+Q^i+P_R^i+Q_R^i\right)
\ee
with the boundary condition
\be
n_it^it^j(1+2Rs^2)\left(\del_j\phi\right)=-n_j\left(P^i+Q^i+P_R^i+Q_R^i\right),
\ee
where
\be
 P_R^i&=P_Rt^i\equiv Rs^2g^{jk}\epsilon_{df}\widetilde{E}^d{}_l\widetilde{E}^f{}_j(D_kt^l)t^i,
\ee
and
\be
Q_R^i&=Q_Rt^i\equiv Rs^2G^{jk}\epsilon_{df}(D_l\widetilde{E}^d{}_j)\widetilde{E}^f{}_kt^lt^i.
\ee
If we decompose the solution into the scalar field $\phi_\mathrm{inv}$ and the rest $\phi_R$,
\be
\phi(x^1,x^2)\equiv\phi_\mathrm{inv}(x^1,x^2)+\phi_R(x^1,x^2),
\ee
some of the terms in the differential equation cancel due to (\ref{julediff}) and we can add the remaining terms to the right-hand side. We then find
\be
D_i\left(t^it^j(1+2Rs^2)(\del_j\phi_R)\right)=-D_i\left(P_R^i+Q_R^i+t^it^j2Rs^2(\del_j\phi_\mathrm{inv})\right)
\ee
with the boundary condition
\be
t^j(1+2Rs^2)\left(\del_j\phi_R\right)=-n_i\left(P_R^i+Q_R^i+t^it^j2Rs^2(\del_j\phi_\mathrm{inv})\right).
\ee

%

The solutions to these equations then give us the scalar field $\phi_R(x^1,x^2)$ that takes the curvature into account and determines our new Vielbein. These equations are in principle solvable numerically and this research is ongoing, to be completed in the near future.

\section{Discussion and Conclusions}

In this chapter we presented the Lagrangian which defines the Vielbein field which is invariant under the averaging process and proved this invariance for the case of the perturbed two-sphere. Based on this Lagrangian we suggested a Lagrangian defining a Vielbein which differs from the invariant Vielbein provided that the Ricci scalar does not vanish. Since the modification term in the new Lagrangian includes the Ricci scalar, we expect the associated Vielbein to be a rotation of the invariant Vielbein about an angle which is larger in regions of higher curvature. The spread of the Vielbeins which are parallel transported to a common point in the averaging process is therefore also larger in these regions and thus the average is sensitive to the local curvature. For dimensional reasons the modification term in the Lagrangian also includes a factor of $s^2$. This factor vanishes at the origin and therefore we do not expect any issues with discontinuities. For more distant points the factor grows larger but its influence is counterbalanced by the averaging function in the process, which is required to weigh the contribution of Vielbeins from more distant points at a smaller rate. Future research will show if the curvature Lagrangian defines the Vielbein fields that it was designed for and if the averaging process based on the new Vielbein field leads to the expected results. 

%% file: concl.tex
\chapter{Conclusions}
\label{conclusions}

In this thesis we have addressed two aspects of the averaging problem in
general relativity. The first is the cosmological backreaction arising
from the non-commutativity of volume averaging with the construction of
the Einstein tensor and the second is the problem of constructing a
generally covariant average metric. Both the average of a metric and the
average of the Einstein tensor are of great importance in a variety of
situations and cosmology forms a useful testbed for both.\\

It has been suggested in the literature that cosmological backreaction can
account for the observed acceleration of the universe but there have been
few quantitative studies. We review the standard cosmological model and
some previous work in this area in chapter~\ref{problem}. In chapter~\ref{paper} we considered a linearly-perturbed
Friedmann-Lema\^itre-Robertson-Walker universe. Previous work has tended
to be in synchronous gauge which contains unphysical gauge modes. We chose
instead to work in Newtonian gauge. Employing the Buchert formalism and
the cmbeasy Boltzmann code we calculated the backreaction terms
numerically for the WMAP concordance cosmology and for a toy Einstein de
Sitter model. In both cases we showed that the backreaction of linear
perturbations is too small to account for the observed acceleration of the
universe and that the corrections act as a dark matter and not a dark
energy. This result agrees with the na\"ive expectations. Using the
Halofit code we extended our analysis to mildly nonlinear scales. The
backreaction from quasilinear scales remains of the order of $10^{-5}$ for
both the $\Lambda$CDM and EdS models and the corrections continue to act
as a dark matter. However, although the backreaction effect is small it is
nevertheless present and is expected to be significantly larger on smaller
scales that we did not consider. On such scales vector and tensor perturbations
become significant and for an unambiguous answer to whether backreaction
effects significantly influence the evolution of the universe the Einstein
equation will have to be averaged in a covariant manner. While our result
is small it does not rule out a significant backreaction. It is also a
prediction of the backreaction from viable cosmological models.\\

The other side of the averaging problem concerns the metric. In chapter
\ref{process} we presented a generally covariant averaging process which can be
used to smooth metrics within the framework of general relativity. This
process involves the decomposition of the metric into a specific choice of
tetrads, namely the maximally smooth tetrad field. Within the geodesic
region to be averaged across this tetrad field is determined by minimising
an action whose Lagrangian describes the variation of the covariant
derivative of the tetrads. The maximally smooth tetrads are then parallel
transported to a common point using a connector. This connector is a
general relativistic reformulation of the Wegner-Wilson operator from QCD.
At the common point the average can be unambiguously taken. The averaged
metric is then obtained from the averaged tetrads.\\

As a first example, we applied the process in chapter \ref{twodimensions} to the
two-dimensional spaces of constant curvature, the two-sphere, the
two-plane and the two-hyperboloid. These spaces provide a convenient
and simple tool to visualize the averaging process. Moreover, any
reasonable averaging process should leave a space of constant curvature
invariant and so they also act as a consistency test. Finally, the
three-dimensional spaces of constant curvature form the basis of the FLRW
models and so an understanding of averaging on such surfaces has an
immediate cosmological application. Considering first the two-sphere we
found that all constituents of the process, especailly the geodesic bundle
through the origin, can be expressed in an elegant way after the
applications of stereographic projection. It was shown that the averaging
indeed leaves the two-sphere unaffected as we required. The averaging
process was then applied to the hyperbolic plane and the flat plane. Both
were also invariant.\\

In chapter \ref{twosphere} we extended the analysis to a slightly
perturbed sphere. This serves as the first vital step towards a covariant
average in three space dimensions as would be required to reconstruct the
FLRW metric. The Lagrangian formulation was shown to yield a partial
differential equation of Neumann type for the maximally smooth dyad field.
This equation is not analytically tractable but can be evaluated
numerically.\\

In chapter \ref{threesphere} we showed that the process can in principle be
applied to space of more than two space dimensions by considering the
three-sphere. This space of constant curvature corresponds to a closed
FLRW universe along a hypersurface of constant time. We showed that the
three-sphere is also left invariant by the averaging procedure as should
be expected but that the differential equations for the maximally smoothed
triad gains a complicated structure that will be difficult to solve. In
more convenient coordinates it will be solveable numerically.\\

In chapter \ref{gauss} we turned to solving the equations for
the perturbed two-sphere. We specified the perturbation function of the
two-sphere to be a Gaussian function and solved the partial differential
equation with the aid of the numerical toolkit Gascoigne. This led to the
conclusion that the averaging process, as originally formulated, does not
smooth the perturbation function in the desired way.\\

The reason for this was analysed in chapter \ref{julelagrangian}, where we
concluded that the Lagrangian generating the maximally smooth dyad field
is similar to the Lagrangian that defines the dyad field which is
completely invariant under the averaging process. This problem was
overcome by adding a curvature-dependent term to the Lagrangian. By
including the Ricci scalar at a general point in the Lagrangian we take
into account the curvature of the manifold and directly link its
fluctuations to the average.\\

This averaging process can now be applied to different perturbation
functions to study their interaction with each other and with the
background sphere. Being applied to a three-dimensional sphere the outcome
of the averaging would be of cosmological relevance for a closed FLRW
model. Considering instead the flat three-plane with perturbations would
correspond to a flat FLRW model. The immediate prospect is then of
applying the formalism developed in chapter \ref{threesphere}, along with the
experience gained in chapters \ref{gauss} and
\ref{julelagrangian} to linearly-perturbed FLRW models and so test to some
degree the cosmological principle. A more general study would require us
to consider fluctuations of arbitrary size. Concerning the extension to
four dimensions would involve an analysis of the boundary conditions on
the congruence of light-like geodesics.\\

The averaging problem in general relativity is one of the greatest
unresolved problems in classical physics. In this thesis we have
considered two aspects of it. We have calculated the corrections to the
averaged Einstein equation compared to that for a smooth model that arise
inevitably from the current concordance model and from a toy competitor
model without dark energy. The impacts from the scales our approach
considered are small and resemble a dark matter but indicate that
backreaction remains a physical effect. We have then proposed a generally
covariant averaging process for metrics based on parallel transport and
much simpler than other existing procedures. We have applied this
procedure to test metrics and analysed the perturbed two- and
three-spheres and demonstrated its applicability.

%% file: append.tex
\begin{appendix}

\chapter{Fundamental Concepts of General Relativity}

Throughout the thesis we use units in which $c=1$.\\

In the general theory of relativity gravity is described as an effect of the curvature of spacetime, and the geometry of spacetime in turn is related to the energy and momentum content. \\

Spacetime $(\mathcal{M},{\bf g})$ is described as a four-dimensional real smooth manifold with a symmetric nondegenerate metric ${\bf g}$ of Lorentzian signature $(+,-,-,-)$. In a coordinate basis we can expand the metric in terms of its covariant components $g_{\mu\nu}$ as
\be
{\bf g}=g_{\mu\nu}~dx^\mu\otimes dx^\nu.
\ee
In abstract index notation the metric tensor is expressed as the line element
\be
ds^2=g_{\mu\nu}dx^\mu dx^\nu.
\ee
The contravariant components of the metric are defined by
\be
g^{\mu\nu}g_{\nu\rho}=\delta^\mu_\rho.
\ee
If we change coordinate systems the contravariant components $A^\mu$ and the covariant components $A_\mu$ of a vector ${\bf A}$ will transform according to
\be
A'^\mu=\frac{\del x'^\mu}{\del x^\nu}A^\nu\qquad\mathrm{and}\qquad A'_\mu=\frac{\del x^\nu}{\del x'^\mu}A_\nu.
\ee 
A geodesic ${\cal C}_{x_0x_1}$ is the path of extremal length between two spacetime points $x_0$ and $x_1$. If we choose an explicit parametrization for the geodesic
\be
{\cal C}_{x_0x_1}=\{(z^{\mu})|z^{\mu}=z^{\mu}(\tau),~\tau_0\leq\tau\leq\tau_1,~z^{\mu}(\tau_0)=x_0^{\mu},~z^{\mu}(\tau_1)=x_1^{\mu}\}.
\ee
and set the variation of the arc length $s({\cal C})$ 
\be
s({\cal C})=\int_{\tau_0}^{\tau_1}d\tau\sqrt{g_{\mu\nu}(z(\tau))\frac{\del z^\mu}{\del\tau}(\tau)\frac{\del z^\nu}{\del\tau}(\tau)}
\ee
with respect to $\tau$ equal to zero, we find the general geodesic equation
\be
\frac{d}{d\tau}\left(\frac{1}{\frac{ds}{d\tau}}g_{\mu\rho}(z(\tau))\frac{dz^\mu}{d\tau}(\tau)\right)-\frac{1}{\frac{ds}{d\tau}}\frac{1}{2}\frac{\del g_{\mu\nu}}{\del z^\rho}(z(\tau))\frac{dz^\mu}{d\tau}(\tau)\frac{dz^\nu}{d\tau}(\tau)=0,
\label{pertgeoeq}
\ee
provided the variation vanishes at the end points $x_0$ and $x_1$. Here, the arc length $s$ as a function of $\tau$ is given by
\be
s(\tau)=\int_{\tau_0}^{\tau}d\tau'\sqrt{g_{\mu\nu}(z(\tau'))\frac{\del z^\mu}{\del\tau'}(\tau')\frac{\del z^\nu}{\del\tau'}(\tau')}.
\ee
For $ds^2>0$ we can use the arc length $s$ to parametrize the curve and therefore set $\tau=s$. Then the geodesic equation reduces to 
\be
\frac{d^2z^\rho}{ds^2}+\Gamma^\rho_{\mu\nu}\frac{dz^\mu}{ds}\frac{dz^\nu}{ds}=0
\label{geodef}
\ee
with the Christoffel symbols
\be
\Gamma^\rho_{\mu\nu}=\frac{1}{2}g^{\rho\lambda}\left(\frac{\del g_{\lambda\mu}}{\del x^\nu}+\frac{\del g_{\lambda\nu}}{\del x^\mu}-\frac{\del g_{\mu\nu}}{\del x^\lambda}\right).
\label{christoffeldef}
\ee
The covariant derivative $D_\mu$ of a contravariant and a covariant vector are defined by
\be
D_\mu A^\rho=\del_\mu A^\rho+\Gamma^\rho_{\mu\nu}A^\nu\qquad\mathrm{and}\qquad D_\mu A_\rho=\del_\mu A_\rho-\Gamma^\nu_{\mu\rho}A_\nu,
\label{defcov}
\ee
respectively, with analogous definitions for higer order tensors. The covariant derivative of the metric vanishes,
\be
D_\rho g_{\mu\nu}=0.
\label{defcovdermetric}
\ee
The Lie derivative $\pounds_A$ evaluates the change of a tensor field $T^\mu{}_\nu$ along the flow of vector field $A^\rho$. The coordinate expression of the Lie derivative is
\be
\pounds_A T^\mu{}_\nu=A^\rho(D_\rho T^\mu{}_\nu)-T^\rho{}_\nu(D_\rho A^\mu)+T^\mu{}_\rho(D_\nu A^\rho).
\label{liedef}
\ee
The Riemann curvature tensor is given by
\be
R^\mu{}_{\nu\rho\lambda}=\del_\rho \Gamma^\mu_{\nu\lambda}-\del_\lambda \Gamma^\mu_{\nu\rho}+\Gamma^\mu_{\rho\sigma}\Gamma^\sigma_{\nu\lambda}-\Gamma^\mu_{\lambda\sigma}\Gamma^\sigma_{\nu\rho}.
\label{riemanndef}
\ee
The Ricci Tensor is the contraction of the Riemann curvature tensor,
\be
R_{\nu\rho}=R^\mu{}_{\nu\rho\mu}.
\label{riccidef}
\ee
Contracting once more yields the scalar curvature, sometimes also referred to as the Ricci scalar,
\be
R=g^{\nu\rho}R_{\nu\rho}.
\label{curvaturedef}
\ee

The fundamental relation between the spacetime metric $g_{\mu\nu}$ and the energy momentum tensor $T_{\mu\nu}$ is given by Einstein's famous equation, which in standard notation is
\be
G_{\mu\nu}=R_{\mu\nu}-\frac{1}{2}g_{\mu\nu}R=8\pi G T_{\mu\nu}+\Lambda g_{\mu\nu}.
\label{einsteindef}
\ee
Whether or not the cosmological term with the cosmological constant $\Lambda$ needs to be included is still the subject of debate.\\

The energy momentum conservation in general relativity is expressed as the vanishing of the divergence of the energy momentum tensor,
\be
D_\mu T^{\mu\nu}=0.
\label{conservationdef}
\ee

\chapter{The Coordinate Transformations for the Sphere}
\label{sphereappend}

Let the coordinates of the sphere be called ${\bf x}=(x,y,z)$ and the ones of the stereographic projection plane at the south pole ${\bf x}_S=(0,0,-a)$ be $(x^1,x^2)$. Let furthermore the coordinates of the sphere after the rotation of the point ${\bf x}_0=(x_0,y_0,z_0)$ $=(a\sin\theta_0\cos\phi_0,a\sin\theta_0\sin\phi_0,a\cos\theta_0)$ to the south pole ${\bf x}_S$ be called $\widetilde{{\bf x}}=(\widetilde{x},\widetilde{y},\widetilde{z})$ and the ones from the new stereographic projection plane $(\widetilde{x}^1,\widetilde{x}^2)$. Recall that in this case $L^2=4a^2+(x^1)^2+(x^2)^2$.\\

The inverse stereographic projection from the coordinates $(x^1,x^2)$ to the coordinates of the sphere ${\bf x}=(x,y,z)$ is
\be
x=\frac{4 a^2x^1}{L^2},\qquad y=\frac{4 a^2x^2}{L^2},\qquad z=a\left(1-\frac{8 a^2}{L^2}\right).
\ee

The rotation matrix is given by $D=D_zD_yD_z^{-1}$ and we find
\be
D&=\begin{pmatrix}\cos\phi_0&-\sin\phi_0&0\\\sin\phi_0&\cos\phi_0&0\\0&0&1\end{pmatrix}\begin{pmatrix}\cosh\theta_0&0&\sinh\theta_0\\0&1&0\\\sinh\theta_0&0&\cosh\theta_0\end{pmatrix}\begin{pmatrix}\cos\phi_0&\sin\phi_0&0\\-\sin\phi_0&\cos\phi_0&0\\0&0&1\end{pmatrix}\\
&=\begin{pmatrix}\cosh\theta_0\cos^2\phi_0+\sin^2\phi_0&\sin\phi_0\cos\phi_0(\cosh\theta_0-1)&\sinh\theta_0\cos\phi_0\\\sin\phi_0\cos\phi_0(\cosh\theta_0-1)&\cosh\theta_0\sin^2\phi_0+\cos^2\phi_0&\sinh\theta_0\sin\phi_0\\\sinh\theta_0\cos\phi_0&\sinh\theta_0\sin\phi_0&\cosh\theta_0\end{pmatrix}.
\ee
After the rotation we apply the stereographic projection to the rotated sphere with the coordinates $\widetilde{{\bf x}}=(\widetilde{x},\widetilde{y},\widetilde{z})$ and find the transformation to the coordinates $(\widetilde{x}^1,\widetilde{x}^2)$ of the new projection plane to be
\be
\widetilde{x}^1=\frac{2a\widetilde{x}}{(a-\widetilde{z})},\qquad\widetilde{x}^2=\frac{2a\widetilde{y}}{(a-\widetilde{z})}.
\ee

The explicit transformation of the coordinates $(x^1,x^2)$ to the coordinates $(\widetilde{x}^1,\widetilde{x}^2)$ and the first derivatives of this transformation are

\be
\widetilde{x}^1&=\frac{(-\cos\theta_0\cos^2\phi_0+\sin^2\phi_0)8a^2x^1-\sin\phi_0\cos\phi_0(\cos\theta_0+1)8a^2x^2}{L^2+\sin\theta_0\cos\phi_04ax^1+\sin\theta_0\sin\phi_04ax^2+\cos\theta_0(L^2-8a^2)}\\
&\qquad +\frac{\sin\theta_0\cos\phi_02a(L^2-8a^2)}{L^2+\sin\theta_0\cos\phi_04ax^1+\sin\theta_0\sin\phi_04ax^2+\cos\theta_0(L^2-8a^2)}\\
\widetilde{x}^2&=\frac{-\sin\phi_0\cos\phi_0(\cos\theta_0+1)8a^2x^1+(-\cos\theta_0\sin^2\phi_0+\cos^2\phi_0)8a^2x^2}{L^2+\sin\theta_0\cos\phi_04ax^1+\sin\theta_0\sin\phi_04ax^2+\cos\theta_0(L^2-8a^2)}\\
&\qquad +\frac{\sin\theta_0\cos\phi_02a(L^2-8a^2)}{L^2+\sin\theta_0\cos\phi_04ax^1+\sin\theta_0\sin\phi_04ax^2+\cos\theta_0(L^2-8a^2)},
\label{coordtrans}
\ee
\be
\frac{\del\widetilde{x}^1}{\del x^1}&=\frac{\del\widetilde{x}^2}{\del x^2}=-\frac{8a^2(4a^2(\cos\theta_0\!-\!1)\!-\!4ax^2\sin\phi_0\sin\theta_0\!-\!2((x^1)^2\!-\!(x^2)^2)\cos^2\phi_0\cos^2(\theta_0/2))}{(L^2+4a\sin\theta_0(x^1\cos\phi_0+x^2\sin\phi_0)+(L^2-8a^2)\cos\theta_0)^2}\\
&-\frac{8a^2(\sin^2\phi_0(1\!+\!\cos\theta_0)((x^1)^2\!-\!(x^2)^2)\!-\!4x^1\cos\phi_0(x^2\sin\phi_0(\cos\theta_0\!+\!1)\!+\!a\sin\theta_0))}{(L^2+4a\sin\theta_0(x^1\cos\phi_0+x^2\sin\phi_0)+(L^2-8a^2)\cos\theta_0)^2},\\
\frac{\del\widetilde{x}^1}{\del x^2}&=-\frac{\del\widetilde{x}^2}{\del x^1}=32a^2\cos(\theta_0/2)(x^2\cos\phi_0-x^1\sin\phi_0)\\
&\times\frac{(x^1\cos\phi_0\cos(\theta_0/2)+x^2\sin\phi_0\cos(\theta_0/2)+2a\sin(\theta_0/2))}{(L^2+4a\sin\theta_0(x^1\cos\phi_0+x^2\sin\phi_0)+(L^2-8a^2)\cos\theta_0)^2}.
\label{coordder2}
\ee
The inverse transformation from the coordinates $(\widetilde{x}^1,\widetilde{x}^2)$ to the coordinates $(x^1,x^2)$ and the first partial derivatives are
\be
x^1&=\frac{(-\cos\theta_0\cos^2\phi_0+\sin^2\phi_0)8a^2\widetilde{x}^1-\sin\phi_0\cos\phi_0(\cos\theta_0+1)8a^2\widetilde{x}^2}{\widetilde{L}^2-\sin\theta_0\cos\phi_04a\widetilde{x}^1-\sin\theta_0\sin\phi_04a\widetilde{x}^2+\cos\theta_0(\widetilde{L}^2-8a^2)}\\
&\qquad -\frac{\sin\theta_0\cos\phi_02a(\widetilde{L}^2-8a^2))}{\widetilde{L}^2-\sin\theta_0\cos\phi_04a\widetilde{x}^1-\sin\theta_0\sin\phi_04a\widetilde{x}^2+\cos\theta_0(\widetilde{L}^2-8a^2)}\\[5mm]
x^2&=\frac{-\sin\phi_0\cos\phi_0(\cos\theta_0+1)8a^2\widetilde{x}^1+(-\cos\theta_0\sin^2\phi_0+\cos^2\phi_0)8a^2\widetilde{x}^2}{\widetilde{L}^2-\sin\theta_0\cos\phi_04a\widetilde{x}^1-\sin\theta_0\sin\phi_04a\widetilde{x}^2+\cos\theta_0(\widetilde{L}^2-8a^2)}\\
&\qquad -\frac{\sin\theta_0\cos\phi_02a(\widetilde{L}^2-8a^2))}{\widetilde{L}^2-\sin\theta_0\cos\phi_04a\widetilde{x}^1-\sin\theta_0\sin\phi_04a\widetilde{x}^2+\cos\theta_0(\widetilde{L}^2-8a^2)},
\label{coordtrans2}
\ee
\be
\frac{\del x^1}{\del \widetilde{x}^1}&=\frac{\del x^2}{\del\widetilde{x}^2}=-\frac{8a^2(4a^2(\cos\theta_0\!-\!1)\!+\!4a\widetilde{x}^2\sin\phi_0\sin\theta_0\!-\!2((\widetilde{x}^1)^2\!-\!(\widetilde{x}^2)^2)\cos^2\phi_0\cos^2(\theta_0/2))}{(L^2-4a\sin\theta_0(\widetilde{x}^1\cos\phi_0+\widetilde{x}^2\sin\phi_0)+(L^2-8a^2)\cos\theta_0)^2}\\
&-\frac{8a^2(\sin^2\phi_0(1\!+\!\cos\theta_0)((\widetilde{x^1})^2\!-\!(\widetilde{x}^2)^2)\!-\!4\widetilde{x}^1\cos\phi_0(\widetilde{x}^2\sin\phi_0(\cos\theta_0\!+\!1)\!-\!a\sin\theta_0))}{(L^2-4a\sin\theta_0(\widetilde{x}^1\cos\phi_0+\widetilde{x}^2\sin\phi_0)+(L^2-8a^2)\cos\theta_0)^2},\\
\frac{\del x^1}{\del \widetilde{x}^2}&=-\frac{\del x^2}{\del\widetilde{x}^1}=32a^2\cos(\theta_0/2)(\widetilde{x}^2\cos\phi_0-\widetilde{x}^1\sin\phi_0)\\
&\times\frac{(\widetilde{x}^1\cos\phi_0\cos(\theta_0/2)+\widetilde{x}^2\sin\phi_0\cos(\theta_0/2)-2a\sin(\theta_0/2))}{(L^2-4a\sin\theta_0(\widetilde{x}^1\cos\phi_0+\widetilde{x}^2\sin\phi_0)+(L^2-8a^2)\cos\theta_0)^2}.
\label{coordder}
\ee

\chapter{The Coordinate Transformations for the Hyperbolic Plane} 
\label{hyperappend}

Let the coordinates of the hyperbolic plane be called ${\bf x}=(x,y,z)$ and the ones of the stereographic projection plane at the minimum point of the upper sheet ${\bf x}_M=(0,0,a)$ be $(x^1,x^2)$. Let furthermore the coordinates of the hyperbolic plane after the rotation of the point ${\bf x}_0=(x_0,y_0,z_0)$ $=(a\sinh\theta_0\cos\phi_0,a\sinh\theta_0\sin\phi_0,a\cosh\theta_0)$ to the minimum point ${\bf x}_M$ be called $\widetilde{{\bf x}}=(\widetilde{x},\widetilde{y},\widetilde{z})$ and the ones from the new stereographic projection plane $(\widetilde{x}^1,\widetilde{x}^2)$. Recall that in this case $L^2=4a^2-(x^1)^2-(x^2)^2$.\\

The inverse stereographic projection from $(x^1,x^2)$ to ${\bf x}=(x,y,z)$ is
\be
x=\frac{4 a^2x^1}{L^2},\qquad y=\frac{4 a^2x^2}{L^2},\qquad z=a\left(\frac{8 a^2}{L^2}-1\right).
\label{newstereo2}
\ee
The rotation matrix of the hyperbolic plane is given by $D=D_zD_yD_z^{-1}$ and we find 
\be
D&=\begin{pmatrix}\cos\phi_0&-\sin\phi_0&0\\\sin\phi_0&\cos\phi_0&0\\0&0&1\end{pmatrix}\begin{pmatrix}\cosh\theta_0&0&\sinh\theta_0\\0&1&0\\\sinh\theta_0&0&\cosh\theta_0\end{pmatrix}\begin{pmatrix}\cos\phi_0&\sin\phi_0&0\\-\sin\phi_0&\cos\phi_0&0\\0&0&1\end{pmatrix}\\
&=\begin{pmatrix}\cosh\theta_0\cos^2\phi_0+\sin^2\phi_0&\sin\phi_0\cos\phi_0(\cosh\theta_0-1)&\sinh\theta_0\cos\phi_0\\\sin\phi_0\cos\phi_0(\cosh\theta_0-1)&\cosh\theta_0\sin^2\phi_0+\cos^2\phi_0&\sinh\theta_0\sin\phi_0\\\sinh\theta_0\cos\phi_0&\sinh\theta_0\sin\phi_0&\cosh\theta_0\end{pmatrix}.
\label{dmatrix}
\ee
After the rotation we apply the stereographic projection to the rotated hyperbolic plane with the coordinates $\widetilde{\bf x}=(\widetilde{x},\widetilde{y},\widetilde{z})$ and find the transformation to the coordinates $(\widetilde{x}^1,\widetilde{x}^2)$
\be
\widetilde{x}^1=\frac{2a\widetilde{x}}{(a-\widetilde{z})},\qquad\widetilde{x}^2=\frac{2a\widetilde{y}}{(a-\widetilde{z})}.
\ee

The explicit transformation of the coordinates $(x^1,x^2)$ to the coordinates $(\widetilde{x}^1,\widetilde{x}^2)$ and the first derivatives of this transformation are
\be
\widetilde{x}^1=\frac{(\cosh\theta_0\cos^2\phi_0+\sin^2\phi_0)8a^2x^1+\sin\phi_0\cos\phi_0(\cosh\theta_0-1)8a^2x^2}{L^2-\sinh\theta_0\cos\phi_04ax^1-\sinh\theta_0\sin\phi_04ax^2+\cosh\theta_0(8a^2-L^2)}\\
\D\qquad -\frac{\sinh\theta_0\cos\phi_02a(8a^2-L^2)}{L^2-\sinh\theta_0\cos\phi_04ax^1-\sinh\theta_0\sin\phi_04ax^2+\cosh\theta_0(8a^2-L^2)}\\[2mm]
\D \widetilde{x}^2=\frac{\sin\phi_0\cos\phi_0(\cosh\theta_0-1)8a^2x^1+(\cosh\theta_0\sin^2\phi_0+\cos^2\phi_0)8a^2x^2}{L^2-\sinh\theta_0\cos\phi_04ax^1-\sinh\theta_0\sin\phi_04ax^2+\cosh\theta_0(8a^2-L^2)}\\
\D\qquad -\frac{\sinh\theta_0\sin\phi_02a(8a^2-L^2)}{L^2-\sinh\theta_0\cos\phi_04ax^1-\sinh\theta_0\sin\phi_04ax^2+\cosh\theta_0(8a^2-L^2)}.
\label{newcoordtrans}
\ee
\be
\frac{\del\widetilde{x}^1}{\del x^1}&=\frac{\del\widetilde{x}^2}{\del x^2}=\frac{8a^2(4a^2(1+\cosh\theta_0)+2x^1x^2(2\cosh\theta_0\sin\phi_0\cos\phi_0-\sin(2\phi_0)))}{(L^2-4a\sinh\theta_0(x^1\cos\phi_0+x^2\sin\phi_0)+(8a^2-L^2)\cosh\theta_0)^2}\\
&+\frac{8a^2(-4a\sinh\theta_0(x^1\cos\phi_0+x^2\sin\phi_0)+\cos(2\phi_0)(1-\cosh\theta_0)((x^2)^2-(x^1)^2))}{(L^2-4a\sinh\theta_0(x^1\cos\phi_0+x^2\sin\phi_0)+(8a^2-L^2)\cosh\theta_0)^2},\\
\frac{\del\widetilde{x}^1}{\del x^2}&=-\frac{\del\widetilde{x}^2}{\del x^1}=-32a^2\sinh(\theta_0/2)(x^2\cos\phi_0-x^1\sin\phi_0)\\
&\times\frac{(2a\cosh(\theta_0/2)-\sinh(\theta_0/2)(x^1\cos\phi_0+x^2\sin\phi_0))}{(L^2-4a\sinh\theta_0(x^1\cos\phi_0+x^2\sin\phi_0)+(8a^2-L^2)\cosh\theta_0)^2}.
\label{newcoordder2}
\ee
The inverse transformation from the coordinates $(\widetilde{x}^1,\widetilde{x}^2)$ to the coordinates $(x^1,x^2)$ and the first partial derivatives are
\be
x^1=\frac{(\cosh\theta_0\cos^2\phi_0+\sin^2\phi_0)8a^2\widetilde{x}^1+\sin\phi_0\cos\phi_0(\cosh\theta_0-1)8a^2\widetilde{x}^2}{\widetilde{L}^2+\sinh\theta_0\cos\phi_04a\widetilde{x}^1+\sinh\theta_0\sin\phi_04a\widetilde{x}^2+\cosh\theta_0(8a^2-\widetilde{L}^2)}\\
\D\qquad +\frac{\sinh\theta_0\cos\phi_02a(8a^2-\widetilde{L}^2))}{\widetilde{L}^2+\sinh\theta_0\cos\phi_04a\widetilde{x}^1+\sinh\theta_0\sin\phi_04a\widetilde{x}^2+\cosh\theta_0(8a^2-\widetilde{L}^2)}\\[2mm]
\D x^2=\frac{\sin\phi_0\cos\phi_0(\cosh\theta_0-1)8a^2\widetilde{x}^1+(\cosh\theta_0\sin^2\phi_0+\cos^2\phi_0)8a^2\widetilde{x}^2}{\widetilde{L}^2+\sinh\theta_0\cos\phi_04a\widetilde{x}^1+\sinh\theta_0\sin\phi_04a\widetilde{x}^2+\cosh\theta_0(8a^2-\widetilde{L}^2)}\\
\D\qquad +\frac{\sinh\theta_0\sin\phi_02a(8a^2-\widetilde{L}^2))}{\widetilde{L}^2+\sinh\theta_0\cos\phi_04a\widetilde{x}^1+\sinh\theta_0\sin\phi_04a\widetilde{x}^2+\cosh\theta_0(8a^2-\widetilde{L}^2)}.
\label{newcoordtrans2}
\ee

\be
\frac{\del x^1}{\del \widetilde{x}^1}&=\frac{\del x^2}{\del\widetilde{x}^2}=\frac{8a^2(4a^2(1+\cosh\theta_0)+2\widetilde{x}^1\widetilde{x}^2(2\cosh\theta_0\sin\phi_0\cos\phi_0-\sin(2\phi_0)))}{(\widetilde{L}^2+4a\sinh\theta_0(\widetilde{x}^1\cos\phi_0+\widetilde{x}^2\sin\phi_0)+(8a^2-\widetilde{L}^2)\cosh\theta_0)^2}\\
&+\frac{8a^2(4a\sinh\theta_0(\widetilde{x}^1\cos\phi_0+\widetilde{x}^2\sin\phi_0)+\cos(2\phi_0)(1-\cosh\theta_0)((\widetilde{x}^2)^2-(\widetilde{x}^1)^2))}{(\widetilde{L}^2+4a\sinh\theta_0(\widetilde{x}^1\cos\phi_0+\widetilde{x}^2\sin\phi_0)+(8a^2-\widetilde{L}^2)\cosh\theta_0)^2},\\
\frac{\del x^1}{\del \widetilde{x}^2}&=-\frac{\del x^2}{\del\widetilde{x}^1}=32a^2\sinh(\theta_0/2)(\widetilde{x}^2\cos\phi_0-\widetilde{x}^1\sin\phi_0)\\
&\times\frac{(2a\cosh(\theta_0/2)+\sinh(\theta_0/2)(\widetilde{x}^1\cos\phi_0+\widetilde{x}^2\sin\phi_0))}{(\widetilde{L}^2+4a\sinh\theta_0(\widetilde{x}^1\cos\phi_0+\widetilde{x}^2\sin\phi_0)+(8a^2-\widetilde{L}^2)\cosh\theta_0)^2}.
\label{newcoordder}
\ee

\end{appendix}

%% file: acknowledgement.tex
\thispagestyle{empty}

\begin{center}
{\sc\Huge Acknowledgements}\\[1.3cm]
\end{center}

 I am very grateful to Prof.\ Dr.\ Otto Nachtmann for his guidance over the years and his patience during the times when the loss of my father distracted me from work. \\

I am also very grateful to Prof.\ Dr.\ Iring Bender for agreeing to co-referee this thesis.\\

I would like to acknowledge the partial funding of my PhD by the Landesgraduierten- f\"orderung Baden-W\"urttemberg.\\

This thesis would have never been possible without Iain Brown. Not only because of the collaboration on the cosmological backreaction, but also because he encouraged me and spent much time on proofreading the whole thesis. Moreover, he stayed awake with me during night-shifts, contributed through valuable discussions and proved to be a good friend. \\

In this context I would also like to thank Thomas Richter who spent much time on the solution of my differential equations and who never complained when I suddenly changed everything or asked for completely different computations. \\

I would also like to acknowledge Georg Robbers and Andreas von Manteuffel who always took the time to help me with my computer problems, and Georg for his collaboration on the cosmological backreaction.\\

It is hard to leave an institute like the Philosophenweg 16 and 19. I really enjoyed working here and I always felt at home. Many of my colleagues became really good friends and I hope nobody is offended if I don't try to list all their names. Someone, however, needs to be mentioned personally, since he has such a strong influence on the atmosphere in this institute, and that is Eduard Thommes. Even in the most stressful times he would never send someone away who knocked on his door for help and I often made use of that.\\

Finally, I want to thank my mother Barbara Behrend and my sister Ulrike Gr\"ozinger and her family for their confidence and support and Hendrik Kasten for all his love.